\documentclass{cimento}

\usepackage{graphicx}  
\usepackage{fancyhdr}
\usepackage[noadjust]{cite}
\def\lesssim{\mathrel{\hbox{\rlap{\hbox{\lower3pt\hbox{$\sim$}}}\hbox{\raise2pt\hbox{$<$}}}}}
\def\gtrsim{\mathrel{\hbox{\rlap{\hbox{\lower3pt\hbox{$\sim$}}}\hbox{\raise2pt\hbox{$>$}}}}}
\def\gsimeq{\,\raise0.14em\hbox{$>$}\kern-0.76em\lower0.28em\hbox
{$\sim$}\,}
\def\lsimeq{\,\raise0.14em\hbox{$<$}\kern-0.76em\lower0.28em\hbox
{$\sim$}\,}

\title{Supernova Neutrinos: Production, Oscillations and Detection}
\shorttitle{Supernova Neutrinos}
\author{A.~Mirizzi\from{ins:b}\from{ins:bb},
 I.~Tamborra\from{ins:e},
\ H.-Th.~Janka\from{ins:a}, 
 N.~Saviano\from{ins:c},
 K.~Scholberg\from{ins:d},
 R.~Bollig\from{ins:a}\from{ins:f},
 L.~H\"udepohl\from{ins:g}, \atque
 S.~Chakraborty\from{ins:h}
}
\instlist{
\inst{ins:b} Dipartimento Interateneo di Fisica ``Michelangelo Merlin,'' Via Amendola 173, 70126 Bari, Italy.
\inst{ins:bb} Istituto Nazionale di Fisica Nucleare - Sezione di Bari,
Via Amendola 173, 70126 Bari, Italy.
\inst{ins:e}  GRAPPA Institute, University of Amsterdam, Science Park 904, 1098 XH Amsterdam, The Netherlands.
\inst{ins:a}
Max-Planck-Institut f\"ur Astrophysik, Karl-Schwarzschild-Str. 1, 85748 Garching, Germany.
\inst{ins:c} Institute for Particle Physics Phenomenology, Department of Physics,
Durham University,\\ Durham DH1 3LE, United Kingdom.
\inst{ins:d} Department of Physics, Duke University, Durham, NC, 27708, USA.
\inst{ins:f}  Physik Department, Technische Universit\"at M\"unchen, James-Franck-Str. 1, 85748 Garching, Germany.
\inst{ins:g}  Max Planck Computing and Data Facility (MPCDF), Gie{\ss}enbachstr.~2,85748 Garching, Germany.
\inst{ins:h} Max-Planck-Institut f\"ur Physik (Werner-Heisenberg-Institut),
 F\"ohringer Ring 6, 80805 M\"unchen, Germany.}

\PACSes{\PACSit{14.60.Pq, 14.60.Lm, 97.60.Bw}\ }

\begin{document}

\maketitle

\begin{abstract}
Neutrinos play a crucial role in the collapse and explosion of massive stars,  governing
 the infall dynamics of the stellar core,  triggering and fueling the explosion
and driving the cooling and deleptonization of the newly formed neutron star. 
Due to their role neutrinos carry information from the heart of the explosion and, due to their weakly interacting nature, offer the only direct probe of the dynamics and thermodynamics at the center of a supernova. In this paper, we review the present status of modelling  the neutrino physics
and signal formation in collapsing and exploding stars. We assess the capability of
current and planned large underground neutrino detectors to yield faithful
information of the time and flavor dependent neutrino signal from a future Galactic 
supernova. We show how the observable neutrino burst would provide a benchmark for fundamental supernova physics with 
unprecedented richness of detail. Exploiting the treasure of the measured
neutrino events requires a careful discrimination of source-generated properties from
signal features that originate on the way to the detector. As for the latter, we 
discuss  self-induced flavor conversions associated with 
neutrino-neutrino interactions that occur in the deepest stellar regions; matter 
effects that modify the pattern of flavor conversions in the dynamical stellar
envelope;  neutrino-oscillation signatures that result from structural features 
associated with the shock-wave propagation as well as turbulent mass motions in 
post-shock layers. Finally, we highlight our current understanding of the formation
of the diffuse supernova neutrino background and we analyse the perspectives for a 
detection of this relic signal that integrates the 
contributions from all past core-collapse supernovae in the Universe.
\end{abstract}


\section{Introduction}

 Core-collapse supernovae  are among the most powerful sources of neutrinos
 in our Universe. During a supernova explosion,  99$\%$ of the emitted energy
 ($\sim 10^{53}$~erg) is released by neutrinos and antineutrinos of 
all  flavors, with energy of several MeV, which play the role of ``astrophysical messengers,''
escaping almost unimpeded from the supernova core.
The supernova neutrino flux  has been extensively
studied as a probe of both fundamental neutrino properties and core-collapse
physics. Therefore, supernova neutrinos
represent a truly interdisciplinary research field at the interface
 between particle physics,
nuclear physics and astrophysics.

While Galactic supernovae are rare, existing or proposed large neutrino
detectors will allow collection of a high-statistics
neutrino signal from the next Galactic explosion.
The supernova neutrino detection  will be crucial to test the explosion mechanism  and
 thus  to compare current supernova models with direct empirical information from the supernova core.
Originating  from deep inside
the core,  neutrinos are affected by 
flavor conversions in  the dense supernova matter on their way through
the stellar mantle and envelope.
Therefore, the neutrino fluxes reaching
the detectors will carry intriguing signatures of oscillation effects
in the deepest supernova regions, depending on the unknown neutrino mass
hierarchy.
In this sense, the dense supernova interior represents a unique laboratory to probe neutrino  flavor mixing under
high-density conditions.

Matter effects in a supernova can be truly dramatic as neutrinos propagate
through a dense turbulent environment. Furthermore
in the deepest supernova regions, the neutrino density is so high
 that it dominates the flavor evolution, producing a fascinating
\emph{collective} behavior associated with  $\nu$-$\nu$ 
interactions~\cite{Pantaleone:1992xh,Pantaleone:1994ns,Pantaleone:1992eq}.
In the recent past, our description of self-induced neutrino oscillations has seen
substantial progress. The seminal studies started almost a decade ago~\cite{Duan:2005cp,Duan:2006an,Duan:2006jv,Hannestad:2006nj,Fogli:2007bk,Duan:2007mv}
 stimulated a still-ongoing torrent of analytical
and numerical works to clarify several aspects of this unusual
flavor dynamics (see~\cite{Duan:2010bg} for a review).

\begin{figure}[t!]
\begin{center}
 \includegraphics[angle=0,width=0.7\textwidth]{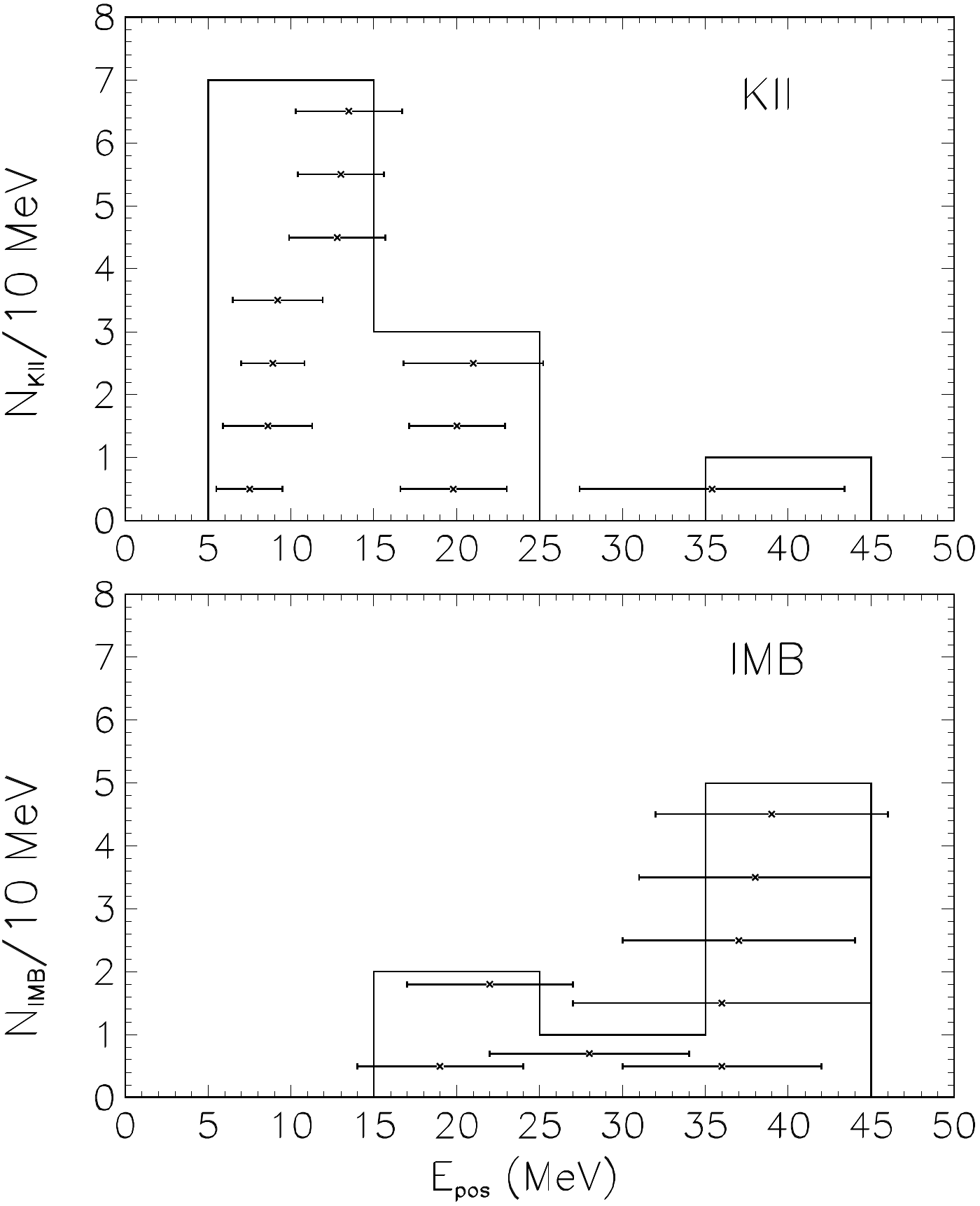} 
   \end{center}
\caption{Positron spectra detected at Kamiokande-II (upper panel) and
IMB (lower panel) in connection with SN 1987A.
(Reprinted figure  from~\cite{Mirizzi:2005tg}; copyright (2005) by the American Physical Society.)
 \label{sn1987a}} 
\end{figure}

It is thus evident that
the physics potential of a supernova neutrino detection is enormous.
Stellar collapse neutrinos were observed
for the first (and so far only)  time in 1987.
Within a decade since the advent of large underground neutrino detectors, Nature was kind enough to provide a  neutrino burst associated with the collapse of Sanduleak -69$^\circ$ 202 in the Large Magellanic Cloud, 51~kpc away. Two kiloton-scale water Cherenkov detectors, one in Japan, Kamiokande-II~\cite{Hirata:1987hu},
and one in the United States, the Irvine-Michigan-Brookhaven (IMB) experiment~\cite{Bionta:1987qt},
observed about 20 events within 13 seconds, with timing consistent with the optical observation of the SN 1987A explosion.
The positron spectra produced by inverse beta decays of 
supernova ${\bar \nu}_e$ in the two detectors are represented in Fig.~\ref{sn1987a}.
Two smaller scintillator detectors,
Baksan~\cite{Alekseev:1987ej} and the Mont Blanc Liquid Scintillation Detector (LSD)~\cite{Aglietta:1987it} also reported several
 events~\footnote{The reported LSD burst is more controversially associated with the supernova, because the events were recorded some hours earlier than the others.}.

The detection of SN 1987A neutrinos has long been taken as a confirmation of the salient
features of our physical comprehension  of the core-collapse supernova
phenomenon and  of the associated neutrino emission~\cite{Vissani:2014doa}.
This observation allowed us to put strong constraints on exotic neutrino properties (e.g., decays, neutrino charge) 
that would have altered the supernova neutrino emission.
Most importantly, the total energy of ${\bar \nu}_e$ and the inferred cooling time scale of a few seconds
of the proto-neutron star  put severe limits on  non-standard
cooling mechanisms associated with 
new particles emitted from the supernova core, notably  right-handed  neutrinos
and axions (see~\cite{Raffelt:1990yu} for a review).
At the same time, with the poor
 statistics of the SN 1987A events, determining both supernova and
neutrino parameters is impossible, although some hints could be obtained, e.g., 
 the signal of SN~1987A 
was analysed in the light of neutrino oscillations~\cite{Minakata:1988cn,Jegerlehner:1996kx,Vissani:2014doa}.

The Galactic supernova rate can be estimated with different techniques (see, e.g., Table X in~\cite{Raffelt:2012kt}).
Typically one expects 1-3 core-collapse supernovae per century.
However, 
except for SN 1987A in the Large Magellanic Cloud,
no stellar collapse has been observed over more than 30 years of neutrino experiments. 
This absence of a signal allowed different neutrino experiments to place non-trivial  upper bounds to the
 rate of collapses and failed supernovae~\cite{Alekseev:2002ji,Ikeda:2007sa,Agafonova:2014leu},
which confirm that a Galactic supernova explosion is a rare event. 

\begin{figure}[t!]
\begin{center}
 \includegraphics[angle=0,width=0.7\textwidth]{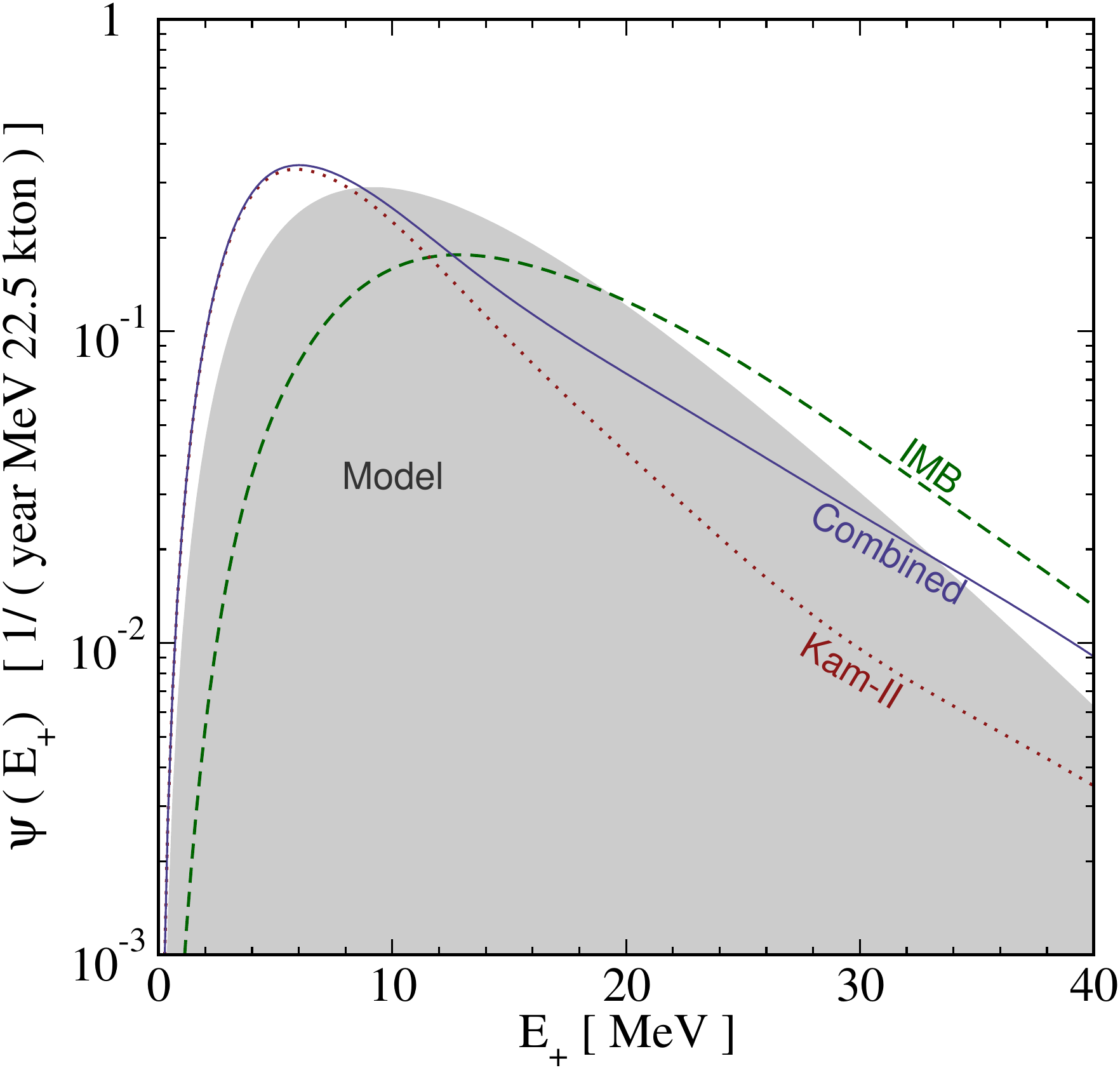} 
    \end{center}
\caption{The  DSNB  detection  spectra  based  on  the  neutrino  spectra inferred from either the Kamiokande-II or IMB data sets
alone or their
combination, compared to a model (shaded shape) with
canonical neutrino emission parameters. 
(Reprinted figure with permission from~\cite{Yuksel:2007mn}; copyright (2007) by the American Physical Society.)
 \label{dsnb1987}} 
\end{figure}

However, there are  $\sim 10$ supernova explosions per second in the visible Universe. The cumulative
emission of neutrinos from all the past core-collapse supernovae produces a cosmic background
 of (anti)neutrinos, the so-called Diffuse Supernova Neutrino Background (DSNB), whose 
existence was predicted  already before the observation of SN 1987A~\cite{Krauss:1983zn}.
 Although weak, the DSNB is a ``guaranteed'' signal that
 can also
probe physics different  from a Galactic explosion, including processes which occur
on cosmological scales in time or space. 
The DSNB covers a wide range of physics, including
the cosmic star formation rate, the stellar dynamics and fundamental
neutrino properties.
Forecasts of the DSNB can be obtained using the neutrino spectra predicted by supernova simulations 
or the ones reconstructed from SN 1987A data~\cite{Yuksel:2007mn}, as shown in Fig.~\ref{dsnb1987}.
The stringent observational upper limit on the DSNB 
flux,
obtained by Super-Kamiokande~\cite{Iida:2009qnz,Bays:2012wty}, is only a factor of $\sim 2$ higher than 
typical theoretical estimates~\cite{Horiuchi:2008jz}. This suggests an imminent detection of the DSNB in current and planned detectors.

 Nearly three decades after SN 1987A, 
we are eager for the next supernova neutrino signal.
Meanwhile
our understanding of supernova neutrinos has grown significantly. 
Remarkably, our modeling of the supernova neutrino emission and of flavor conversions
in the stellar matter   has
experienced several breakthroughs. On the other side there is  vivid experimental activity
in low-energy neutrino astronomy, whose main goal is the detection of Galactic and diffuse supernova neutrino signals.
Given the recent advances both on the theoretical and experimental fronts, 
in this review we aim to present state-of-the-art supernova neutrino physics and astrophysics.
Our plan is as follows.

In Section~\ref{sec:SNmodels} we review the neutrino-emission phases of core-collapse
supernovae, the characteristic signal properties expected during these
phases, and our current understanding of the delayed neutrino-driven
mechanism based on multi-dimensional modeling of stellar collapse and
explosion. Besides reporting new and unexpected neutrino-emission 
features associated with the nonradial hydrodynamic flows during the
accretion phase, we also present long-timescale proto-neutron star cooling
calculations with state-of-the-art treatment of neutrino transport, in
which ---for the first time--- convective transport effects were taken
into account by a mixing-length description for spherically symmetric 
simulations.
In Section~\ref{sec:detection}  we discuss the detectability of (extra)galactic supernova neutrinos
in current and planned detectors. We describe the physics potential of the different detection techniques proposed to
 measure the   neutrino burst and present the most relevant
applications of  a supernova  detection  to neutrino 
 astronomy and astrophysics. 
Section~\ref{sec:oscillations} is devoted to the characterization of supernova neutrino flavor conversions. Self-induced effects associated
with the neutrino-neutrino interactions in the deepest supernova regions and matter effects in the dynamical
supernova environment are discussed. Finally, observable signatures of flavor conversions imprinted on the neutrino burst
are presented. 
In Section~\ref{sec:DSNB} we focus on the DSNB. 
 We discuss our current knowledge of the cosmic supernova rate and present estimates of  the expected DSNB
signal in large underground neutrino detectors by taking into account the effects
of  flavor conversions as well as delicate background issues, and including core-collapse and invisible supernova progenitors.
Finally, conclusions and future perspectives are discussed in Section~\ref{sec:conclusions}.

We remark that in the current review we assume a standard three-neutrino oscillation framework.
 We neglect exotic neutrino properties
such as a neutrino magnetic moment, neutrino decays or extra sterile neutrino states that would have a potential  impact on the supernova neutrino signal~\cite{Raffelt:1990yu}.
Also, aspects of neutrino-induced and neutrino-affected nucleosynthesis will only be mentioned in passing. This field is 
linked to a large diversity of questions that reach far beyond the scope of the topic of our review and which are addressed
in Ref.~\cite{Martinez-Pinedo:2014}.

\newpage

\section{Neutrino Signals from Stellar Collapse and Supernova Explosions}
\label{sec:SNmodels}
\noindent {\it Authors: H.-Th.~Janka, R.~Bollig, L.~H\"udepohl}
\\

In a sequence of hydrostatic nuclear burning stages massive stars
build up degenerate iron or oxygen-neon-magnesium cores, whose
gravitational collapse terminates the stellar life. These events 
are the strongest cosmic sources of MeV neutrinos, comparable to 
neutron-star mergers, but roughly 1000 times more frequent. An intense
burst of order $10^{58}$ neutrinos is released on a time scale of 
several seconds when a neutron star (NS) or black hole (BH) is born, possibly
accompanied by a supernova (SN) explosion that expels several solar masses 
of stellar debris with velocities up to some ten percent of the speed
of light.

Neutrinos are the main agents for the transport of energy and lepton
number during the infall of the stellar core and the formation of its
compact relic. Therefore they play a decisive role during all stages of
such an event. Electron neutrinos ($\nu_e$) are produced by electron
captures on nuclei and on free protons and thus accelerate the initial 
implosion.
Their continuous release drives the evolution from the lepton-rich 
post-collapse configuration to the final deleptonized NS,
while neutrinos and antineutrinos of all flavors carry away the 
gravitational binding energy of the assembling remnant. Neutrino-energy
deposition behind the stalled bounce shock can revive the shock and can
thus initiate the SN explosion. The interaction of electron
neutrinos and antineutrinos with the neutrino-heated outflow sets the
neutron-to-proton ratio and hence determines the nucleosynthesis in the
innermost SN ejecta. The intense neutrino radiation that
escapes from the central regions can trigger nuclear spallations, whose
free nucleons can seed interesting chemical element formation even 
in the outer stellar shells. Matter and neutrino-induced flavor oscillations
inside the exploding star can affect these processes, and vacuum oscillations 
as well as the matter effects in the Earth contribute in shaping the neutrino
signal that is detectable in experimental facilities on earth.

In this Section we summarize the present status of the numerical modeling of
stellar core collapse and explosion and of the associated neutrino emission
by spherically symmetric and multi-dimensional simulations. The effects
of neutrino flavor oscillations, which are usually ignored in the source
modeling and only computed in a post-processing treatment, will be
discussed in Section~\ref{sec:oscillations}.

\subsection{Neutrinos from supernovae and SN~1987A in retrospect}
\label{sec:SN1987A}

The Nobel prize awarded detection of neutrinos from SN 1987A 
on February 23, 1987, sets a landmark for the beginning of extragalactic 
neutrino astronomy. It is
the first direct empirical proof of the formation of a hot NS
 in the gravitational collapse of the core of a massive, evolved
star. The overall properties of the measured neutrinos 
\cite{Hirata:1987hu,Bionta:1987qt,Alekseev:1987ej}, i.e., their individual
particle energies, signal duration, and total signal energy, were compatible
with theoretical predictions of the neutrino emission from the birth of
NSs \cite{Burrows:1986}. Although the limited statistics
of only two dozen registered events did not allow for high significance,
numerous relevant constraints on particle properties and fundamental physics
could be derived on the basis of this neutrino observation in the months and
years after SN~1987A \cite{Raffelt:1990yu}. 

The lack of signal statistics also appears as the most likely explanation of 
some puzzling features of the neutrino detection, in particular the gap
of roughly 7\,s between $\sim$2\,s and $\sim$9\,s in the Kamiokande-II data,
and the clear excess of events pointing away from the source. Moreover, the 
Kamiokande and IMB measurements are only marginally consistent with each
other. A joint analysis of both experiments also yields a best-fit value 
for the spectral temperature of the detected electron antineutrinos 
($\bar\nu_e$) that 
is considerably lower than expected for the time-integrated neutrino 
emission on grounds of the most detailed models existing at the time of 
SN~1987A. This tension is even enhanced given the neutrino mixing 
parameters indicated by the current phenomenology 
(cf.\ Sec.~\ref{sec:oscillations}) and the high mean energies that had
been predicted for the radiated heavy-lepton neutrinos 
($\nu_x = \nu_\mu,\,\bar\nu_\mu,\,\nu_\tau,\,\bar\nu_\tau$) by those
early models \cite{Jegerlehner:1996kx}.

Meanwhile, however, this conflict has disappeared because 
state-of-the-art simulations of the Kelvin-Helmholtz cooling of nascent NSs
 yield significantly softer heavy-lepton neutrinos and a smaller
difference of their spectrum compared to that of electron antineutrinos
\cite{Keil:2002in,Buras:2006b,Huedepohl:2009wh,Fischer:2009af}.
These trends are along the lines of model adjustments that had been
identified in Ref.~\cite{Jegerlehner:1996kx} 
as a necessary implication of the SN~1987A neutrino data if 
solar neutrino experiments would bear out a large mixing angle between $\bar\nu_e$
and a heavy-lepton neutrino. The improvements in modern treatments of 
neutrino transport in SNe are connected to the introduction of 
nucleon-nucleon bremsstrahlung as the most important pair-production process
\cite{Hannestad:1998,Thompson:2000}, whose relevance
for neutrino-spectra formation in NS cooling models had first
been pointed out in Ref.~\cite{Suzuki:1989}. The similarity of $\bar\nu_e$ and
$\nu_x$ spectra is further enhanced by the inclusion of energy transfers in 
neutrino-nucleon scatterings \cite{Raffelt:2001,Keil:2002in}.
A detailed reanalysis of the SN~1987A neutrino
detections on the basis of signal predictions by modern hydrodynamical models of
SN~1987A, also applying our current knowledge of SN neutrino oscillations,
still needs to be performed. This would remove some of the uncertainties in
the two-phase parametrizations used in 
Refs.~\cite{Vissani:2009,Pagliaroli:2009,Loredo:2002} and would allow
for a closer assessment of the remaining discrepancies between model 
predictions and SN~1987A neutrino measurements.

One of the biggest unsolved riddles in connection to SN~1987A is a cluster
of 5 neutrinos of 7--11\,MeV registered within 7\,s by the LSD scintillator
experiment in the Mont Blanc Laboratory roughly 4.5 hours before the Kamiokande II,
IMB and Baksan events occurred \cite{Aglietta:1987it}. The fact that none of
these other experiments reported any significant detection at the time of the
Mont Blanc measurement might be understood by the relatively low energies of these 
neutrinos and the higher detection thresholds of the experiments, whereas the
small active mass of only 90 tons of the LSD detector has been used as an
argument why this facility did not see any neutrinos when the other three
experiments captured their events. However, the detection of five low-energy 
neutrinos by the small LSD mass requires a total energy in the neutrino burst
that is several times higher than the typical energy release from NS
formation \cite{Schramm:1990}. No truly convincing scenario has been 
proposed as an explanation so far. In Refs.~\cite{Imshennik:2004,Imshennik:2011} 
a two-stage collapse scenario was suggested in which the fragmentation of
the collapsing stellar core leads to the formation of a binary NS
whose inspiral is driven by gravitational radiation and whose final merger 
causes the second neutrino burst. While core fragmentation seems possible
in the presence of extreme rotation \cite{Rampp:1998},
the long delay to the second burst can hardly be attributed to gravitational-wave
emission, because the collapse fragments are embedded in the dense, infalling 
gas of the SN progenitor and must be expected to dissipate the energy
of their orbital motion by hydrodynamic effects (pressure waves and shocks)
much faster. 

Another long-standing question concerns the possible formation of a 
BH in SN~1987A. Considering the fact that recent observations
set the lower limit of the maximum mass of nonrotating, cold NSs  
to more than 2\,$M_\odot$ \cite{Demorest:2010,Antoniadis:2013},
BH formation from the collapse of an 18--20\,$M_\odot$ 
progenitor of SN~1987A \cite{Arnett:1989} appears very unlikely
from the theory perspective. Neither the stellar iron core of such a 
star nor the fallback mass expected for an explosion energy of more
than $10^{51}$\,erg, which was deduced from SN~1987A observations,
are sufficiently large
to push the NS beyond the BH threshold after the emission
of the detected neutrino burst. From the observational perspective there is
also no real problem with the lack of clear evidence for a pulsar so far.
The current luminosity of the ejecta cloud of this
SN ($\sim 10^{36}$\,erg\,s$^{-1}$; \cite{Larsson:2011,Seitenzahl:2014})
is not compatible with the spin-down power of a bright pulsar like Crab
($\sim$4.5$\times 10^{38}$\,erg\,s$^{-1}$).
However, the ejecta emission is strongly 
affected by the shock interaction with the circumstellar
medium, and the corresponding brightening is well able to cover the 
radiation of a thermally cooling NS like the compact object in 
Cassiopeia~A with a present bolometric luminosity of about 
$7\times 10^{33}$\,erg\,s$^{-1}$ \cite{Ho:2009} and 
an initial luminosity of at most
$\sim$10$^{35}$\,erg\,s$^{-1}$ (e.g., \cite{Page:2006,Shternin:2011}).
Interestingly, very
recent spectral and morphological analysis of the remnant of
SN~1987A with ALMA and ATCA data seems to indicate the possible 
presence of a diffuse synchrotron
source in the unshocked ejecta at a westward offset from the SN
position. If caused by the energy release of a pulsar wind nebula
it would set an upper limit to the pulsar spin-down power of
$\sim$10$^{35}$\,erg\,s$^{-1}$ \cite{Zanardo:2014}. The pulsed 
emission of such a spinning NS would so far have remained 
undetected because of its relative faintness and the obscuration of the
scattering and absorbing stellar debris around the SN site.

While the neutrinos from SN 1987A were able to yield basic
confirmation of the theory of NS formation, their
event statistics was too poor to allow for useful insights into
the dynamics of the beginning explosion and thus into the still
debated explosion mechanism of core-collapse SNe. 
A next Galactic SN, however, is likely to provide a 
high-statistics neutrino signal, depending on the availability
of running experimental facilities. It will therefore 
carry valuable information about the properties of the neutrinos 
themselves (e.g., \cite{Raffelt:2011}), and it will, 
in particular, help us unravelling
one of the most nagging problems of stellar astrophysics, namely
why and how massive stars achieve to reverse their catastrophic
collapse to the gigantic blast of the SN. But even if a 
massive star in our Milky Way ends its life without a brilliant
explosion, which is a possibility that could happen in a fair 
fraction of up to several 10\% of all stellar core collapses
\cite{Horiuchi:2011zz,Ugliano:2012kq,Horiuchi:2014ska,Kochanek:2013yca},
the formation of a BH will be
accompanied by a luminous outburst of neutrinos. In this case
neutrinos (and possibly gravitational waves)
might be the only messengers of the ``silent''
stellar death. The magnitude of the neutrino luminosity will be
a measure of the rate of mass infall from the collapsing stellar 
core to the accreting NS, and large-amplitude luminosity
variations will reflect the dynamics of the (unsuccessful) 
SN shock, caused by shock expansion or contraction 
episodes, which modulate the
NS accretion rate. The final termination of the
neutrino emission will mark the collapse of the transiently
stable NS to a BH and thus might set constraints
to the uncertain equation of state (EoS) at supernuclear densities. 

A high-statistics measurement of neutrinos from a future 
Galactic SN will therefore be of paramount 
importance not only for astrophysics and neutrino physics but
for nuclear physics, too. Also the measurement of the
diffuse supernova neutrino background (DSNB) as an integral
signal of stellar core collapses in the past will bear a high
potential for interesting astrophysics. It will be able to 
set limits on the total rate of stellar core-collapse events
in the local Universe. If the rate of BH formation
cases accounts for a considerable fraction of these events, they
might leave characteristic fingerprints in the DSNB spectrum
because of pronounced differences between the neutrino
luminosities and spectra radiated by NS and
BH formation. Improved predictions of the properties of
the DSNB are an important task for future studies to be performed
when more detailed stellar collapse and SN models with 
sophisticated neutrino transport have become available for large
sets of progenitors.

\subsection{Numerical methods for supernova modeling in this work}

All simulations by the Garching group reported in this article,
spherically symmetric (1D) as well as multi-dimensional, 
were performed with the \textsc{Prometheus-Vertex} SN code
\cite{Rampp:2002}. It utilizes a two-moment scheme
for the transport of neutrinos and antineutrinos of all three
flavors, which employs a Boltzmann closure for the variable Eddington 
factor and accounts for the full energy and velocity (to order
$(v/c)$) dependence of the transport in the comoving frame
of the fluid. For multi-dimensional problems a ``ray-by-ray-plus'' 
approximation is applied \cite{Rampp:2002,Buras:2006a}. 
Gravity is generalized beyond the Newtonian description by an effective
relativistic potential \cite{Rampp:2002}, adopting ``Case~A''
of Ref.~\cite{Marek:2006}, and relativistic reshifting is taken
into account in the neutrino transport. The code
includes the whole set of neutrino opacities as detailed
in Ref.~\cite{Rampp:2002} with the improvements of Ref.~\cite{Buras:2006a};
see also the summaries in Refs.~\cite{Mueller:2012is,Janka:2012a}.
Recent upgrades of the code include\footnote{The implementations of
points (ii) and (iii) in the simulation code in an early, preliminary 
version were assisted by Bernhard M\"uller.}:
\begin{itemize}
\item[(i)]
The possibility to treat
all individual types of muon and tau neutrinos and antineutrinos as 
separate species instead of clustering them together into one 
representative heavy-lepton neutrino \cite{Bollig:2013};
\item[(ii)]
the implementation of self-energy shifts of unbound neutrons and 
protons in their charged-current $\beta$-reactions with neutrinos;
\item[(iii)]
an optional treatment of quasi-stationary convection in 1D
simulations by a mixing-length description 
(see Sect.~\ref{sec:mixlength}).
\end{itemize}
These upgrades will be mentioned when they are applied in the
models described in this section.

Although \textsc{Prometheus-Vertex} makes use of an approximate
description of the effects of general relativistic gravity (the 
hydrodynamics solver is Newtonian), it permits simulations up
to the onset of BH formation. Tests did not only show
good compatibility with fully relativistic calculations
during the whole cooling evolution of nascent NSs
\cite{Groote:2014}. Also for proto-neutron stars that 
approach gravitational instability by continuous accretion, 
test calculations in comparison to cases in the
literature \cite{Fischer:2008rh} revealed an almost
perfect match of the time scale to reach the collapse of
the NS and very good agreement with respect to the
neutrino-emission properties
(with minor differences for $\nu_e$ and $\bar\nu_e$
and maximal differences of 10--20\% for $\nu_x$;
\cite{Huedepohl:2013}).

\begin{figure}[!t]
\centering
\includegraphics[width=0.325\textwidth]{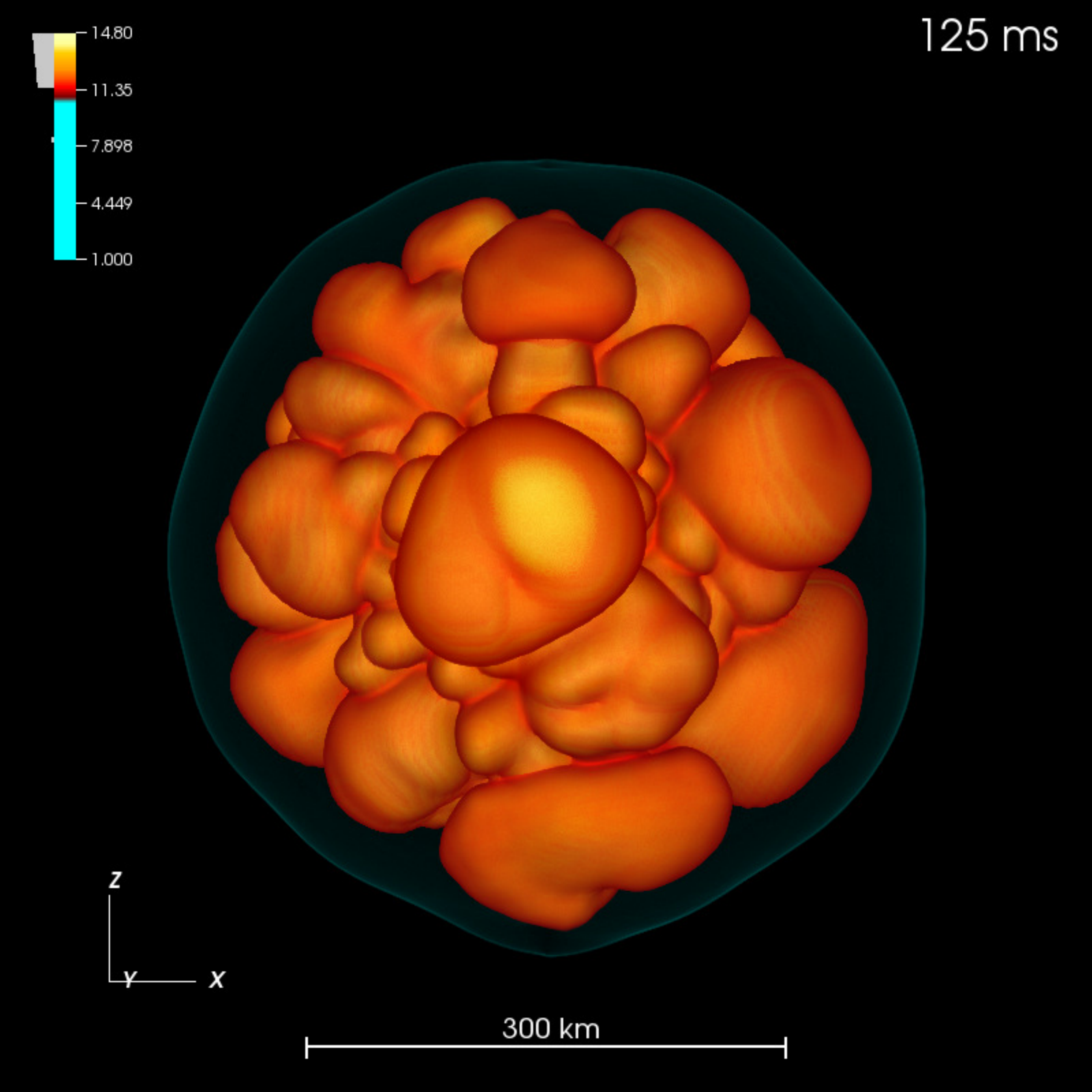} 
\includegraphics[width=0.325\textwidth]{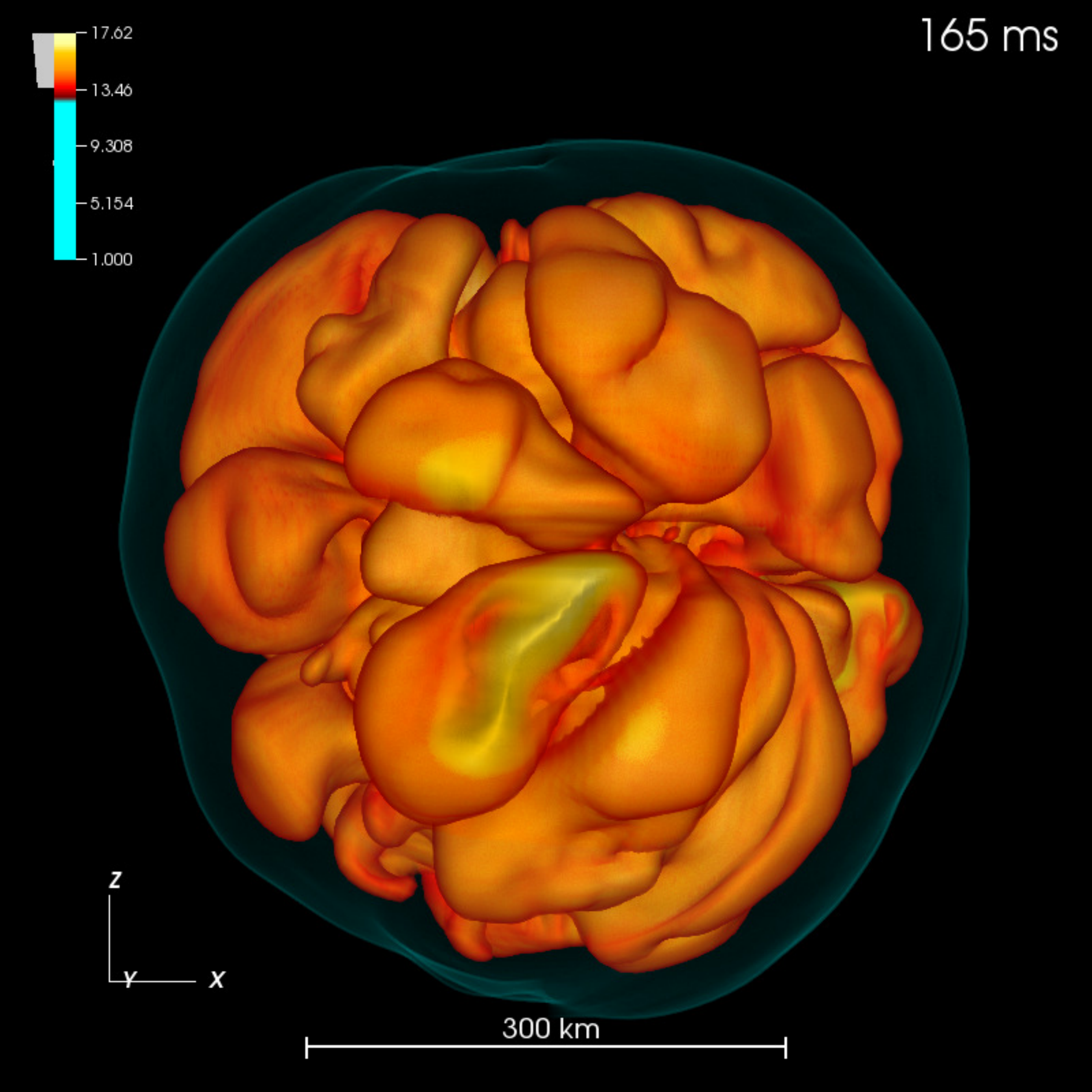} 
\includegraphics[width=0.325\textwidth]{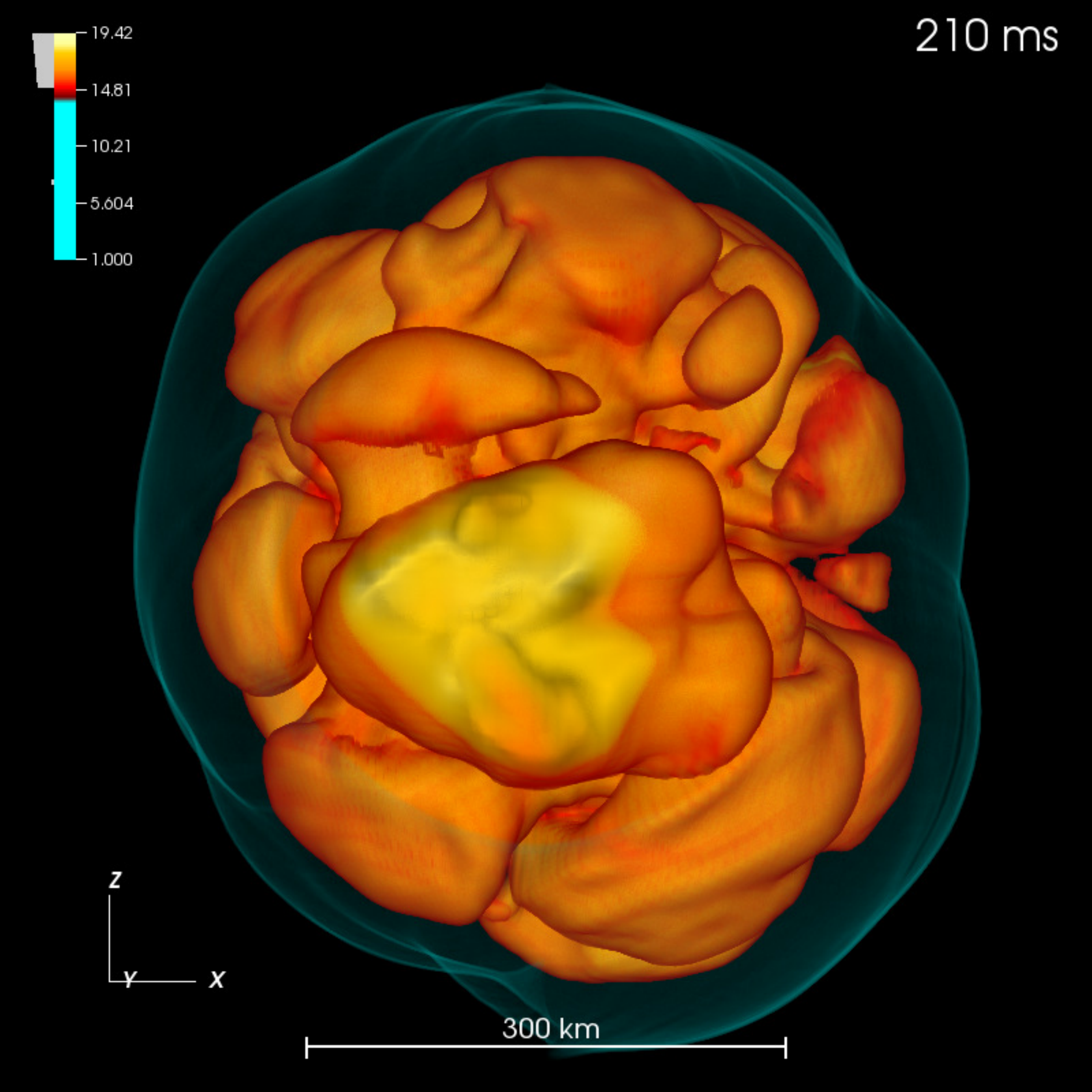} \\ \vspace{2pt}
\includegraphics[width=0.325\textwidth]{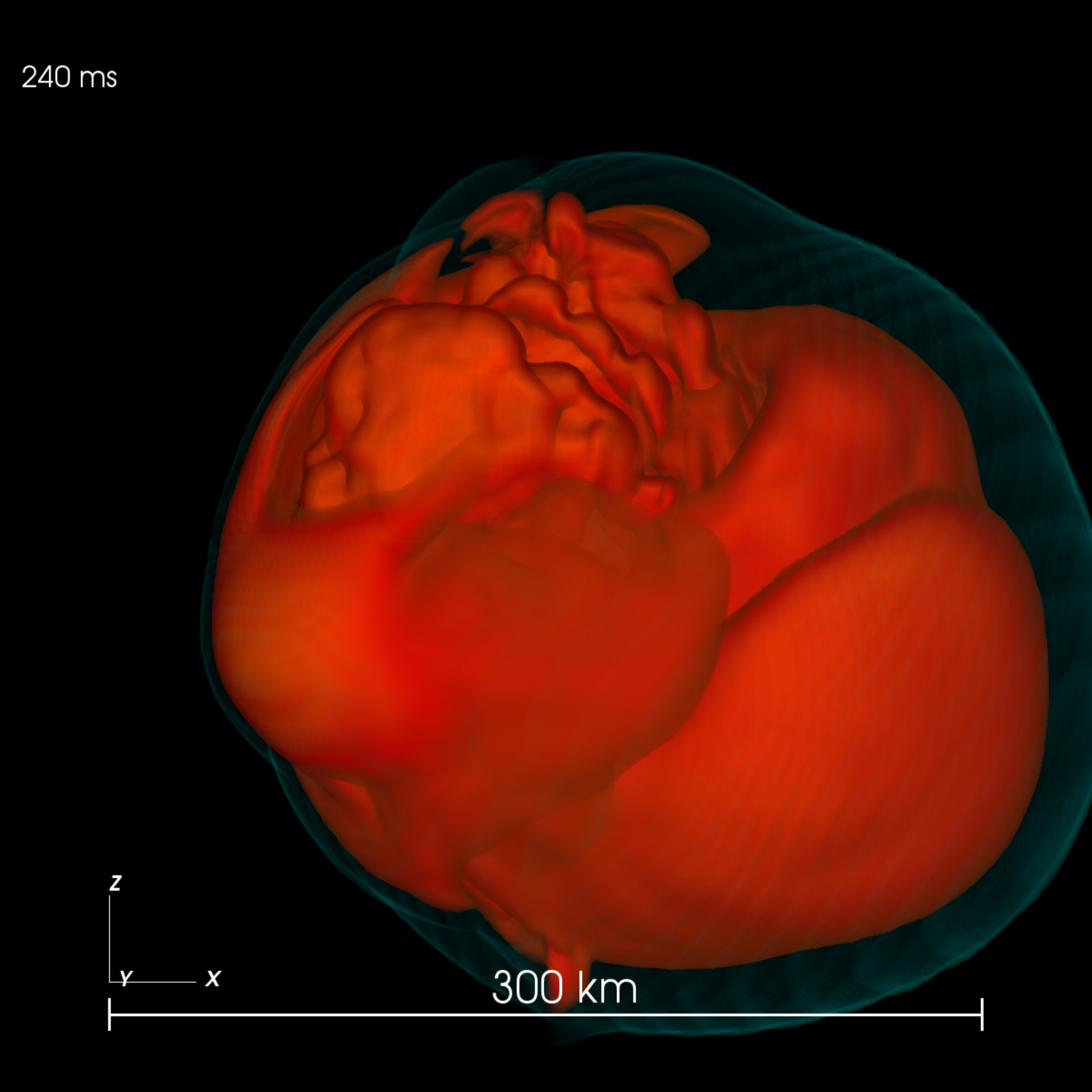} 
\includegraphics[width=0.325\textwidth]{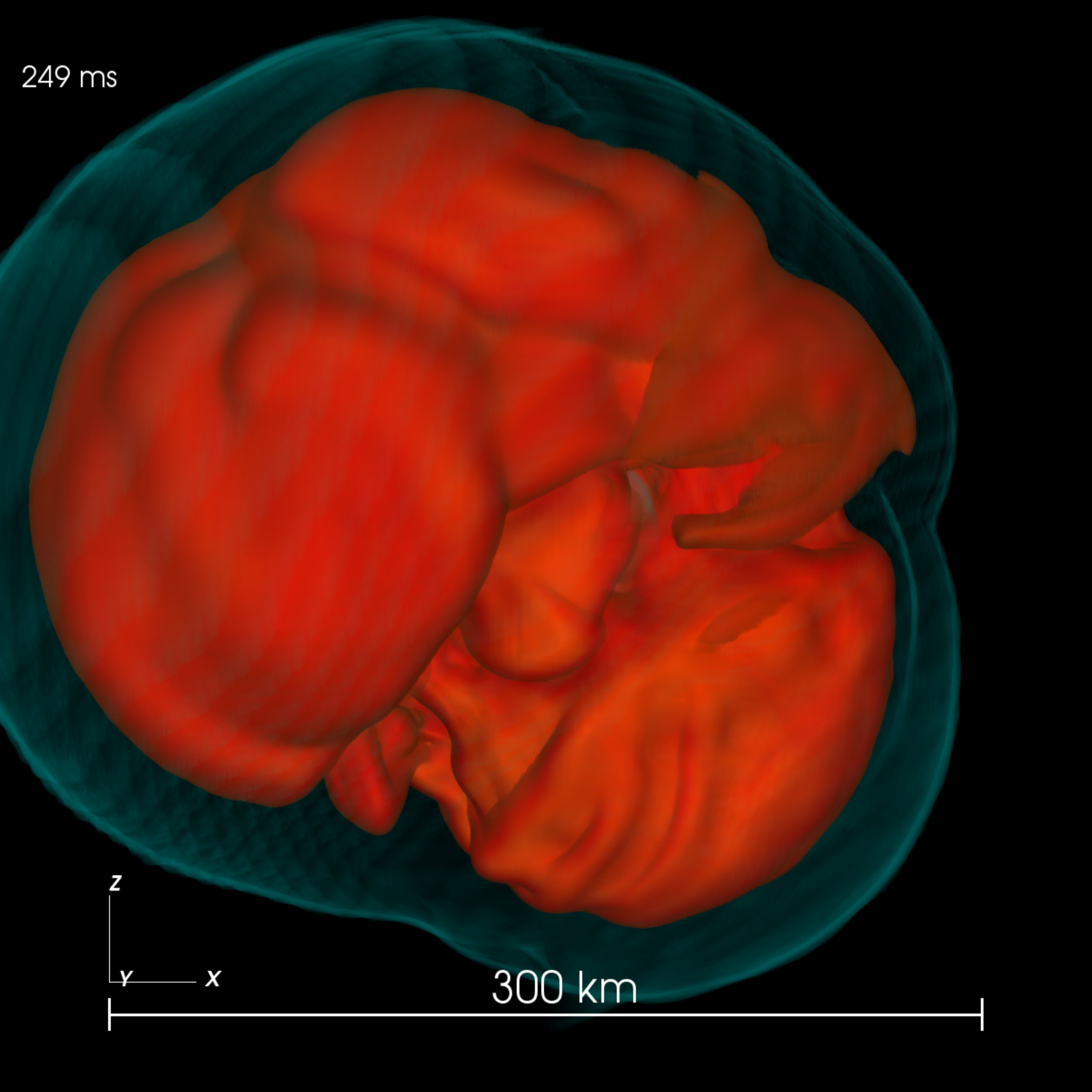} 
\includegraphics[width=0.325\textwidth]{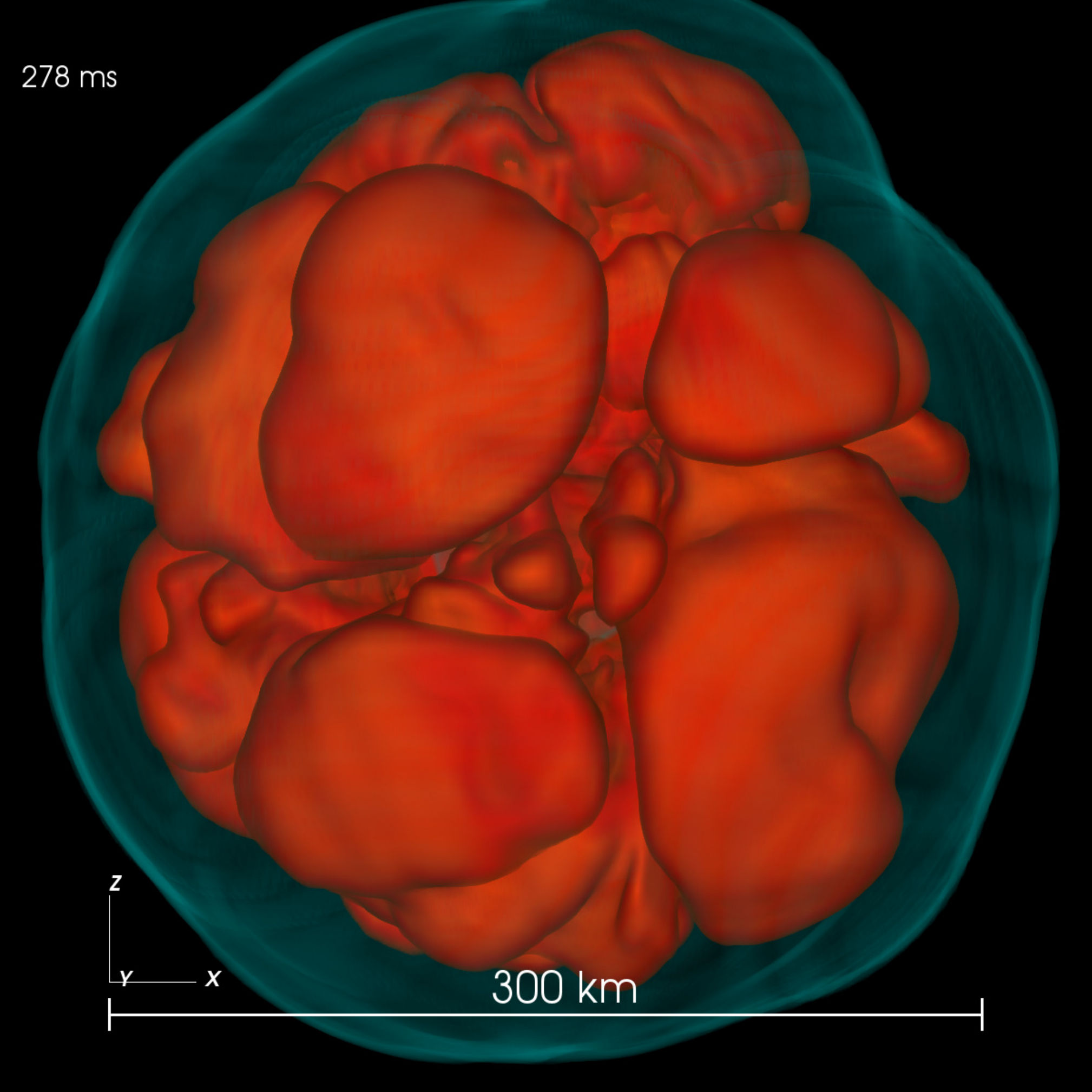}
\caption{{\em Top:} Three post-bounce snapshots (125, 165, 210\,ms p.b.) 
of a 3D simulation of an 11.2\,$M_\odot$ star, whose postshock accretion
flow is characterized by convection. (Image from Ref.~\cite{Tamborra:2014a}; 
copyright (2014) by the American Astronomical Society.)
{\em Bottom:} Three images (240, 249, 278\,ms p.b.) from a 27\,$M_\odot$ 3D
simulation, in which episodes of strong activity by the standing accretion 
shock instability (SASI) alternate with convection-dominated periods.
(Image from Ref.~\cite{Hanke:2013}; copyright (2013) by the American 
Astronomical Society.) Surfaces of constant entropy are displayed
in yellow and red; the SN shock is visible as a bluish,
semi-transparent envelope. SASI sloshing or spiral motions show up by
large-amplitude unipolar or dipolar deformations, whose orientation
flips between the hemispheres on time scales of milliseconds.
\label{fig:3D-snaps}}
\end{figure}

\subsection{Multi-dimensional phenomena in supernova cores and 
explosion mechanism}
\label{sec:explmech}

\subsubsection{\em The neutrino-driven mechanism}

When nuclear-matter densities are reached at the center of a 
gravitationally imploding
stellar core, a NS begins to assemble from the infalling
matter. The stiffening of the EoS due to repulsive forces
between the nucleons leads to an abrupt bounce that sends a strong 
hydrodynamical shock wave into the supersonically collapsing outer
material of the stellar core. Only milliseconds after core bounce,
however, the newly formed SN shock is weakened
by iron photo-disintegration and a prompt burst 
of electron neutrinos, which drain energy from the postshock material
and reduce the pressure behind the shock. 
Therefore the shock stagnates and becomes
an accretion shock with mass infall instead of expansion characterizing
the flow in the downstream region. Increasingly more sophisticated
numerical models have consolidated this nowadays generally accepted
failure of the prompt bounce-shock mechanism (see \cite{Janka:2012a} for a
review). On a much longer time scale of tens to hundreds of milliseconds
after core bounce, however,
neutrinos leaving the nascent NS deposit energy in the 
gain layer between the so-called gain radius and the shock, mainly 
through charged-current reactions with free nucleons:
\begin{eqnarray}
\nu_e + n &\longrightarrow& p + e^- \,,\label{eq:nuabs} \\
\bar{\nu}_e + p &\longrightarrow& n + e^+\,.
\label{eq:nubarabs}
\end{eqnarray}
As time goes on, the conditions
for this neutrino heating become continually more favorable because the
neutrinos are radiated with increasingly harder spectra as the 
neutrinospheric temperature rises in the contracting and 
compressionally heated NS.

If the energy transfer by neutrinos is strong enough, it can raise the
postshock pressure to trigger the reacceleration of
the SN blast wave and to thus initiate a successful
explosion \cite{Bethe:1985}. 
At later times after bounce, not only the rising spectral temperatures
of the radiated neutrinos improve the conditions for this shock revival.
Also the rate at which matter of the collapsing stellar core falls into
the shock gradually declines and with it the ram pressure of the infalling
material. This also favors the rejuvenation of the stalled shock
when the mass accretion rate drops below the critical value at which
the thermal pressure created by neutrino heating can overcome the ram
pressure of matter ahead of the shock. The explosion therefore sets in
with a considerable time delay after the initial shock formation
at the moment of core bounce. 

For a period of 30 years, Wilson's ``delayed neutrino-driven 
mechanism'' has survived as the standard paradigm for explaining how
massive stars achieve to explode, despite the fact that 
numerical modeling repeatedly experienced setbacks in its efforts
to demonstrate the viability of the mechanism and
despite the nagging lack of an ultimately
convincing theoretical or observational confirmation, both of which still 
nourish scepticism and criticism. However, in the absence of rapid 
rotation and very strong magnetic fields, whose relevance for
ordinary SNe is disfavored by current stellar evolution models
\cite{Heger:2005}, neutrinos are the most efficient
way to extract energy from the rich reservoir of the hot, newly 
formed NS and to transfer parts of this energy to the 
overlying stellar shells in order to reverse their infall to an explosion.
In fact, neutrinos
carry away the huge gravitational binding energy of the compact remnant,
several $10^{53}$\,erg, and less than one percent of this energy is
well sufficient to account for the observed power of a typical 
SN.

\begin{figure}[!t]
\centering
\includegraphics[width=1.0\textwidth]{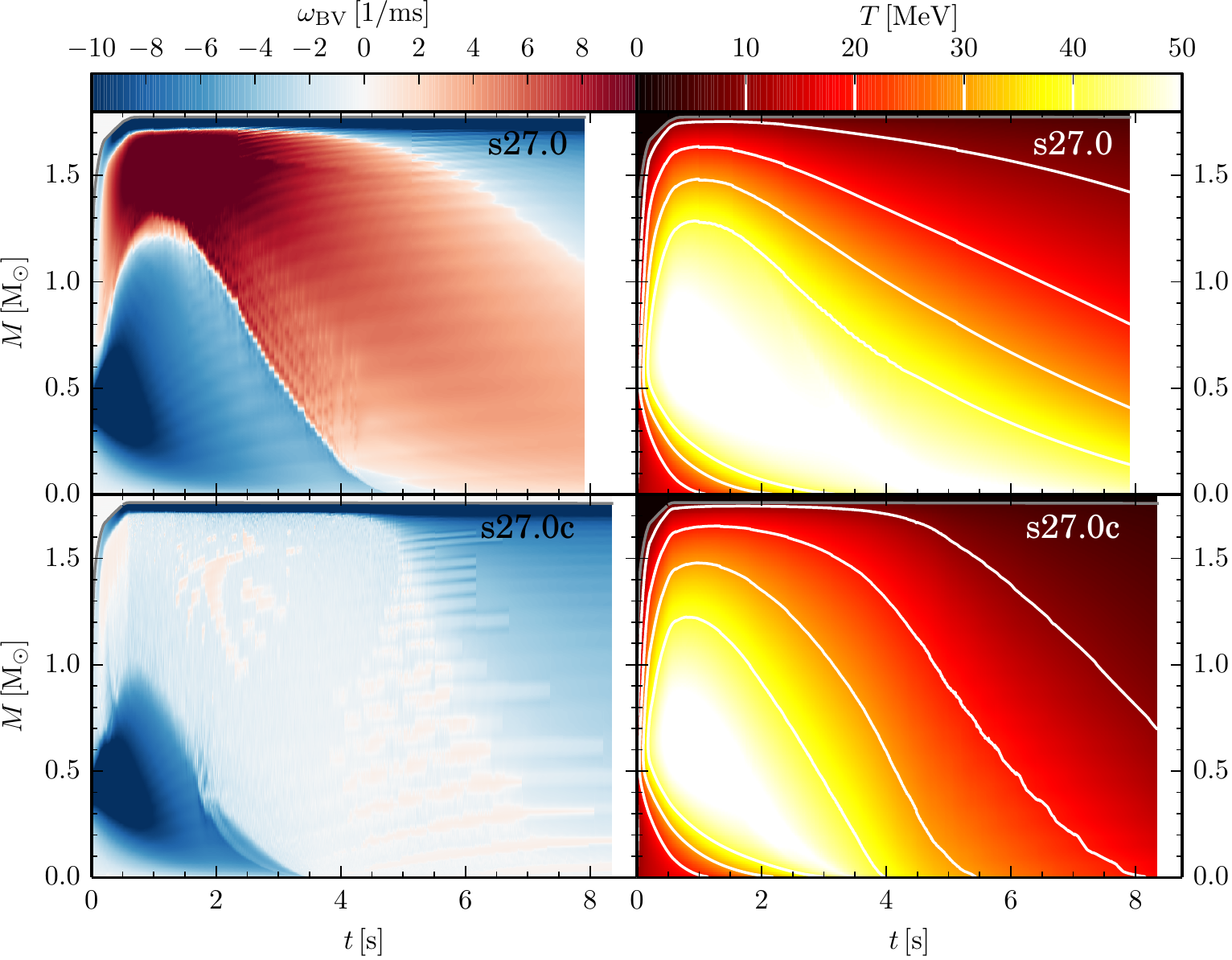}
\caption{Evolution of the convective region in the newly formed
NS of a collapsing and exploding 27\,$M_\odot$ star 
\cite{Huedepohl:2013} using the
EoS of Lattimer \& Swesty \cite{Lattimer:1991nc} with a
nuclear incompressibility modulus of $K = 220\,$MeV. The NS 
 has a baryonic (final gravitational) mass of 1.776 (1.592)\,$M_\odot$.
Time is normalized to bounce. The {\em upper panels} show a 1D simulation
without convection, the {\em lower panels} with a mixing-length treatment
of convection. The {\em left column} displays the evolution of the
Brunt-V\"ais\"al\"a frequency, positive values of which indicate
instability for Ledoux convection. In the model that includes
convective energy and lepton-number transport,
the convectively unstable regions (red in the upper left plot)
become convectively neutral (whitish in the lower left plot).
The {\em right panels} display the
temperature evolution. The cooling (and deleptonization) of the
nascent NS is considerably faster when convective
fluxes accelerate the neutrino loss.
\label{conv-s27}}
\end{figure}

The basic functioning of the neutrino-driven explosion mechanism can 
be understood as a global runaway instability of the postshock 
accretion layer caused by the neutrino-energy deposition. According
to the ``critical luminosity condition'' \cite{Burrows:1993}
the explosion is launched when the neutrino luminosity exceeds a
critical threshold value that depends on the mass-accretion rate of
the stalled shock. This concept is now widely accepted and basically
consistent with many parametric numerical models (e.g., 
\cite{Janka:1996,Murphy:2008,Nordhaus:2010,Fernandez:2012,Hanke:2012})
and analytic studies (e.g., \cite{Yamasaki:2006,Pejcha:2012}),
although many details are still disputed, e.g. which
parameters define the critical condition most accurately and can be
used as most reliable indicator of the threshold for runaway (e.g.,
\cite{Fernandez:2012,Pejcha:2012,Mueller:2015}).

\subsubsection{\em Importance of non-radial hydrodynamic instabilities}

Hydrodynamic simulations with an increasingly more sophisticated 
description of the crucial neutrino physics and transport, however,
have shown that a state-of-the-art treatment of the microphysics
(in particular of the neutrino interactions and EoS)
does not allow for explosions in spherical symmetry (i.e., in 
one dimension; 1D). An exception to this are 
the lowest-mass progenitors of SNe, i.e.,
$\sim$9--10\,$M_\odot$ stars with oxygen-neon-magnesium cores
\cite{Kitaura:2006,Dessart:2006,Fischer:2009af} 
or small iron cores \cite{Melson:2015a},
which are surrounded by extremely dilute overlying shells with 
very low gravitational binding energy. But even in these cases the 
neutrino-heated postshock layer is convectively unstable and
high-entropy, neutrino-heated plasma becomes buoyant and rises 
in Rayleigh-Taylor plumes (see Fig.~\ref{fig:3D-snaps}, upper 
row). These multi-dimensional flows assist neutrinos in triggering
the explosion earlier, accelerate the expansion of the SN
shock, enhance the explosion energy, create asymmetries in the 
explosion ejecta, and modify the conditions for element formation in
the neutrino-processed gas as well as in the layers of explosive 
nucleosynthesis \cite{Wanajo:2011,Melson:2015a}.

For progenitor stars that are more massive than about 10\,$M_\odot$,
the support by multi-dimensional hydrodynamic instabilities is 
indispensable to reach the critical condition for a neutrino-driven
explosion. There is a variety of effects by which non-radial flows
in the postshock layer can aid the onset of the SN explosion
\cite{Herant:1994,Burrows:1995,Janka:1996}.
Non-radial convective flows stretch the residence time
of matter in the heating layer and thus enhance the energy transfer
by neutrinos (e.g., \cite{Buras:2006b,Murphy:2008,Marek:2009a}).
The buoyant rise of neutrino-heated matter 
also reduces the energy loss by the re-emission of neutrinos (mostly
through the inverse reactions of 
Eqs.~(\ref{eq:nuabs},\ref{eq:nubarabs})). Moreover, the expanding
plumes push the shock to larger radii and thus increase the mass and
the volume of the neutrino-heated layer. With the larger optical
depth of the gain layer the efficiency of neutrino heating grows
even more, creating ideal conditions for a runaway. Some authors have 
attributed the combination of these explosion-assisting effects, 
which are associated with the non-radial mass motions in
the postshock layer, to the generation of turbulent pressure
behind the shock \cite{Murphy:2013,Couch:2015a}.
The corresponding turbulent energy scales with the neutrino-energy
deposition rate \cite{Thompson:2000,Mueller:2015},
and the turbulent pressure in addition to the gas pressure has been
shown to effectively reduce the critical
neutrino luminosity for shock revival \cite{Mueller:2015}.
Although this conceptual picture seems to be able to capture some 
basic features found in multi-dimensional explosion models, the
actual, highly time-dependent gas dynamics in collapsing stellar
cores is probably too complex to be compatible in all phases and
in all aspects with the properties of fully developed turbulent
flows at steady-state conditions.

\begin{figure}[!t]
\centering
\includegraphics[width=0.44\textwidth]{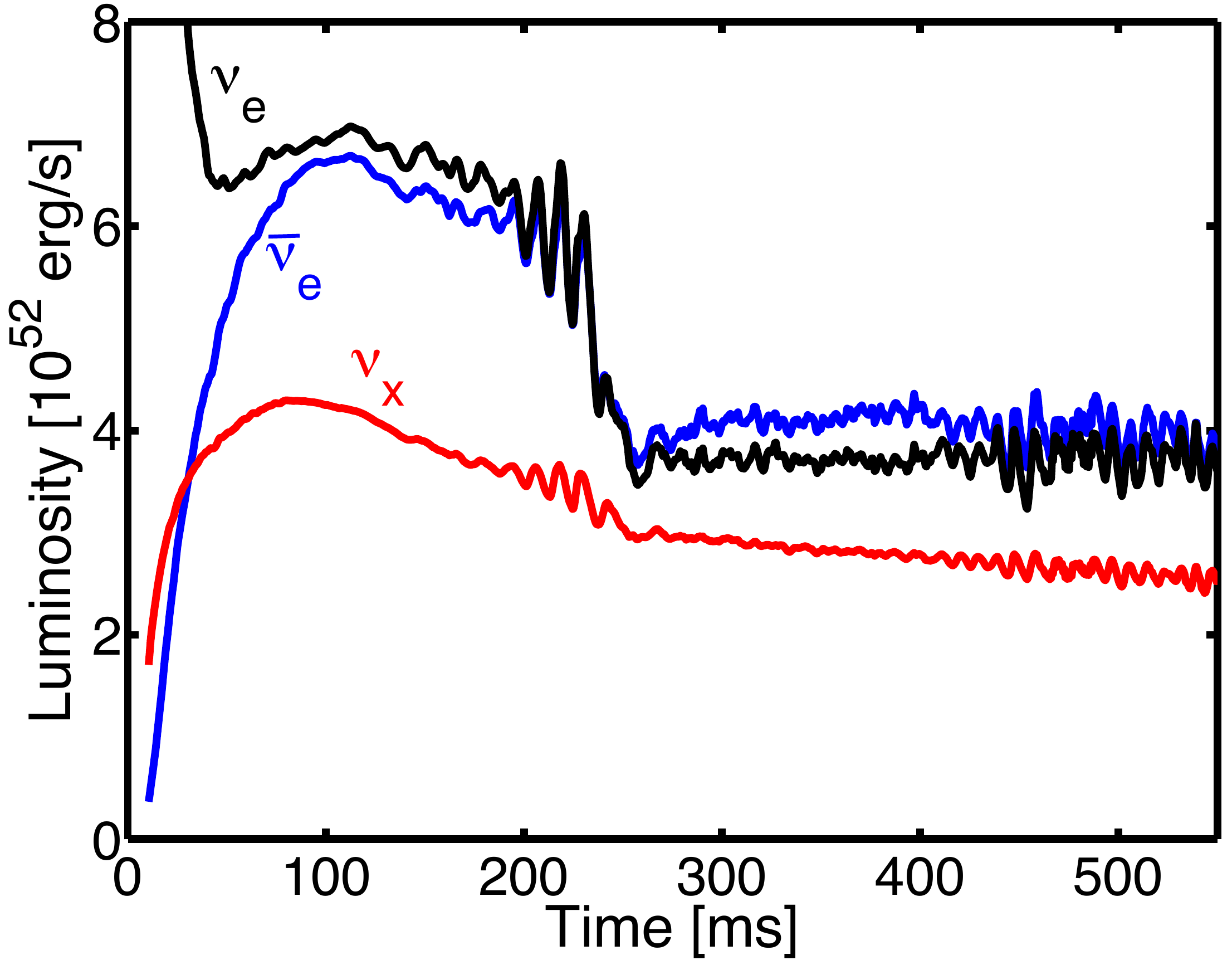}
\includegraphics[width=0.55\textwidth]{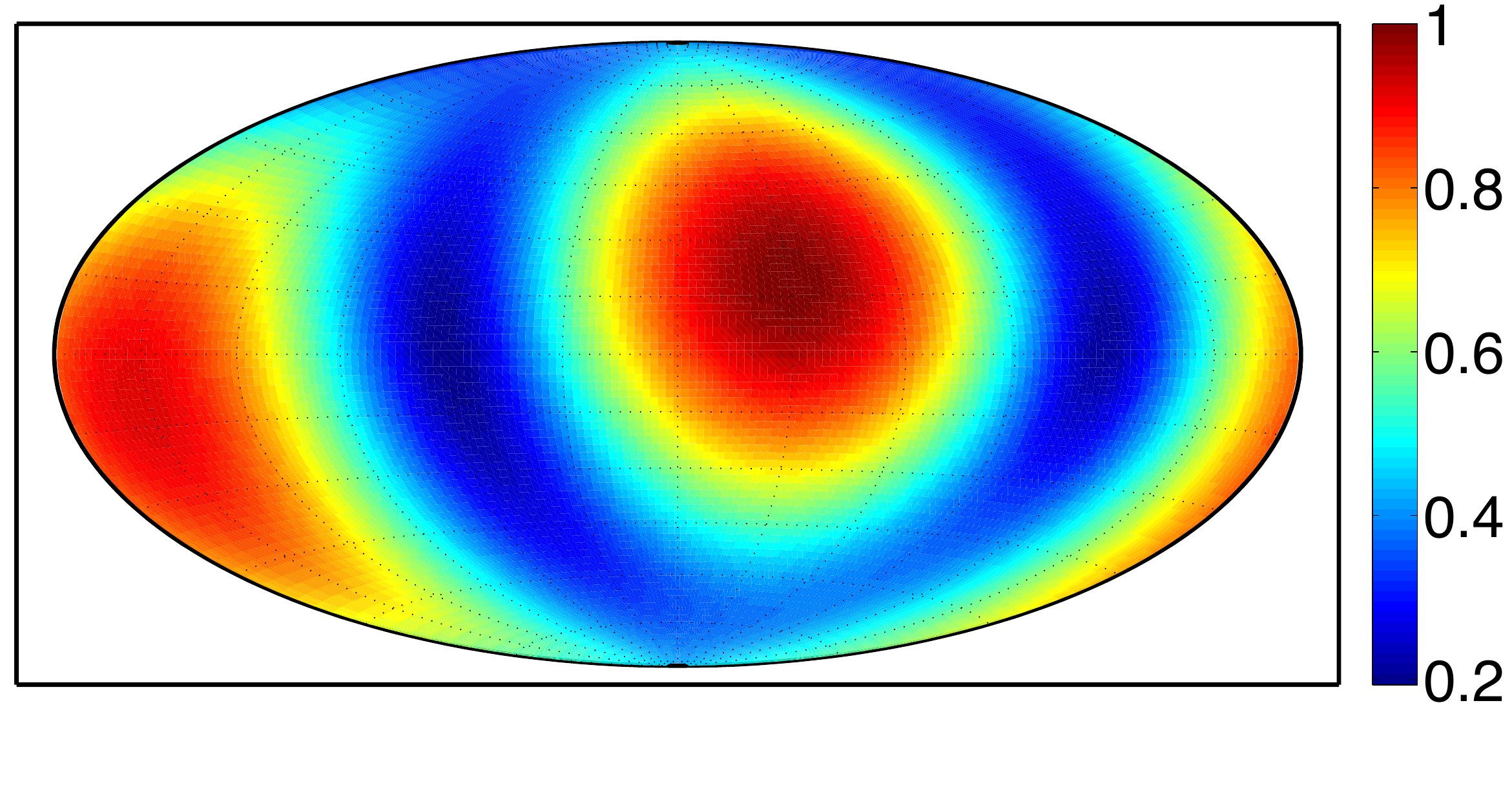}
\caption{{\em Left:} Post-bounce luminosities of $\nu_e$, $\bar\nu_e$,
and $\nu_x$ as measured by a distant observer
for a 3D collapse simulation of a 27\,$M_\odot$ star. The observer
position is close to the plane of a spiral SASI shock-oscillation
mode between $\sim$120\,ms and $\sim$250\,ms. For the displayed
luminosities the neutrino emission was integrated over the hemisphere
facing the observer with limb-darkening effects taken into account
as described in the appendix of Ref.~\cite{Tamborra:2014b},
from where the figure was taken. (Image copyright (2014) by the 
American Physical Society.) The SASI modulations of
the mass-accretion of the NS imprint large-amplitude,
near-sinusoidal variations on the neutrino emission.
There is a second phase of SASI activity starting at $\sim$400\,ms,
whose orientation is inclined to the direction of the observer.
Therefore the plot does not reflect the full amplitude of these later 
modulations. 
{\em Right:} Relative amplitude of the $\bar\nu_e$ signal variations
on a ``sky-plot'' of all observer directions during the first SASI phase
(120--250\,ms p.b.) in the 27\,$M_\odot$ model. (Image from 
Ref.~\cite{Tamborra:2013}; copyright (2013) by the American Physical 
Society.)
\label{fig:SASI-s27}}
\end{figure}

Theoretical models therefore led to the conclusion that the 
neutrino-driven SN mechanism
is a generically multi-dimensional phenomenon. This important
insight is in line with observations of SN~1987A and of other
relatively nearby and well studied SNe, and it is further 
supported by the morphological properties seen in young SN
remnants. All of these events
show large-scale asymmetries, non-spherical deformation, and extended
radial mixing of the chemical elements ejected during the explosion,
none of which could be understood if the onion-shell stratification
of the pre-collapse star was preserved during the SN blast.

The existence of convectively unstable regions in the
SN core was recognized soon after the first detailed 
hydrodynamic simulations had revealed the evolution and structure of 
the newly formed NS and its surrounding layers.
Besides the negative entropy gradient that is built up by neutrino 
heating behind the stalled shock, and which is unstable to convective
overturn \cite{Bethe:1990} setting in typically 80--100\,ms after 
bounce, the decelerating and weakening bounce
shock also leaves a negative entropy profile behind, which can decay
in an early post-bounce phase of prompt convection lasting some ten
milliseconds \cite{Bethe:1987,Burrows:1992,Burrows:1993b,Mueller:1997}.
Moreover, in the hot and
still lepton-rich proto-neutron star the combination of a negative 
entropy gradient and an unstable lepton-number 
gradient drives Ledoux convection in a thick shell interior to the
neutrinosphere \cite{Epstein:1979,Bethe:1987}. Over a period of seconds
this convective activity penetrates deeper and deeper towards the center
and accelerates the cooling and deleptonization of the nascent
NS (Fig.~\ref{conv-s27}; 
\cite{Burrows:1986,Burrows:1988,Keil:1996,Janka:2001}).

\begin{figure}[!t]
\centering
\includegraphics[width=1.0\textwidth]{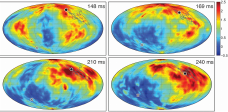}
\caption{Time evolution of the lepton-number ($\nu_e$ minus $\bar\nu_e$)
flux density normalized by the average value over all directions in a 3D
core-collapse simulation of an 11.2\,$M_\odot$ progenitor.
The panels show ``all-sky'' images for
the indicated post-bounce times. One can see the emergence of a clear
dipole pattern with strong excess of the $\nu_e$ emission in one
hemisphere and a reduced $\nu_e$ emission or even excess flux of
$\bar\nu_e$ in the opposite hemisphere. This phenomenon has been
found in all 3D simulations of the Garching group and was termed
LESA for lepton-emission self-sustained asymmetry. The black dot marks
the maximum of the dipole, the cross the anti-direction. The gray line
indicates the path described by a slow drift of the dipole direction.
(Image from Ref.~\cite{Tamborra:2014a}; copyright (2014) by the American
Astronomical Society.) 
\label{fig:LESA-4pi}}
\end{figure}

\begin{figure}[!t]
\centering
\includegraphics[width=1.0\textwidth]{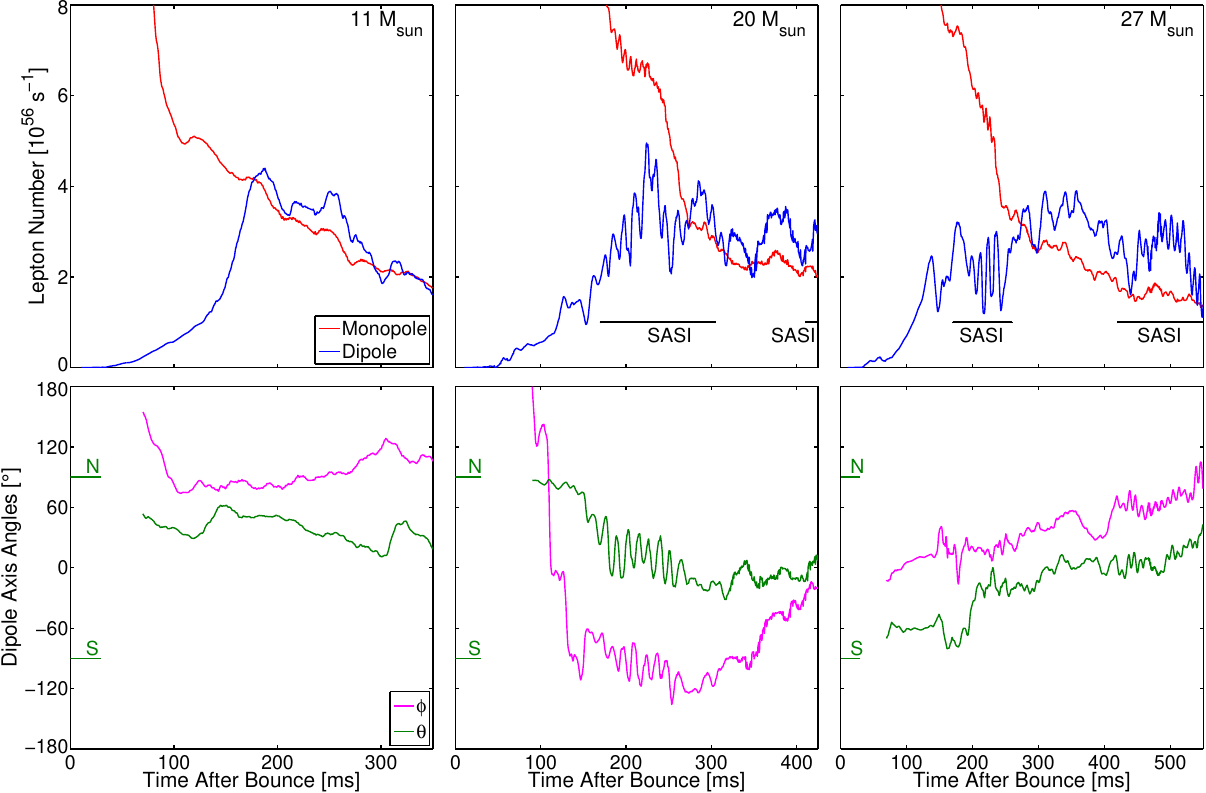}
\caption{Post-bounce evolution of the lepton-number emission in 3D
core-collapse simulations of 11.2\,$M_\odot$ ({\em left}),
20\,$M_\odot$ ({\em middle}) and 27\,$M_\odot$ ({\em right}) progenitors.
The upper panels show the monopole, i.e., the total lepton-number flux
(red curve), and the dipole component (blue curve). The bottom panels
display the polar angles $\theta$ and $\phi$ defining the dipole direction
(i.e., the direction of the maximum excess of the $\nu_e$ emission
relative to the $\bar\nu_e$ emission) in the polar coordinate grid of
the star (north and south pole directions are indicated by ``N'' and ``S'',
respectively). The monopole and dipole amplitudes $A_\mathrm{mon}$ and
$A_\mathrm{dip}$ are normalized such that the lepton-number flux is given
by $A_\mathrm{mon} + A_\mathrm{dip}\cos\vartheta$ in coordinates aligned
with the dipole direction, if the flux distribution contains only
monopole and dipole terms. 
(Image from Ref.~\cite{Tamborra:2014a}; copyright (2014) by the American
Astronomical Society.) 
\label{fig:LESA-3models}}
\end{figure}

In addition to being stirred by convective mass motions, the 
accretion shock and the whole postshock accretion flow
are unstable to global, non-radial deformation modes, which
do not allow the accretion shock to remain spherical and at
a stationary radius. The possibility of such a non-convective
instability in the SN core was first discussed in
Ref.~\cite{Blondin:2003} and was termed ``standing accretion
shock instability'' (SASI). It leads to the oscillatory growth 
of an initially small seed perturbation with the 
largest growth rates being found for the spherical harmonics modes
of lowest order, i.e., the dipolar and quadrupolar modes
\cite{Blondin:2006,Foglizzo:2006,Foglizzo:2007}.
This instability leads to large-amplitude shock sloshing and
spiral motions \cite{Blondin:2007} (see
Fig.~\ref{fig:3D-snaps}, bottom row). The underlying 
growth mechanism for this instability is a so-called advective-acoustic
cycle in the accretion flow between stagnant shock and proto-neutron
star \cite{Foglizzo:2002,Scheck:2008,Guilet:2012} and can also be 
studied in an inexpensive water experiment that shares basic
features with the accretion flow in a SN core 
(\cite{Foglizzo:2012}; for a nice review with focus on this
instability and a more complete collection of relevant references,
see Ref.~\cite{Foglizzo:2015}).

Both instabilities, postshock convection and the SASI, can provide 
crucial support to the onset of neutrino-driven explosions. A 
growing number of groups with energy-dependent neutrino-transport 
schemes of different levels of sophistication (and quite a variety
of differences in other important numerical and physical 
aspects of the hydrodynamical 
modeling) have meanwhile obtained successful neutrino-driven 
explosions of progenitors above $\sim$10\,$M_\odot$ in self-consistent,
first-principle simulations in two dimensions (2D), i.e., in simulations
that were artifically constrained to axisymmetry (e.g., 
\cite{Buras:2006b,Marek:2009a,Suwa:2010,Suwa:2013,Mueller:2012is,Mueller:2013,Mueller:2014,Janka:2012b,Bruenn:2013,Pan:2015}). This suggests
the basic viability of the neutrino-driven mechanism, although
there is still no unanimous agreement \cite{Dolence:2015}.
Simulations in 2D, however, are problematic because of the artifically
imposed axial symmetry, which attributes a toroidal geometry to all
structures, and because of the inverse direction of the
turbulent energy cascade (from small to large scales) compared to 
the realistic three-dimensional (3D) situation. It is therefore
indispensable to confirm the functioning of the mechanism by 3D
simulations, which have recently become possible because of the 
increasing power of modern supercomputers.

\subsubsection{\em Status of explosion modeling in 3D}

The first self-consistent 3D simulations of stellar core collapse
and explosion were performed with a relatively schematic neutrino
treatment by grey diffusion \cite{Fryer:2002,Fryer:2004}.
More recently, first-principle 3D simulations with energy-dependent
neutrino transport (however applying transport schemes with largely
differing degrees of sophistication) have also obtained successful
explosions, although the revival of the stalled shock happens 
somewhat later than in the corresponding 2D models
\cite{Takiwaki:2012,Takiwaki:2014,Melson:2015b,Lentz:2015}.
The longer delay of the onset of the explosion may be problematic 
because later explosions tend to be weaker and might be incompatible 
with observed SN energies.
However, the detailed effects in 3D and the
differences between 2D and 3D models are still a matter of intense
research, and the current models do not yet provide final answers,
in particular concerning the role of the SASI in 3D 
(cf.~\cite{Fernandez:2015}) and concerning the energetics 
of the explosions (cf.~\cite{Mueller:2015b}).
Moreover, more studies are needed to investigate
the question of numerical convergence in describing the potentially
turbulent postshock flow 
(e.g., \cite{Couch:2014,Abdikamalov:2014,Radice:2015})
and to explore the influence of so far poorly understood perturbations
and non-sphericities in the convective silicon and oxygen burning layers
of the pre-collapse star (e.g., \cite{Couch:2013,Couch:2015b,Mueller:2015}).

Also the details of the microphysics in the newly formed NS
remain a matter of concern. Interestingly, in Ref.~\cite{Melson:2015b}
it was found that a modest reduction of the neutrino-nucleon 
neutral-current scattering opacity by just 10--20\%, e.g.\ due to
possible effects of strange-quark spins in the axial-vector structure
factor of the scattering cross section, is sufficient to convert a
3D simulation of a 27\,$M_\odot$ by the Garching group from failure
to a successful explosion. This result clearly demonstrates the 
proximity of current state-of-the-art 3D simulations of stellar core
collapse to explosions, and it emphasizes the sensitivity of the outcome
of these simulations to the detailed input in the neutrino-opacity 
sector. A similar sensitivity exists with respect to the NS
EoS, because it has been known for a while already that 
``softer'' EoSs are favorable for the possibility of neutrino-driven
explosions \cite{Marek:2009a,Janka:2012b,Suwa:2013}.  
Both effects, reduced neutral-current neutrino-nucleon scattering rates
as well as a soft EoS, lead to a faster contraction of the newly formed,
hot NS in the first $\sim$0.5\,s after bounce\footnote{Note
that here a ``softer'' EoS is defined by the faster contraction of the 
newly formed NS during the early post-bounce accretion phase
and {\em not} by a lower maximum mass of a cold, non-rotating NS,
which is often considered as the criterion for discriminating soft from
stiff EoSs. It is important to realize that neither the radius nor
the maximum mass of a cold NS are of direct relevance for the 
properties of the hot, accreting remnant immediately after core bounce 
and therefore for the conditions that influence the onset of the 
SN explosion during this phase.}. In the case of a reduced 
neutrino-scattering opacity the faster contraction is a consequence
of the enhanced emission of heavy-lepton neutrinos. A faster 
NS contraction causes more rapid compressional heating of 
the neutrinospheric layer and thus allows for higher accretion
luminosities and a steeper rise of the average energies of the
radiated neutrinos with time. Both higher luminosities and higher
mean energies enhance the neutrino heating behind the shock and are
therefore supportive for neutrino-driven explosions.

Non-radial hydrodynamic instabilities in the SN core are not
only of crucial importance for the neutrino-driven mechanism; they
also impose asymmetries on the beginning explosion and therefore
set the conditions for NS acceleration through the 
``gravitational tug-boat mechanism'', which can yield NS
kick velocities in agreement with those observed for young pulsars
(e.g., \cite{Wongwathanarat:2010,Wongwathanarat:2013}).
Moreover, they seed the observable ejecta asymmetries that develop
by the interaction of the primary non-sphericities with secondary 
mixing instabilities that grow during the first day of the 
SN blast (e.g., \cite{Wongwathanarat:2015}).
The violent hydrodynamic flows associated with convective overturn
and the SASI in the postshock layer also produce characteristic 
imprints on the neutrino signal emitted during the shock-stagnation
phase after bounce. While convection leads to small-amplitude,
high-frequency ($\gsimeq$100\,Hz) fluctuations of the neutrino 
luminosities and mean energies, SASI sloshing and spiral motions
create quasi-periodic variations of much bigger amplitudes (up to
10--20 percent in the neutrino luminosities, cf.\ the left panel
of Fig.~\ref{fig:SASI-s27}, and 1--2\,MeV in the mean neutrino 
energies; \cite{Tamborra:2013,Tamborra:2014b}) and with typical
frequences of $\lsimeq$100\,Hz, because the
large-scale shock expansion and contraction phases associated 
with the SASI modulate the mass-accretion flow towards the nascent
NS massively. During episodes of enhanced accretion
the emission of neutrinos is boosted by additional accretion
luminosity, and the neutrinos escape with higher mean energies
because of the compressional heating of the accretion layer.

Measurements of neutrinos from a future Galactic SN
with Cherenkov telescopes will well be able to detect
these signal modulations 
\cite{Lund:2010,Lund:2012,Tamborra:2013,Tamborra:2014b}
(see Sect.~\ref{sec:SASI} for more information).
The observation of SASI modulations for SN neutrinos
would provide a very important confirmation of our present,
purely theoretical picture of the shock dynamics in the SN 
core. The presence of one or more SASI episodes is more probable
in cases of more massive progenitor stars (maybe with masses
beyond $\sim$15\,$M_\odot$), where the shock expansion comes
to a longer halt and even shock contraction can occur before
the onset of runaway. However, the possibility to see the 
corresponding neutrino signal modulations will depend on the
viewing angle relative to the main direction or the plane of
the SASI sloshing or spiral motions. Observers with positions
close to the SASI sloshing axis or near the main plane of the
SASI spiral mode will receive bigger modulation amplitudes 
(Fig.~\ref{fig:SASI-s27}, right panel; see also Sect.~\ref{sec:SASI}) 
and will have
the better chance to diagnose them in the detection.

\subsubsection{\em LESA: A dipolar neutrino-emission asymmetry as
new phenomenon}

Another interesting and novel phenomenon was recently discovered
in the neutrino emission of the first 3D stellar core-collapse
models with state-of-the-art multi-group three-flavor neutrino 
transport. Instead of persistently picturing the high-order 
multipole pattern of buoyant plumes and sinking downdrafts
in the convection zones in the interior of the proto-neutron star
and in the neutrino-heated postshock layer, the neutrino emission
develops a strong dipolar asymmetry on a time scale of 
about 100--150\,ms after bounce
(Fig.~\ref{fig:LESA-4pi}; \cite{Tamborra:2014a}). 
The observable hemispheric luminosity difference of electron 
neutrinos, $\nu_e$, and electron anti-neutrinos, $\bar\nu_e$,
individually can reach up to $\sim$20\% and even higher for
the lepton-number ($\nu_e$ minus $\bar\nu_e$) luminosity
(Fig.~\ref{fig:LESA-4pi}),
whereas it is only of order 1--2\% for the sum of $\nu_e$ and
$\bar\nu_e$ and for heavy-lepton neutrinos, $\nu_x$. This
astonishing and completely unexpected phenomenon was found in
all 3D SN simulations conducted by the Garching group
so far, i.e., in a convection-dominated 11.2\,$M_\odot$ model
as well as in SASI-dominated 20 and 27\,$M_\odot$ cases 
(see Fig.~\ref{fig:LESA-3models}) and also in a 
more recently published model of a 9.6\,$M_\odot$ progenitor,
in which the explosion starts on a short post-bounce time scale
of only $\sim$130\,ms \cite{Melson:2015a}. 
The development of a strong lepton-emission
dipole therefore seems to be a generic instability of the 
neutrino transport in the convectively stirred proto-neutron
star. The exact growth conditions of this instability are not
yet understood, but the main contribution to the emission 
dipole builds up in the convective layer well inside of the 
neutrinosphere, while the outer accretion develops a dipolar
asymmetry, too, and enhances the hemispheric difference
of the lepton-number emission. In fact, the accretion asymmetry
might trigger the growth of the dipole and seems to stabilize
its existence over periods of at least hundreds of milliseconds.
The phenomenon was therefore named ``lepton-emission 
self-sustained asymmetry'' or LESA (see Ref.~\cite{Tamborra:2014a}
for a detailed discussion).

After its steep growth phase until roughly 200\,ms post bounce,
the amplitude of the LESA lepton-number emission dipole can
become even larger than the corresponding monopole
(Fig.~\ref{fig:LESA-3models}). The dipole direction is
amazingly stable if one compares its slow drifting
with the much shorter life time of convective cells in
the inner and outer convection regions, which collapse
and are regenerated on a time scale of typically $\sim$10\,ms
only. SASI and LESA are two different phenomena, which can
be well distinguished also with respect to their effects
on the neutrino emission \cite{Tamborra:2014b}. SASI
modulates the $\nu_e$ and $\bar\nu_e$ emission coherently,
whereas LESA amplitudes of the $\nu_e$ and $\bar\nu_e$
fluxes have opposite signs. The
LESA dipole direction and SASI mass motions can have arbitrary 
orientations relative to each other, for example the LESA
dipole and normal vector of the plane of SASI spiral mode
are roughly parallel during the first SASI episode in the 
20\,$M_\odot$ model and nearly perpendicular in the 27\,$M_\odot$ 
case. The LESA amplitude and dipole orientation can reflect
the SASI-imposed modulations of the neutrino emission
(see Fig.~\ref{fig:LESA-3models}), but the presence of the
LESA is not overruled by the SASI, and the LESA dipole direction
is also remarkably stable during the phases of violent 
SASI sloshing and spiralling activity.

The LESA phenomenon is a new type of instability, which
occurs as a combined effect of multi-dimensional hydrodynamics
and neutrino transport. Presently, however, it cannot be 
rigorously excluded that this stunning phenomenon is a 
numerical artifact, e.g.\ connected to the ``ray-by-ray-plus''
approximation used for the neutrino transport in the 
multi-dimensional simulations \cite{Dolence:2015}.
It is also possible that the growth of the LESA dipole is 
fostered by some
special (but not necessarily realistic) aspects of the employed
microphysics, in particular of the neutrino opacities. 
Nevertheless, even in these cases a lepton-number emission 
dipole would still be an interesting result of the 3D 
simulations. 

If happening in nature, LESA would have far-reaching 
observable consequences for SNe. It would not 
only imply that the detectable signal of electron antineutrinos
is a function of the viewing direction during
the LESA-active phase of the SN evolution.
Depending on the duration of the asymmetry,
a neutrino-emission dipole (as sum over all neutrino species)
with an amplitude of about one percent could also lead to 
a NS kick of tens of kilometers per second. Such 
velocities, however, are far too low to account for the typical 
space velocities observed for young pulsars, and they are
subdominant compared to the velocities that can be achieved
by the ``gravitational tug-boat mechanism'' associated with
asymmetries of the mass ejection in SN explosions
(\cite{Wongwathanarat:2010,Wongwathanarat:2013} and references 
therein). Also the nucleosynthesis conditions in the innermost,
neutrino-processed ejecta, whose neutron-to-proton ratio is set by
the reactions of Eqs.~(\ref{eq:nuabs}) and (\ref{eq:nubarabs}),
will depend on the direction, with proton excess being present
on the side of the stronger $\nu_e$ emission and neutron excess
in the opposite hemisphere. 3D models that track the explosion
considerably beyond the onset of shock revival are needed
to study the exact implications of these possibilities for
a varity of progenitors.

\begin{figure}[!t]
\centering
\includegraphics[width=1.0\textwidth]{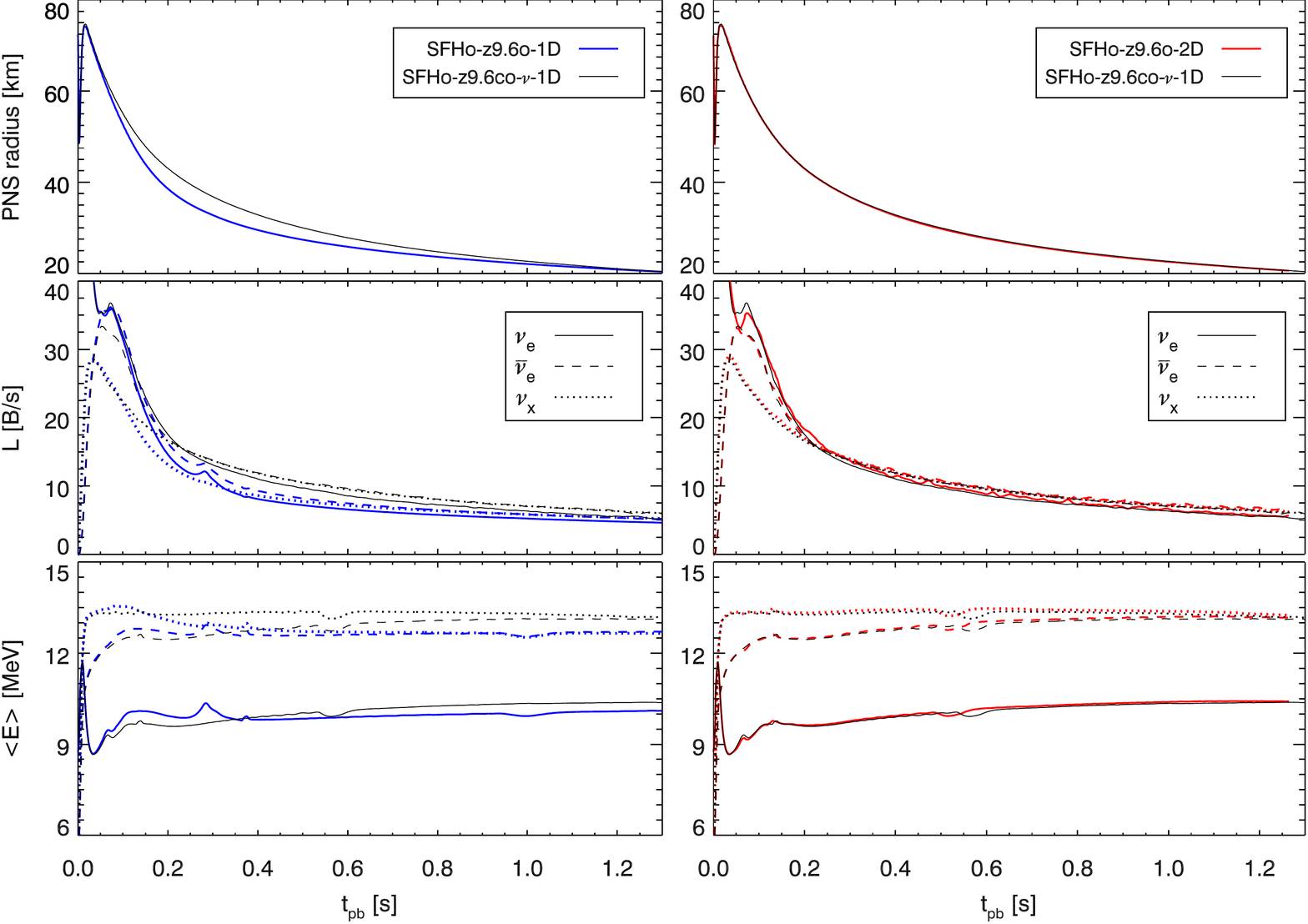}
\caption{Post-bounce evolution of the NS radius
(defined by a density of $10^{11}$\,g\,cm$^{-3}$; {\em top panels}),
luminosities of $\nu_e$, $\bar\nu_e$, and (a single representative of)
$\nu_x$ (given in units of B\,s$^{-1} = 10^{51}$\,erg\,s$^{-1}$;
{\em middle panels}) and corresponding mean energies
(defined as the ratio of energy flux to number flux;
{\em bottom panels}) in the
laboratory frame for the proto-neutron star formed during the
SN explosion of a 9.6\,$M_\odot$ progenitor star.
The baryonic (final gravitational) mass of the compact remnant is
$\sim$1.363\,(1.252)\,$M_\odot$. The {\em left panels} show a
comparison of 1D models with mixing-length treatment of
NS convection (SFHo-z9.6co-$\nu$-1D; thin black lines) 
and without (SFHo-z9.6o-1D; thick blue lines). The {\em right panels}
demonstrate the excellent agreement between the 1D model
with convection and a corresponding 2D simulation, in which
the convection is treated hydrodynamically. Both models explode
self-consistently at nearly the same time in the 1D and 2D cases,
which offers ideal conditions for a clean and accurate comparison.
\label{conv-tests-z9.6}}
\end{figure}

\begin{figure}[!t]
\centering
\includegraphics[width=1.0\textwidth]{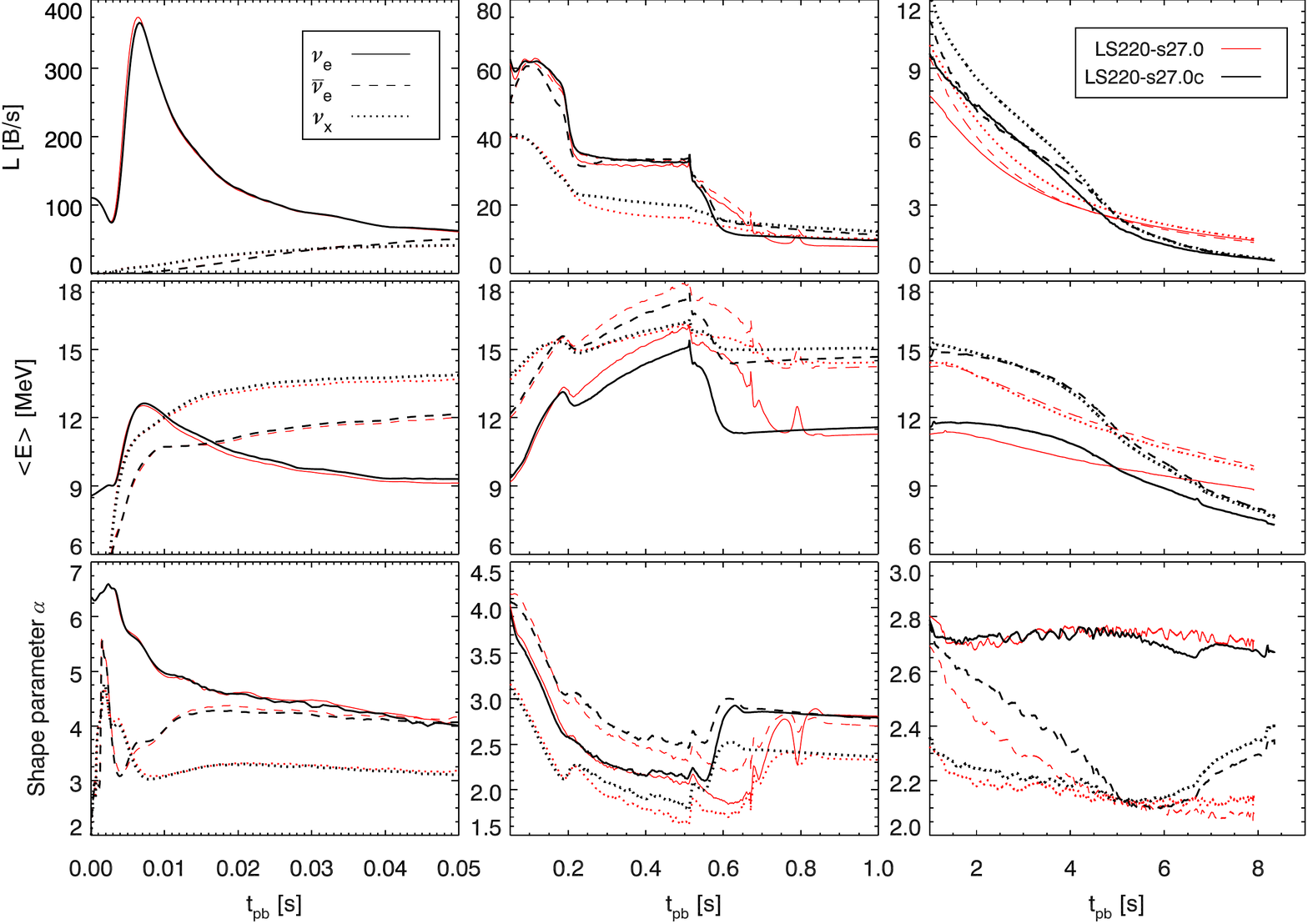}
\caption{Neutrino-signal evolution for the $\sim$1.776
($\sim$1.592)\,$M_\odot$ baryonic (final gravitational) mass
 NS formed in the explosion of a 27\,$M_\odot$
progenitor. The red lines correspond to a 1D simulation without
convection, the black lines to a 1D model with a mixing-length
treatment of proto-neutron star convection (indicated by the 
letter ``c'' in the model name). The {\em left column} shows the 
phase of the $\nu_e$-burst at shock-breakout, the {\em middle 
column} displays the accretion phase, the {\em right column}
the Kelvin-Helmholtz cooling phase of the proto-neutron star. 
The {\em top panels} display the neutrino luminosities for $\nu_e$,
$\bar\nu_e$ and a single species of $\nu_x$ (given in units of
B\,s$^{-1} = 10^{51}$\,erg\,s$^{-1}$), 
the {\em middle panels} show the radiated mean energies
(ratios of energy fluxes to number fluxes), and the {\em bottom
panels} the shape parameters $\alpha$ of the neutrino spectra,
all given as functions of time after bounce and as measured
by an observer in the laboratory frame. Note that the explosions
of the displayed 1D models were artificially triggered at 
$\sim$0.5\,s p.b.\ (in contrast to the explosions of the 
low-mass progenitor of Fig.~\ref{conv-tests-z9.6}, which
developed self-consistently). The initiation of the explosion
was done slightly differently in both models, which explains
the differences that are visible between $\sim$0.5\,s and 
$\sim$0.8\,s during the transition from the accretion to the
cooling phase.
\label{neutrinos-convection-s27}}
\end{figure}

\subsection{Neutrino signals from proto-neutron star cooling}
\label{sec:protoneutronstars}

Full 3D modeling of SN explosions has only just
begun and many aspects of the neutrino-driven mechanism are
still heavily debated and not generally agreed on (see 
Sect.~\ref{sec:explmech}). In particular, it 
is presently still uncertain which progenitor stars explode,
when exactly the explosions set in, which energies
they develop, and which stars 
collapse to give birth to BHs. Parametric explosion 
simulations are therefore useful to explore the landscape of
possibilities and to predict progenitor dependent variations
of the neutrino signal as input to the investigation of neutrino
oscillations in SNe (Sect.~\ref{sec:oscillations}), to
calculations of the DSNB (Sect.~\ref{sec:DSNB}), and to 
studies of neutrino detection in connection with future Galactic
SNe (Sect.~\ref{sec:detection}).

Spherically symmetric, i.e.\ one-dimensional (1D), simulations
are a very efficient way to achieve this goal, in particular
if the long-time cooling evolution of the newly formed NS 
over seconds to tens of seconds shall be followed, too.
However, 1D simulations do not explode in a self-acting way
(except for the lightest progenitor stars of core-collapse
SNe, cf.\ Sect.~\ref{sec:explmech}) and therefore the
explosions have to be triggered artificially. Moreover, 1D
models lack in the generically multi-dimensional effects 
in the neutrino emission associated with convection, the SASI
and the LESA during the accretion phase that precedes the runaway
acceleration of the shock. Also the transition from the 
accretion phase to the Kelvin-Helmholtz cooling phase of the
remnant cannot be reliably 
represented by 1D models, because such models are unable to 
describe the simultaneous presence of shock expansion and
accretion that continues even after the revival of the stalled
SN shock. This post-runaway accretion phase might last 
for hundreds of milliseconds \cite{Marek:2009a,Mueller:2015b}
and will not only exhibit an
enhanced level of $\nu_e$ and $\bar\nu_e$ emission due to
the persistent accretion contribution to the neutrino
luminosity; it may also show short, spike-like eruptions
caused by bursts of $\nu_e$ and $\bar\nu_e$ production
when particularly massive accretion downflows hit the 
proto-neutron star surface and dissipate their kinetic
energy in shock-heated plasma \cite{Mueller:2014}.

\subsubsection{\em Mixing-length treatment of proto-neutron star
convection}
\label{sec:mixlength}

After accretion has ended, the proto-neutron star is
left behind as a quasi-spherical object (if rotation does not
play an important role). However, as discussed already in 
Sect.~\ref{sec:explmech}, a convective layer eats  
deeper and deeper into the NS and accelerates the
transport of energy and lepton number compared to neutrino
diffusion on its own. In contrast to the highly time-dependent
and aspherical mass flows associated
with convective overturn and the SASI 
in the neutrino-heating layer, the quasi-stationary convection
inside of the proto-neutron star offers the possibility to 
describe the effects of convective energy and lepton-number 
transport in 1D simulations by
a mixing-length approximation (e.g., \cite{Weiss:2004}),
because the time scale of hydrostatic equilibration is much
smaller than the time scale of convective transport,
which again is much smaller than the time scale of evolutionary
changes of the NS.
For the convective lepton-number and energy fluxes one can 
thus write \cite{Huedepohl:2013}:
\begin{eqnarray}
F_\mathrm{conv}^\mathrm{lep} &=& \rho 
v_\mathrm{c}\lambda_\mathrm{c}\,
\frac{\mathrm{d}Y_\mathrm{lep}}{\mathrm{d}r}\,,\label{eq:convlep}\\ 
F_\mathrm{conv}^\mathrm{erg} &=& \rho
v_\mathrm{c}\lambda_\mathrm{c}\,
\left(\frac{\mathrm{d}\varepsilon}{\mathrm{d}r} +
P\frac{\mathrm{d}(\rho^{-1})}{\mathrm{d}r}\right)\,,\label{eq:converg}
\end{eqnarray}
where $r$ is the radius, $\rho$ the baryonic-mass density, $P$ the
pressure and $\varepsilon$ the specific internal energy. In the 
neutrino-trapping regime at densities above the neutrinospheres,
these thermodynamic quantities include
the contributions from neutrinos. Correspondingly, $Y_\mathrm{lep}$
is the total electron-lepton number including electron neutrinos
and antineutrinos. The mixing length is coupled to the pressure
scale height, 
$\lambda_\mathrm{c} = \zeta P|\mathrm{d}P/\mathrm{d}r|^{-1}$,
where $\zeta$ is a dimensionless free parameter of order
unity. The convective velocity is determined as
$v_\mathrm{c} = \sqrt{2g\Delta\rho\,\rho^{-1}\lambda_\mathrm{c}}$
with $g > 0$ being the local gravitational acceleration, and
$\Delta\rho = \lambda_\mathrm{c}C_\mathrm{Ledoux}$ is the density
contrast between surrounding medium and convective fluid elements
after travelling a distance $\lambda_\mathrm{c}$ when $C_\mathrm{Ledoux}$ 
is the Ledoux-convection criterion given by
\begin{eqnarray}
\protect\label{eq:ledoux}
C_\mathrm{Ledoux} = \frac{\rho}{g}\,\omega_\mathrm{BV}^2 
&=& \left(\frac{\partial\rho}{\partial s}\right)_{\!\! P,Y_\mathrm{lep}}
\frac{\mathrm{d}s}{\mathrm{d}r} +
\left(\frac{\partial\rho}{\partial Y_\mathrm{lep}}\right)_{\!\! P,s}
\frac{\mathrm{d}Y_\mathrm{lep}}{\mathrm{d}r} \\ \nonumber
&=& \frac{\mathrm{d}\rho}{\mathrm{d}r} -
\frac{1}{c_\mathrm{s}^2}\frac{\mathrm{d}P}{\mathrm{d}r} \\ \nonumber
&=& -\,\frac{\rho}{\Gamma_\rho}\left(\Gamma_s\frac{\mathrm{d}\ln s}{\mathrm{d}r}
+ \Gamma_{Y_\mathrm{lep}}\frac{\mathrm{d}\ln Y_\mathrm{lep}}{\mathrm{d}r}
\right)\,.
\end{eqnarray}
Here, $s$ is the entropy per baryon, $c_\mathrm{s} = 
\sqrt{(\partial P/\partial\rho)_{s,Y_\mathrm{lep}}}$ is the adiabatic sound
speed, and the thermodynamic $\Gamma$-indices are given by the
derivatives
\begin{equation}
\Gamma_\rho =\left(\frac{\partial\ln P}{\partial\ln\rho}\right)_{s,Y_\mathrm{lep}}\,,
\quad\quad
\Gamma_s =\left(\frac{\partial\ln P}{\partial\ln s}\right)_{\rho,Y_\mathrm{lep}}\,,
\quad\quad
\Gamma_{Y_\mathrm{lep}}=\left(\frac{\partial\ln P}{\partial\ln Y_\mathrm{lep}}\right)_{\rho,s}\,.
\label{eq:gammas}
\end{equation}
Instability to Ledoux convection requires $C_\mathrm{Ledoux}>0$,
in which case the Brunt-V\"ais\"al\"a frequency $\omega_\mathrm{BV}$ is
real and $\omega_\mathrm{BV} = 
\mathrm{sign}(C_\mathrm{Ledoux})\sqrt{g\rho^{-1}|C_\mathrm{Ledoux}|}>0$ 
denotes the growth rate of convective perturbations
in the linear regime. In the numerical simulations a value of $\zeta=1$
is used, but tests showed no noticeable changes for values between
1 and 10. 

Figure~\ref{conv-tests-z9.6} testifies the good agreement that can be
achieved between 1D simulations utilizing this mixing-length treatment
of convection and corresponding 2D calculations. Results of a 
proto-neutron star formed in the collapse of
a 9.6\,$M_\odot$ progenitor are displayed. The compact remnant 
assembles a baryonic mass of about 1.363\,$M_\odot$,
corresponding to a final gravitational mass of $\sim$1.252\,$M_\odot$
for the nuclear EoS used in the simulation. Proto-neutron star
convection sets in typically 30--50\,ms after bounce, which is
the time when differences between convective and non-convective
models become visible first. The NS radius shows the usual
increase associated with interior convection, and the radiated neutrino 
luminosities and mean energies reveal the well known time and 
neutrino-species dependent relative changes compared to the 1D,
non-convective case (for a detailed discussion of the pre-explosion
effects, see Ref.~\cite{Buras:2006b}; for post-explosion differences,
see Refs.~\cite{Burrows:1986,Keil:1996,Janka:2001}).
The neutrino luminosities during the
displayed phase of the first 1.3\,s after bounce are enhanced 
by convective energy and lepton-number transport by
up to about 20\%, and the mean neutrino energies are initially reduced
and at later times increased by up to nearly 2\,MeV with the largest
effects for heavy-lepton neutrinos. The 2D model
and the 1D simulation with convection exhibit basically identical
time evolutions of the NS radius and neutrino parameters
(right panels of Fig.~\ref{conv-tests-z9.6}).

\subsubsection{\em Neutrino-emission phases and characteristics}

The effects of post-bounce accretion on the nascent NS
are not very prominent in the neutrino signal emitted from 
SNe of low-mass progenitors like the 9.6\,$M_\odot$ star 
of Fig.~\ref{conv-tests-z9.6}. In these events the 
shock moves outward essentially in a continuous expansion and
the explosion sets in early after bounce. In the 9.6\,$M_\odot$
star this happens already
at $\sim$130\,ms in the convective models and at roughly
300\,ms p.b. in the non-convective case
\cite{Melson:2015a}. The accretion phase is therefore
very short in such a low-mass progenitor, and the mass delivered 
on the accreting NS during this time interval is 
relatively small. 

The influence of longer and stronger accretion and of
convection in the proto-neutron star interior on the neutrino 
emission of a more massive compact remnant 
can be seen in Fig.~\ref{neutrinos-convection-s27}.
The NS there is born in the collapse of a
27\,$M_\odot$ progenitor and has a baryonic mass of
1.776\,$M_\odot$ and a final gravitational mass of 
$\sim$1.592\,$M_\odot$. Note that in contrast to the 9.6\,$M_\odot$ 
model the explosion of the 27\,$M_\odot$ star cannot be obtained 
self-consistently in 1D but requires artificial initiation. This
was done at 0.5\,s after bounce in both of the models shown in 
Fig.~\ref{neutrinos-convection-s27} by applying slightly different
methods. This is the reason for the differences in the neutrino
emission between 0.5\,s and $\sim$0.8\,s after bounce. The
features seen during this time and the differences between 
both models should not be interpreted by physics.

Figure~\ref{neutrinos-convection-s27}
displays the neutrino signal properties for the three main
phases of neutrino production after core bounce, which are
\begin{itemize}
\item[(a)] 
the high-luminosity $\nu_e$ burst at shock breakout from the 
neutrino-trapping regime,
\item[(b)]
the subsequent accretion phase of the stalled shock until
the beginning of the explosion,
\item[(c)]
the long-time Kelvin-Helmholtz cooling of the newly formed
NS on its way to a final, deleptonized and cold
compact remnant.
\end{itemize}
The neutronization burst radiates away the electron neutrinos
that are abundantly produced by rapid electron captures on 
protons when the hot postshock matter becomes transparent
to neutrinos. It exhibits universal properties concerning
its width (several milliseconds) and height (around
$4\times 10^{53}$\,erg\,s$^{-1} = 400\,\mathrm{bethe\,s}^{-1}
= 400$\,B\,s$^{-1}$), which depend only
weakly on the progenitor star and uncertain nuclear physics
in the SN core (\cite{Kachelriess:2004ds}; 
cf.\ Fig.~\ref{rise} in Sec.~\ref{sec:oscillations}).
Also the faster rise of the $\nu_x$ luminosities compared
to the slower increase of the $\bar\nu_e$ emission in the
first 10--20\,ms after bounce is a generic feature of 
state-of-the-art neutrino transport simulations. It
can be understood by the high electron and $\nu_e$ 
degeneracy in the postshock material right after shock
breakout, which suppresses the rapid production of 
$\bar\nu_e$. These neutrino-emission features,
which are predicted to be independent of the progenitor
star, offer interesting possibilities for
neutrino diagnostics and neutrino astronomy (see
Sec.~\ref{sec:oscillations}).
Differences of the neutrino emission from different
collapsing stars grow as time after core bounce 
progresses. This is a consequence of the gradual
infall of the stellar shells that surround the 
initially collapsing inner core and whose density profiles
vary greatly between progenitors of different masses.

In contrast to the 9.6\,$M_\odot$ model of 
Fig.~\ref{conv-tests-z9.6}, the 27\,$M_\odot$ star shows
evidence for prompt post-bounce convection, which causes
small differences in the neutrino luminosities and energies
even during the first $\sim$50\,ms after bounce. Towards
the end of this phase the convective activity interior to the 
neutrinosphere begins to develop. Since the explosion in
1D simulations of such a massive star has to be triggered
artificially, the duration of the accretion phase is uncertain
(in fact, it is prescribed by the modeler),
and the transition from the accretion to the cooling phase is
not reliable. Moreover, neither the accretion phase nor the
transition to the cooling phase can exhibit special features 
of the neutrino emission associated with multi-dimensional
flows in the accretion region between the stalled shock and
the NS (e.g., convective overturn, SASI, or sporadic,
post-explosion accretion episodes), which were discussed in 
Sect.~\ref{sec:explmech} and at the beginning of 
this section. Nevertheless, the neutrino signal
of 1D models still reflects some generic properties of the
accretion phase. These include the following aspects:
\begin{itemize}
\item
The luminosities of heavy-lepton neutrinos can rather
well be expressed in terms of a Stefan-Boltzmann relation,
\begin{equation}
L_{\nu_x} = 4\pi\phi\sigma_{\nu}T_{\nu}^4 R_\nu^2 \,,
\label{eq:SBnux}
\end{equation}
where $\sigma_{\nu}=4.751\times 10^{35}$\,erg\,MeV$^{-4}$cm$^{-2}$s$^{-1}$.
The luminosity, $L_{\nu_x}$, effective spectral temperature,
$T_\nu$ (measured in MeV), and neutrinospheric
radius, $R_\nu$, are measured at infinity, and the radiated
neutrino spectrum is assumed to be represented by a Maxwell-Boltzmann
distribution, in which case $\left<E_\nu\right> = 3T_\nu$. 
The ``greyness factor'' $\phi$ is determined by numerical 
simulations to range between $\sim$0.4 and $\sim$0.8 
\cite{Huedepohl:2009wh,Mueller:2014}.
\item
The luminosities of $\nu_e$ and $\bar\nu_e$ are enhanced by
an accretion component that strongly depends on the progenitor
star through the mass-accretion rate of the infalling stellar
core and the mass and radius of the assembling proto-neutron
star (see Fig.~\ref{rise}). The sum of $\nu_e$ and $\bar\nu_e$ 
luminosities can be expressed as
\begin{equation}
L_{\nu_e} + L_{\bar\nu_e} = 2\beta_1 L_{\nu_x} +
\beta_2\,\frac{G M_\mathrm{ns}\dot M}{R_\mathrm{ns}} \,.
\label{eq:acclum}
\end{equation}
The first term on the r.h.s.\ represents the ``core component''
of the luminosity carried by neutrinos diffusing out from the
high-density inner region of the proto-neutron star. This 
component can be assumed to be similar to the heavy-lepton
neutrino luminosity, because the core radiates neutrinos in
roughly equal numbers from a reservoir in thermal equilibrium,
which is confirmed by the close similarity of the luminosities
of all neutrino species found after the end of accretion.
The second term on the r.h.s.\ of Eq.~(\ref{eq:acclum}) stands for 
the accretion component of neutrinos lost from the hot,
inflated accretion mantle around the neutrinosphere. It can
be approximately represented by the product of mass accretion
rate, $\dot M$, and Newtonian surface gravitational potential
of the NS, $\Phi = GM_\mathrm{ns}/R_\mathrm{ns}$,
with neutron-star mass $M_\mathrm{ns}$ and radius $R_\mathrm{ns}$.
By a least-squares fit to a large set of 1D results for the 
post-bounce accretion phase of different progenitor stars,
the values $\beta_1\approx 1.25$ and $\beta_2\approx 0.5$ were
deduced in Ref.~\cite{Huedepohl:2013}.
These values apply later than about 150\,ms p.b., when
the postshock accretion layer has settled into a quasi-steady
state, and they depend only weakly on the nuclear EoS.
In Ref.~\cite{Mueller:2014} the authors used a form similar to 
Eq.~(\ref{eq:acclum}) with $\beta_1 = 1$; they
found $\beta_2\approx 0.5$--1 for the pre-explosion 
phase of the 2D models they investigated.
\item
During the accretion phase the mean energies of all neutrino
species show an overall trend of increase, which is
typically steeper for $\nu_e$ and $\bar\nu_e$ than for $\nu_x$.
Interestingly, for progenitors more massive than 
$\sim$10\,$M_\odot$, which possess extended accretion phases, this
leads to a violation of the canonical hierarchy of the mean 
neutrino energies (defined as ratios of energy to number
fluxes), i.e., one observes a moment when the usual order
$\left<E_{\nu_e}\right> < \left<E_{\bar\nu_e}\right>
< \left<E_{\nu_x}\right>$ changes to
$\left<E_{\nu_e}\right> < \left<E_{\nu_x}\right>
< \left<E_{\bar\nu_e}\right>$ (see Ref.~\cite{Marek:2009b}).
The reason for this behavior is a local temperature maximum somewhat
inside of the neutrinospheres of $\nu_e$ and $\bar\nu_e$, which 
forms because of the compressional heating of the growing accretion 
layer, whose cooling becomes ineffective when the radius of the
proto-neutron star shrinks and the density profile becomes steeper
\cite{Huedepohl:2013,Mueller:2014}. 
The hierarchy inversion is enhanced and shifted to a considerably
earlier time when energy transfer in neutrino-nucleon scattering
is taken into account. Non-isoenergetic neutrino-nucleon scattering
reduces the mean energies of $\nu_x$ in a ``high-energy filter''
layer between the $\nu_x$ number sphere and the $\nu_x$ scattering
sphere \cite{Raffelt:2001,Keil:2002in}. The corresponding
energy transfer to the stellar medium also raises the luminosities
and mean energies of $\nu_e$ and $\bar\nu_e$. 
Different from the mean energies, $\left<E_\nu\right>$, 
the rms energies, $\sqrt{\left<E_\nu^2\right>}$, always
obey the canonical hierarchy.
\item
The secular rise of the mean energies of the radiated 
$\nu_e$ and $\bar\nu_e$ during the accretion phase is rather
well captured by the proportionality $\left<E_\nu\right>
\propto M_\mathrm{ns}(t)$ between the mean energies and the growing
mass of the proto-neutron star. The proportionality constant
depends on the neutrino type but
is only slightly progenitor-dependent \cite{Mueller:2014}.
\end{itemize}

During the accretion phase, convection inside of the massive
NS of Fig.~\ref{neutrinos-convection-s27}
causes differences compared to the non-convective case
similar to those already mentioned for the low-mass remnant of
a 9.6\,$M_\odot$ star above. It is stressed again that the 
transition from accretion to proto-neutron star cooling
exhibits features that differ between the two models shown
in Fig.~\ref{neutrinos-convection-s27}, because in both cases the
artificial initiation of the explosion was handled slightly
differently. During the Kelvin-Helmholtz cooling the influence
of convection leads to growing effects and prominent differences
in the long-time evolution of the neutrino emission properties.
Convection enhances the luminosities over the first few seconds
and thus reduces the late-time luminosities 
(Fig.~\ref{neutrinos-convection-s27}, right upper panel), because
the NS cooling and deleptonization are accelerated by
convective transport effects; see also Fig.~\ref{conv-s27} for
the faster evolution of the temperature profile in the convective
model. Correspondingly, the mean neutrino energies are 
increased by up to $\sim$1.5\,MeV in the first seconds, followed
by a faster decline because of the cooler NS at late
times (Fig.~\ref{neutrinos-convection-s27}, right middle panel).

\subsubsection{\em Spectral properties of the neutrino emission}

The spectrum of radiated neutrinos is usually somewhat different
from a thermal spectrum. Since neutrino-matter interactions are 
strongly energy dependent, neutrinos of different
energies decouple from the background medium at different radii
with different temperatures of the stellar plasma. Nevertheless,
the emitted neutrino spectrum can still be fitted by a 
Fermi-Dirac distribution, 
$f(E) \propto E^2[1+\exp{(E/T-\eta)}]^{-1}$,
with fit temperature $T$ and effective degeneracy parameter 
$\eta$ \cite{Janka:1989}.
A mathematically more convenient representation was introduced
in Ref.~\cite{Keil:2002in}, who proposed the following
dimensionless form for the energy distribution at a large
distance from the radiating source:
\begin{equation}
f_\alpha(E)=\left(\frac{E}{\langle E\rangle}\right)^\alpha
e^{-(\alpha+1)E/\langle E\rangle} \,,
\label{eq:fitspectrum}
\end{equation}
where 
\begin{equation}
\langle E\rangle = 
\frac{\int_0^\infty\mathrm{d}E\,E f_\alpha(E)}{\int_0^\infty\mathrm{d}E\,f_\alpha(E)}
\label{eq:meane}
\end{equation}
is the average energy\footnote{With $\alpha$ being determined,
a normalization factor $A$ relates the spectrum to the neutrino
luminosity: $L = 4\pi r^2 A\int_0^\infty\mathrm{d}E Ef_\alpha(E)
= 4\pi r^2 A \langle E\rangle \int_0^\infty\mathrm{d}E f_\alpha(E)$.}.
The parameter $\alpha$ represents the amount of spectral
pinching and can be computed from the two lowest energy 
moments of the spectrum, $\langle E\rangle$ and
$\langle E^2\rangle$, by
\begin{equation}
\frac{\langle E^2\rangle}{\langle E\rangle^2}
= \frac{2+\alpha}{1+\alpha} \,.
\label{eq:alpha}
\end{equation}
Higher energy moments $\langle E^\ell\rangle$ for $\ell > 1$
are defined analogue to Eq.~(\ref{eq:meane}) with $E^\ell$
under the integral in the numerator instead of $E$.
Besides its analytic simplicity, this functional form has the
advantage to also allow for the representation of a wider range
of values for the spectral (anti)pinching than a Fermi-Dirac fit.
A Fermi-Dirac spectrum with vanishing degeneracy parameter
($\eta=0$) corresponds to $\alpha\approx 2.3$,
a Maxwell-Boltzmann spectrum to $\alpha=2$, and
$\alpha \gsimeq 2.3$ yields pinched spectra whereas 
$\alpha \lsimeq 2.3$ gives antipinched ones.
Comparing to high-resolution transport results, the authors
of Ref.~\cite{Tamborra:2012ac} showed that these ``$\alpha$-fits''
also reproduce the high-energy tails of the radiated 
neutrino spectra very well.

\begin{figure}[!t]
\centering
\includegraphics[width=1.0\textwidth]{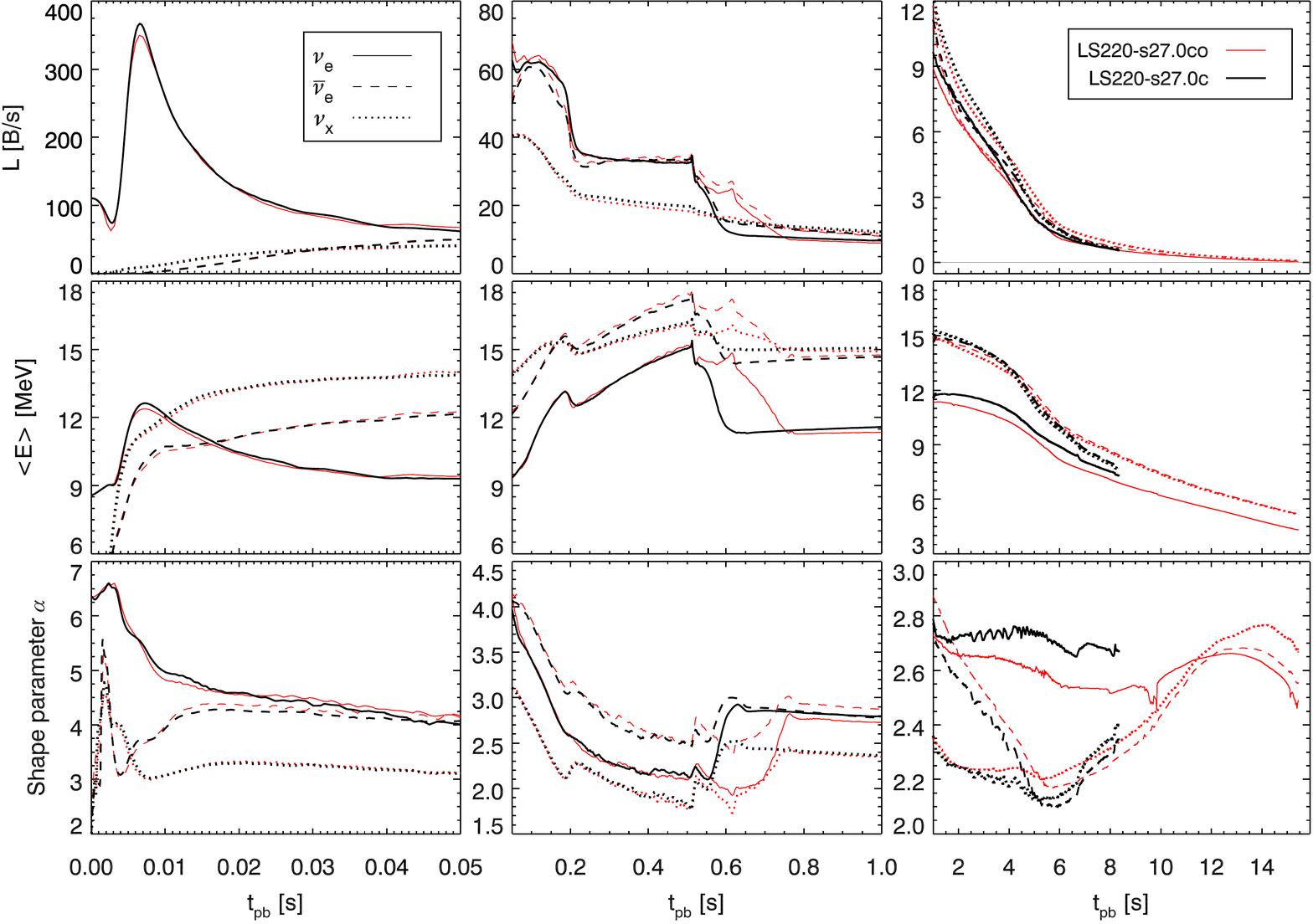}
\caption{Similar to Fig.~\ref{neutrinos-convection-s27} but for
1D collapse and explosion simulations of the 27\,$M_\odot$
star that include nucleon self-energy shifts in the charged-current
neutrino-nucleon interactions in one case (model LS220-s27.0co)
while ignoring these self-energy potentials in the other case (model 
LS220-s27.0c). Both models were performed with a mixing-length
treatment of convection. Note that similar to the models shown in
Fig.~\ref{neutrinos-convection-s27}, also here the explosions
of the two cases were triggered slightly differently, for which
reason the simulations exhibit artificial differences in the
time interval from $\sim$0.5\,s to $\sim$0.8\,s p.b.\ without
any deeper physical meaning and without any significant perturbing
consequences for the subsequent evolution. 
The effects of nucleon self-energy shifts
during the $\nu_e$-burst and accretion phases are small as pointed out
before \cite{Martinez-Pinedo:2012}. During the proto-neutron
star cooling evolution, there are visible differences of up to about
10\% in the neutrino luminosities. The main, persistent
effect of the nucleon mean-field potentials is a reduction of the
average energy of the
emitted $\nu_e$ by up to $\sim$0.7\,MeV. At very late times
($t_\mathrm{pb}\gsimeq 6$\,s) the relative enhancement of the
luminosity of $\nu_x$ for the case with nucleon self-energies
increases and the mean energies of $\bar\nu_e$ and $\nu_x$ become
$\sim$1\,MeV higher than in the other model.}
\label{neutrinos-potential-s27}
\end{figure}

In Fig.~\ref{neutrinos-convection-s27} the shape parameters
$\alpha$ as functions of time are given in the bottom panels.
The evolution is similar in the convective and non-convective
models, with moderate quantitative differences in some phases.
The modest effects of convection are not astonishing, because 
the convective layer is located well inside of the
neutrinospheres and the neutrinospheric layer itself is
convectively stable. The region where the radiated spectrum
is shaped is therefore dominated by neutrino radiative
transport and not by convective transport. In particular 
neutrinos of higher energies, around and beyond the spectral
peak, which have the strongest influence on the higher
energy moments $\langle E^\ell\rangle$ with $\ell\ge 2$,
are still diffusively coupled to the stellar medium outside
of the convective layer. Therefore a major change of the 
shape and time evolution of the emitted spectra due to 
convection cannot be expected.

Figure~\ref{neutrinos-convection-s27} shows that the $\nu_e$
spectrum during the shock breakout burst has a considerable
pinching ($\alpha$ up to nearly 7), but the pinching
decreases with time. The opposite trend occurs for $\bar\nu_e$
during the early post-bounce evolution 
($t_\mathrm{pb}\gsimeq 3$\,ms), while heavy-lepton neutrinos
have spectra closest to the Maxwell-Boltzmann form. 
As the growing and contracting accretion mantle of the
proto-neutron star heats up during the accretion phase,
the pinching of the spectra of all types of neutrinos
decreases. During the subsequent cooling phase, $\nu_e$
and $\nu_x$ exhibit fairly stable shape parameters 
($\alpha\approx 2.7$ for $\nu_e$ and $\alpha\approx 2.2$ 
for $\nu_x$), whereas the spectra of $\bar\nu_e$ become
more similar to those of $\nu_x$ as time goes on. 
In the convective model, different from the non-convective
one, the pinching of the $\bar\nu_e$ and $\nu_x$ spectra
increases again at late times. This reflects the much 
faster progression of cooling in the convective model
(cf.\ Fig.~\ref{conv-s27}), where the late evolution 
towards more pinching of the $\bar\nu_e$ and $\nu_x$ 
spectra occurs earlier than in the non-convective case.

\subsubsection{\em Nucleon self-energy shifts in $\beta$-reactions}

Besides convection in newly formed NSs, the 
EoS of hot NS matter and the mean-field potentials of
the nucleons play an important role for the neutrino emission 
and cooling evolution of the compact remnant. The importance
of the EoS-dependent nucleon self-energies has been pointed 
out in the recent past 
\cite{Roberts:2012a,Martinez-Pinedo:2012,Horowitz:2012,Fischer:2014,Hempel:2015}
after the corresponding physics had been ignored in earlier studies
of proto-neutron star cooling 
\cite{Huedepohl:2009wh,Fischer:2009af,Fischer:2012}
and of proto-neutron star accretion to BH formation
\cite{Fischer:2008rh}.
Considering neutrons and protons as gases of nonrelativistic
quasi-particles that move in a single-particle mean-field 
potential, $U$, one can write their energy-momentum relation
$E_i(\bf{p}_i)$ in generalization of the
non-interacting case as
\begin{equation}
E_i({\bf p}_i) = \frac{{\bf p}_i^2}{2m_i^\ast} + 
m_ic^2 + U_i
\quad \mathrm{for}\quad i = n, p \,,
\label{eq:enmomrel}
\end{equation}
with $m_i$ and $m_i^\ast$ being the particle rest masses and 
the (Landau) effective masses, respectively. Both the masses 
$m_i^\ast$ and
the potentials $U_i$ depend on density, temperature, and the 
neutron-to-proton ratio (or the electron-to-nucleon ratio $Y_e$).
The electron and positron energies of the reactions of 
Eqs.~(\ref{eq:nuabs}) and (\ref{eq:nubarabs}) are therefore 
related with the
energies of their corresponding $\nu_e$ and $\bar\nu_e$ by
\begin{eqnarray}
E_{\nu_e} &=& E_{e^-} + \frac{{\bf p}_p^2}{2m_p^\ast}
- \frac{{\bf p}_n^2}{2m_n^\ast} - (m_n - m_p)c^2 - (U_n - U_p) \,,
\label{eq:nuerg} \\
E_{\bar\nu_e} &=& E_{e^+} + \frac{{\bf p}_n^2}{2m_n^\ast}
- \frac{{\bf p}_p^2}{2m_p^\ast} + (m_n - m_p)c^2 + (U_n - U_p) \,.
\label{eq:nubarerg}
\end{eqnarray}
The potential difference $\Delta U = U_n-U_p$ is directly related 
to the nuclear symmetry energy \cite{Hempel:2015}:
\begin{equation}
\Delta U \cong 4(1-2Y_e) E_\mathrm{sym}^{\mathrm{int},0}(\rho)\,,
\label{eq:deltau}
\end{equation}
where $E_\mathrm{sym}^{\mathrm{int},0}$ is the interaction part
of the symmetry energy at zero temperature, and the r.h.s.\ ignores 
terms of higher than linear order in $Y_e$.
Since $\Delta U$ is positive at the very neutron-rich conditions 
in forming NSs, the contribution from the
mean-field potentials reduces the energy of $\nu_e$ and
increases the energy of $\bar\nu_e$ created in the $\beta$-reactions.
Moreover, the mean-free path for $\nu_e$ absorption on neutrons is 
decreased because the extra energy associated with the potential difference
$\Delta U$ diminishes the importance of Pauli phase-space blocking in the
final state of the electron, whereas the mean-free path for $\bar\nu_e$
is increased at low energies and decreased at high energies.

\begin{figure}[!t]
\centering
\includegraphics[width=1.0\textwidth]{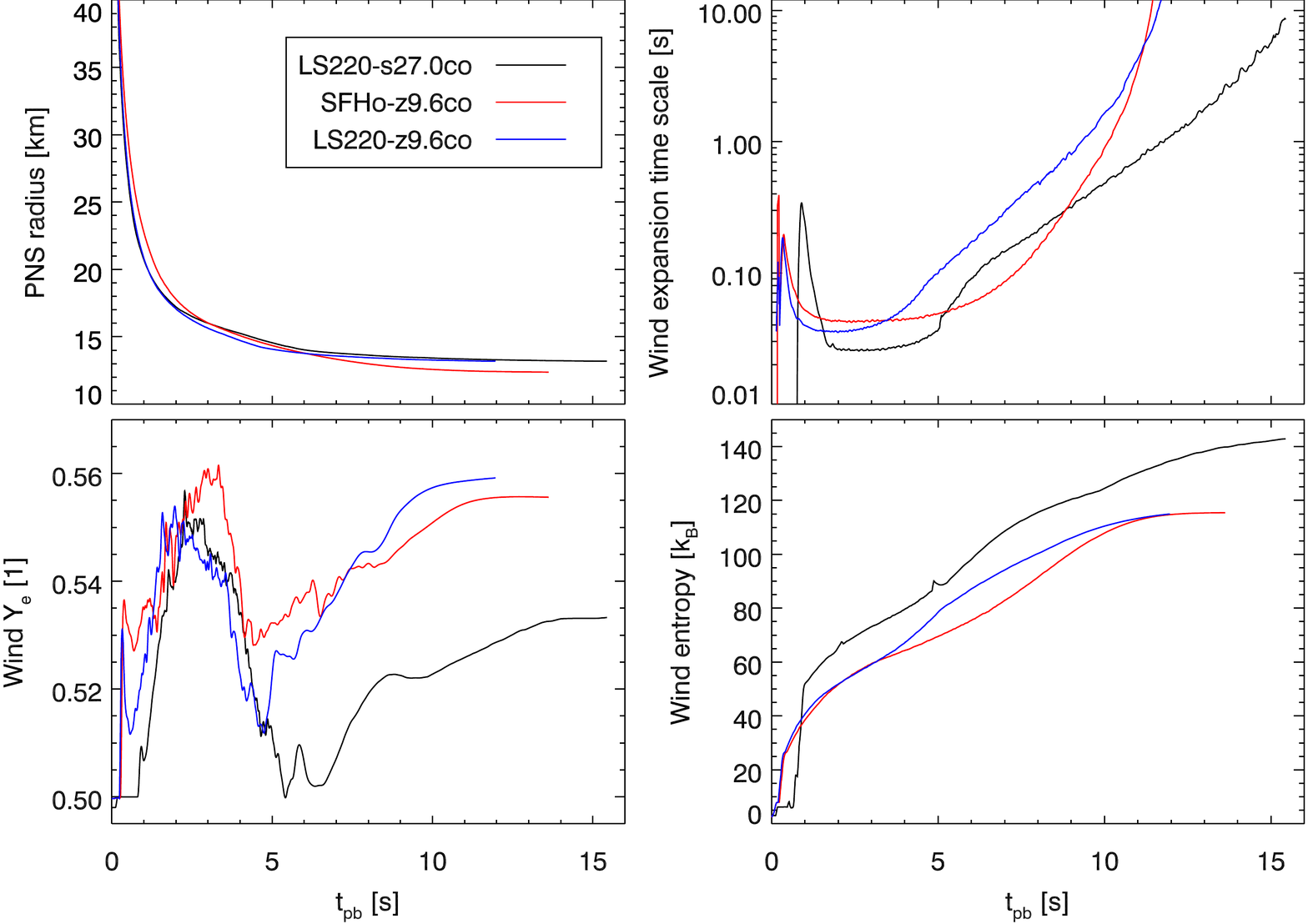}
\caption{Evolution of the proto-neutron star radius ({\em upper left}) and
of characteristic parameters of the neutrino-driven wind: expansion
time scale from $5\times 10^9$\,K to $2\times 10^9$\,K ({\em upper right}),
asymptotic electron fraction, $Y_{e,\mathrm{wind}}$ ({\em bottom left}),
and asymptotic entropy ({\em bottom right}); the last two quantities 
were evaluated at a radius of 1000\,km. The two 1D simulations for the
9.6\,$M_\odot$ progenitor (LS220-z9.6co and SFHo-z9.6co) were performed
with different hadronic EoSs and produced explosions self-consistently,
whereas the explosion of the 27.0\,$M_\odot$ star in 1D was triggered
artificially at 0.5\,s after bounce. All simulations were performed with
a mixing-length treatment of proto-neutron star convection and including
nucleon self-energy shifts in the charged-current neutrino-nucleon reactions.
All wind ejecta are proton-rich.
\label{wind-z9.6+s27}}
\end{figure}

\begin{figure}[!t]
\centering
\includegraphics[width=1.0\textwidth]{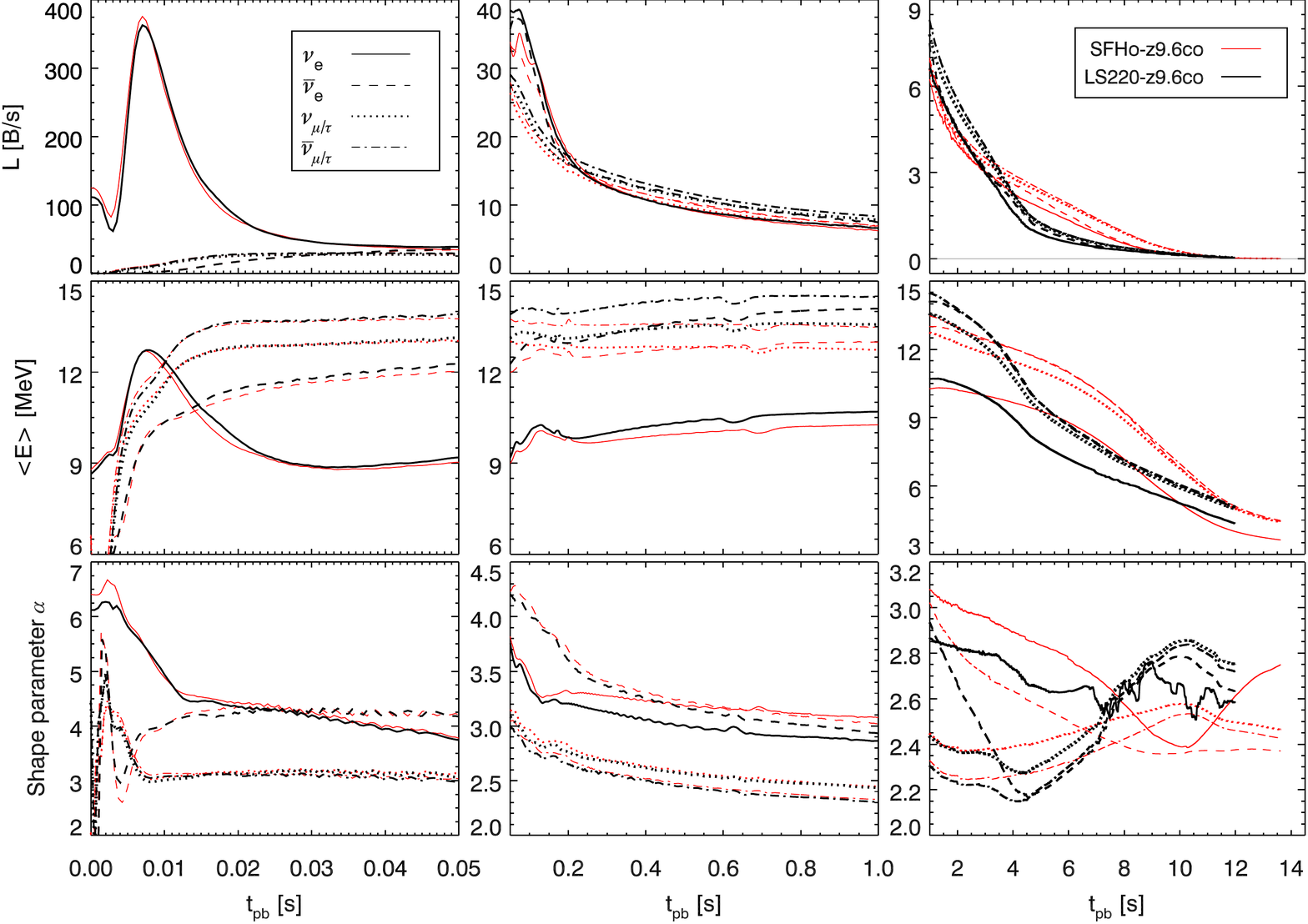}
\caption{Similar to Fig.~\ref{neutrinos-convection-s27} but for
two 1D collapse and explosion simulations of a 9.6\,$M_\odot$
star with two different nuclear EoSs. Model LS220-z9.6co uses the
EoS of Ref.~\cite{Lattimer:1991nc} with a
nuclear incompressibility modulus of $K = 220\,$MeV, whereas
model SFHo-z9.6co employs the SFHo hadronic SN EoS of
Ref.~\cite{Steiner:2013}. The cooling evolution of the
proto-neutron star is clearly different for both EoSs.
Both models were performed with a mixing-length
treatment of convection and in both cases the explosions developed
self-consistently within 170--200\,ms after bounce. Note that
different from Figs.~\ref{conv-tests-z9.6}--\ref{neutrinos-potential-s27},
the neutrino emission properties are shown here for $\nu_e$, $\bar\nu_e$,
one species of $\nu_\mu$ or $\nu_\tau$, and one species of $\bar\nu_\mu$
or $\bar\nu_\tau$, because the neutral-current scattering cross sections
of neutrinos and antineutrinos differ in the sign of the weak-magnetism
corrections \cite{Horowitz:2002}. The slightly lower nucleon-scattering
opacity of $\bar\nu_{\mu,\tau}$ leads to slightly higher
luminosities and up to $\sim$1\,MeV higher mean energies.
In all other simulations discussed in this Section, the four heavy-lepton
neutrinos were treated equally by one representative species $\nu_x$, 
whose emission properties turned out to agree well with the average 
of those of $\nu_{\mu,\tau}$ and $\bar\nu_{\mu,\tau}$ \cite{Bollig:2013}.
\label{neutrinos-z9.6}}
\end{figure}

As a consequence of these changes, the authors of 
Ref.~\cite{Martinez-Pinedo:2012} reported a strong impact of
the nucleon self-energy shifts on the luminosities and spectra
of all neutrinos, but with an important difference between pre-explosion
and post-explosion phases. While they found the nucleon mean-field 
potentials in the charged-current reactions to have a negligible
influence during the accretion phase prior to explosion, they obtained 
reduced luminosities for all neutrino species and a considerable
decrease of the mean energy of the emitted $\nu_e$ during the 
Kelvin-Helmholtz cooling of the proto-neutron star. They also
observed values for the electron fraction in 
the neutrino-driven wind (see Sect.~\ref{sec:neutrinowind})
that were significantly lower (down to $Y_{e,\mathrm{wind}}\sim 0.45$) 
than in the simulation without nucleon mean-field potentials.
Using a more accurate determination of $Y_{e,\mathrm{wind}}$,
Ref.~\cite{Martinez-Pinedo:2014} obtained 
less neutron-rich ejecta with $Y_{e,\mathrm{wind}}\sim 0.48$
despite having replaced the TM1 EoS \cite{Shen:1998gq}
of the previous work by a DD2 EoS table provided by 
M.~Hempel \cite{Fischer:2014}, 
which predicts higher values of $\Delta U$ between
the neutrinospheric layer and the nuclear saturation density,
from where most of the neutrino emission originates. 

Using the STOS TM1 EoS \cite{Shen:1998gq} and the LS220 EoS
\cite{Lattimer:1991nc}, Ref.~\cite{Huedepohl:2013} confirmed
the basic trends of the results of Ref.~\cite{Martinez-Pinedo:2012}.
Similar absolute shifts were obtained for both 
investigated EoSs, although the symmetry energy of the STOS
EoS is larger.
Ref.~\cite{Huedepohl:2013} also found that nucleon self-energy
potentials lead to reduced luminosities for all neutrino
species with the biggest effects for $\nu_e$. While the
average energy of the emitted $\nu_e$ is lowered by roughly
0.75\,MeV compared to the case without self-energy shifts, 
the effect is about one third of that for $\nu_x$ 
and even smaller for $\bar\nu_e$, in contrast to the
slight rise seen in Ref.~\cite{Martinez-Pinedo:2012}\footnote{Our
simulations differ from those of
Ref.~\cite{Martinez-Pinedo:2012} in many aspects of 
the neutrino opacities. In particular, our models take into
account weak magnetism and recoil effects in neutrino-nucleon
interactions, while these relevant corrections were ignored
in the calculations discussed in Ref.~\cite{Martinez-Pinedo:2012} 
(Meng-Ru Wu; private communication).}.
More important, however, is the fact that mixing-length 
convection shrinks the differences between the results with 
and without nucleon self-energies to about half of the 
values obtained without convection. Because convection
speeds up the transport of electron-lepton number out of
the NS, it enhances the $\nu_e$ emission and partly
compensates the effects of the mean-field potentials in the
neutrino opacities, at least as long as the deleptonization
of the proto-neutron star is proceeding.
Figure~\ref{neutrinos-potential-s27} displays two simulations
for our 27\,$M_\odot$ progenitor for comparison. As pointed
out before in Ref.~\cite{Martinez-Pinedo:2012}, the nucleon
self-energy shifts are not relevant during the shock-breakout 
and accretion phases. However, even during the proto-neutron
star cooling the differences are small until very late times,
$t_\mathrm{pb}\gsimeq 6$\,s. The main consequences are a 
persistent reduction of the mean energy of the radiated
$\nu_e$ by at most $\sim$0.7\,MeV and no late-time convergence
of all mean energies as observed in the case without nucleon
self-energies. Also the spectral pinching of the $\nu_e$ is
moderately lowered when the mean-field potentials are taken
into account. 

\begin{figure}[!t]
\centering
\includegraphics[width=1.0\textwidth]{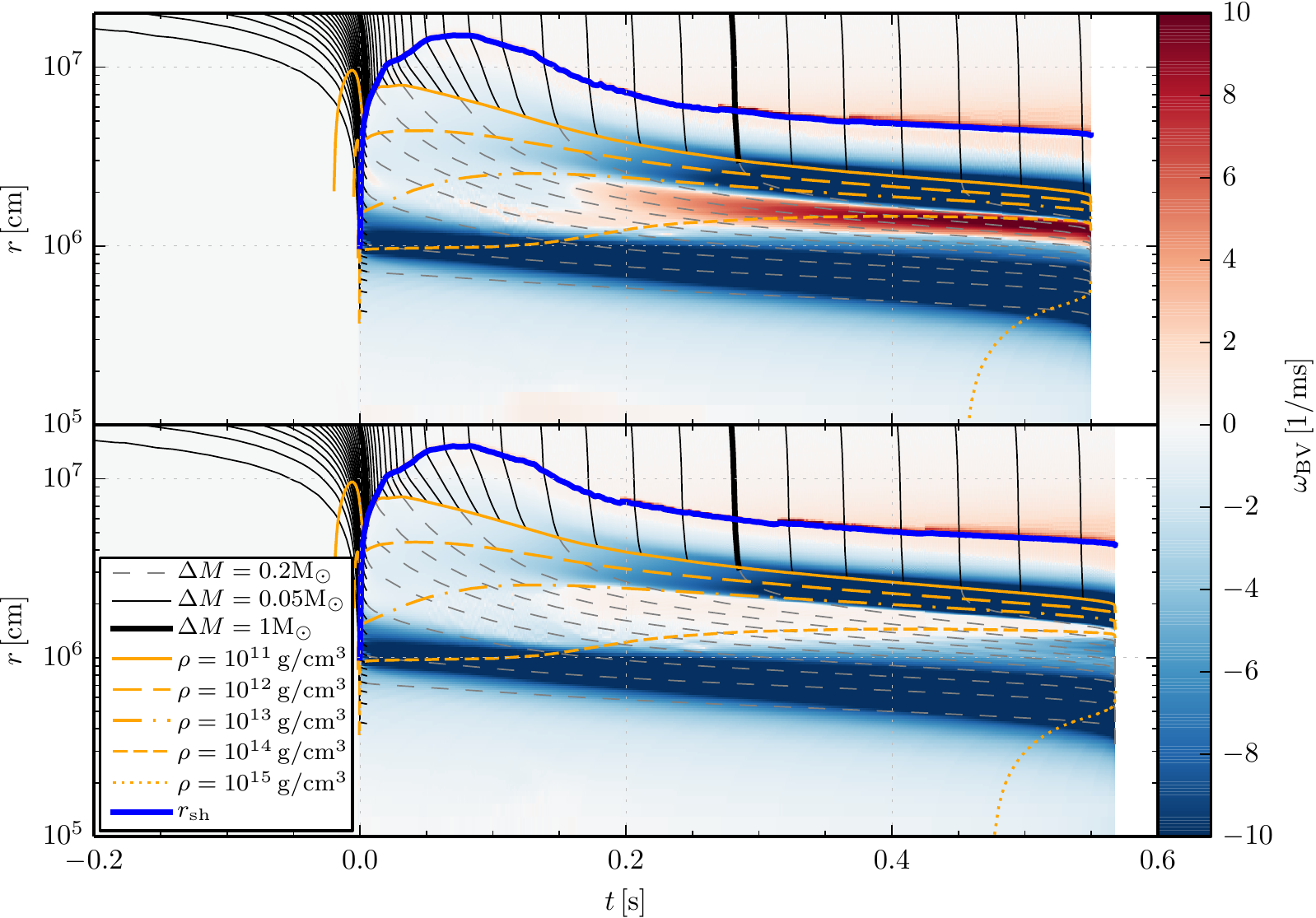}
\caption{Mass-shell plots for the evolution of a collapsing progenitor
star with a birth mass of 40\,$M_\odot$ 
(model s40s7b2 of \cite{Woosley:1995}) towards BH formation
without mixing-length treatment of proto-neutron star convection ({\em upper
panel}) and including such a treatment ({\em lower panel}). The simulations
made use of the LS220 model of the nuclear EOS (\cite{Lattimer:1991nc} with
incompressibility modulus of $K = 220\,$MeV). The color coding shows the
Brunt-V\"ais\"al\"a frequency as defined in Eq.~(\protect\ref{eq:ledoux}).
The different lines are explained in the legend of the plot. The thick blue,
solid line is the stalled and retreating shock, the ingoing solid lines
indicate continuous accretion of infalling matter, the thin dashed lines
the contraction of the proto-neutron star. Convection slightly stretches
the evolution time until the proto-neutron star becomes unstable to BH
formation, because the accelerated deleptonization leads to a faster heating
of the proto-neutron star interior and therefore enhanced thermal support.
\label{conv-s40s7b2}}
\end{figure}

\begin{figure}[!t]
\centering
\includegraphics[width=1.0\textwidth]{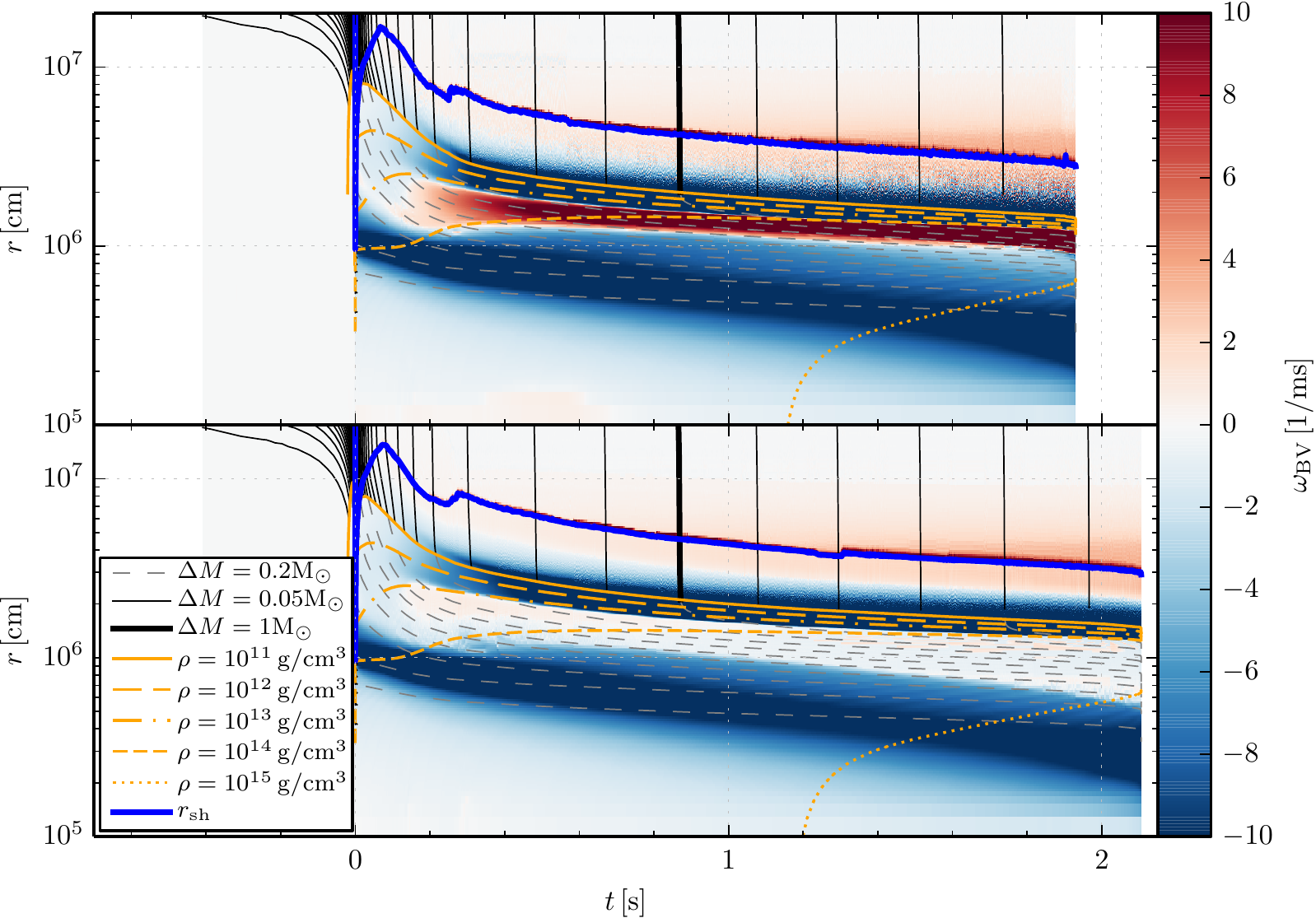}
\caption{Same as Fig.~\ref{conv-s40s7b2}, but for the 40\,$M_\odot$ progenitor
s40.0 of \cite{Woosley:2002zz}, which has a lower mass-accretion rate and
therefore needs a much longer evolution until BH formation.
\label{conv-s40}}
\end{figure}

\subsubsection{\em Neutrino-driven outflows from proto-neutron stars}
\label{sec:neutrinowind}

Neutrinos leaving the hot proto-neutron star deposit energy 
just outside of the neutrinosphere and thus launch a baryonic mass
outflow, the so-called ``neutrino-driven wind'' (e.g., 
\cite{Qian:1996,Thompson:2001} and references therein). This wind
is discussed as a potentially important site of heavy-element
formation (for recent reviews, see 
Refs.~\cite{Arnould:2007,Arcones:2013}). The neutron-to-proton
ratio in the ejected matter is set by $\nu_e$ and $\bar\nu_e$
absorption via the processes of Eq.~(\ref{eq:nubarabs}). If 
neutrons dominate over protons, the wind environment may provide
favorable conditions for the formation of neutron-rich nuclei in
the rapid-neutron capture process (r-process), depending on 
sufficiently high entropies and/or sufficiently short expansion
time scales of the ejecta (cf., e.g., 
\cite{Woosley:1994,Takahashi:1994,Hoffman:1997}). If the outflow
material develops a proton excess, nuclei on the proton-rich
side of the valley of stability might form through the 
neutrino-proton process \cite{Froehlich:2006,Pruet:2006}.

The neutrino-driven winds computed in our models are proton-rich
during the whole proto-neutron star cooling evolution.
Figure~\ref{wind-z9.6+s27} displays
two simulations for the 9.6\,$M_\odot$ progenitor and one for
the 27\,$M_\odot$ star, which bracket the conditions that can be
expected for proto-neutron stars in the range of baryonic 
masses of 1.36--1.8\,$M_\odot$ (final gravitational masses
of 1.25--1.6\,$M_\odot$), which are representative for the far 
majority of core-collapse SNe.
All three displayed simulations include the mixing-length
treatment of convection and the mean-field potentials of nucleons
in the $\beta$-processes. The lower-mass case was computed with 
two different nuclear EoSs, both yielding qualitatively similar
time evolutions of the wind parameters. Instead of an early 
wind phase of about 3\,s with $Y_{e,\mathrm{wind}}<0.5$, which 
was reported as a consequence of the
nucleon self-energy shifts in Ref.~\cite{Martinez-Pinedo:2012},
the simulations shown in Fig.~\ref{wind-z9.6+s27} 
exhibit a prominent maximum of $Y_{e,\mathrm{wind}}>0.5$ during the
early cooling phase. This maximum is a consequence of proto-neutron
star convection, which was not included in the models of 
Ref.~\cite{Martinez-Pinedo:2012}. Correspondingly, the 
maximum is also absent in simulations that we performed without the
mixing-length treatment of convection or for conditions (e.g.,
nuclear EoSs) that yield
only weak convection, in which cases $Y_{e,\mathrm{wind}}$ increases
monotonically during the proto-neutron cooling \cite{Huedepohl:2013}. 
Convective enhancement
of the lepton-number transport inside of the nascent NS
increases the $\nu_e$ number flux relative to that of $\bar\nu_e$
and, in particular, leads to increased pinching of the $\bar\nu_e$
spectrum (see Fig.~\ref{neutrinos-convection-s27}), thus reducing
$\bar\nu_e$ captures on protons compared to $\nu_e$ absorptions on
neutrons in the outflowing wind material. This favors higher 
values of $Y_{e,\mathrm{wind}}$ until the main period of 
deleptonization is over and the proto-neutron star 
convection becomes weaker at around 3--4\,s after bounce (cf.\
Fig.~\ref{conv-s27}), at which time $Y_{e,\mathrm{wind}}$ declines
from a local maximum. Only subsequently, after having gone through
a local minimum on the proton-rich side, $Y_{e,\mathrm{wind}}$ 
resumes the trend found in Ref.~\cite{Martinez-Pinedo:2012}
with increasing values towards later times.
The entropies in the neutrino-driven outflows rise continuously
until they reach maximum values of 120--140\,$k_\mathrm{B}$ per 
nucleon with the higher values for the more massive NS.
Such conditions of entropy and proton excess clearly disfavor the
formation of r-process elements in this environment, 
confirming previous results of less complete models
\cite{Huedepohl:2009wh,Fischer:2009af}.

Active-active as well as active-sterile neutrino-flavor oscillations
modify the neutrino emission from the proto-neutron star and
can thus have an impact on the neutron-to-proton ratio in the 
neutrino-driven wind. This requires that the flavor-changing 
effects occur interior to the neutrinosphere or in the close
vicinity of the nascent NS, i.e., in the region where
neutrino interactions set the electron fraction of the outflowing
matter. While quite a number of studies have
explored the implications of matter-enhanced neutrino conversions
to active (e.g., \cite{Qian:1993}) and sterile channels
(e.g., \cite{McLaughlin:1999,Fetter:2003,Balantekin:2004ug,Wu:2013gxa,Wu:2014kaa,Tamborra:2011is,Pllumbi:2014saa})
in relevant parts
of the space of possibilities for mixing parameters and scenarios, 
the implications of additional neutrino-background induced flavor
conversions are not yet satisfactorily understood. First 
investigations have not been able to reveal promising conditions
for r-processing in the wind ejecta
\cite{Tamborra:2011is,Pllumbi:2014saa}, but more
studies that account for the growing insight into the complex 
physics of collective flavor transformations are urgently
needed (see also Sec.~\ref{sec:oscillations}).

\subsubsection{\em Equation-of-state dependence of the neutrino emission}

The neutrino signal from the proto-neutron star cooling phase
of a future Galactic SN explosion has the potential to
provide us with extremely valuable information about the nuclear
EoS that describes the physics in the NS interior.
Figure~\ref{neutrinos-z9.6} shows interesting EoS dependent
effects of the neutrino signal radiated by the NS
forming in the collapsing and exploding 9.6\,$M_\odot$ progenitor.
Both displayed simulations were carried out with different models
of the high-density EoS above $10^{11}$\,g\,cm$^{-3}$ and both include
a mixing-length treatment of NS convection as well as
the nucleon self-energy effects in the $\beta$-processes. During
the shock-breakout and the post-bounce accretion phase, which in
this low-mass model lasts only until $\sim$200\,ms after bounce,
the differences of the neutrino luminosities and mean energies
are still modest, with slightly higher values for the LS220 EoS.
At later times, however, the differences become considerable
and the simulation with the LS220 EoS yields 10--20\% higher
neutrino luminosities and mean energies until $\sim$3\,s after
bounce. This can be understood as a consequence of the more
rapidly contracting proto-neutron star and the smaller NS
radius over a period
of $\sim$6\,s in the case of the LS220 EoS (see left upper panel
of Fig.~\ref{wind-z9.6+s27}). Also differences in the efficiency
of convective transport in the NS interior play a role,
because the region of convective instability is sensitive to
the properties of the nuclear EoS, in particular to the symmetry
energy of the matter composed of neutrons and protons 
\cite{Roberts:2012b}. At intermediate times during
the cooling evolution,
$3\,\mathrm{s}\lsimeq t_\mathrm{pb}\lsimeq 10$\,s, 
the order of the luminosities
and mean energies for the two investigated EoSs is reversed.
Now the NS computed with the SFHo EoS exhibits 
considerably higher values of the neutrino-emission parameters,
because the remnant with the LS220 EoS has cooled faster and
its temperatures have become correspondingly lower.
As a consequence of the accelerated cooling of the latter model,
the time $t_{\nu,90}$ at which 90\% of the gravitational binding
energy of the cold, compact remnant have been radiated away
is considerably shorter for LS220. The effect is so extreme that 
for our $\sim$1.36\,$M_\odot$ (baryonic mass) NS
computed with the SFHo EoS as well as for our $\sim$1.8\,$M_\odot$ 
(baryonic mass) NS described by the LS220 EoS,
$t_{\nu,90}\approx 6$\,s is roughly the same, although
the total release of gravitational binding energy is 
$198\,\mathrm{B} = 1.98\times 10^{53}$\,erg in the former case and 
$335\,\mathrm{B} = 3.35\times 10^{53}$\,erg in the latter case.

Such differences indicate the paramount importance of the 
high-density EoS model for reliable calculations of the neutrino-cooling
signal of nascent NSs. They also suggest that highly valuable
constraints of the still incompletely understood properties of
matter in the interior of
NSs could be deduced from the measurement of the neutrino 
signal from a future Galactic SN. If the distance to the 
SN is well determined, the energy-loss time scale and
calorimetric arguments can yield useful information. But even
if the SN distance is not known, the characteristic time
structure of the signal \cite{Roberts:2012b} and the
spectral evolution can provide crucial evidence.
This will require most accurate predictions of the SN
neutrino emission (employing consistent microphysics of the EoS
and neutrino opacities) and
the combined information from detections in all existing, upcoming, 
and planned experimental facilities. All together this will allow
one to interpret the detailed features of a high-statistics
measurement of the time structure and spectral evolution of the 
neutrino emission during all phases of stellar collapse and
explosion, despite the complexities of the SN physics and
of various neutrino oscillation phenomena on the way from the 
SN core
to the detectors on earth. Because of the stochasticity of
multi-dimensional hydrodynamics phenomena that assist shock revival
and because of the limited sensitivity of the early-time neutrino
emission to the high-density EoS, the 
post-explosion emission, in particular the late-time neutrino signal,
will be the most reliable probe of the state of matter at nuclear
and supernuclear densities in the deep interior of the NS.

\begin{figure}[!t]
\centering
\includegraphics[width=0.495\textwidth]{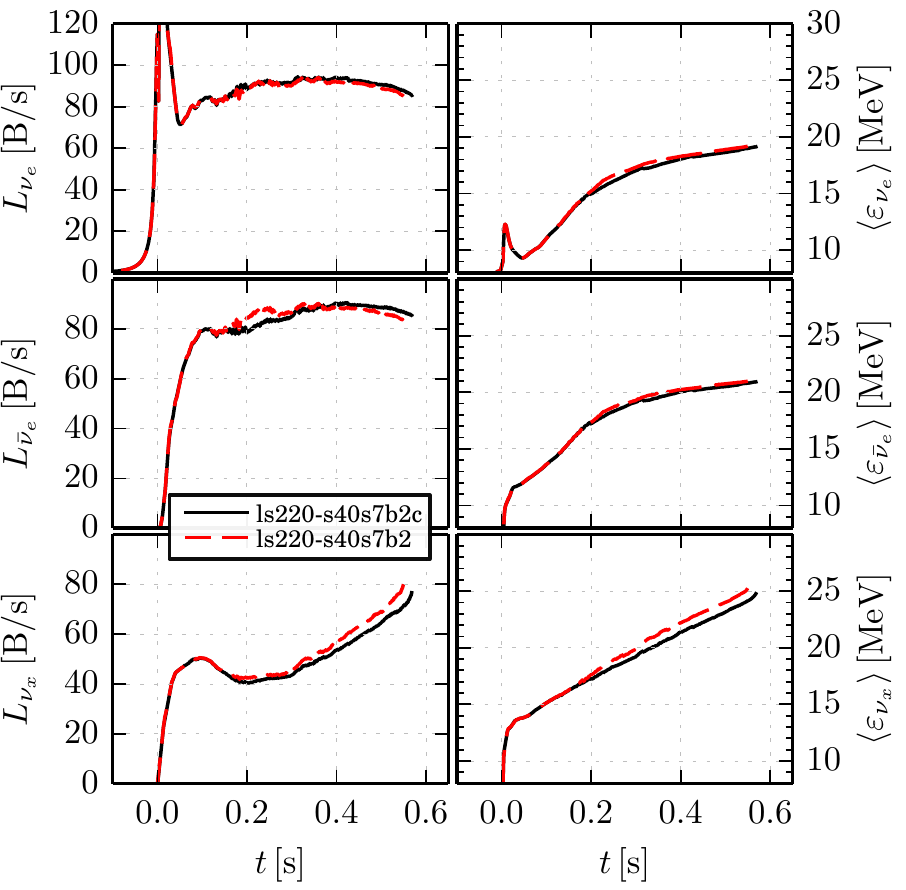}
\includegraphics[width=0.495\textwidth]{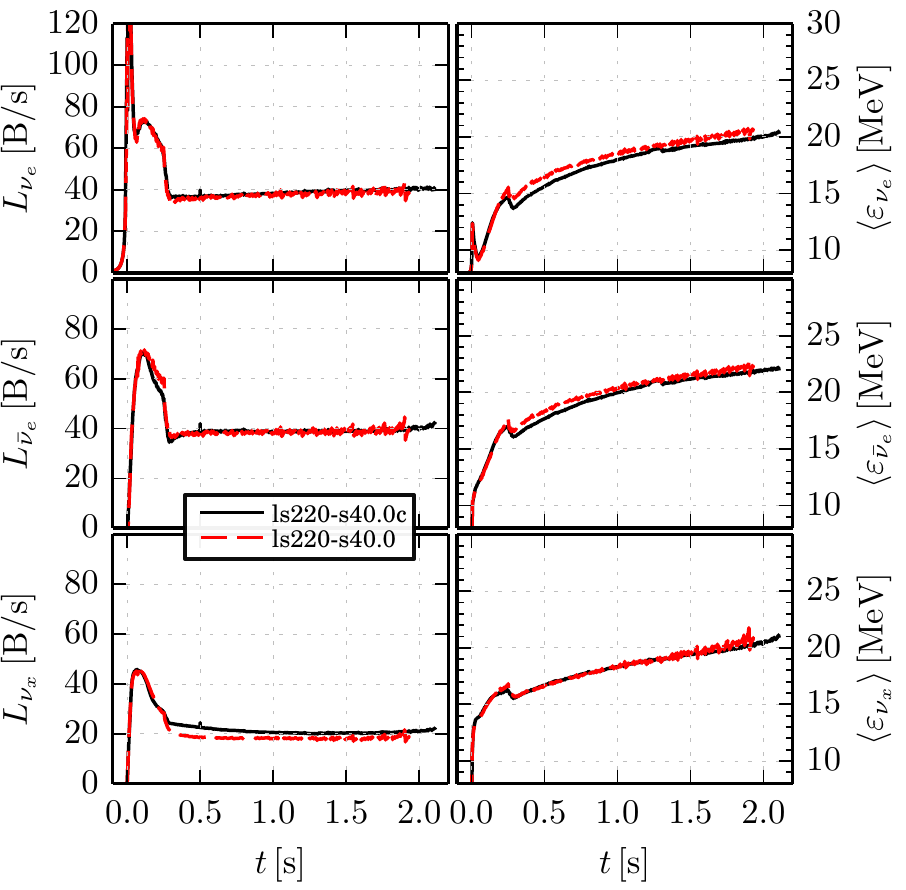}
\caption{Neutrino signals (luminosities and mean energies for an
observer in the laboratory frame) for $\nu_e$ ({\em top}), $\bar\nu_e$ 
({\em middle}), and one species of $\nu_x$ ({\em bottom}) for the 40\,$M_\odot$
BH formation simulations shown in Fig.~\ref{conv-s40s7b2}
({\em left panel}) and Fig.~\ref{conv-s40} ({\em right panel}). 
The results of the
models including mixing-length convection in the proto-neutron star 
(black solid lines) exhibit only relatively small differences compared to the
models without convection (red dashed lines). In the case of the s40s7b2
progenitor ({\em left}) convection mainly reduces the luminosities and
mean energies of heavy-lepton neutrinos, whereas in the s40.0 case 
({\em right}) it reduces the mean energies of $\nu_e$ and $\bar\nu_e$ by
up to $\sim$1\,MeV. 
\label{neutrinos-BH}}
\end{figure}

\subsection{Neutrino signals from black hole formation}
\label{sec:blackholes}

A considerable fraction of all stellar core collapses, possibly 
similarly frequent as SNe, might fail to explode. In these
cases the newly formed NS continues to accrete matter
from the infalling stellar core until it reaches the mass limit for 
BH formation and gravitational instability sets in.
Modern calculations of the neutrino emission of such events,
performed with energy-dependent neutrino transport schemes,
were carried out, e.g., in 
Refs.~\cite{Sumiyoshi:2006id,Sumiyoshi:2007pp,Sumiyoshi:2008zw,Nakazato:2008vj,Nakazato:2011vd,Fischer:2008rh}
for different EoSs of hot 
NS matter. However, these models did not
include convection in the NS and also ignored many aspects
of the neutrino interactions, e.g., did not take into account
neutrino-pair conversion between different flavors \cite{Buras:2003},
energy transfers in neutrino-nucleon interactions, and
nucleon-correlations in the dense medium.

Here we present models of the Garching group \cite{Huedepohl:2013}
that include all of these effects. Figures~\ref{conv-s40s7b2} and
\ref{conv-s40} show mass-shell plots of two 40\,$M_\odot$ progenitors
(s40s7b2 of Ref.~\cite{Woosley:1995} and s40.0 of 
Ref.~\cite{Woosley:2002zz}, respectively), both of which were computed
with the LS220 EoS. The upper panels of the figures display results
of simulations without mixing-length treatment of convection, the
lower ones with convection inside of the proto-neutron stars.
Figure~\ref{neutrinos-BH} provides the time evolution of the 
corresponding neutrino luminosities and mean energies. The 
differences between convective and non-convective models are
relatively small because the convective shell is buried below
a hot, convectively stable accretion layer that is growing in mass 
and whose neutrino production dominates the properties of the 
radiated neutrinos. The models with active proto-neutron star 
convection exhibit slightly lower values of some quantities of 
the neutrino emission. Most remarkably, the convective models
tend to collapse to BHs a bit later than their
counterparts without mixing-length convection, despite the higher
values of the integrated energy and lepton-number loss through
neutrinos. This delay is explained by the stabilizing influence of 
thermal pressure, because the accelerated deleptonization 
associated with convective lepton-number transport leads to a 
faster rise and higher values of the temperature in the deep
interior of the NS.

The two considered 40\,$M_\odot$ stars differ significantly
with respect to the duration of their evolution until BH
formation. While it takes $\sim$0.57\,s after bounce for model 
s40s7b2 to reach this moment, the gravitational instability in
model s40.0 occurs as late as $\sim$2.1\,s after bounce, because
the mass accretion rate of the collapsing stellar core in this
latter case is much lower after a composition-shell interface
has passed the stalled shock at $\sim$0.3\,s after bounce.
Correspondingly, the neutrino luminosities in the s40.0 
run are only roughly half of those in the s40s7b2 model; also the 
mean energies of the radiated neutrinos rise more slowly and 
remain lower in model s40.0 when the same times are compared. 
For heavy-lepton neutrinos
these differences in the time-dependent behavior are most 
prominent. While their luminosity and mean energy grow steeply
in the s40s7b2 simulation, $L_{\nu_x}$ in s40.0 is basically
constant for $t_\mathrm{pb}\gsimeq 0.3$\,s and 
$\langle E_{\nu_x}\rangle$ increases only with a flatter
gradient (Fig.~\ref{neutrinos-BH}). These differences 
reflect the compression and corresponding heating of the rapidly
mass-accumulating accretion layer in model s40s7b2, whereas 
in s40.0 neutrino cooling can well keep pace with the
inflow of energy by the accretion of fresh gas, for which reason
the mantle layers of the newly formed NS in the
latter model remain cooler.

Failed SNe, i.e., ``direct'' BH 
formation events in which the transiently formed NS  is pushed across the BH formation limit by permanent
post-bounce accretion, therefore produce neutrino signals with
distinct features like a continuous hardening of the emitted 
spectra before the emission terminates abruptly.
The two discussed simulations show that the 
detailed properties of the neutrino radiation
depend strongly on the survival time of the NS.
Better predictions of the integrated neutrino emission
of such events, whose contribution might imprint characteristic
structures on the DSNB, will therefore require systematic 
calculations of failed explosions for large model sets of 
progenitor stars, taking into account the growing understanding
of stellar core-collapse cases that are not likely to blow up
as SNe 
\cite{O'Connor:2010tk,Ugliano:2012kq,Horiuchi:2014ska,Pejcha:2015,Ertl:2015rga}.

One should keep in mind, however, that spherically symmetric
simulations (even if they include a mixing-length treatment of 
convection interior to the neutrinosphere) are unlikely to
capture all effects of relevance for the neutrino emission.
When the stalled shock retreats in reaction to the 
contraction of the NS, favorable growth conditions
of the SASI are established and, correspondingly, episodes of
violent SASI sloshing and spiral-mode activity are found in
3D simulations that follow the post-bounce accretion of 
non-exploding massive stars over hundreds of milliseconds
\cite{Hanke:2013,Tamborra:2013,Tamborra:2014b,Couch:2014,Abdikamalov:2014}.
It must be
expected that large-amplitude neutrino-emission variations
caused by SASI modulations of the NS accretion
(cf.\ Sect.~\ref{sec:explmech}) accompany the whole evolution
at least until the collapse of the NS to a BH
sets in. 
Future 3D simulations will also have to clarify the long-time
development of the LESA dipole that was found for the 
neutrinos radiated during the accretion phase 
\cite{Tamborra:2014a}. It is presently
unclear at which level these and other multi-dimensional 
phenomena (e.g.\ connected to stellar core rotation, whose
effects potentially increase on the way to NS
collapse) can affect the predictions of the DSNB from failed 
SNe (see Sect.~\ref{sec:DSNB}). Although most
(maybe all) of the direction dependences and short-time signal
modulations of the observable neutrino emission will average
away in the time and direction-integrated signals that are relevant
for the DSNB, the interpretation of a Galactic BH formation
event will greatly benefit from a precise knowledge of the signal 
structure until the emission is terminated.

\subsection{Conclusions}
\label{sec:SN-conclusions}

In this Section we discussed the current status of modeling
the neutrino emission from stellar core collapse events. 
Hydrodynamical simulations with most sophisticated
neutrino transport are needed to calculate detailed signal
properties, which are necessary for
studies of neutrino oscillations in supernovae,
for investigations of neutrino-dependent and neutrino-induced
nucleosynthesis, and for the exploration of detection
possibilities of the DSNB and of neutrinos from 
a future Galactic supernova.

The neutrino emission from core-collapse supernovae is strongly
affected by multi-dimensional hydrodynamic instabilities inside
and around the newly formed NS. Convective overturn
and the SASI (standing-accretion shock instability) 
in the postshock accretion layer support the initiation
of the explosion and impose large-amplitude time variations and
large-scale directional asymmetries on the neutrino emission
during the accretion phase (Sec.~\ref{sec:explmech}; 
\cite{Tamborra:2013,Tamborra:2014b}). Proto-neutron star convection 
is active interior to the neutrinospheres during this phase as
well as during the post-explosion Kelvin-Helmholtz cooling phase
of the compact remnant.

Recent 3D simulations with detailed neutrino transport have
led to the discovery of the growth of hemispheric differences
in the strength of proto-neutron star convection on a time scale
of about 150\,ms after core bounce (Sec.~\ref{sec:explmech}). 
These hemispheric differences
produce a large dipolar asymmetry of the lepton-number flux with
relative enhancement of the $\nu_e$ emission compared to $\bar\nu_e$
in one hemisphere and increased $\bar\nu_e$ emission on the
opposite side. The hemispheric differences of the lepton-number 
($\nu_e$ minus $\bar\nu_e$) flux normalized to the average can be
factors of a few, corresponding to dipole amplitudes of the
individual emission of $\nu_e$ and $\bar\nu_e$ of 10--20\%,
whereas the $\nu_x$ emission dipole is only of order percent
\cite{Tamborra:2014a}. While this unexpected and
stunning new phenomenon is not yet fully understood and still
needs further exploration and confirmation by independent 
results, the first successful explosions in 3D simulations with 
sophisticated, energy-dependent neutrino transport have become 
available \cite{Takiwaki:2014,Melson:2015a,Melson:2015b,Lentz:2015}.
They provide support for the
viability of the neutrino-driven mechanism for driving and
powering supernova explosions.

During the Kelvin-Helmholtz cooling phase convection interior
to the neutrinospheres accelerates the lepton-number and energy
loss of the nascent NS. The basic effects of this
quasi-stationary convective transport can be well described in
spherically symmetric simulations by a mixing-length treatment,
whose results we showed to compare nicely with those of
direct hydrodynamic modeling of convection in 2D calculations
(Sec.~\ref{sec:protoneutronstars}).
We also presented here the first time-dependent, hydrodynamical
simulations with state-of-the-art spectral neutrino transport,
in which a mixing-length treatment was applied during the 
whole cooling evolution of proto-neutron stars. The calculations
were performed with different nuclear equations of state and 
include the recently intensely discussed effects of nucleon
self-energy shifts in the charged-current $\beta$-reactions of 
free nucleons. 

Convection interior to the neutrinospheres facilitates the faster
deleptonization of the high-density core of the NS 
and thus enhances
the emission of $\nu_e$ relative to $\bar\nu_e$ during the
first seconds, thus leading to more proton-rich conditions in the
neutrino-driven wind. The possibility of a $\nu$-p process 
\cite{Froehlich:2006,Pruet:2006} during the
early phase of proto-neutron star cooling has to be reinvestigated.
However, with the neutrino physics used (which excludes effects
of active or sterile neutrino-flavor oscillations in the hydrodynamic 
modeling), neutron-rich conditions in the wind are not found at any
time during the cooling evolution of compact remnants that bracket 
the mass range of NSs formed in the far majority
of all supernova explosions;
models with NS baryonic (final gravitational) masses 
of $\sim$1.363\,(1.252)\,$M_\odot$ and 1.776 (1.592)\,$M_\odot$
were discussed. Spherically symmetric simulations of proto-neutron
star formation, however, require artificial initiation of the 
supernova explosion. Even in stars near the low-mass end
of supernova progenitors, where multi-dimensional hydrodynamic 
effects are not crucial for the onset of the explosion 
\cite{Kitaura:2006,Melson:2015a}, convective overturn
stirs the postshock accretion layer. One-dimensional models therefore
cannot correctly describe the transition from the accretion phase
to the explosion and proto-neutron star cooling phase, because
simultaneous inflows and outflows coexist in the postshock layer
during this transition.
Two-dimensional explosion models show the ejection of
matter with moderate neutron excess in rapidly expanding plumes
of buoyant plasma during the early expansion phase of the
shock front in particular in low-mass oxygen-neon-magnesium
and iron-core progenitors \cite{Wanajo:2011,Janka:2012b}.
These ejecta might provide the conditions for a weak r-process
with production of neutron-rich nuclei up to mass numbers of
$A\sim 100$--110 \cite{Wanajo:2011}.

Continuing accretion of matter onto the nascent NS in the case of unsuccessful explosions unavoidably leads
to BH formation. Recent insights suggest that these 
failed supernovae might be more
common than previously thought 
(e.g., \cite{O'Connor:2010tk,Horiuchi:2011zz,Kochanek:2013yca,Ugliano:2012kq,Horiuchi:2014ska,Pejcha:2015,Ertl:2015rga})
and might therefore account for a significant contribution to the 
DSNB. Neutrino signals from such events up to the moment when the
NS becomes gravitationally unstable were computed
with energy-dependent neutrino transport before 
\cite{Sumiyoshi:2006id,Sumiyoshi:2007pp,Sumiyoshi:2008zw,Nakazato:2008vj,Nakazato:2011vd,Fischer:2008rh}.
Here we presented state-of-the-art 1D hydrodynamic
simulations that include proto-neutron star
convection by a mixing-length treatment and the full set of 
currently discussed neutrino interactions of relevance in the
core of collapsing stars (Sec.~\ref{sec:blackholes}). While 
convection inside of the neutrinospheres does not have a major
impact on the calculated neutrino emission (except for slightly
stretching the life time of the transiently
existing NS), it must be expected
that violent SASI sloshing and spiral-mode activity induces
large-amplitude modulations on the neutrino luminosities and
spectra radiated by the 
accreting NS all the way until BH formation.
Three-dimensional simulations of the evolution from the
onset of stellar core collapse to the gravitational instability
of the NS will therefore have to be performed also
for obtaining reliable predictions of the neutrino
(and gravitational-wave) signals from failed supernovae in order
to get prepared for the ---not unlikely--- case of a future
Galactic event.

Neutrino data and structural profiles from the Garching simulations 
can be made available upon request for download at 
{\tt www.mpa-garching.mpg.de/ccsnarchive}.
\newpage
\section{Perspectives for Future Supernova Neutrino Detection}\label{sec:detection}
\noindent {\it Authors: K.~Scholberg, A.~Mirizzi, I.~Tamborra}
\\

In this Section, we will consider the
physics of neutrino detection in the regime relevant for SN neutrinos, and discuss the interaction channels for relevant detector types.  We will survey main existing and planned detectors sensitive to SN neutrinos, and touch on some topics related to real-time SN neutrino astronomy and astrophysics.

\subsection{Supernova neutrino detection}

Few detectors are constructed with the primary aim of SN  neutrino detection; most neutrino detectors are built for studies of neutrino oscillation with solar, atmospheric, reactor and beam neutrinos, or for high-energy astrophysical neutrino observation, or for nucleon decay searches.  Happily, however,  many large neutrino detectors --- especially those sited underground --- have excellent capabilities for the capture of a core-collapse SN neutrino burst.

Detectors suitable for SN neutrinos must in general be sensitive to the products of interactions of neutrinos in the few to few tens of MeV range.  Typically the final-state particle energies are also in that energy range, although sometimes can be lower.  Neutrinos interact with detector materials via both charged-current (CC) and neutral-current (NC) channels.  For standard-model CC processes, the resulting lepton type depends on the incoming neutrino flavor, and the lepton charge is negative for a neutrino and positive for an antineutrino.  Supernova neutrino spectra are, however, almost entirely below threshold for CC production of a muon or a tau particle (the thresholds are 110~MeV and 3.5~GeV respectively for quasi-elastic production).  Therefore CC interactions are accessible only for $\nu_e$  and $\bar{\nu}_e$ from core collapse.  The muon and tau components of the flux are accessible only via NC interactions.  

The neutrino interaction products depend also on nature of the target; the observability of the interaction products depends on the nature of the detection technology.  The following subsections describe the different channels, and relevant detector technologies, for the different components of the SN flux. 
Detection of SN neutrinos is reviewed in~\cite{Scholberg:2012id}.  

\subsubsection{\emph{Neutrino interactions with matter}}

The interaction products of SN neutrinos are detected using conventional means, typically via the energy loss of the primary particle out of the interaction, or via energy loss of secondary particles.  
Detectors have different capabilities for energy resolution, angular resolution, time resolution, and particle identification.  They are limited usually by some combination of the intrinsic physics of the particle energy loss, the detection mechanism, and the parameters of the detector technology (i.e., by technical sophistication and cost).

Real detectors are plagued by backgrounds.  For surface or near-surface detectors, cosmogenic backgrounds are a serious issue, typically inducing event rates at or above SN neutrino energies at rates far higher than the rate expected from a burst from all but the nearest SNe.  Cosmogenics include direct cosmic rays (muons being the most penetrating) and secondaries, some of which can be long-lived and produce radioactive decays in the SN event energy range (e.g., ~\cite{Abe:2009aa,Li:2014sea, Li:2015kpa}). Nevertheless, some surface detectors can expect to collect useful SN burst events when triggered by some external observation.  With even modest overburden, cosmogenic rates can be reduced significantly, to the point where the expected background counts within a typical few-tens-of-second burst are negligible.  Cosmogenic background rates depend on overburden, detector size and geometry, and penetrating particle veto capability.  Intrinsic detector background from radioactivity from detector components and nearby materials can also be a problem, especially at low energy.  In the regime below about 5-10 MeV of energy deposition, radioactivity can be dominant.  Radioactive background rates depend sharply on the composition of the nearby environment and the care taken with detector radio-cleanliness.  Finally, instrumental backgrounds (electronic noise, calibration sources, etc.) can sometimes be a problem in practice.  In general, underground detectors designed to measure low-energy neutrinos from steady-state sources such as the Sun, the atmosphere, or reactors, are able to achieve background levels low enough that a Galactic SN burst will have negligible background.  Backgrounds become more of an issue for detection beyond the Milky Way, and for the DSNB (see Sec.~\ref{sec:DSNB}).

Different detector materials have different flavor sensitivities in part by virtue of the relative interaction rates of CC and NC channels in the material.
The ability of a detector to determine flavor content by distinguishing different interactions on an event-by-event basis (``tagging capability'') also affects the information content that will be possible to extract from a burst observation.  
We summarize here the main detection channels for different flavors.

\begin{itemize}
\item{\textit{Electron antineutrinos:}}
The detectors that were running during the SN 1987A neutrino burst, as well as most current detectors, are sensitive primarily to the $\bar{\nu}_e$ component of the neutrino flux.  The reason is that the main detector materials for large underground detectors, water and liquid scintillator (C$_n$H$_{2n}$) are rich in free protons, which have a rather large, and well known, cross section for interaction with $\bar{\nu}_e$ via inverse beta decay (IBD)~\cite{Strumia:2003zx}, $\bar{\nu}_e + p \rightarrow n + e^+$.  This interaction has a threshold of $E_{\nu_{\rm thr}}=1.8$~MeV.  
For typical expected SN neutrino spectra, IBD dominates in water and scintillator.   The IBD interaction products are detected via the energy loss of the positron.  In some cases, the neutron can be detected after moderation and capture (with well-defined delay on the tens or few hundred microsecond scale, depending on materials available for capture).   For example, a neutron capture on a free proton, $n+p \rightarrow d+\gamma$, produces a 2.2-MeV gamma-ray, which can be detected via its Compton-scattering energy loss.   Other nuclei (notably Gd) can also capture neutrons with high cross section, and the subsequent deexcitation gammas can be detected.   If the time-delayed neutron can be detected via capture, it often provides an experimentally-useful tag of the IBD interaction.  It is also possible to observe the energy loss of Compton-scattered 0.511-MeV gamma-rays resulting from annihilation of the positron.

Charged-current $\bar{\nu}_e$ interactions on protons bound in nuclei of common detector materials (carbon, oxygen, argon, iron, lead) can also occur
($\nu_e+(N,Z)\rightarrow (N-1,Z+1) +e^-$), and may result in ejected nucleons and nuclear deexcitation gamma-rays in addition to the positron (e.g.,~\cite{Kolbe:2003ys,Volpe:2000zn}).
However CC $\bar{\nu}_e$ interactions on nuclei are usually subdominant, as the production of a neutron in the final state is often Pauli-suppressed.

\item{\textit{Electron neutrinos:}}
Electron neutrinos interact via  CC channels on neutrons in nuclei (alas, no free-neutron neutrino target exists),  according to
$\bar{\nu}_e+(N,Z)\rightarrow (N+1,Z-1) +e^+$.
This interaction will occur for oxygen in water, and on $^{12}$C in liquid scintillator, and will represent an observable component of a large-statistics sample; however IBD will still be dominant for these detector materials, and it may be difficult to do an event-by-event tagging.
As for CC $\bar{\nu}_e$ interactions, prompt and delayed final-state particles besides the electron may be produced, depending on the nature of the target nucleus.
One of the most promising targets for clean $\nu_e$ detection is liquid argon.  This has a reasonably high-cross-section CC $\nu_e$ channel
$\nu_e + {}^{40}{\rm Ar} \rightarrow e^- + {}^{40}{\rm K}^*$~\cite{Raghavan:1986hv, Bhattacharya:1998hc}.  In a time projection chamber,  the resulting electron track can be reconstructed. Furthermore, the resulting characteristic cascade of gamma rays, for which Compton scatters can be detected,  can serve to tag the channel as $\nu_e$.  This channel should dominate in argon for the expected core-collapse spectrum.

Lead and iron also have a large cross sections (especially lead) for CC $\nu_e$ interaction
\cite{Fuller:1998kb,Volpe:2001gy,Kolbe:2000np,Kolbe:2002gk,Toivanen:2001re,Langanke:1995he}, for which observable products are (in principle) the lepton, but also single or double emitted neutrons (for which there are different thresholds).

\item{\textit{Muon and tau neutrinos:}}
As noted above, in order to access the $\nu_\mu$ and $\nu_\tau$ (and antineutrino) components of the signal, NC sensitivity is required.  Because  NC reactions are flavor-blind, all flavors will be entangled in the observed NC signal; nevertheless, since under many scenarios the mu/tau component is hotter (and furthermore represents two-thirds of the luminosity), the NC signal will have a significant $\nu_\mu$, $\nu_\tau$ component.
NC interaction channels occur in every detector type, although detectors differ in their ability to tag the channel.   The NC interaction results in a scattered nuclear target, and is either elastic or inelastic.  
Inelastic NC scattering on nuclei produces excited final-state nuclei which can deexcite via gamma emission, or ejection of protons or neutrons.  These deexcitation products are the experimental observables.  
While neutrino differential spectral information is lost,  some integral spectral information may be available by counting observed events over threshold. 
In water, resulting protons and neutrons will always be below Cherenkov threshold; gammas may however produce Compton scatters that are observable.   Neutrons may also be captured on protons or dopants such as Gd to produce above-Cherenkov-threshold light.   In scintillator, all ejected products are in principle observable, and neutrons are generally visible via the gamma-rays emitted following their captures. A particularly attractive 15-MeV NC-induced deexcitation gamma (the cross section for which has been measured~\cite{Armbruster:1998gk}) occurs for $^{12}$C.  

In argon, some NC deexcitation channels may be observable~\cite{Hayes}; however this is  currently rather unexplored in the literature. Lead also has prominent neutron-ejection NC channels which can be exploited by detectors.
(Note there is also a NC component to elastic scattering on electrons, which is discussed separately below.)

Elastic NC scattering on protons or nuclei is also a possibility for detectors with low energy thresholds.  Elastic scattering produces simple recoil spectra, the shapes of  which retain source spectral information~\cite{Dasgupta:2011wg}.  Elastic  scattering on protons~\cite{Beacom:2002hs}, which produces hundreds-of-keV recoils,  is observable in scintillator, and coherent elastic neutrino-nucleus scattering (CE$\nu$NS)~\cite{Freedman:1977xn,Drukier:1983gj,Horowitz:2003cz}, will also yield events with recoils in the tens to hundreds of keV scale.  This regime is out of reach for very large detectors, but is attainable by detectors designed to search for weakly-interacting dark matter.  

\item{\textit{Elastic scattering on electrons:}}
A final category of interaction, the simplest physically, is sensitive to all flavors of SN burst neutrinos; this is elastic scattering on electrons (ES). It proceeds via both CC and NC channels~\cite{Marciano:2003eq}.    The detectable signature is the energy loss of scattered electron. The ES cross section is low compared to cross sections on nuclei, typically representing a few percent of a SN burst signal.
This interaction has directional sensitivity--- the electron is scattered with respect to the incoming neutrino direction within an (energy-dependent) angle of about 30 degrees.
Because all materials have electrons, this interaction will occur in all detector types, although detectors vary in the efficiency for observing the electron and in ability to track the scattered electron direction.

\end{itemize}

We note that although the cross sections for the simplest targets --- elastic scattering on electrons and inverse beta decay of $\bar{\nu}_e$ on protons--- are well understood, and have been employed as low-energy detection interactions for decades with e.g., solar and reactor neutrinos, the interactions of neutrinos on heavier nuclei are quite poorly understood.  At present, the only other interactions studied experimentally with better than $\sim$ 10\% precision is $^{12}$C~\cite{Armbruster:1998gk,Auerbach:2002iy}.  The targets of interest for SN neutrino detection have never been irradiated with a well-understood neutrino source to determine their cross sections and the properties of the interaction products (particle types, angular distributions, etc.). Furthermore, there are relatively few theoretical studies in the relevant energy regime.   We note promising opportunities to study these cross sections using a pion decay-at-rest source, such as the Spallation Neutron Source at Oak Ridge National Laboratory in Tennessee, within the next decade~\cite{Bolozdynya:2012xv}.

\subsubsection{\emph{Main detector types and considerations}}

There are several kinds of relevant detector technologies, some currently running and others planned for the future.  The most important  and the main instances of these are described below.

\begin{itemize}

\item \textit{Water Cherenkov (WC):}  These employ water in liquid or ice form as the detector material.   Optical-frequency Cherenkov radiation from particles moving faster than the speed of light in water is collected by photomultiplier tubes.  The Super-Kamiokande  detector, a 50-kton detector in Japan~\cite{Ikeda:2007sa} is the main instance of this class. Among detectors running at the time of this review, Super-Kamiokande will collect the largest number of individually-reconstructed SN neutrino events.   A possible enhancement to Super-Kamiokande currently under investigation is the potential loading with Gd to improve the neutron capture cross section for better tagging of $\bar{\nu}_e$~\cite{Beacom:2003nk} (and $\nu_e$~\cite{Laha:2013hva}). 
The tagging efficiency for $\bar{\nu}_e$ via neutron capture on free protons is about 18\%. A Gd tagging efficiency of at least 67\% has been demonstrated~\cite{Watanabe:2008ru}, but it can likely be improved beyond that. 
We will discuss the impact of Gd tagging in the background reduction for DSNB in Sec.~\ref{sec:DSNB}.
 The next-generation anticipated large WC detector is Hyper-Kamiokande, with half a Mton of mass~\cite{Abe:2011ts}.

An important special case of WC detector for SN detection is the ``long-string'' type.  This type of detector comprises a very large volume of water or ice (part of a natural body), instrumented with strings of photomultiplier tubes.  These kinds of detectors are nominally intended for ultra-high-energy neutrino detection ($>$GeV), and they are too sparsely instrumented to resolve individual SN neutrino interactions.  However, they are still able to observe a SN neutrino burst as a diffuse glow of Cherenkov photons in the ice or water, via coincident increase in count rate of all the photomultiplier tubes within the time window of the burst~\cite{Halzen:1994xe,Halzen:1995ex}.   Although event-by-event energy, angle and flavor information is not available, the photon statistics are high enough to provide good time resolution for the overall time structure of the burst~\cite{Halzen:2009sm}.
Instances of this class are IceCube, which has a SN trigger; in its complete configuration, IceCube
has 5160 optical modules~\cite{Abbasi:2011ss} and about 3 Mton effective detection mass, representing the largest current detector for SN neutrinos. Others, which are noisier~\cite{Ageron:2011nsa}, may still retain information.   Increased photomultiplier tube density, such as that proposed for the PINGU infill~\cite{Aartsen:2014oha}, will improve SN sensitivity by enhancing the probability to for multiple photons per interaction to be recorded.

\item \textit{Liquid scintillator:} These detectors employ organic hydrocarbons in liquid form. Like water, the material has a large fraction of free protons.  Scintillator detectors are therefore also primarily sensitive to IBD and hence to $\bar{\nu}_e$.   Scintillators emit photons in response to energy loss of charged particles. 
Rather large light yields per energy loss ($\sim$50 times higher than light yields for Cherenkov radiation) can be obtained, resulting in good event-by-event energy resolution and potentially low energy thresholds.    Typically neutrons can be detected via capture on free protons with high efficiency;
gadolinium doping is often employed to enhance capture rate and deexcitation gamma energy release.
In order to take advantage of low energy thresholds, scintillator detectors must tackle the difficulties of radioactive contamination which are typically fierce below a few MeV.

The currently-running instances of large scintillator detectors are LVD~\cite{Agafonova:2014leu,Aglietta:1992dy,Agafonova:2007hn}, Borexino~\cite{Cadonati:2000kq,Monzani:2006jg}, and KamLAND~\cite{Eguchi:2002dm}; the SNO+~\cite{Kraus:2010zzb} experiment should turn on in the near future.  
A future planned 20-kton detector is JUNO~\cite{Li:2014qca,An:2015jdp}; another well-developed concept is the 50-kton  LENA detector~\cite{Wurm:2011zn}. 
Surface reactor-neutrino scintillator detectors (all Gd-loaded),  although relatively small (tens of ton scale) and unprotected by significant overburden, also represent a SN burst detection opportunity.  Daya Bay in particular can defeat cosmogenic background by high IBD selection efficiency and by the requirement of coincidence between multiple, spatially-separated detectors~\cite{HanyuWeifortheDayaBay:2013oma}.
Because scintillation light is emitted isotropically,  the direction sensitivity of these detectors is limited (although see Sec.~\ref{sec:directionality}).
Thanks to the low energy threshold, scintillator also has sensitivity to NC elastic scattering of neutrinos on protons.  ``Quenching'' of scintillation light from heavy recoil products suppresses observability of this channel; nevertheless there should be some capability for existing and future detectors.  Furthermore, information about the neutrino spectrum is available via the proton recoil spectrum~\cite{Dasgupta:2011wg}.

\item \textit{Liquid argon (LAr):}  An extremely interesting possibility for the future is liquid argon time projection chamber (LAr TPC) technology.  These detectors drift ionization charge from energy loss of charged particles in an electric field, and collect this charge on a planar anode (via different mechanisms-- either collection or induction wires, or multiplication at a liquid-gas interface).   Two dimensions of the track can be reconstructed from collected-charge projection on the anode plane, and the third dimension is determined from drift at known constant velocity; therefore three-dimensional tracks can be reconstructed.   Products of low-energy SN neutrino interactions can be in general reconstructed as short ($\sim$10-cm scale) lepton tracks. In principle energy loss from Compton scatters from deexcitation and bremsstrahlung gamma rays can also be recorded, and used to tag interaction channels.
The primary interaction from a SN neutrino burst will be $\nu_e$ absorption which gives liquid argon excellent sensitivity to the electron flavor 
component of the flux~\cite{GilBotella:2004bv}.  The detector also has intrinsic directional capability for anisotropic interaction products.
Liquid argon time projection chamber technology has been demonstrated for the Icarus~\cite{Amoruso:2003sw} detector, as well as smaller prototypes such as ArgoNEUT (although with little emphasis on low-energy events).  MicroBooNE, a 200-ton fiducial surface detector at Fermilab~\cite{Soderberg:2009rz}, will be online soon.  The prospects for a large detector, DUNE, underground in the United States are very good~\cite{Adams:2013qkq}.

\item \textit{Lead:} Another possible method of SN neutrino detection is to employ a heavy material, such as lead or iron.  These materials have a high cross section for neutrino interactions in the few tens of MeV range.  In principle, detectors can observe energy loss of leptons, gammas, or other interaction products.  However a potentially inexpensive mode of operation is to exploit the high probability for ejection of neutrons subsequent to a CC or NC neutrino interaction; counting single and double neutrons gives crude spectral information~\cite{Vaananen:2011bf}. The ejected neutrons can be detected with dedicated neutron-detection technology.  There is currently one operating instance of a lead-based SN neutrino detector, the HALO experiment at SNOLAB~\cite{Duba:2008zz}, which incorporates 76 ton of lead in conjunction with leftover $^3$He neutron counters from the SNO experiment.  The HALO-2 detector of kton scale is an envisioned future upgrade.

\item \textit{Dark matter detectors:}  Another category of detector will have some sensitivity to SN neutrinos via CE$\nu$NS, the neutral-current interaction of a neutrino with an entire nucleus~\cite{Horowitz:2003cz}.   The cross section for this process is relatively high, producing a handful of events per ton of detector material for a 10-kpc core collapse.  However recoil energies are very low (few tens to hundreds of keV), making this channel experimentally challenging.  Dark matter direction detection detectors are optimized to observe these low energy recoils and hence will have sensitivity to the total 
SN flux~\cite{Chakraborty:2013zua}.  Current-generation detectors of less than ton scale will observe small statistics, but with potential scale-ups to multi-ton scales, this category of detectors will provide useful information about a SN burst.

\end{itemize}

\subsection{Detector summary and future prospects}\label{sec:detectionprospects}

\begin{figure}[t!]
\begin{center}
\includegraphics[angle=0,width=0.6\textwidth]{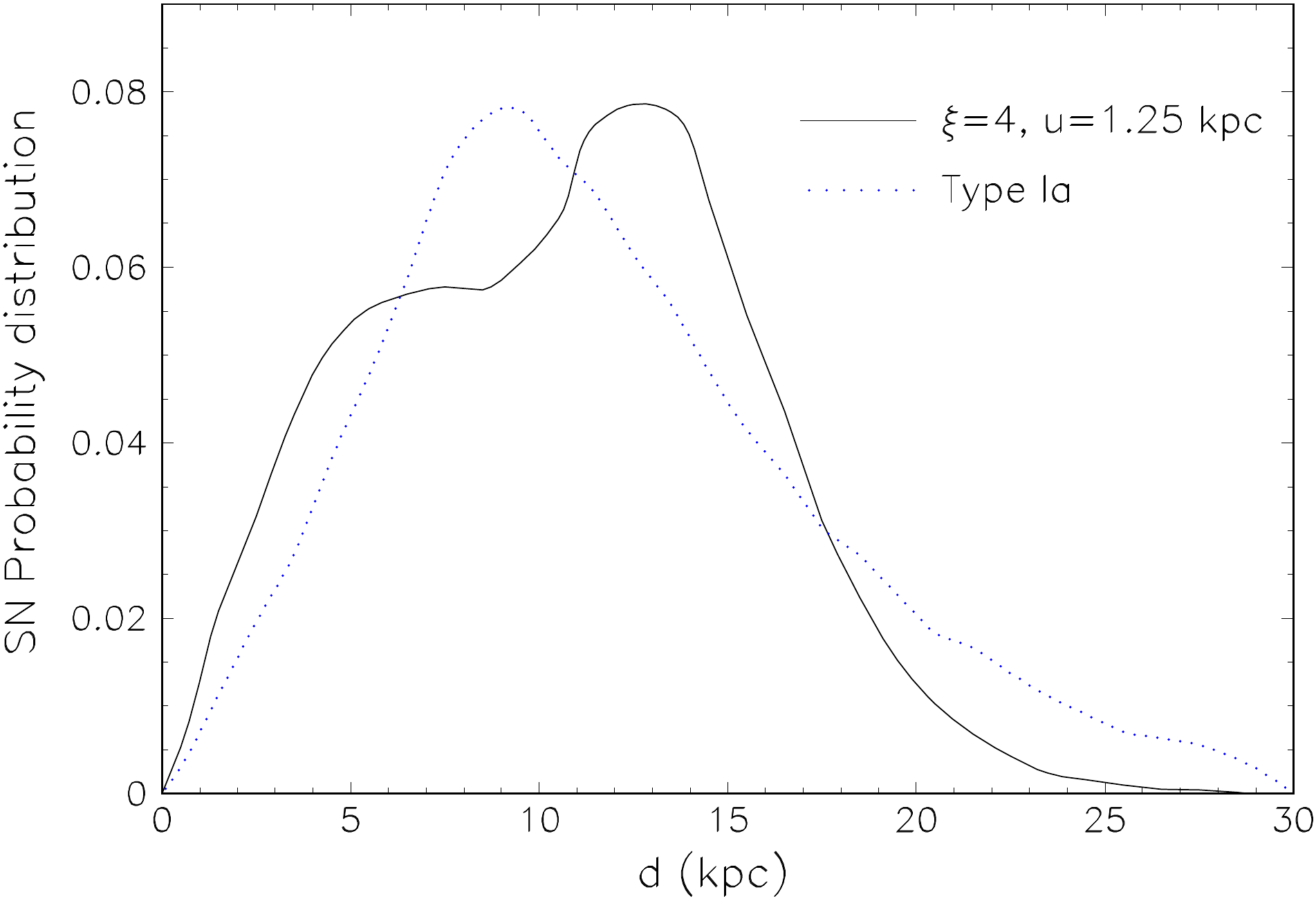} 
    \end{center}
\caption{SN probability vs. distance from the Sun for a simple model of progenitor distribution
(continuous curve).
In comparison also the SN distribution for Type Ia SNe is shown (dotted curve).
(Reprinted figure  from~\cite{Mirizzi:2006xx}; copyright (2003) by the Institute Of Physics Publishing.)
\label{Sndistrib}}
\end{figure}

\begin{figure}
\includegraphics[width=6in]{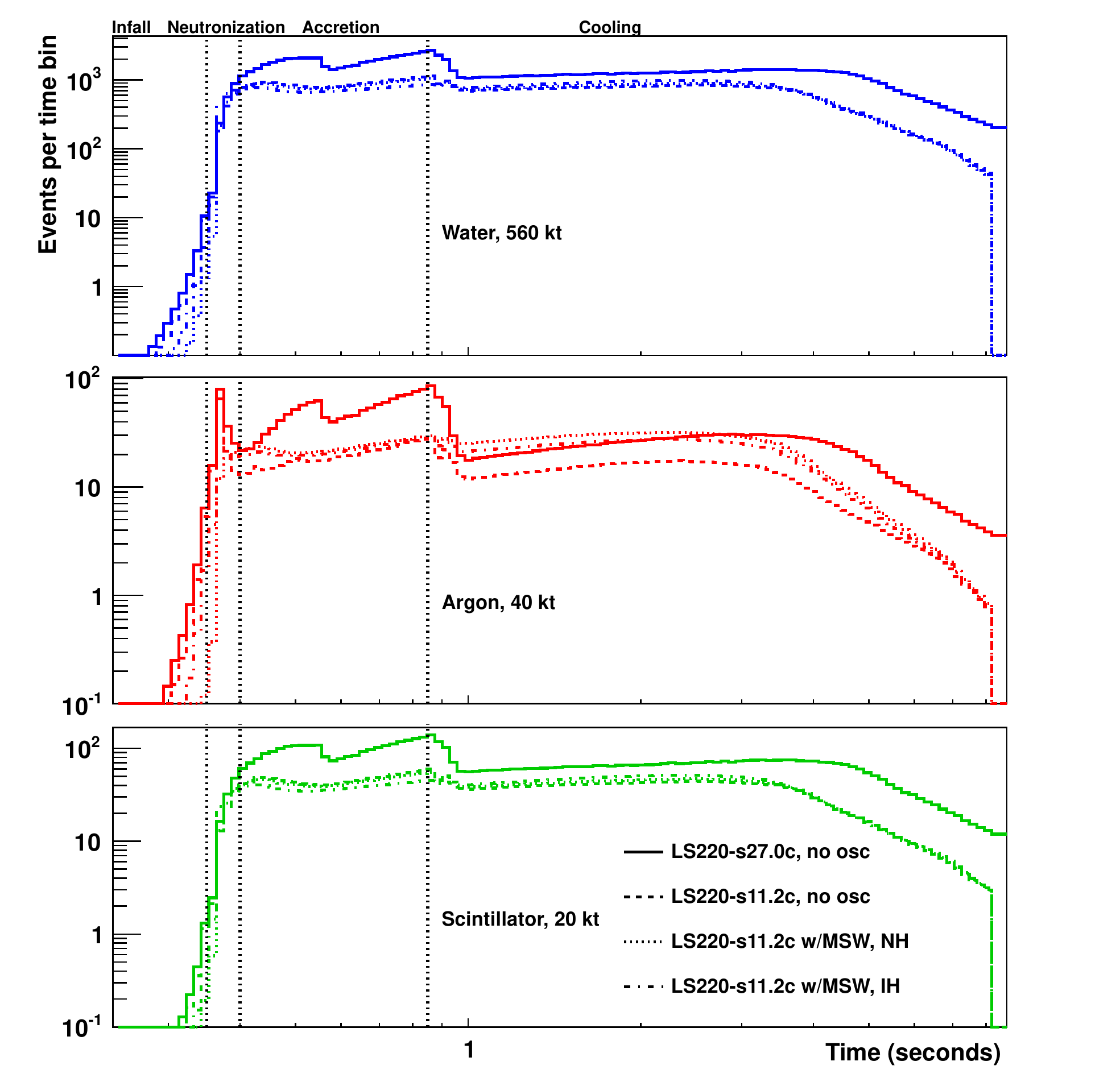}     
\caption{Examples of total event rates per time bin for   $27\ M_\odot$ and $11\ M_\odot$ SN progenitors (see Sec.~\ref{sec:protoneutronstars}), for
water, argon and scintillator detectors, respectively from the top to the bottom panel, assuming a Galactic SN at $d=10$~kpc.   Core bounce is at 0.35 seconds. The dotted (dashed dotted) lines are rates for a simplified MSW oscillation assumption for the normal (inverted) hierarchy case (see Sec.~\ref{sec:signatures}). Note that expected events in non-equal logarithmic time bins are plotted, not rates.}\label{fig:timedep}
\end{figure}

\begin{figure}
\includegraphics[width=5.5in]{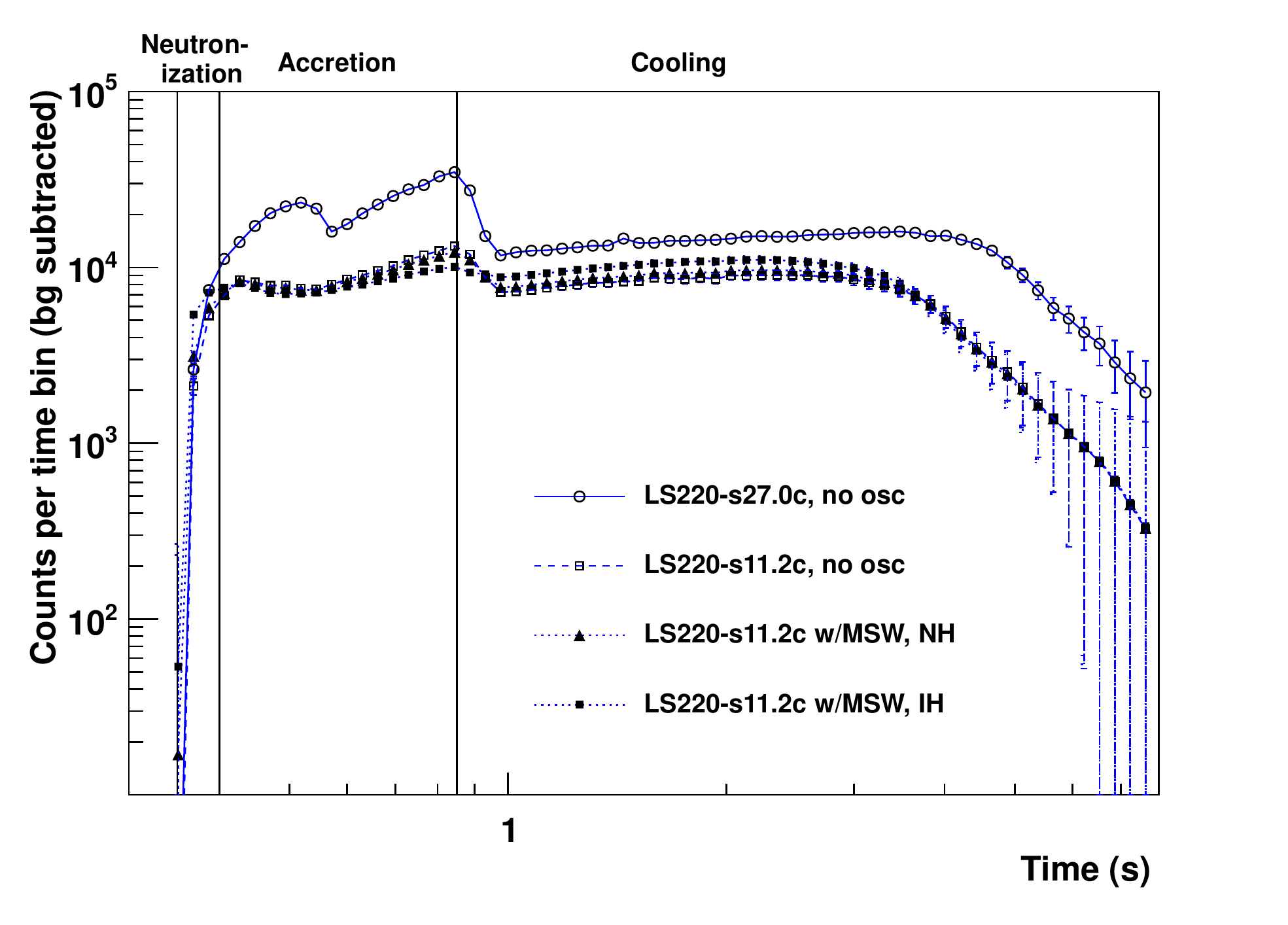}     
\caption{Expected background-subtracted counts per time bin for IceCube, for IBD events (following~\cite{Serpico:2011ir}), for the same models as in Fig.~\ref{fig:timedep}. Error bars are statistical. Again note non-equal logarithmic time bins.} \label{fig:ictimedep}
\end{figure}

The possible distribution of core-collapse SNe in the Galaxy must follow the regions
of star formation, notably in the spiral arms.  The expected distance distribution for a
simple model is shown in Fig.~\ref{Sndistrib}. 
One realizes that 
the distribution is very broad and that 10 kpc is probably a reasonable benchmark
value. However, every distance between 2 and 20~kpc has similar probability. Therefore, one would expect a factor of $\sim$100 for variations
in the predicted neutrino event rate for a future Galactic SN explosion.
\begin{table}[t]
\caption{Current and proposed SN neutrino detectors as of the time of this writing.
Event rate estimates are approximate for 10~kpc; note there may be significant variation by SN model.  The ``Flavors'' column indicates the dominant flavor sensitivity (note that other flavor components may be detectable, with varied tagging quality).
Not included are smaller detectors
(e.g., reactor neutrino scintillator experiments) and detectors sensitive primarily to coherent elastic neutrino-nucleus scattering (e.g., WIMP dark matter search detectors).  
An asterisk indicates a surface detector, which may not be self-triggering due to background.  Numbers in parentheses indicate that individual events will not be reconstructed; see text.
}
\begin{tabular}{@{}ccccccc@{}}%
\hline
Detector&Type &Mass (kt) &Location & Events & Flavors & Status \\ \hline

Super-Kamiokande & H$_2$O& 32 & Japan& 7,000 & $\bar{\nu}_e$ & Running\\    
LVD & C$_n$H$_{2n}$& 1 & Italy& 300 & $\bar{\nu}_e$ & Running\\
KamLAND & C$_n$H$_{2n}$& 1 & Japan& 300& $\bar{\nu}_e$ & Running \\
Borexino& C$_n$H$_{2n}$& 0.3 & Italy& 100 & $\bar{\nu}_e$ & Running \\ 
IceCube & Long string& (600) & South Pole & ($10^6$) & $\bar{\nu}_e$ & Running \\  
Baksan & C$_n$H$_{2n}$  & 0.33 & Russia & 50 & $\bar{\nu}_e$ & Running\\  
MiniBooNE$^*$ &  C$_n$H$_{2n}$ & 0.7 & USA & 200 & $\bar{\nu}_e$ & (Running) \\ 
HALO &  Pb & 0.08 & Canada & 30 & ${\nu}_e, \nu_x$ & Running \\   
Daya Bay &  C$_n$H$_{2n}$ & 0.33 & China & 100 & $\bar{\nu}_e$ & Running \\   
NO$\nu$A$^*$ &  C$_n$H$_{2n}$ & 15  & USA &  4,000& $\bar{\nu}_e$ & Turning on \\ 
SNO+ & C$_n$H$_{2n}$& 0.8 & Canada& 300 & $\bar{\nu}_e$ & Near future \\    
MicroBooNE$^*$ &  Ar & 0.17 & USA & 17 & ${\nu}_e$ & Near future \\ 
DUNE&  Ar & 34 &USA  & 3,000  & ${\nu}_e$ & Proposed \\
Hyper-Kamiokande &  H$_2$O & 560 & Japan & 110,000 & $\bar{\nu}_e$ & Proposed \\ 
JUNO &   C$_n$H$_{2n}$& 20 &  China & 6000 & $\bar{\nu}_e$ & Proposed \\  
RENO-50 &   C$_n$H$_{2n}$& 18 &  Korea & 5400 & $\bar{\nu}_e$ & Proposed\\   
LENA &   C$_n$H$_{2n}$& 50 &  Europe & 15,000 & $\bar{\nu}_e$ & Proposed\\   
PINGU & Long string& (600) & South Pole & (10$^6$) & $\bar{\nu}_e$ & Proposed\\ 
\hline
\end{tabular}

\label{tab:detectors}

\end{table}
For definiteness,  
Table~\ref{tab:detectors} summarizes neutrino event rates for current and future detectors assuming  a typical SN at $d=10$~kpc.
 Despite the scaling of the event rates with $1/d^2$, most of the existing detectors would be massive enough
to assure a  sensitivity 
to a SN event in the whole Galaxy. Moreover, a 0.5 Mton water Cherenkov detector would collect $\sim 10$ events from an
extragalactic SN in nearby galaxies at $d\sim 1$~Mpc, such as M31 and M33.
The prospects for detecting  mini-neutrino bursts from nearby galaxies is quite exciting, since
about one SN per year is expected within distances up to 10~Mpc~\cite{Ando:2005ka,Kistler:2008us}. 
In this regard, the patient accumulation of data on an event-by-event
basis from SN in nearby galaxies (up to 10 Mpc) would be an additional possibility
for study of the average supernova neutrino spectrum~\cite{Ando:2005ka,Kistler:2008us}.

   Figure~\ref{fig:timedep} shows some examples of expected event rates as a function of time for several models in  proposed future large detectors for a $11\ M_\odot$ and a $27\ M_\odot$ SN progenitors with LS EoS (see Sec.~\ref{sec:protoneutronstars}), and Fig.~\ref{fig:ictimedep} shows the time profile for the same models observable by IceCube.
Among the current detectors IceCube is the one having the largest volume for SN neutrinos. Moreover,  before two decades have passed we will possibly have a 40-kton argon detector (DUNE), a 560-kton water detector (Hyper-K) and a 20-kton scintillator detector (JUNO).  For water and scintillator, the event rates are dominated by $\bar{\nu}_e$, and for argon, the event rates are dominated by $\nu_e$.
This figure shows that all detectors can clearly distinguish the accretion phase, resulting in a ``hump'' in the neutrino signal, lasting
$\simeq 0.5$~s and the following cooling on a timescale of $\sim 10$~s.  In the $\nu_e$ signal collected by the argon detector one would also clearly see
the neutronization peak, lasting $\simeq 50$~ms just after the core-bounce, unless suppressed by oscillations (see Sec.~\ref{sec:oscillations}).  We note that for large detectors, provision must be made when designing electronics and data acquisition to prevent data loss for the very high event rates expected for core collapses at $\lesssim$ kpc distances.

\subsection{Neutrino astronomy and astrophysics}

The neutrino burst from a core collapse is also useful to astronomers and astrophysicists. In the following we present some of the applications discussed in literature. 

\subsubsection{\emph{Finding the supernova: Early alert}}

Because neutrinos are released from the stellar collapse on a timescale of seconds, starting within milliseconds of infall, detection of a SN burst offers an opportunity to provide astronomers with an early alert of a SN occurrence.   The relative delay between the neutrino wavefront reaching Earth and the first photons from shock breakout is expected to be on the timescale of hours~\footnote{The delay of three hours between the $\bar\nu_e$ burst and the optical
signal from SN 1987A  implies that the  velocity of electron antineutrinos is equal to that of light with an 
accuracy $\sim 2 \times 10^{-9}$~\cite{Stodolsky:1987vd,Longo:1987ub}.
This bound represents an accurate test   of special relativity.}.
The SuperNova Early Warning System (SNEWS)~\cite{Antonioli:2004zb,Scholberg:2008fa} is a world-wide network designed to alert astronomers, or any other interested scientists, of  a burst.
At the time of this review, the SNEWS network involves Super-Kamiokande, LVD, Borexino, KamLAND, IceCube and Daya Bay; sensitivity is to core collapses within the Milky Way.
A central computer at Brookhaven National Lab (with backup at Bologna) receives datagrams sent by individual experiments and sends an automated alert to a mailing list if a coincidence is found within ten seconds according to the burst UT time stamps.  
The SNEWS network has been operational for an automated alert since 2005, although no coincidences have been detected. 
It is worth noting that gravitational waves also potentially provide a prompt signal of core collapse, and correlation with neutrino burst can potentially enhance the early alert~\cite{Halzen:2009sm,Pagliaroli:2009qy}.
Moreover, assuming that  the angular location of the SN burst can be determined quickly, there is a very high probability for detection of the SN shock breakout in the infrared~\cite{Adams:2013ana}. Such cross-correlations would be extremely useful against false triggers. 

It has been estimated  that the silicon-burning phase preceding a SN explosion can release an energy of about
$5 \times 10^{50}$~erg  for a few days
in neutrino-antineutrino pairs with average energy $\sim 2$~MeV~\cite{Odrzywolek:2003vn,Misiaszek:2005ax}.
IBD events from silicon burning have very low
(below nominal detection threshold) positron
energies, and the best prospects for detection are via neutron capture by adding Gd,
provided that the events can be statistically distinguished from background fluctuations.
For a very nearby SN at a distance $d=1$~kpc, one would expect
$\sim 500$ events per day in  a 0.5-Mton  WC detector~\cite{Fogli:2004ff}. The kton liquid scintillator detector KamLAND will be 
able to detect pre-SN neutrinos for progenitors at distances within $d=660$~pc in the most optimistic scenario~\cite{Asakura:2015bga}. Detection prospects for
detectors under construction or proposed, e.g., SNO+, JUNO, and LENA, look also excellent~\cite{Laha:2014yua}. 
Such detection would make it possible to foresee the death of a massive star  a few days before the stellar core collapse; eventually this could allow discrimination of the progenitor type~\cite{Kato:2015faa}.
Possible pre-supernova candidates that could explode at unpredictable future times include Betelgeuse, 3 Ceti, Antares, Epsilon Pegasi, Pi Puppis, NS Puppis, and Sigma Canis Majoris.

\begin{figure}[t!]
\begin{center}
\includegraphics[angle=90,width=0.9\textwidth]{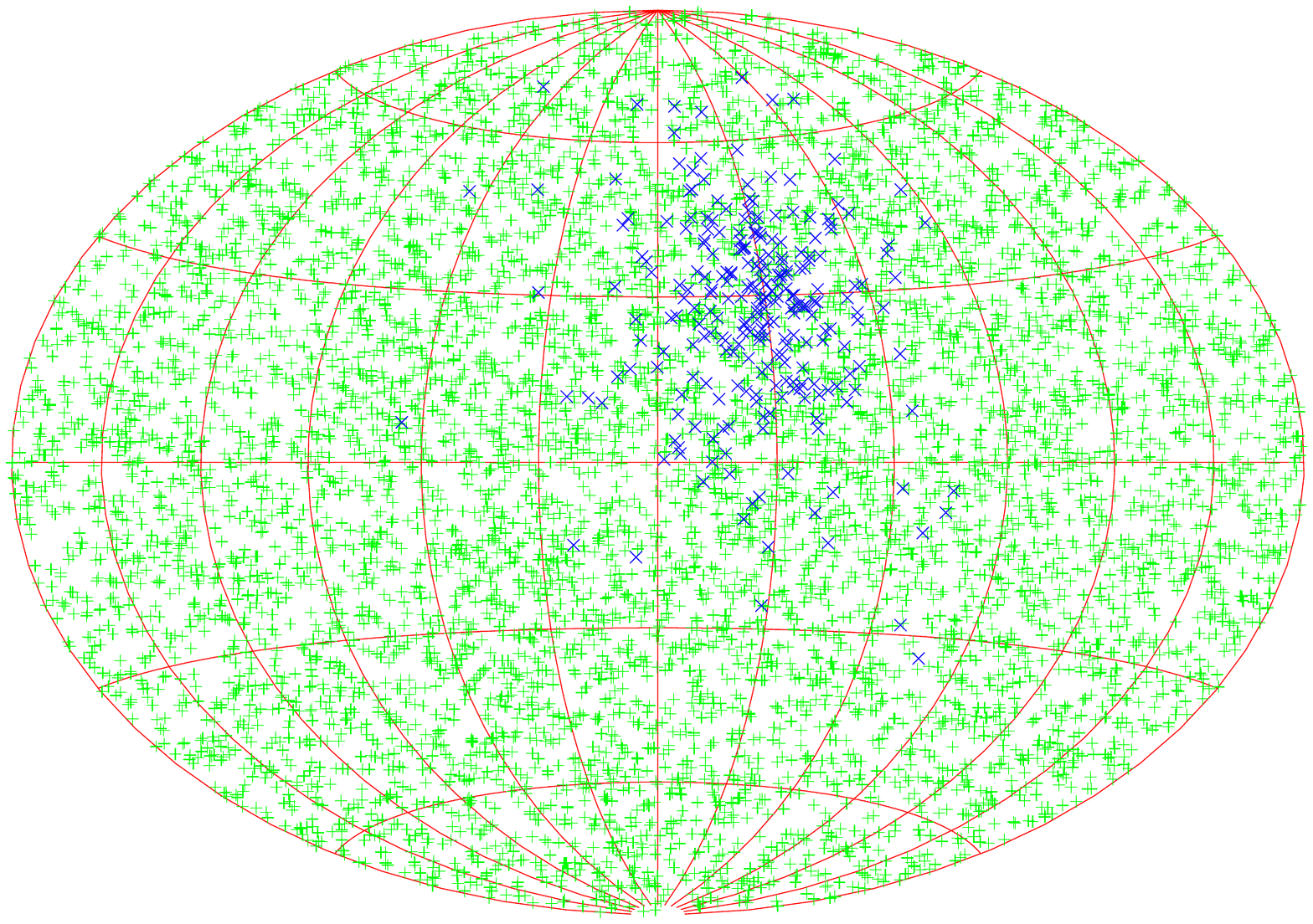} 
    \end{center}
\caption{Angular distribution of IBD events (green) and ES events (blue) of one simulated SN.
(Reprinted figure with permission from~\cite{Tomas:2003xn}; copyright (2003) by the American Physical Society.)
\label{point}}
\end{figure}

\subsubsection{\emph{Pointing to the supernova with neutrinos}}\label{sec:directionality}
It is of clear value to know the position in the sky of the SN, for several reasons.  First, obviously astronomers need to know where to look--- given that the visible SN may not show up in electromagnetic wavelengths for some time, a direction will aid in observing the very early stages.  Furthermore, it is not clear that the core collapse always results in a bright SN, and direction information will help to potentially locate a weak explosion, or even a vanished progenitor.
Knowledge of distance to the progenitor is valuable for precise determination of neutrino luminosity.
Knowledge of the direction will aid as well in interpretation of the signal itself, in that the specific matter effects undergone by the neutrino flux will depend on the pathlength through the Earth~\cite{Mirizzi:2006xx}.

The ability of a detector to determine the direction of a signal depends both on the intrinsic anisotropy of the interaction, and on the capability of tracking the resulting products.  CC and NC interactions with nuclei, including IBD, have some energy-dependent anisotropy (e.g., \cite{Vogel:1999zy}),  but this is generally relatively weak for the most important interactions.  Elastic scattering of electrons is however fairly strongly peaked forward. 
The angular distribution of  IBD and ES events for a simulated SN is shown in Fig.~\ref{point}.
 Of the interactions relevant for current detectors, argon and oxygen interactions do have some potentially observable anisotropy (e.g., \cite{Haxton:1990ks}).

Of the detector types currently available, Cherenkov and LAr tracking detectors are capable of determining the direction of charged particle tracks.
Unfortunately scintillator produces mostly isotropic light with little chance of determining scattered particle direction (some information may be available from reconstructed relative positions of IBD-produced positron and neutron energy loss~\cite{Fischer:2015oma}).
Of detectors currently running, Super-Kamiokande has the only plausible chance of determining the SN direction using ES events (which should constitute a few percent of the observed events). 
The pointing accuracy in the worse case would be $\sim 8^\circ$~\cite{Tomas:2003xn} for a SN at $d=10$~kpc.
This ability should improve with addition of Gd by reducing the near-isotropic background of IBD events, reaching
$\sim 3^\circ$~\cite{Tomas:2003xn}. 
Hyper-Kamiokande's statistics would improve determination even further (with an accuracy as good as $\sim 0.6^\circ$).
LAr should also have decent pointing capabilities via lepton tracking.

Another potential method for source direction determination is via triangulation, using timing of several signals observed in different locations around the Earth~\cite{Beacom:1998fj,Muhlbeier:2013gwa}.  This is challenging, given that the spread over time of the burst is longer than the travel time through the Earth; therefore, detection statistics of the burst need to be such that a sharp feature of the signal, such as the risetime or other feature, can be determined.  IceCube may be the best prospect for contributing to this kind of determination.   With current detectors, triangulation in real time may be difficult, although with large future detectors prospects are more promising.
Another possible pointing method is via the matter oscillation effect pattern, which might be possible to combine with timing information~\cite{Scholberg:2009jr}. However this requires large statistics and good energy resolution, and is unlikely to do better than ES in a single directional detector.

\subsubsection{\emph{Probing supernova hydrodynamical instabilities through neutrinos}}\label{sec:SASI}

As discussed in Sec.~\ref{sec:SNmodels}, the first  hydrodynamical 3D SN simulations  with sophisticated neutrino transport
have recently become available. The more massive cases of the simulated SN progenitors (with $27$ and $20\ M_\odot$ progenitor masses) show pronounced SASI phases interrupted by episodes of dominant  convective overturn activity. The neutrino signal carries imprints of these hydrodynamical instabilities (see Fig.~\ref{fig:SASI-s27}) and  the detection of such modulations
 will therefore offer a unique chance to probe the core-collapse mechanism.
It has been recently shown~\cite{Tamborra:2013,Tamborra:2014b}  that  measurements of neutrinos from a future Galactic SN with neutrino Cherenkov telescopes (i.e., IceCube and Hyper-Kamiokande, see Sec.~\ref{sec:detectionprospects}) will indeed be able to discriminate the SASI neutrino 
 modulation. 

\begin{figure}
\begin{center}
\includegraphics[angle=0,width=0.6\textwidth]{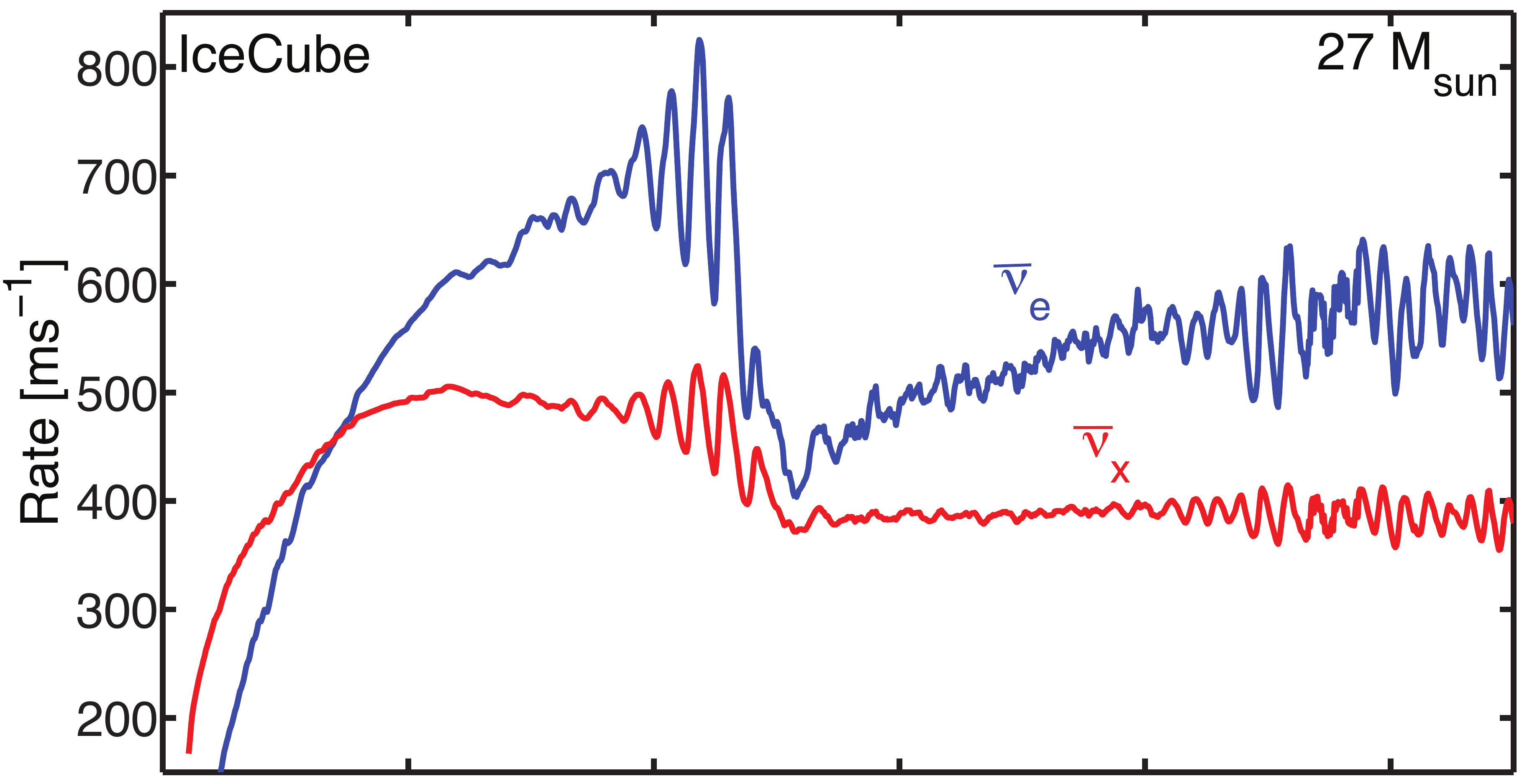}\\
\includegraphics[angle=0,width=0.6\textwidth]{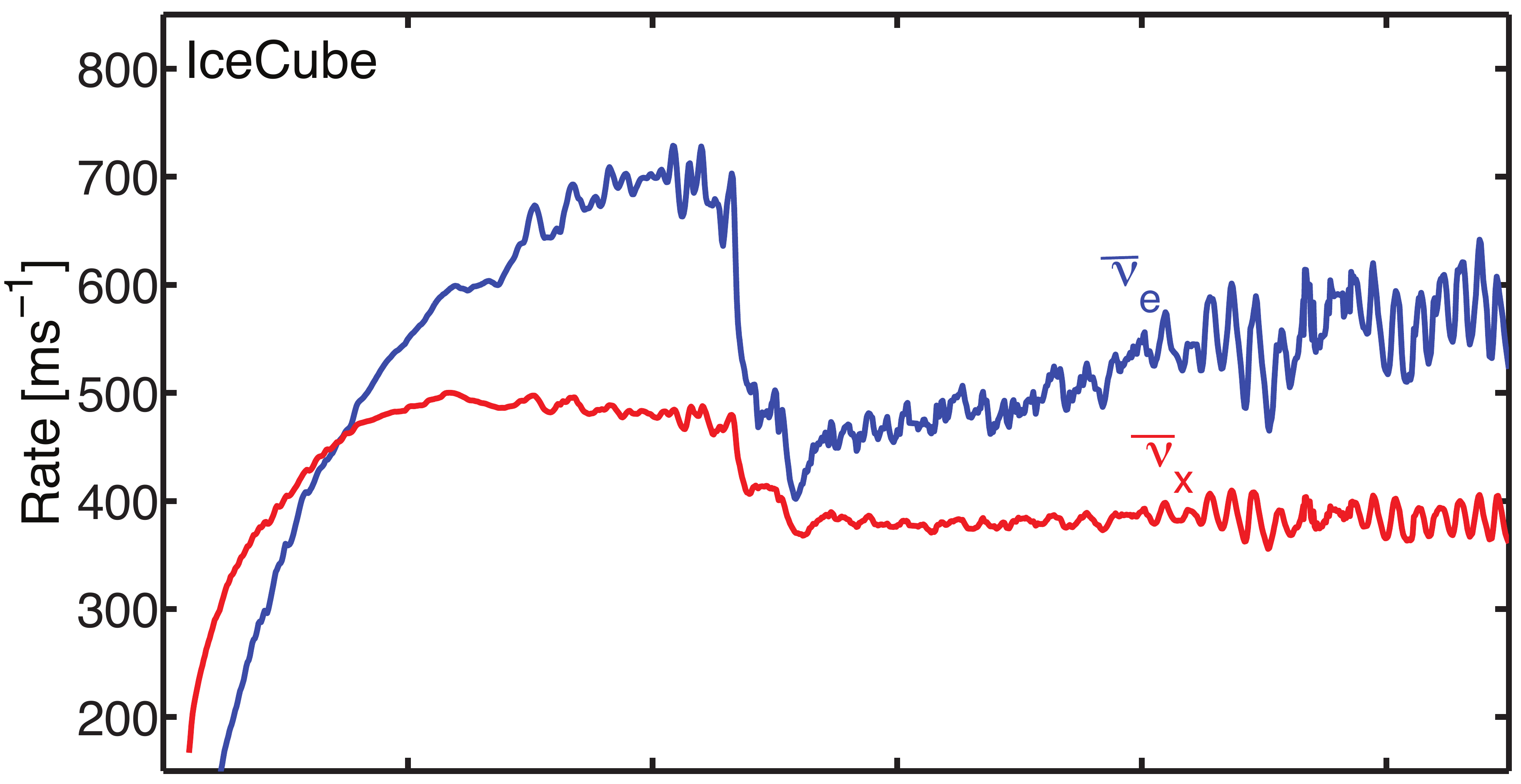}\\
\includegraphics[angle=0,width=0.6\textwidth]{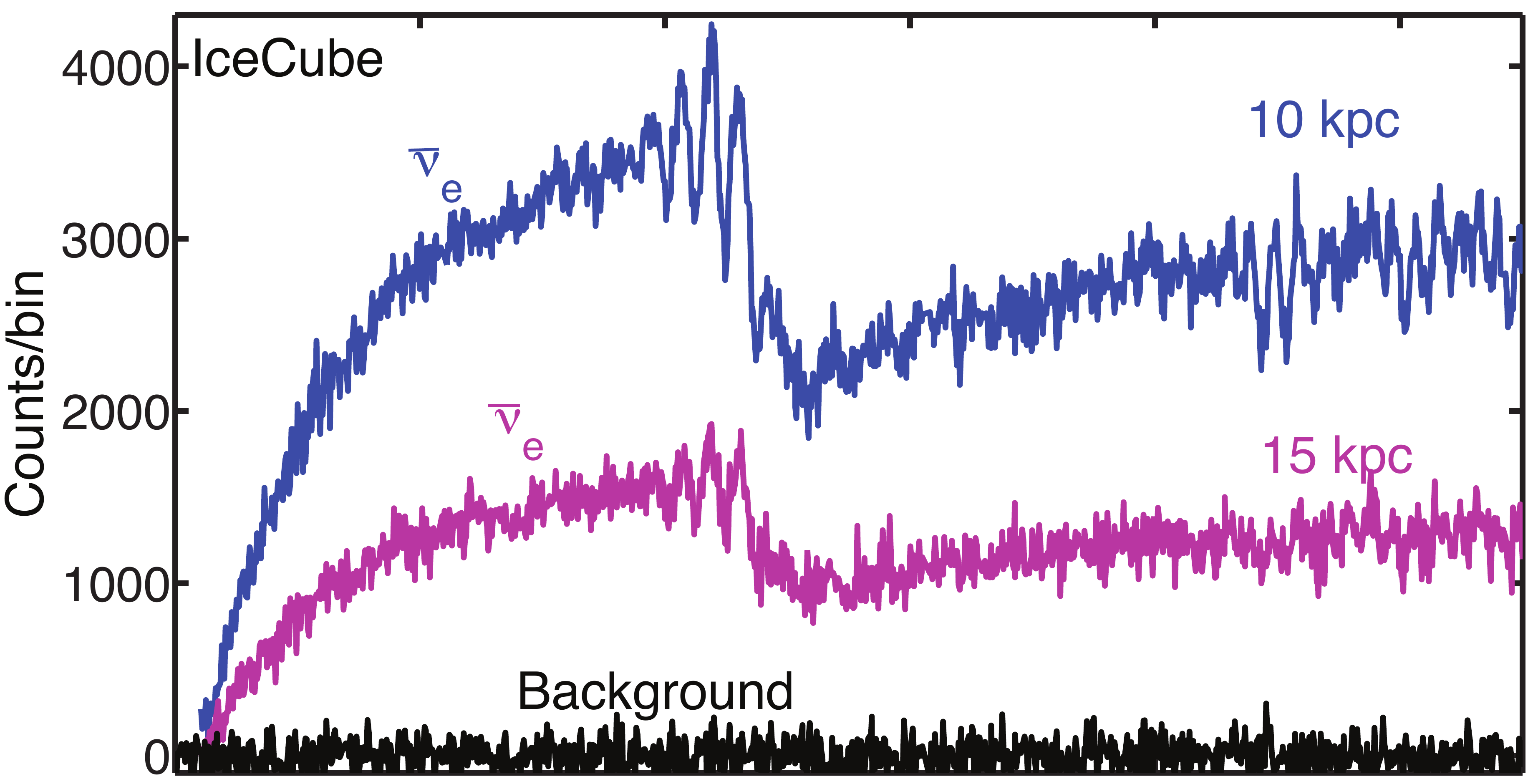}\\
\includegraphics[angle=0,width=0.6\textwidth]{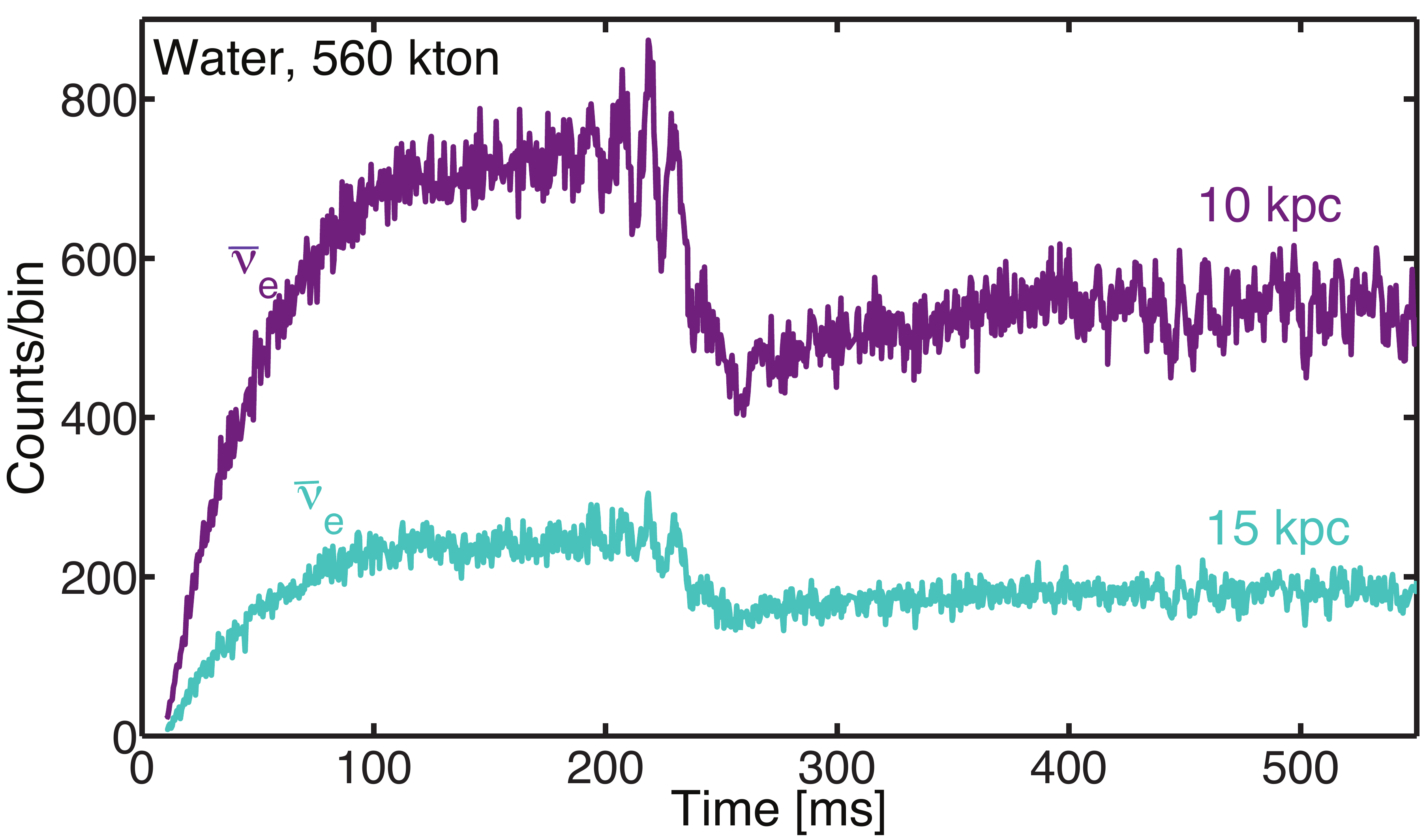}
    \end{center}
\caption{Detection rate in IceCube (three upper panels) and in  a  560-kton WC detector  (bottom panel) for a $27\ M_\odot$ SN 
progenitor at $10$~kpc for $\bar{\nu}_e$ (no flavor conversions) and $\bar{\nu}_x$ (complete flavor swap).  
The observer direction is chosen close to the SASI plane during the first SASI episode ($120$--$260$~ms), where a strong modulation of the neutrino signal is expected, except
for the second panel where a direction of minimal modulation of the neutrino signal is shown. The lower panels include a random shot-noise realisation in 
5 ms bins; the IceCube background fluctuations in the absence of a SN are shown in black. 
(Figure adapted from~\cite{Tamborra:2013}.)
\label{fig:SASI_rate}}
\end{figure}

Figure~\ref{fig:SASI_rate} (top panel) shows the expected IceCube rate as a function of time for an observer located close to the SASI plane\footnote{Note that for all angular locations of the observer,  the 
detected neutrino rate has been computed by integrating the neutrino emission  emitted from the hemisphere facing the observer including 
projection effects associated with limb darkening. We refer the reader to Ref.~\cite{Tamborra:2014b} for details on the projection procedure.} for a 3D SN simulation of a   $27\ M_\odot$ star (see Sec.~\ref{sec:SNmodels}). Large-amplitude sinusoidal modulations of the IceCube event rate signal appear
in correspondence to the first SASI episode at $120$--$260$~ms. After a phase where convective motions dominate, a second SASI episode
begins at about $410$~ms.  Given our scarce knowledge of neutrino flavor oscillations in the presence of SASI, two extreme cases have
been plotted: One in which flavor conversions have been neglected and therefore the signal is caused by the non-oscillated $\bar{\nu}_e$ (blue curve)
and another one where  complete flavor swap was assumed  so that the signal is caused by $\bar{\nu}_x$ (red curve). Both cases exhibit   large-amplitude signal modulations with clear periodicity.

The possibility to detect neutrino signal modulations 
 depends on the viewing angle relative to the  plane of the SASI sloshing or spiral motions.  Observers located along the direction orthogonal to the plane of the first SASI episode will only detect very weak SASI modulations, as shown in the second panel of Fig.~\ref{fig:SASI_rate} (see also right panel of Fig.~\ref{fig:SASI-s27} for an estimation of the favorable locations of the observer with respect to the SASI plane);  observers placed on opposite directions away from the source will detect almost the same signal modulations but with opposite phases.

The third panel of Fig.~\ref{fig:SASI_rate} shows the IceCube signal in $5$~ms time bins, including a random shot noise realization,
and the IceCube background fluctuations are plotted in black for comparison. For a SN up to $15$~kpc, the  SASI modulations 
in the neutrino signal will be clearly visible.  The correspondent 
rate in a 560-kton water Cherenkov detector is plotted in the bottom panel of Fig.~\ref{fig:SASI_rate}. A 560-kton water Cherenkov telescope will have no background in contrast to IceCube and it will be able to collect event-by-event energy
information, but its expected rate should be
about $1/5$ of the IceCube one.  Similar modulations of the neutrino signal due to SASI are also clearly
detectable for the $20\ M_\odot$ SN progenitor~\cite{Tamborra:2013,Tamborra:2014b}.

The Fourier power spectrum of the IceCube rate exhibits  a prominent peak at about $80$~Hz both for the $20\ M_\odot$ and the 
$27\ M_\odot$ SN progenitors. Such a frequency corresponds to the sloshing frequency of the shock front and  is a function of the 
neutron star radius (similar for both progenitors as they were modeled with the same EoS) and of the shock radius
(again comparable for both progenitors as their mass-infall history in the collapsing stellar core is fairly similar)~\cite{Tamborra:2013}. 

In synthesis, the observation of SASI modulations by Cherenkov neutrino telescopes will provide a confirmation of our
current theories of stellar core collapse. However, the possibility to discriminate SASI signatures will be 
dependent on the location of the observer relative to the plane where 
SASI sloshing or spiral motions develop. 

\subsection{Outlook} 
We are reasonably well prepared world-wide for the next SN burst, with a number of detectors online, of which Super-Kamiokande will give the largest event signal, and potentially provide pointing information.
Additional scintillator detectors, some surface detectors, and a small lead detector, will enhance the yield, and IceCube will provide excellent burst time structure information. However, the current sensitivity is overwhelming for the $\bar{\nu}_e$ component of the flux. The planned next generation of very large detectors will provide even more abundant $\bar{\nu}_e$ statistics and, in addition, should also have broader flavor sensitivity via $\nu_e$ detection in argon, and high-statistics NC channels in water and scintillator. We look forward to recording the next burst with these enhanced capabilities.

A high-statistics detection of a Galactic SN $\nu$ burst would allow accurate studies of the signal in energy and time domains and in different
interaction channels. In particular, we have shown how the observable SN $\nu$ light curve would carry interesting information concerning the explosion mechanism
and the hydrodynamical instabilities occurring in the deepest SN regions. Moreover, by reconstructing the duration of the different phases of the $\nu$ burst
one could have a direct empirical test for the predictions of the SN simulations. Furthermore, the duration of the neutrino signal
is an important diagnostic parameter to constrain exotic energy-loss processes in a SN, associated with the emission of novel light weakly-interacting particles~\cite{Raffelt:1990yu}. 
These would shorten the duration of the observable SN neutrino signal.
Finally, the SN $\nu$ light curves and energy spectra would also carry intriguing signatures associated with the flavor conversions occurring in the deepest
stellar regions, as we will discuss in the next Section.

\clearpage

\section{Supernova Neutrino Flavor Conversions}\label{sec:oscillations}
\noindent {\it Authors: A.~Mirizzi, N.~Saviano, I.~Tamborra, S.~Chakraborty}
\\

In this Section, 
we will present an overview  of the neutrino flavor transitions in core-collapse
SNe.  
At first we introduce the equations of motion  (EoMs)
for  SN neutrinos in the formalism of the density matrix, describing the vacuum,  matter and  $\nu$-$\nu$ refractive terms  entering the neutrino Hamiltonian. 
Then the phenomenology of self-induced flavor conversions, associated with the large $\nu$-$\nu$ interaction potential,  is illustrated through representative examples of its rich 
phenomenology.
The Mikheyev-Smirnov-Wolfenstein (MSW) matter effects are also discussed as well as the impact of the SN shock-wave propagation and of matter turbulence on the neutrino
conversion probabilities. 
Finally, observable signatures of the SN neutrino oscillations imprinted on the neutrino burst 
 are characterized. 
 Particular attention is devoted to the sensitivity of these observables to the unknown neutrino mass hierarchy.

\subsection{Three-neutrino oscillation framework}
\label{sec:threenu}

There is compelling 
experimental evidence~\cite{PDG4} that the three known active neutrino states with
definite flavor $\nu_\alpha$ ($\alpha=e,\mu,\tau$)  are linear
combinations of the mass eigenstates  $\nu_i$ ($i=1,2,3$), and
that the Hamiltonian of neutrino propagation in vacuum~\cite{Pontecorvo:1957qd,Pontecorvo:1967fh,Maki:1962mu}
 and
matter~\cite{Matt} does not commute with flavor. The evidence
for flavor non-conservation (i.e., ``neutrino oscillations'') 
comes from a series of experiments performed
in about four decades of research with very different neutrino
beams and detection techniques.
Namely, 
the solar neutrino~\cite{Davis:1968cp,Ba89,BP00} experiments:
Homestake~\cite{Home}, 
SAGE~\cite{SAGE}, GALLEX-GNO~\cite{GALL,GGNO},  Kamiokande~\cite{Kami},
Super-Kamiokande (SK)~\cite{SKso,SK04}, the Sudbury Neutrino Observatory (SNO)~\cite{SNO1,SNO2,SNOL}, and
Borexino~\cite{Arpesella:2007xf}; 
the long-baseline reactor-neutrino 
experiment KamLAND~\cite{Kam1,Kam2};
the atmospheric neutrino  
experiments like
Super-Kamiokande~\cite{SKev,SKLE,AtSK}, 
MACRO~\cite{MACR}, and Soudan-2~\cite{Soud};  the
long-baseline accelerator neutrino  experiments: KEK-to-Kamioka (K2K)~\cite{K2K1},
MINOS~\cite{Adamson:2013whj}, OPERA~\cite{Agafonova:2014bcr}  and 
Tokai-to-Kamioka (T2K)~\cite{Abe:2013fuq, Abe:2014ugx}.

Except for yet controversial results from short-baseline neutrino experiments, that
 point towards the existence of extra sterile neutrino states [with $m \sim {\mathcal O}(1)$~eV]
~\cite{Abazajian:2012ys,Gariazzo:2015rra},
all the data from the above experiments are consistent with the simplest
extension of the standard electroweak model needed to accommodate
nonzero neutrino masses and mixings. Namely, with a scenario where the
three known flavor eigenstates $\nu_{e,\mu,\tau}$ are mixed with the  three
mass eigenstates {$\nu_{1,2,3}$},
through a unitary matrix $U$, which 
in terms of one-particle neutrino states
$|\nu\rangle$,  is defined as (see, e.g., \cite{PDG4})
\begin{equation}
\label{Ustar}
|\nu_{\alpha}\rangle=\sum_{i=1}^3 U_{\alpha i}^* |\nu_{i}\rangle\ .
\end{equation}

A common parametrization  for the matrix $U$ is~\cite{Fogli:2005cq}:
\begin{equation}
\label{UPDG}
U=
\left(
\begin{array}{ccc}
1 & 0 & 0 \\
0 & c_{23} & s_{23}\\
0 & -s_{23} & c_{23}
\end{array}
\right)\left(
\begin{array}{ccc}
c_{13} & 0 & s_{13}e^{-i\delta} \\
0 & 1 & 0\\
-s_{13} e^{i\delta} & 0 & c_{13}
\end{array}
\right)\left(
\begin{array}{ccc}
c_{12} & s_{12} & 0\\
-s_{12} & c_{12} & 0 \\
0 & 0 & 1
\end{array}
\right)
\end{equation}
with $c_{ij}=\cos\theta_{ij}$ and $s_{ij}=\sin\theta_{ij}$, $\theta_{ij}$
being the mixing angles and $\delta \in [0,2\pi]$ being the CP-violating phase%
~\footnote{We neglect here extra-phases possible if neutrinos are Majorana particles,
since they are not relevant in oscillations.}.

The current neutrino phenomenology implies that the three-neutrino mass
spectrum $\{m_i\}_{i=1,2,3}$ is made by a ``doublet'' of relatively close
states and by a third ``lone'' neutrino state, which may be either
heavier than the doublet (``normal hierarchy,'' NH) or lighter
(``inverted hierarchy,'' IH). Typically, the lightest (heaviest) neutrino in the
doublet is called $\nu_1$ ($\nu_2$) and the corresponding mass squared 
difference is defined as
\begin{equation}
\delta m^2=m^2_2-m^2_1>0
\label{eq:solar}
\end{equation}
by convention. 
The $\delta m^2$ is traditionally named the \emph{solar} mass squared difference.
The lone state is then labeled as $\nu_3$, and the physical
sign of 
$ m^2_3-m^2_{1,2} $
 distinguishes NH from IH.
 The second independent squared mass difference $\Delta m^2$, called also
 \emph{atmospheric} mass squared difference, 
 is  
\begin{equation}
\Delta m^2=\left| m^2_3-\frac{m^2_1+m^2_2}{2}\right|\ ,
\label{eq:atm}
\end{equation}
so that the two hierarchies (NH and IH) are simply related by the transformation
$+\Delta m^2\to-\Delta m^2$. Numerically, 
it results that $\delta m^2 \ll \Delta m^2$. 

 The latest
solar, reactor and  long-baseline   neutrino oscillation analyses indicate 
the following $\pm 2\sigma$ ranges for each parameter
(95\% C.L.) taken from~\cite{Capozzi:2013csa}
(see also~\cite{Forero:2014bxa,Gonzalez-Garcia:2014bfa} for other updated analyses) 
\begin{eqnarray}
\Delta m^2  &=& (2.43^{+0.12}_{-0.13})
\times 10^{-3}\mathrm{\ eV}^2\ , \nonumber \\
\delta m^2 &=& (7.54^{+0.46}_{-0.39})\times 10^{-5}\mathrm{\ eV}^2\ , \nonumber \\
\sin^2\theta_{12} &=& (3.08^{+0.17}_{-0.34}) \times 10^{-1}  , \nonumber \\
\sin^2\theta_{23} &=& (4.37^{+1.15}_{-0.44}) \times 10^{-1}\ , \nonumber \\
\sin^2\theta_{13} &= &(2.34^{+0.40}_{-0.39}) \times 10^{-2} \,\ . \nonumber \\
\label{eq:parameters}
\end{eqnarray}
Remarkably,
the  $1-3$ mixing angle
$\theta_{13}$ has been the last one measured 
in 2012 
by the Daya Bay~\cite{An:2012eh} and RENO~\cite{Ahn:2012nd}  reactor experiments. These recent measurements confirmed
and greatly strengthened the significance of early hints suggested by the long-baseline $\nu_{\mu}$-$\nu_e$ 
experiments T2K~\cite{Abe:2011sj} and MINOS~\cite{Adamson:2011qu} as well as by the
Double Chooz reactor experiment~\cite{Abe:2011fz}, especially when analyzed in 
combination with other oscillation data~\cite{Fogli:2008jx,Fogli:2011qn}.
The measurement of a ``large'' value of $\theta_{13}$ has siginifcantly
reduced the ambiguity in characterizing the oscillated SN neutrino oscillations.
Notably, the latest global analysis 
hints towards a non-zero CP-violation around
$\delta \sim 1.4 \pi$ at $\gtrsim 1 \sigma$ level~\cite{Capozzi:2013csa}, 
 see also~\cite{Forero:2014bxa,Gonzalez-Garcia:2014bfa}~\footnote{It has been shown that in the context of SN neutrino oscillations, CP-violation effects 
are negligible~\cite{Akhmedov:2002zj,Balantekin:2007es,Gava:2008rp}.
Therefore, we will not include them in the following discussion.}.

Within the three-neutrino scenario,  the most important unsolved problems
require probing the CP-violating phase $\delta$, the mass hierarchy, and the absolute neutrino masses. 
In this context, a high--statistics detection of Galactic SN
neutrinos
could give a unique help to  solve some of these open issues.

\subsection{Equations of motion for supernova neutrinos}
\label{sec:eom}

The treatment of neutrino mixing is well understood
in terms of the  propagation of a beam of particles in  vacuum or in a medium, and  neutrino 
flavor transitions have  been described by means of a Schr\"odinger-like equation~\cite{Kuo:1989qe}.
Neutrinos free streaming beyond the neutrinosphere also interact among themselves (\emph{neutrino self-interactions}). As pointed out in seminal papers~\cite{Pantaleone:1992xh,Pantaleone:1994ns,Pantaleone:1992eq,Kostelecky:1993yt,Pantaleone:1998xi,%
Samuel:1993uw,Kostelecky:1993ys,Kostelecky:1994dt,Samuel:1995ri,Kostelecky:1996bs,Pastor:2001iu,%
Dolgov:2002ab,Pastor:2002we,Wong:2002fa,Balantekin:2004ug,Sawyer:2005jk}, in the deepest regions of a SN  (as well as in the Early Universe) the neutrino gas is so dense that  neutrinos themselves form a background medium leading to intriguing non-linear effects in the flavor distribution. 
 Neutrino-neutrino interactions  could trigger 
large \emph{self-induced} flavor conversions
 in the deepest SN regions~\cite{Duan:2006an,Duan:2006jv}. Under these circumstances, 
neutrinos emitted with different energies would be locked to oscillate in a collective fashion. 
 
A natural treatment to properly characterize the flavor evolution
in this situation requires the formalism of the neutrino density  matrix~\cite{Sigl:1992fn} 
 \begin{equation}
 \varrho_{\mathbf{p, x}}=
\left(\begin{array}{ccc} 
\varrho_{ee} &  \varrho_{e \mu} & \varrho_{e\tau} \\
\varrho_{\mu e}  & \varrho_{\mu \mu} &  \varrho_{\mu \tau} \\
\varrho_{\tau e} &\varrho_{\tau \mu} &\varrho_{\tau \tau}
\end{array}\right)~.
\label{eq:rho}
 \end{equation}
The diagonal
entries of this matrix are the usual occupation numbers, whereas the
off-diagonal terms encode phase information related to the oscillations.  
An analogous expression holds for the  antineutrino density matrix,
$\overline{\varrho}_{\mathbf{p, x}}$.
See also~\cite{Raffelt:1991ck, Raffelt:1992uj,McKellar:1992ja,Dolgov:1980cq, Barbieri:1990vx}
for similar derivations
and~\cite{Balantekin:2006tg,Cardall:2007zw,Volpe:2013uxl,Vlasenko:2013fja} 
for recent developments on the formalism.

In SNe one is concerned with 
 the spatial evolution of the neutrino fluxes in
a quasi-stationary situation. Therefore, the matrices
$\varrho_{\mathbf{p, x}}$
do not explicitly depend on time, so that the EoMs
reduce to the Liouville term involving only spatial
derivatives~\cite{Sigl:1992fn,Cardall:2007zw}
\begin{eqnarray}
\mathbf{v_p} \cdot \nabla_{\mathbf{x}}  \varrho_{\mathbf{p, x}} &=& -i [\Omega^{\textrm{vac}}_ {\mathbf{p}},  \varrho_{\mathbf{p,x}}]-i[ \Omega^{\textrm{ref}}_ {\mathbf{p,x}},  \varrho_{\mathbf{p,x}}] \,\ , \nonumber \\
\mathbf{v_p} \cdot \nabla_{\mathbf{x}} {\overline\varrho}_{\mathbf{p, x}} &=& +i [\Omega^{\textrm{vac}}_ {\mathbf{p}}, 
{\overline\varrho}_{\mathbf{p, x}}]-i[ \Omega^{\textrm{ref}}_ {\mathbf{p,x}},  {\overline\varrho}_{\mathbf{p, x}}] \,\ .
\label{eq:eom}
\end{eqnarray} 
In the flavor basis the first term in the Hamiltonian at the right-hand-side (r.h.s.) of  Eq.~(\ref{eq:eom})
for ultrarelativistic neutrinos represents the vacuum oscillation term, 
\begin{equation}
\Omega^{\textrm{vac}}_ {\mathbf{p}}=U\frac{M^2}{2p}U^{\dag} \,\ .
\label{eq:vac}
\end{equation}
The $U$ matrix describes the mixing  [Eq.~(\ref{UPDG})] and $M^2$ is the squared mass matrix which, except for a common term  proportional to the identity matrix and irrelevant for oscillations, is parametrized,
in terms of the solar $\delta m^2$ [Eq.~(\ref{eq:solar})] and atmospheric $\Delta m^2$ 
[Eq.~(\ref{eq:atm})]
mass-squared differences,
 as~\cite{Fogli:2005cq}
\begin{equation}
\footnotesize{
M^2=\textrm{diag}\left(m_{1}^2, m_{2}^2, m_{3}^2\right)= \left(-\frac{\delta m^2}{2}, +\frac{\delta m^2}{2}, \pm {\Delta m^2}\right)} \,\ . 
\end{equation}
 The sign $\pm$ in front of $\Delta m^2$ refers to NH (+) and IH (-), respectively.
One can associate two vacuum oscillation frequencies to  these mass-squared differences, i.e.,
\begin{eqnarray}
\omega_L = \frac{\delta m^2}{2 E} \  \mathrm{and}\ 
\omega_H = \frac{\Delta m^2}{2 E}\ . 
\label{eq:vacfreq}
\end{eqnarray}

 The vacuum energy can be affected by a shift (similar to the photon index of refraction) when
neutrinos propagate in a medium. This effect is
 induced by their forward scatterings
with the medium constituents~\cite{Kuo:1989qe}.
 The refractive effect for SN neutrinos is described by  the
second term in the Hamiltonian at the r.h.s. of  Eq.~(\ref{eq:eom})
\begin{equation}\label{eq:ham1}
 \Omega^{\textrm{ref}}_{\mathbf {p, x}}= \lambda_{\mathbf {x}} {\sf L}+\sqrt{2}\,G_{\rm F}
 \int\ \frac{d^3{\mathbf q}}{(2\pi)^3}
 \left(\varrho_{\mathbf{ q, x}}-\overline\varrho_{\mathbf {q, x}}\right)
 (1-{\mathbf v}_{\mathbf q}\cdot{\mathbf v}_{\mathbf p})\,.
\end{equation}
The first term at the r.h.s. in Eq.~(\ref{eq:ham1}) represents the ordinary refractive  matter effect.
In particular, for typical energies of SN neutrinos $[E\sim {\mathcal O}(10)$~MeV] the only relevant process 
is due to 
charged-current interactions of electron neutrinos $\nu_e$ (or antineutrinos $\bar\nu_e$)
with the background  $e^{\pm}$ (the neutral current is 
flavor conserving and therefore equal for all neutrino  flavors).
This is the
well-known MSW effect, first pointed out by Wolfenstein, and by Mikheyev
and Smirnov~\cite{Matt}. 
 This term  is represented by
$\lambda_{\mathbf {x}}=\sqrt{2}\,G_{\rm F}(n_{e^-}-n_{e^+})$ with ${\sf L}={\rm
diag}(1,0,0)$ in the weak interaction basis. 

Neutrino-neutrino interactions, dominant in the deepest SN regions, make an
additional contribution to the refractive energy shift, represented by the second term 
 at the r.h.s. of the Eq.~(\ref{eq:ham1}). This term is proportional to the neutrino density
 matrix $\varrho_{\mathbf{ p, x}}$ that in the presence of mixing has also off-diagonal elements,
 giving rise to ``off-diagonal refractive indices'' as first pointed out by Pantaleone~\cite{Pantaleone:1992xh,Pantaleone:1994ns,Pantaleone:1992eq}. 
 The main complication  in the $\nu$-$\nu$ refractive term
for SN neutrinos is the angular
factor $(1-{\mathbf v}_{\mathbf q}\cdot{\mathbf v}_{\mathbf
p})=(1-\cos\theta_{{\mathbf p}{\mathbf q}})$ coming from the current-current nature
of the weak interactions, where ${\mathbf v}_{\mathbf p}={\mathbf p}/p$ is the
neutrino
velocity. This angular term  averages to zero for an isotropic $\nu$ gas.
However, for the non-isotropic neutrino emission from the SN core
 this velocity-dependent term 
would not average to zero, producing a net current that   leads to a different refractive index for neutrinos that propagate
on different trajectories. This is at the origin of the so-called
 ``multi-angle effects''~\cite{Duan:2006jv,Raffelt:2007yz,EstebanPretel:2007ec,Sawyer:2008zs}.
Remarkably, while in an isotropic neutrino gas the self-induced 
effects would lead to a collective behavior in the flavor evolution~\cite{Hannestad:2006nj},
in an anisotropic case this is not guaranteed.
Indeed the multi-angle term 
in some cases challenges the collective behavior of the flavor evolution,
 leading to   
\emph{flavor decoherence} under certain conditions, with a resultant 
flux equilibration among electron and non-electron (anti)neutrino 
species~\cite{Raffelt:2007yz,EstebanPretel:2007ec,Sawyer:2008zs}.

\begin{figure}[t!]
\begin{center}
 \includegraphics[angle=0,width=0.9\textwidth]{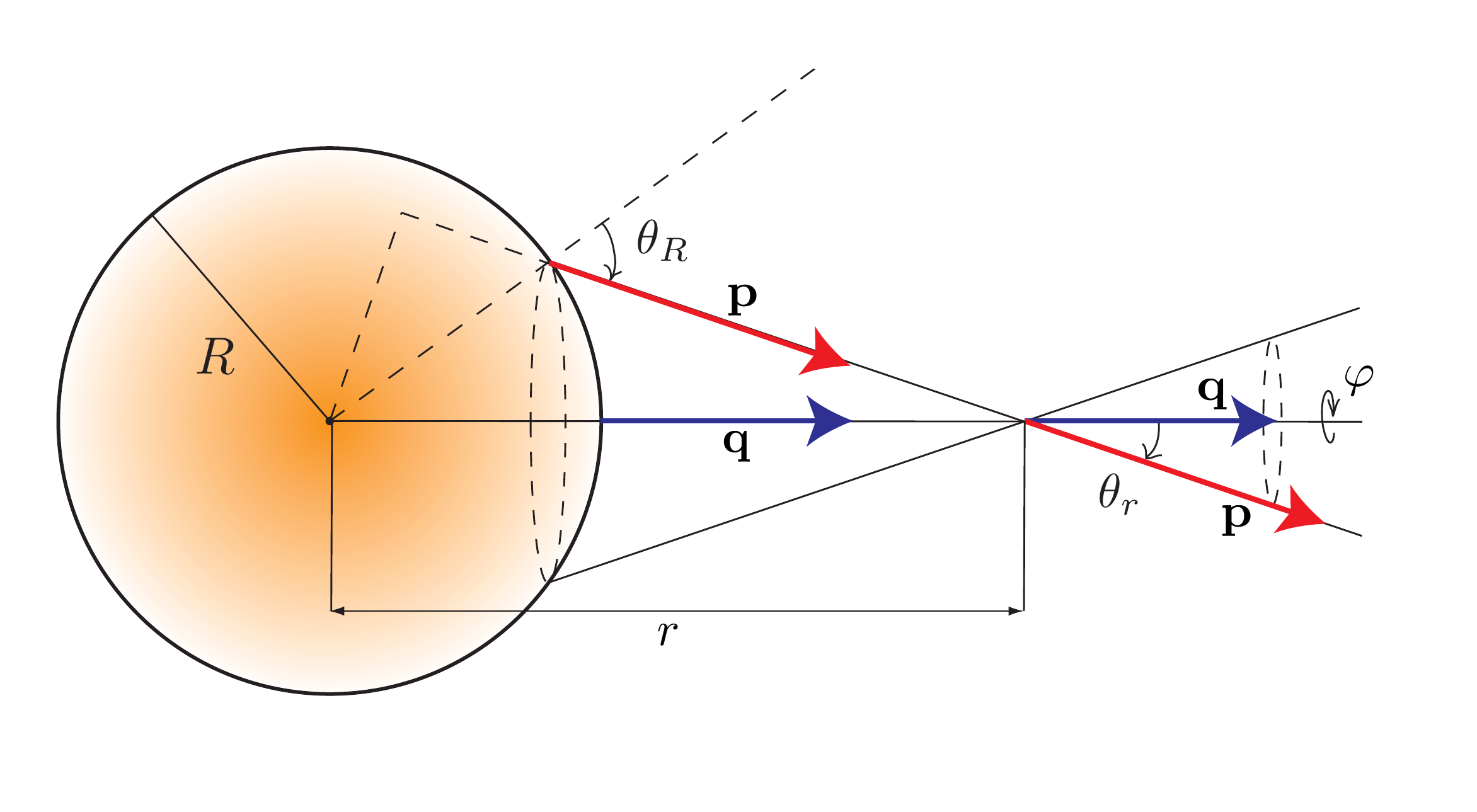} 
    \end{center}
\caption{Geometrical picture of the neutrino bulb model. Courtesy of A.~Marrone. \label{nsphere}} 
\end{figure}

The multi-angle  flavor evolution described by the partial differential
equations, Eq.~(\ref{eq:eom}),  has not been solved  in its full complexity until now.
 However, a few years ago
the first large-scale multi-angle simulations were developed within the so-called  
\emph{``bulb model''}~\cite{Duan:2006an,Fogli:2007bk,EstebanPretel:2007ec}, whose
 geometry is represented in Fig.~\ref{nsphere}.
In this framework it is assumed that 
neutrinos are emitted uniformly and half-isotropically (i.e., with all the outward-going modes occupied, and 
all the backward going modes empty)  from the surface of a spherical neutrinosphere, like in a blackbody.
 Moreover,  it is assumed that there is azimuthal symmetry of the neutrino emission at the neutrinosphere and that
all the  physical conditions in the star only depend on the  distance $r$ from the center
of the star.

Under the assumption of azimuthal symmetry, one can reduce the general partial differential equations [Eq.~(\ref{eq:eom})] 
into  ordinary differential equations, projecting the evolution along the radial direction.
Then
\begin{eqnarray}
{\bf v_p}\cdot \nabla_{\bf x} &\to & v_r \frac{d}{dr} \,\ , \nonumber \\
{\bf p} & \to & (E\simeq |{\bf p}|, u \equiv  \sin^2 \theta_R) \,\ , \nonumber \\
 \varrho_{\mathbf{p, x}} & \to & \varrho_{p,u,r} \,\ ,
\end{eqnarray}
where $\theta_R$ is the emission angle relative to the radial direction at the 
neutrinosphere, $r=R$. 

The distribution matrices $\varrho_{p,\theta,r}$ (or  $\varrho_{p,u,r}$)
are not especially useful to describe a spherically symmetric system because they vary with radius even in  absence of
oscillations. A quantity that is conserved in absence of oscillations is the
total flux matrix~\cite{EstebanPretel:2007ec}
\begin{equation}
{\sf J}_r = 4 \pi r^2 \int \frac{d^3 {\bf p}}{(2\pi)^3} \varrho_{p,\theta, r} v_r \,\ ,
\end{equation}
that in terms of $\varrho_{p,u,r}$ reads
\begin{equation}
{\sf J}_r  = \int_{0}^{\infty} dp \int_{0}^{1} du\ {\sf J}_{p,u,r} =
\int_{0}^{\infty} dp \int_{0}^{1} du\ \frac{\varrho_{p,u,r} p^2 R^2}{2\pi} \,\ .
\end{equation}
This implies that the trace of ${\sf J}_{p,u,r}$ is $r$-independent. 
For the sake of  brevity, we address the interested reader to~\cite{EstebanPretel:2007ec}
for the explicit form of the EoMs for the ${\sf J}_{p,u,r}$ matrices.

\subsection{Polarization vectors and equations of motion for a $2\nu$ system}
\label{sec:polar}
Often SN neutrino flavor  evolution can be  characterized in
a 2$\nu$ flavor scenario, 
associated with $\Delta m^2$ and $\theta_{13}$, in which the electron state
$\nu_e$ mixes with the non-electron one $\nu_x$. 
We will comment later about possible three-flavor effects associated with
$\delta m^2$ and $\theta_{12}$~\cite{Dasgupta:2007ws,Fogli:2008fj}.

In the $2\nu$ framework, one can expand  the $2 \times 2$ matrices in the EoMs in terms of the  unit matrix $\sf{I}$ and the Pauli matrices \hbox{\boldmath$\sigma$}.
One can introduce polarization vectors ${\bf P}_{p,u,r}$ from the expansion of the
flux matrices  ${\sf J}_{p,u,r}$.
  One of the possible definitions
suitable in the SN context is the following~\cite{EstebanPretel:2007ec}
\begin{eqnarray}
{\sf J}_{p,u,r} &=& \frac{F(\nu_e)_{p,u} +F(\nu_x)_{p,u}}{2} + 
\frac{F(\bar\nu_e) -F(\bar\nu_x)}{2} {\bf P}_{p,u,r} \cdot\hbox{\boldmath$\sigma$} \,\ , \nonumber \\
\overline{\sf J}_{p,u,r} &=& \frac{F(\bar\nu_e)_{p,u} +F(\bar\nu_x)_{p,u}}{2} + 
\frac{F(\bar\nu_e) -F(\bar\nu_x)}{2} \overline{\bf P}_{p,u,r} \cdot\hbox{\boldmath$\sigma$}  \,\ ,
\label{eq:polarJ}
\end{eqnarray}
where the quantities $F(\nu_\alpha)_{p,u}$ are the number
flux of $\nu_\alpha$ emitted from the neutrinosphere with
energy $E=|{\bf p}|$ and emission angle $u= \sin^2 \theta_R$. 
The total number flux of flavor $\alpha$ is a simple integral $F(\nu_\alpha)=
\int dp \int d u F(\nu_\alpha)_{p,u}$.
Note that the normalization of
the second term in Eq.~(\ref{eq:polarJ}) is the same for $\nu$ and $\bar\nu$ and it is the difference between the  $\bar\nu_e$'s and
$\bar\nu_x$'s fluxes.
We factorize the flux of each flavor as $F(\nu_\alpha)_{p,u}= N_{\nu_\alpha} \times f_{\nu_{\alpha}}(p) \times \varphi_{\nu_{\alpha}}(u)$.
The $\nu$ number $N_{\nu_\alpha} = L_{\nu_\alpha}/ \langle E_{\nu_\alpha} \rangle$ is expressed in terms of the
$\nu$ luminosity $L_{\nu_\alpha}$ and of the $\nu$ average energy $\langle E_{\nu_\alpha} \rangle$ of the
different species. The function $f_{\nu_\alpha}(p)$ is the normalized neutrino energy spectrum ($\int dp f_{\nu_\alpha}(p) =1$).
An example of a quasi-thermal energy spectrum, widely used in SN literature is given in Eq.~(\ref{eq:fitspectrum}).
Finally, $\varphi_{\nu_\alpha}(u)$ represents the $\nu$ angular distribution that we assume flat in $u$ and flavor and energy independent, 
i.e. $\varphi_{\nu_\alpha}(u)=1$.
 
 One can expand the other quantities in Eqs.~(\ref{eq:eom})-(\ref{eq:vac})-(\ref{eq:ham1}) as
\begin{eqnarray}\label{eq:matrices}
 \Omega^{\textrm{vac}}_{\bf p}&=&{\textstyle\frac{1}{2}}
 \bigl(\omega_0 {\sf I}+\omega_{\bf p}\,{\bf B}
 \cdot\hbox{\boldmath$\sigma$}\bigr)\,,\nonumber\\
  {\sf L} &=&{\textstyle\frac{1}{2}}
 \bigl({\sf I}+\,{\bf L}
 \cdot\hbox{\boldmath$\sigma$}\bigr)\,,
 \end{eqnarray}
where $\omega_0= m^2_{1,2} + m^2_3$, and the vacuum oscillations are determined by
the mass difference $\Delta m^2$ and vacuum mixing angle $\theta_{13}$~\cite{Hannestad:2006nj}:
\begin{eqnarray}\label{eq:vdef}
 \omega_{\bf p}&=& \Delta m^2/2p\,,\nonumber\\
 {\bf B}&=&(\sin2\theta_{13},0, \mp\cos2\theta_{13})\,\ ,
\end{eqnarray}
with the upper (lower) sign refers to normal (inverted) mass hierarchy. For small $\theta_{13}$, ${\bf B} \simeq \mp {\bf z}$. Moreover,
in an ordinary medium composed by electrons and positrons, ${\bf L}= {\bf e}_z$.

Using the definition of  Eqs.~(\ref{eq:polarJ})--(\ref{eq:vdef})  one can write the EoMs 
[Eq.~(\ref{eq:eom})] for the polarization vectors~\cite{EstebanPretel:2007ec} 

\begin{eqnarray}
\label{eq:polarbulb}
\partial_r {\bf P}_{p,u,r} &=& \bigg[\frac{1}{v_{r,u}}(\omega_p {\bf B} + \lambda_r {\bf L})
 \nonumber \\
 &+&  \mu_r \int d q \int du^\prime ({\bf P}_{q,u^\prime,r}-\overline{\bf P}_{q,u^\prime,r}) \bigg(\frac{1}{v_{r,u} v_{r,u^\prime}}-1 \bigg) \bigg]
 \times {\bf P}_{p,u,r} \,\ , \nonumber \\
 \partial_r \overline{\bf P}_{p,u,r} &=& \bigg[\frac{1}{v_{r,u}}(-\omega_p {\bf B} + \lambda_r {\bf L})
 \nonumber \\
 &+&  \mu_r \int d q \int du^\prime ({\bf P}_{q,u^\prime,r}-\overline{\bf P}_{q,u^\prime,r}) \bigg(\frac{1}{v_{r,u} v_{r,u^\prime}}-1 \bigg) \bigg]
 \times \overline{\bf P}_{p,u,r} \,\ ,
\end{eqnarray}
where we introduced the self-interaction potential
\begin{equation}
\mu_r = \sqrt{2}G_F \frac{F(\bar\nu_e) -F(\bar\nu_x)}{4 \pi r^2} \,\ ,
\label{eq:selfpoten}
\end{equation}
and the radial neutrino velocity
\begin{equation}
v_{u,r} = \cos \theta_r = \sqrt{1-u \left(\frac{R}{r} \right)^2} \,\ .
\end{equation}
The  initial conditions
for the polarization vectors at the neutrinosphere are the following
\begin{eqnarray}
P_{p,u}^z (R) &=& \frac{F(\nu_e)_{p,u} -F(\nu_x)_{p,u}}{F(\bar\nu_e) -F(\bar\nu_x)} \,\ , \nonumber \\
\overline{P}_{p,u}^z (R) &=& \frac{F(\bar\nu_e)_{p,u} -F(\bar\nu_x)_{p,u}}{F(\bar\nu_e) -F(\bar\nu_x)} \,\ .
\end{eqnarray}

The presence of the trajectory-dependent factor
$v_{u,r}$ in the neutrino-neutrino interaction term  in  Eq.~(\ref{eq:polarbulb}) leads to the multi-angle effects in the self-induced oscillations. 
In this regard, it has been understood~\cite{Raffelt:2007yz}  that the flavor asymmetry 
is crucial to assess the impact   of multi-angle effects. 
In particular, in the presence of a sufficiently large flux hierarchy among different species, multi-angle
decoherence would be suppressed.

\subsubsection{\emph{Matter suppression of self-induced flavor conversions}}
\label{sec:mattsupp}

Remarkably also the matter term in Eq.~(\ref{eq:polarbulb}) is affected by   multi-angle effects due to the factor $v_{u,r}$.
 It has been shown in~\cite{EstebanPretel:2008ni} that matter effects
play a sub-dominant role in the development of  self-induced flavor conversions  when 
\begin{equation}
n_{e^-}- n_{e^+} \ll n_{\bar\nu_e} - n_{\bar\nu_x}  \,\ .
\end{equation}
 Conversely, if $n_{e^-}- n_{e^+} \gg n_{\bar\nu_e} - n_{\bar\nu_x}$, the multi-angle matter effects
 could produce a large spread in the oscillation frequencies for neutrinos travelling on different trajectories, 
 blocking the self-induced flavor conversions. Finally, when $n_{e^-}- n_{e^+} \sim n_{\bar\nu_e} - n_{\bar\nu_x} $,
   matter-induced multi-angle decoherence in the neutrino ensemble  may occur.
 According to realistic SN models, the matter density is expected to dominate over the neutrino one during the early accretion phase (see Sec.~\ref{sec:profiles}). 
 In this situation, dedicated studies have been performed 
in~\cite{Chakraborty:2011nf,Chakraborty:2011gd,Dasgupta:2011jf,Sarikas:2011am,Saviano:2012yh,Chakraborty:2014nma}  using inputs from  recent   hydrodynamic SN simulations. It has been found that the large matter term  inhibits the 
development of collective oscillations at early times for iron-core SNe.
Only in the case of low-mass  O-Ne-Mg SNe, where the matter density  is never larger than the neutrino one,
the matter suppression is not complete and partial self-induced flavor conversions are possible 
at early times~\cite{Chakraborty:2011nf,Chakraborty:2011gd,Chakraborty:2014nma}.

The matter suppression of self-induced effects for iron-core SNe makes relatively easy the prediction of the oscillation effects during the accretion phase, since
the neutrino fluxes will be processed by the only MSW matter effects. As a consequence, 
as we will discuss in Sec.~\ref{sec:signatures},  the detection of the SN neutrino signal at early times is particularly relevant for 
the mass hierarchy discrimination.
Nevertheless, for all types of SN progenitors
 self-induced effects still remain crucial during the later cooling phase, when the
matter density decreases continuously (at post-bounce times $t_{\rm pb} \gtrsim 1$~s) and becomes sub-dominant with respect to 
the neutrino density.   

 As final caveat, we  remark that
 the  matter suppression of collective oscillations studied in~\cite{Chakraborty:2011nf,Chakraborty:2011gd,Dasgupta:2011jf,Sarikas:2011am,Saviano:2012yh,Chakraborty:2014nma}  has been characterized 
 referring to spherically symmetric one-dimensional SN models.
 However, as discussed in Sec.~\ref{sec:SNmodels},
large deviations from a spherical neutrino emission 
can be generated by  hydrodynamical instabilities such as SASI or LESA especially during the accretion phase.
Before the explosion sets in, during the standing-accretion-shock phase, the  recently discovered
LESA instability~\cite{Tamborra:2014a} is responsible for a  neutrino lepton number flux emerging primarily in one hemisphere (see Sec.~\ref{sec:SNmodels}).
The asymmetry between $\nu_e$ and $\bar{\nu}_e$ can be very small and even negative  in some regions of the star, 
becoming potentially responsible for flavor instabilities~\cite{EstebanPretel:2007ec}.  LESA also influences the electron density and therefore the matter background felt by neutrinos. Moreover, the neutrino flavor ratio can also be affected by SASI that
seems to coexist with LESA in more massive SN progenitors (see Sec.~\ref{sec:SNmodels}) and that is responsible for
wild shock oscillations and a time-dependent directional bias of the  neutrino emission. 
Exploratory work conducted within a simplified setup~\cite{Chakraborty:2014lsa} confirmed the suppression of flavor conversions within the shock front radius in the presence of LESA. However, such results need to be tested within a more realistic scenario and in the presence of SASI. Besides the relevance for phenomenological purposes, the confirmation of the possible suppression of self-induced  flavor conversions will also be  crucial for the modeling of the neutrino propagation in the  hydrodynamical SN simulations.

\subsection{Supernova potential profiles}
\label{sec:profiles}

\begin{figure}[t!]
\begin{center}
 \includegraphics[angle=0,width=0.9\textwidth]{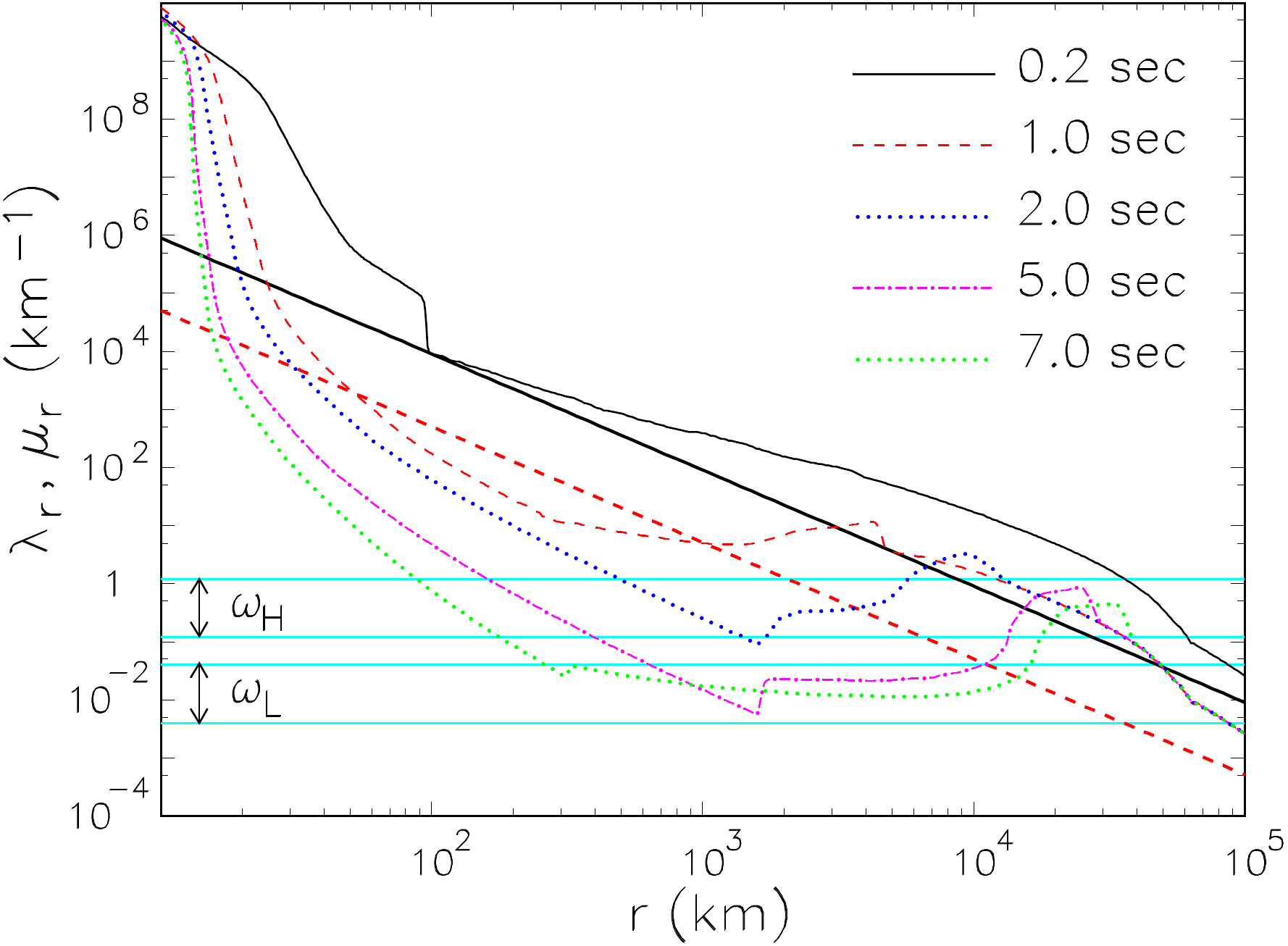} 
    \end{center}
\caption{Snapshots of SN potentials for different post-bounce times
$(1.0-7.0\,$s)
for a $27$ M$_{\odot}$  SN 
progenitor (see Sec.~\ref{sec:SNmodels}). 
The profile at $0.2\,$s is an illustrative case for a typical condition
before shock revival.
The matter potential $\lambda_r$ is drawn with thin curves,
while the neutrino potential $\mu_r$ with thick ones.
The horizontal bands represent the vacuum
oscillation frequencies relevant
for the MSW resonant conversions associated with  $\Delta m^2$ ($\omega_H$) 
and  $\delta m^2$ ($\omega_L$), respectively (see the text for details). \label{densities}} 
\end{figure}

Before discussing the solution of the neutrino EoMs  for the dense SN neutrino gas, 
we find it worthwhile  to show    snapshots of  the interaction potentials appearing in the r.h.s. of Eq.~(\ref{eq:polarbulb}) in Fig.~\ref{densities}. We use the data 
from a  $27$ M$_{\odot}$  SN simulation for different post-bounce times (see Sec.~\ref{sec:SNmodels}). 
As from Eq.~(\ref{eq:selfpoten}), the neutrino-neutrino potential is
\begin{eqnarray}
\mu_r &=&\sqrt{2}G_F \frac{F(\bar\nu_e) -F(\bar\nu_x)}{4 \pi r^2}= \frac{1}{4 \pi r^2} \left(\frac{L_{\bar\nu_e}}{{\langle E_{\bar\nu_e}\rangle}} 
- \frac{L_{{\bar\nu}_x}}{{\langle E_{{\bar\nu}_x}\rangle}}  \right) \, \nonumber \\
&=& 7.0 \times 10^{5} \,\ \textrm{km}^{-1} 
\left(\frac{L_{\bar\nu_e}}{\langle E_{\bar\nu_e}\rangle} -
\frac{L_{\bar\nu_x}}{\langle E_{\bar\nu_x}\rangle}
 \right)
\frac{15 \,\ \textrm{MeV}}{10^{52} \,\ \textrm{erg}/\textrm{s}}
\left(\frac{10 \,\ \textrm{km}}{r} \right)^2 \,\ ,
\label{eq:mur}
\end{eqnarray}
where the number fluxes of the different species $\nu_\alpha$ are expressed in terms
of the neutrino luminosities $L_{\nu_\alpha}$ and of the average energies  
$\langle E_{\nu_\alpha} \rangle$.
As shown in  Fig.~\ref{densities}, the $\nu$-$\nu$  potential decreases as the SN cools.

Matter effects on SN neutrinos [Eq.~(\ref{eq:polarbulb})] depend on the potential 
\begin{eqnarray}
\lambda_r &=& \sqrt{2}G_F  n_{e}(r) = 1.9 \times 10^6  \,\ 
\textrm{km}^{-1} 
\times  \left(\frac{Y_e}{0.5} \right) 
\left(\frac{\rho}{10^{10} \,\ \textrm{g}/\textrm{cm}^{3}} \right) \,\ ,
\end{eqnarray}
encoding the net electron density $n_e\equiv n_{e^-}- n_{e^+}$, where
$Y_e=Y_{e^-} - Y_{e^+}$ is the net electron fraction and
$\rho$ is the matter density. 
 The numerical values of $\mu_r$ and $\lambda_r$ from the previous two expressions
 are quoted in km$^{-1}$.

Figure~\ref{densities} shows that the SN electron density profile is non-monotonic, time-dependent, and presents an abrupt discontinuity,
corresponding to the position of the \emph{forward shock-front}. In fact, the
shock-wave, while propagating outwards at supersonic speed, leaves behind
a rarefaction zone, and creates a high-density front with a sharp density drop
(down to the static value) at 
\footnote{Note that in hydrodynamical simulations, like the ones shown in Fig.~\ref{densities}, 
the forward shock-front is typically  softened because of the limited numerical
resolution. Therefore it has to be steepened by hand when adopted for studying the neutrino
flavor evolution.}
\begin{equation}
r_{\rm fwd} = \textrm{forward shock radius} \,\ .
\end{equation}
We also note that additional features can
appear such as the \emph{reverse}  shock propagation~\cite{Tomas:2004gr}. For example, after the
explosion, a neutrino-driven baryonic wind can develop and collide with the
(slower) SN ejecta, thus triggering a second (reverse) shock at
\begin{equation}
r_{\rm rev} = \textrm{reverse shock radius} \,\ ,
\end{equation}
which propagates (at lower velocity) behind the forward one~\cite{Tomas:2004gr}. Neutrinos
may thus encounter two subsequent density discontinuities, leading to
significantly different spectral features with respect to the case of a single
discontinuity.

One expects that the matter term would lead to \emph{resonant} flavor conversions 
via the  MSW effect~\cite{Matt} when the matter potential is of the order of the vacuum
oscillation frequencies [Eq.~(\ref{eq:vacfreq})]  $\omega_{H,L}$~\cite{Di00}, i.e.
\begin{equation}
\lambda_r \simeq \omega_{L,H} \,\ .
\end{equation}
These oscillation frequencies are represented by the two horizontal strips in Fig.~\ref{densities} 
for a neutrino energy range $E \in [5-50]$ MeV. Note that resonant flavor conversions should be expected for
$r \in [10^3, 10^5]$~km.

Comparing the neutrino and matter density profiles, we realize  that in the deepest SN regions 
$n_\nu \gg n_e$ ($r <10^3$ km, see $t=1.0$~s in Fig.~\ref{densities}), except during the accretion phase ($t=0.2$~s in Fig.~\ref{densities}).
When  the neutrino density dominates over the matter one (as during the cooling phase), 
self-induced flavor conversions would develop without any hindrance.
This schematic investigation suggests that intriguing effects should be expected in the SN neutrino flavor
evolution, with an  interplay between self-induced and matter terms. 
Moreover, self-induced oscillations usually develop at lower $r$ than the MSW flavor conversions, so that
their effects effects could be studied separately for this progenitor.

\subsection{Self-induced flavor conversions: Single-angle approximation}
\label{sec:single-angle}

In order to simplify the numerical treatment of the EoMs 
 [Eq.~(\ref{eq:polarbulb})] and to develop
analytical  interpretations of  neutrino self-interactions, 
it has been often adopted in the literature the so-called \emph{``single-angle''} approximation~\cite{Duan:2006an,Fogli:2007bk,EstebanPretel:2007ec}. 
The main idea is to assume a single angular mode representative of all the neutrino ensemble. For a
blackbody emission, in which all of the angular modes are equally occupied,
it is natural to take neutrinos emitted at $\theta_R= \pi/4$ (i.e., $u_0 = 0.5$) as the representative
ones%
~\footnote{A formal derivation of the single-angle approximation has been presented in~\cite{Dasgupta:2008cu}.}.
 In this case, there is no integration over $u$ and all the radial
velocities appearing in the EoMs are set to  
\begin{equation}
v_{r,u_0} =\sqrt{1-\frac{R^2}{2r^2}} \,\ .
\end{equation}

The EoMs in the single-angle approximation read
\begin{eqnarray}
\partial_r {\bf P}_p =\left[+ \omega_p {\bf B}+ \lambda_r {\bf L} + \mu^\ast_r {\bf D} \right]
\times {\bf P}_p \,\ , \nonumber \\
\partial_r \overline{\bf P}_p =\left[- \omega_p {\bf B}+ \lambda_r {\bf L} + \mu^\ast_r {\bf D} \right]
\times \overline{\bf P}_p \,\ ,
\label{eq:singleanglepolar}
\end{eqnarray}
where the self-interaction term only depends on 
 the radial neutrino self-interaction
strength
\begin{eqnarray}
\mu^\ast_r = \sqrt{2}G_F \frac{F(\bar\nu_e) -F(\bar\nu_x)}{4 \pi R^2}
\frac{R^4}{2 r^4}\frac{1}{1-\frac{R^2}{2 r^2}} 
= \mu_r \frac{R^2}{2 r^2} \frac{1}{1-\frac{R^2}{2 r^2}} \,\ ,
\label{eq:muast}
\end{eqnarray}
that at large distances from the core  declines as
$\mu^\ast_r \sim R^4/2r^4$,
and on the difference of the total polarization vectors
 ${\bf D} = {\bf P} - \overline{\bf P}$, where ${\bf P} = \int dp {\bf P}_p$
(and analogously for  $\overline{\bf P}$). In order to simplify the notation we  remove
the subscript  $r$ in the polarization vectors.

Assuming that self-induced effects are 
not suppressed by the multi-angle effects associated with a dominant matter term,
it has been shown that in the single-angle approximation, when $\lambda_r \gg \omega_p$,   the EoMs can be studied in a co-rotating frame with angular
 velocity $\lambda_r {\bf L}$~\cite{Hannestad:2006nj,Fogli:2007bk,Duan:2007mv}. 
After this rotation,
the only notable matter effect would be the suppression of the in-medium mixing angle with respect to the vacuum one. 

The EoMs [Eq.~(\ref{eq:singleanglepolar})] for $\lambda_r=0$ imply conservation of the scalar 
product
\begin{equation}
{\bf D}\cdot{\bf B}=\mathrm{const}={\bf D}^i\cdot{\bf B}\simeq
\mp{\bf D}^i\cdot{\bf z}=\mp \frac{F_{\nu_e} - F_{\overline{\nu}_e}}{F_{\bar\nu_e}- F_{\bar\nu_x}} 
\equiv \mp \epsilon \,\ ,
\label{eq:lepton}
\end{equation}
where $\epsilon$ indicates the flavor asymmetry. 
Equation~(\ref{eq:lepton})
corresponds to the conservation of the  lepton number~\cite{Hannestad:2006nj},
and implies \emph{pair conversions}  $\nu_e\overline\nu_e\to \nu_x\overline \nu_x$~\cite{Hannestad:2006nj}. 
In the co-rotating frame ${\bf D}\cdot{\bf L}$ is the corresponding conserved quantity.

\subsection{Synchronized vs. bipolar oscillations and spectral splits}
\label{sec:selfinduc}

\begin{figure}[t!]
\begin{center}
 \includegraphics[angle=0,width=0.9\textwidth]{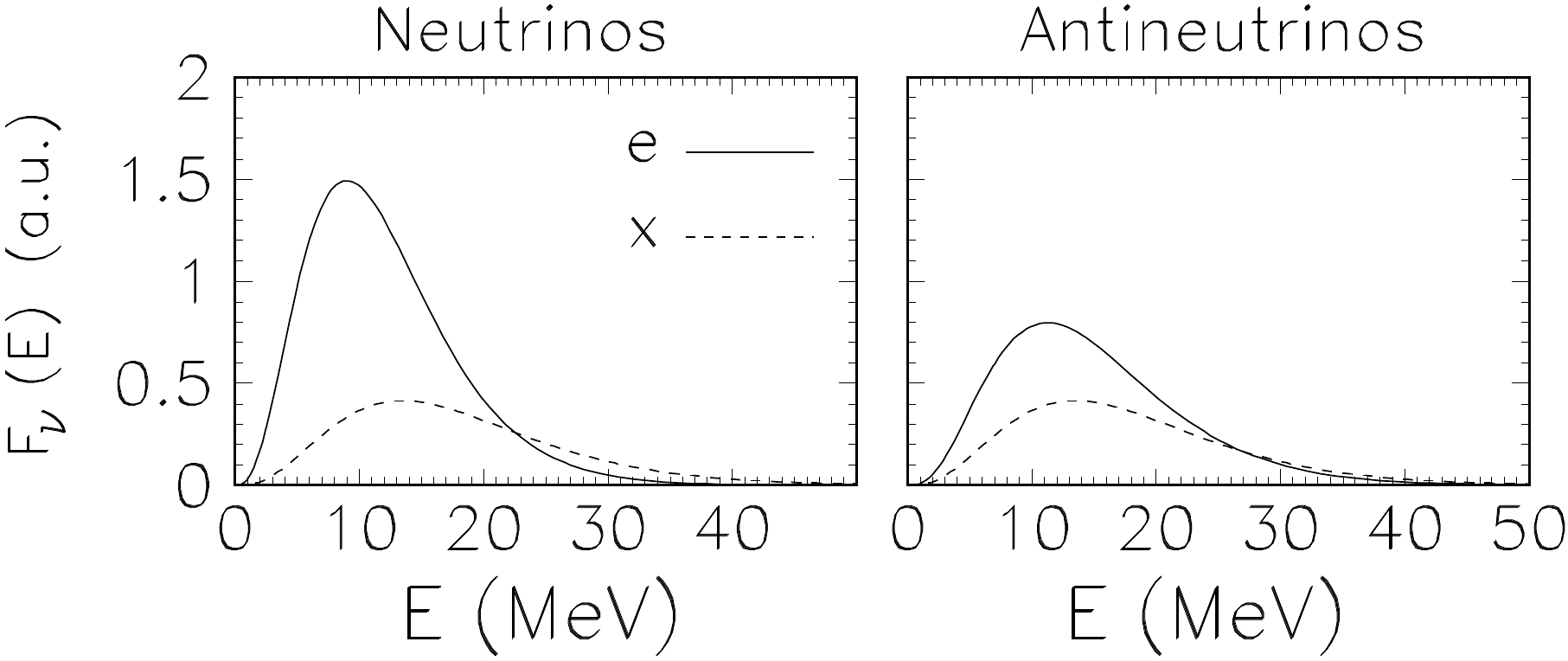} 
    \end{center}
\caption{Initial SN neutrino and antineutrino fluxes with 
$F^0_{\nu_e}:F^0_{\bar\nu_e}:F^0_{\nu_x}=2.40:1.60:1.0$. Neutrinos are plotted in the panel on the left, 
while antineutrinos are on the right panel. Electron species ($e$) are plotted with continuous curves, while
non-electron species ($x=\mu,\tau$) with dashed curves. 
 \label{spectraaccr}} 
\end{figure}

We will start to characterize the  SN neutrino flavor conversion phenomenology, first discussing 
 self-induced effects associated with the neutrino-neutrino interactions 
in the single-angle approximation [Eq.~(\ref{eq:singleanglepolar})]. Assuming to be in a co-rotating frame 
such that $\lambda_r=0$ (see Sec.~\ref{sec:single-angle}), 
the EoMs for the flavor evolution of the dense SN neutrino gas, even in the single-angle approximation,
 present a rich phenomenology  since neutrino-neutrino
interactions are strongly  dependent
on the ordering and hierarchy of  the SN neutrino fluxes of the different species. Since the ordering of the SN fluxes
 changes significantly during the different post-bounce phases, one should expect  a rich phenomenology
 as a function of the time after the bounce~\cite{Duan:2006an,Fogli:2007bk,Fogli:2008fj,Dasgupta:2009mg,Friedland:2010sc,Dasgupta:2010ae,Dasgupta:2010cd,Fogli:2009rd}.  

In this Section, we  take as illustrative example   a ``classical'' case, where the flavor
dynamics of the polarization vectors under the self-induced effects is   analytically well-understood 
through  an analogy with  a
\emph{gyroscopic pendulum in flavor space}~\cite{Hannestad:2006nj,Fogli:2007bk,Duan:2007mv,Duan:2007fw}.
 This behavior is realized for cases presenting a large excess of electron neutrinos over
non-electron species, i.e. $F^0_{\nu_e} \gg F^0_{\bar\nu_e} \gg F^0_{\nu_x}$ (as expected, e.g.,
 during the accretion phase). 
In the pendulum analogy, the generic motion 
of the polarization vectors in flavor space is a combination
of nutation and precession. The neutrino-antineutrino asymmetry $\epsilon$ [Eq.~(\ref{eq:lepton})]  acts as a ``spin'' for
the system with a large asymmetry inducing a precession-like motion around $ \omega_p {\bf B}$,
leaving only the nutation motion when the asymmetry is vanishing~\cite{Hannestad:2006nj}.
The role of the inertia moment in the pendulum analogy is played by the inverse strength of
neutrino-neutrino interaction $\mu^\ast_r$ [Eq.~(\ref{eq:muast})].
The neutrino mass hierarchy sets the behavior of the system.
In NH, the pendulum starts in downward (stable) position and,
assuming a small mixing angle, stays nearby, i.e. only manifests small flavor changes.
Conversely, the pendulum starts in upward (unstable) position in IH---the misalignment being of $\mathcal{O}(\theta_{13})$---and it is subject
to the maximal flavor reversal  \emph{pair-conversions} $\nu_e\overline\nu_e\to \nu_x\overline \nu_x$~\cite{Hannestad:2006nj}, that conserve
the lepton number [Eq.~(\ref{eq:lepton})]. 
It has been shown that  large flavor asymmetry could block possible multi-angle decoherence
effects~\cite{EstebanPretel:2007ec}. Therefore, the single-angle approximation 
is well justified in this case~\cite{Fogli:2007bk,Fogli:2008pt}.
Figure~\ref{spectraaccr} shows  an example of this flux configuration, obtained considering representative neutrino fluxes with the following spectral ordering at the neutrinosphere: $F^0_{\nu_e}:F^0_{\bar\nu_e}:F^0_{\nu_x}=2.40:1.60:1.0$.

\begin{figure}[t!]
\begin{center}
 \includegraphics[angle=0,width=0.75\textwidth]{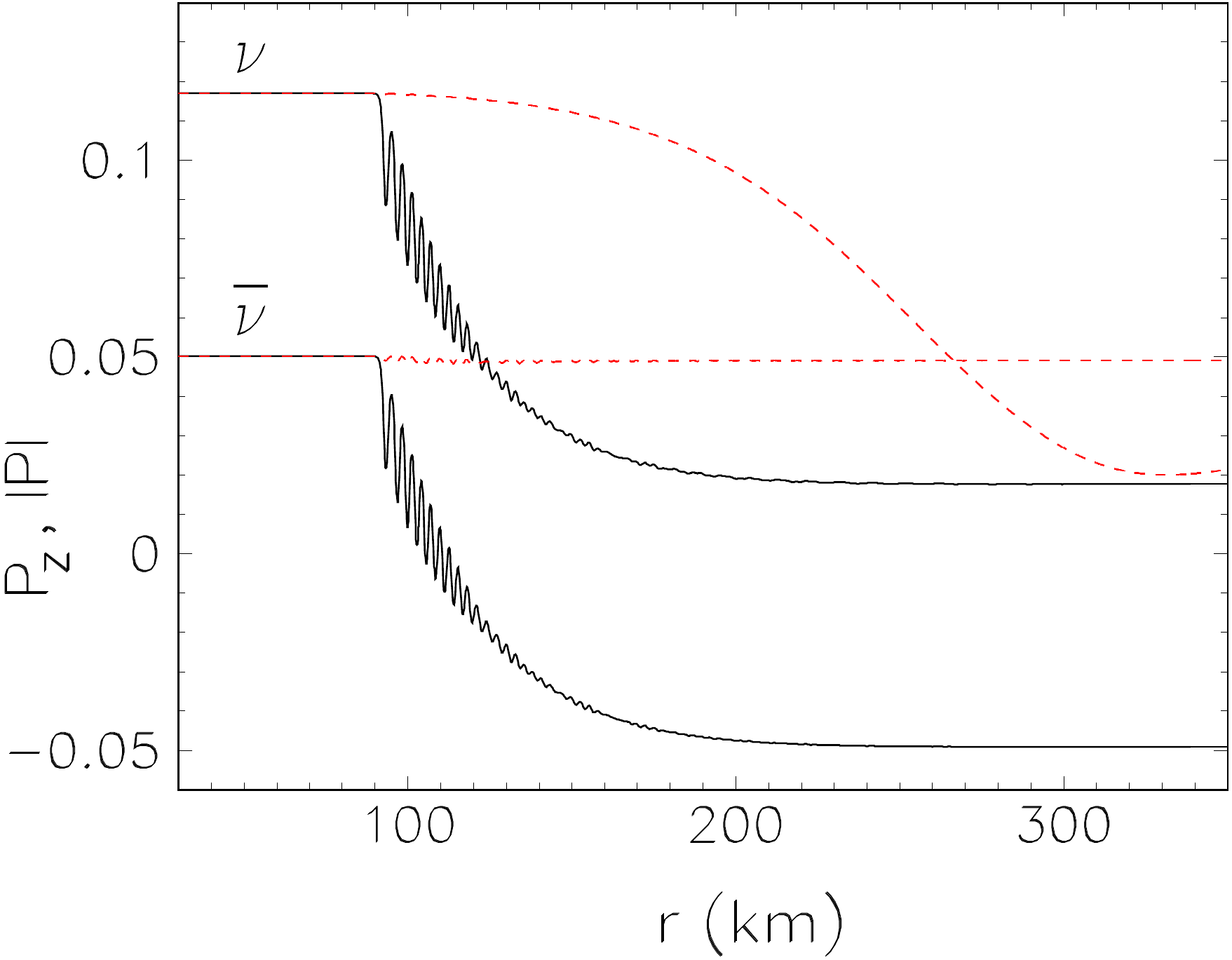} 
    \end{center}
\caption{
Single-angle simulation in inverted mass hierarchy for an initial flux  ordering
$F^0_{\nu_e}:F^0_{\bar\nu_e}:F^0_{\nu_x}=2.40:1.60:1.0$. Modulus (dashed curves) and $z$-component of the global polarization vector
(continuous curves)
of neutrinos $(\bf P)$ and antineutrinos $(\overline{\bf P})$, as a function of radius. 
\label{globalpol}}
\end{figure}

In order to illustrate the self-induced flavor dynamics, 
Fig.~\ref{globalpol}  shows the radial evolution of the modulus $P=|{\bf P}|$ and
$z$-component $P_z$ of the global neutrino polarization vector
$\bf P$ (and analogously for the antineutrino vector $\overline{\bf P}$)
in the nontrivial case of IH for the fluxes shown in Fig.~\ref{spectraaccr} (no significant effect occurs in NH).
The radial evolution of the polarization vectors can be interpreted as follows.  
Up to 
$\sim 90$~km,
it is $P=P_z$ and $\overline P=\overline P_z$: all polarization
vectors are ``glued'' along the vertical axis. In the gyroscopic pendulum analogy,
this corresponds to a precessing top  with a huge spin and
negligible displacement from the vertical $z$-axis. 
It  just spins in the upward position without falling.
 This behavior is named
 \emph{synchronized oscillation} regime~\cite{Hannestad:2006nj,Pastor:2001iu}.
At $r\sim 90$~km, the pendulum falls for the first time
and nutations appear, marking the transition from
synchronized to the so-called \emph{bipolar oscillation}  regime%
~\footnote{Notably, matter effects   delay this 
transition  by a few 
nutation periods with respect to what expected in the presence of 
 $\nu$-$\nu$ interactions only~\cite{Hannestad:2006nj}.}.
The nutation amplitude gradually decreases
as $\sim {\mu^\ast_r}^{1/2}$ being driven down by the decreasing
$\nu$-$\nu$ potential strength.
In the spinning
top analogy, the relaxation of the pendulum to its downward
rest position as kinetic energy is extracted by the reduction of the neutrino-neutrino
interaction potential and thus the increase of the pendulum moment of inertia.
This would lead to large self-induced flavor transitions, occurring in the form of  \emph{pair-conversions} $\nu_e\overline\nu_e\to \nu_x\overline \nu_x$~\cite{Hannestad:2006nj} that conserve
the lepton number [Eq.~(\ref{eq:lepton})].

As a consequence of the self-induced flavor dynamics, 
antineutrinos tend to completely
reverse the polarization vector $(\overline{\bf P}\to -\overline{\bf P})$,
so that $\overline P_z\simeq-\overline P$ asymptotically.
Neutrinos also try to invert their global polarization vector. Then,  $|{\bf P}|$ decreases for $r \gtrsim 150$~km.
However, the inversion cannot be complete, because of the  
lepton number conservation $D_z=P_z-\overline P_z$ at any $r$ [see Eq.~(\ref{eq:lepton})]. 
We remark that the inversion of the polarization vectors corresponds to large flavor conversions occurring
at low-radii ($r < 200$~km) where the system would not have exhibited any flavor change in 
the presence of matter effects only.

\begin{figure}[t!]
\begin{center}
 \includegraphics[angle=0,width=0.9\textwidth]{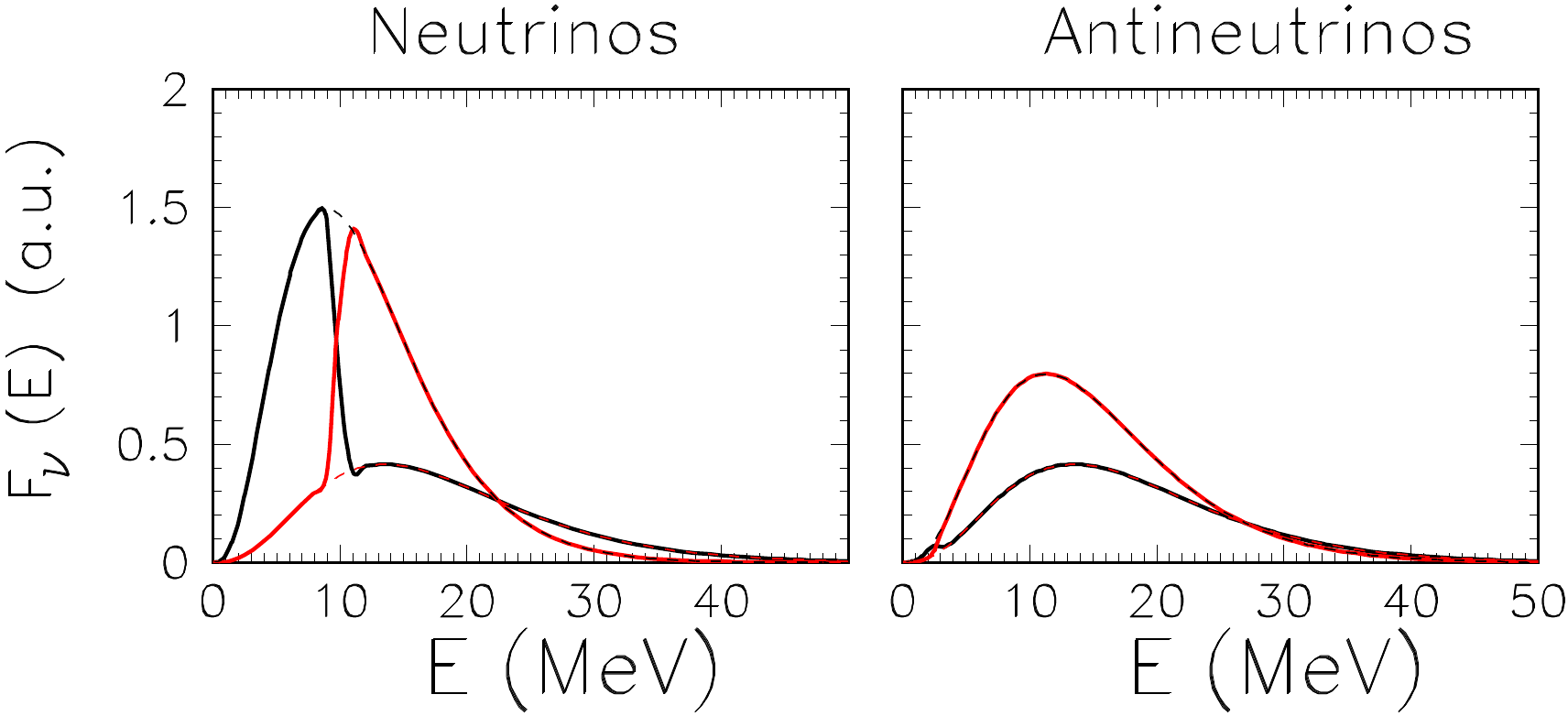} 
    \end{center}
\caption{Two-flavor
single-angle simulation in inverted hierarchy for   an initial flux  ordering
$F^0_{\nu_e}:F^0_{\bar\nu_e}:F^0_{\nu_x}=2.40:1.60:1.0$.   Initial energy spectra  for
$\nu_e$ (black dashed curves) and $\nu_x$ (light dashed curves)
and after collective oscillations for $\nu_e$ (black continuous curves)
and $\nu_x$ (light continuous curves) at $r=350$~km.
 \label{splitaccr}} 
\end{figure}

A generic outcome of the self-induced flavor conversions is the development of
``spectral swaps'' among the fluxes of different flavors, marked by
 ``spectral splits'' at the boundary features  of each
swap interval~\cite{Friedland:2010sc,Dasgupta:2010cd,Duan:2007bt,Raffelt:2007cb,Raffelt:2007xt,Duan:2008za,Dasgupta:2008cd,%
Galais:2011gh,Choubey:2010up}. 
In order to elucidate these effects with the specific example we are discussing, we refer to the 
flavor evolution of the total polarization vectors ${\bf P}$ and  $\overline{\bf P}$ shown in Fig.~\ref{globalpol},
for the initial fluxes of Fig.~\ref{spectraaccr}.
One would not expect  complete flavor conversions in both
neutrino and antineutrino sectors, due to the partial inversion of  ${\bf P}$.  
The initial spectra shown in Fig.~\ref{spectraaccr} give
a larger number of $\nu_e$ over $\bar\nu_e$ and $\nu_x$. 
Therefore,  if all antineutrinos are swapped, the conservation of the net
lepton number~\cite{Fogli:2007bk}
\begin{equation}
\int_{E_{\rm split}}^\infty dE (F_{{\nu_e}}-F_{{\nu_x}})
= \int_{0}^{\infty} dE (F_{{\bar\nu_e}}-F_{{\bar\nu_x}}) \,\ ,
\label{eq:splitcons}
\end{equation}
prevents the same thing happening to the neutrino channel.
This explains the appearance of the  spectral neutrino splits
in the oscillated fluxes. 
The previous equation allows us to determine the ``spectral split'' energy
${E_{\rm split}}$ separating the swapped part of the spectrum from the unswapped one.
This is shown in Fig.~\ref{splitaccr}, where we plot the initial energy spectra  for
$\nu_e$ (black dashed curves) and $\nu_x$ (light dashed curves)
and after collective oscillations for $\nu_e$ (black continuous curves)
and $\nu_x$ (light continuous curves) at $r=350$~km.
Neutrino oscillated spectra (left panel)  clearly show the  split
effect and  also the corresponding sudden swap of $\nu_e$ and $\nu_x$ fluxes above
$E_{\rm split} \simeq 10$~MeV. The antineutrino spectra (right panel) are almost completely swapped with respect to the
initial ones. However, they also present a spectral split 
 at low-energy $E_{\rm split} \simeq 3$~MeV. This feature is not explained by the conservation law expressed
 by Eq.~(\ref{eq:splitcons}) and its nature is associated with non-adiabatic features of the flavor evolution,
 as discussed in~\cite{Fogli:2008pt}.
We address the interested reader to~\cite{Raffelt:2007cb,Raffelt:2007xt,Duan:2008za} for a detailed theoretical explanation  of the spectral splits
through the adiabatic solution of the EoMs.

\subsection{Self-induced spectral splits and multi-angle effects}
\label{sec:difffluxes}

For a   long time the results shown  in Fig.~\ref{splitaccr}---an almost complete flavor exchange in
the antineutrino sector and a spectral split in the neutrino channel in IH,  no conversion
in NH, and sub-leading multi-angle effects---have been considered the paradigm of the 
self-induced effects in SN neutrino flavor conversions. 
However, later it has been realized that this is not the most general situation~\cite{Dasgupta:2009mg}. 
Indeed,  when $\nu$'s  have  a moderate 
 flavor hierarchy of fluxes and spectral
energies are not large, as expected during the cooling phase, more complicated conversion patterns 
could be associated with the self-induced effects~\cite{Fogli:2009rd}.
 \emph{Multiple} spectral splits are possible  in both
NH and IH~\cite{Dasgupta:2009mg,Fogli:2009rd} for neutrinos and antineutrinos.
Under such conditions, also multi-angle and three-flavor effects may play a crucial role. 

\begin{figure}[!h]
\centering
\includegraphics[width=0.52\textwidth]{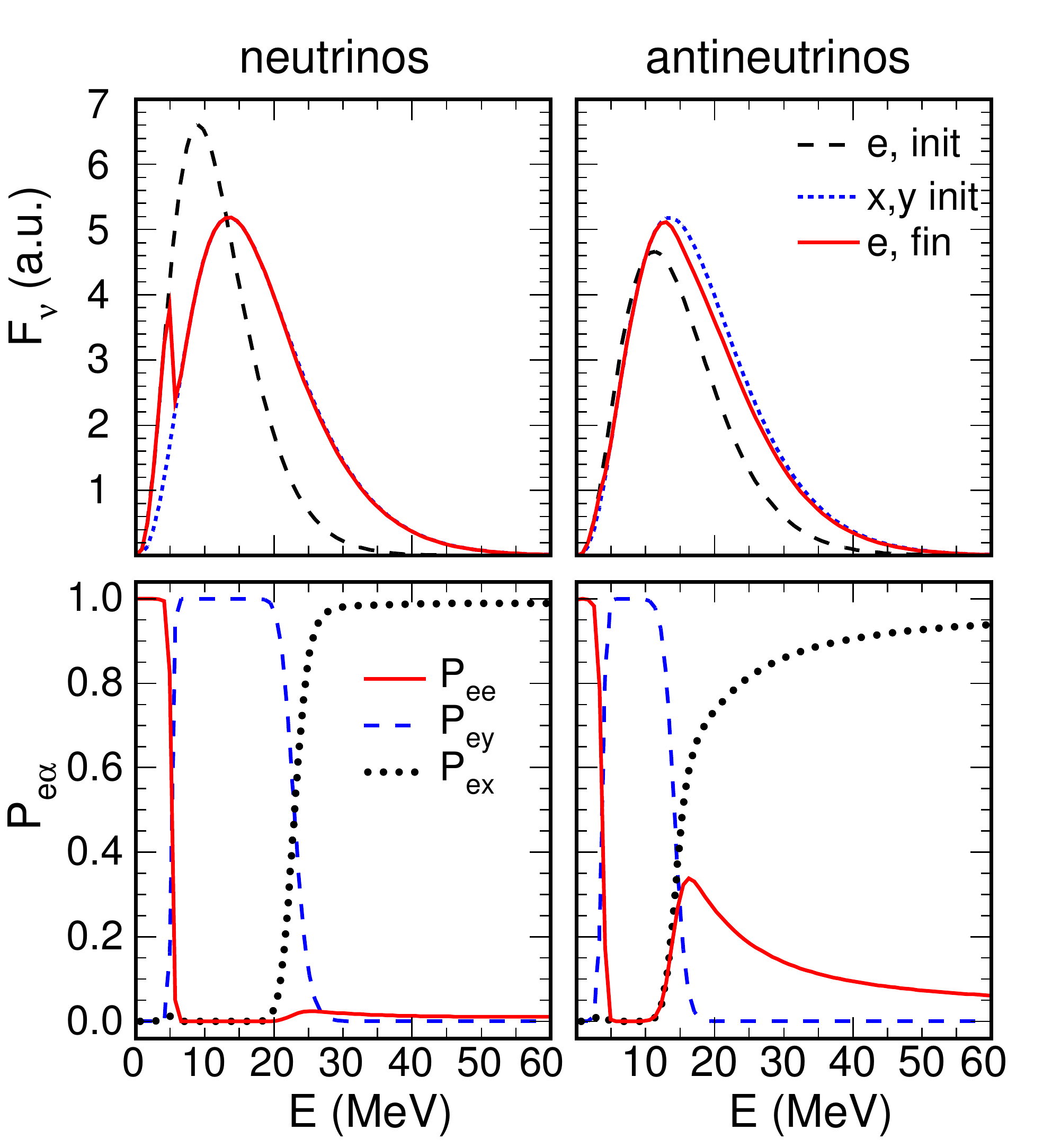} \caption{
Case with $F^0_{\nu_e}: F^0_{\bar\nu_e}: F^0_{\nu_x}=0.85:0.75:1.0$.
Three-flavor
 evolution in the single-angle case and in inverted mass hierarchy  for neutrinos (left panels) and antineutrinos (right panels). Upper panels: Initial energy spectra for $\nu_e$  (long-dashed curve) and $\nu_{x,y}$ (short-dashed curve) and for $\nu_e$ after collective oscillations (solid curve). Lower panels: Probabilities $P_{ee}$ (solid red curve), $P_{ey}$ (dashed blue curve), $P_{ex}$ (dotted black curve).
(Reprinted figure from~\cite{Mirizzi:2010uz}; copyright (2011) by the American Physical Society.)
\label{coolsingl}}
\vskip 1pt
\includegraphics[width=0.52\textwidth]{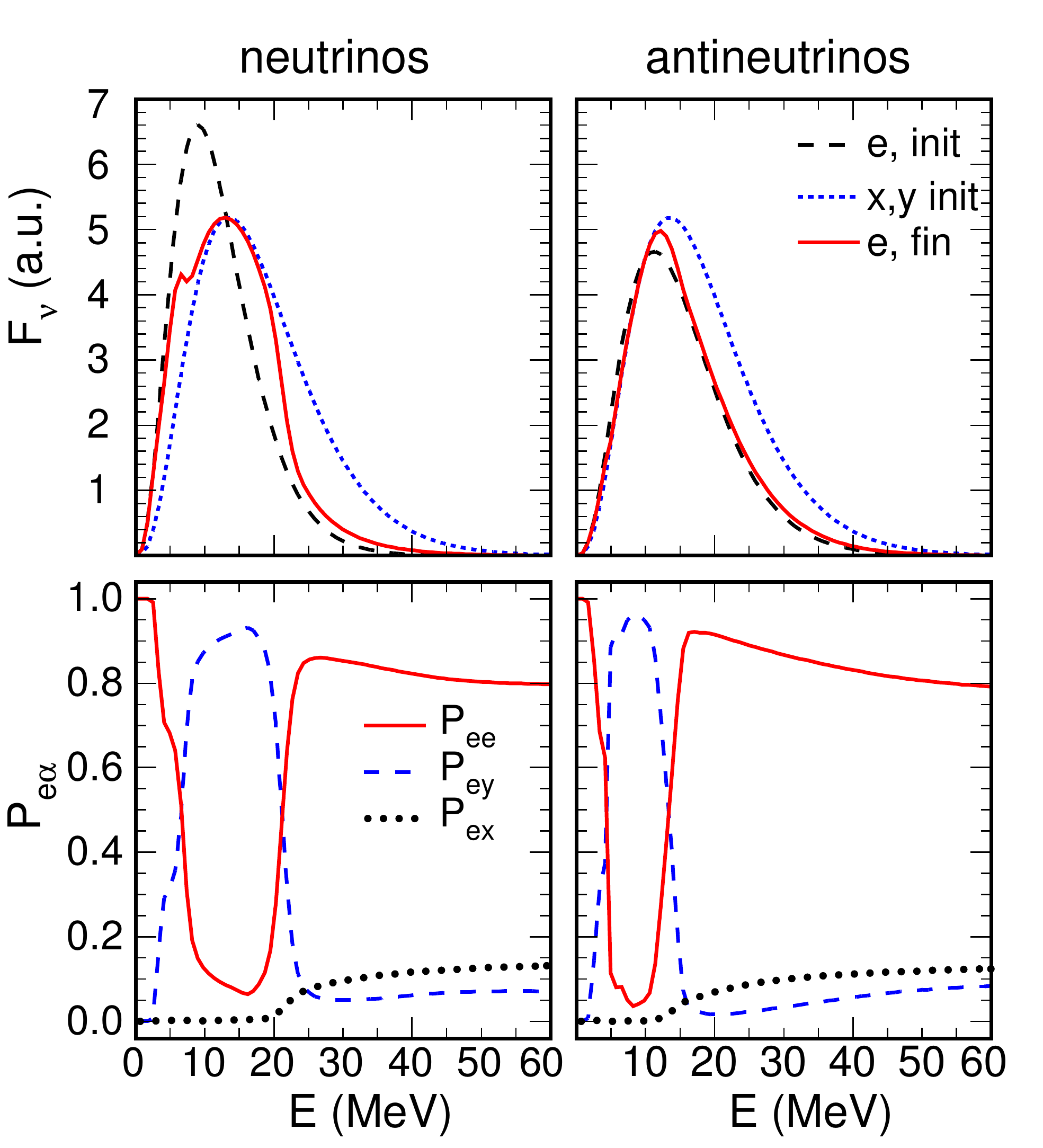}
\caption{The same as Fig.~\ref{coolsingl} but in the multi-angle case. 
(Reprinted figure  from~\cite{Mirizzi:2010uz}; copyright (2011) by the American Physical Society.)
\label{coolmulti}}
\end{figure}

The first large-scale multi-angle simulations have been performed  
in 2006 in Ref.~\cite{Duan:2006an}, adopting the neutrino bulb model. 
 After that, different groups developed independent multi-angle simulations (see, e.g.~\cite{Fogli:2007bk,EstebanPretel:2007ec,Fogli:2008pt,Duan:2010bf,Mirizzi:2010uz,Cherry:2010yc}) that extended the seminal findings
 of~\cite{Duan:2006an} and explored the dependence of the flavor evolution on the initial SN neutrino fluxes. 
 Surprises and unexpected results were found with a strong dependence 
on many details (e.g. neutrino flavor asymmetries~\cite{EstebanPretel:2007ec,Fogli:2009rd}, 
angular distributions~\cite{Mirizzi:2011tu,Mirizzi:2012wp}, 
three-flavor effects~\cite{Dasgupta:2007ws,Fogli:2008fj,Dasgupta:2010ae,Dasgupta:2010cd,Mirizzi:2010uz}). 
Despite the numerous studies on the subject,
at the moment a complete 
picture of the self-induced flavor conversions under multi-angle effects in SNe is still missing. 
Here we show a few  examples of the possible behavior due to self-induced flavor conversions in a $3\nu$ scenario in the presence of multi-angle effects
during  different post-bounce phases:
\begin{itemize}
\item[-] $F^0_{\nu_e} \gg F^0_{\nu_x} \gg F^0_{\bar\nu_e}$. This is the typical flux ordering 
expected during the neutronization phase (see Fig.~\ref{neutrinos-convection-s27}), where the $\nu_e$ flux
is strongly enhanced with respect to $\nu_x$, while the ${\bar\nu}_e$ flux is strongly suppressed.
In this case, bipolar flavor conversions,
proceeding through pair productions of $\nu_e {\bar\nu}_e \to \nu_x {\bar \nu_x}$, are not possible~\cite{Hannestad:2006nj}. Therefore, 
only synchronized oscillations occur, with no relevant effect of flavor conversion, since the in-medium mixing angle is small. 
Multi-angle effects are also negligible \footnote{
A different scenario could be encountered in the case of low-mass SNe with an oxygen-neon-magnesium core.
For these SN progenitors, the matter density profile  can be very steep. The usual MSW matter effect occurs within the region of  high neutrino densities
close to the neutrino sphere. Therefore, self-induced flavor conversions will be possible at low-radii~\cite{Dasgupta:2008cd,Duan:2007sh}.}.

 \begin{figure}[!h]
\centering
\includegraphics[width=0.52\textwidth]{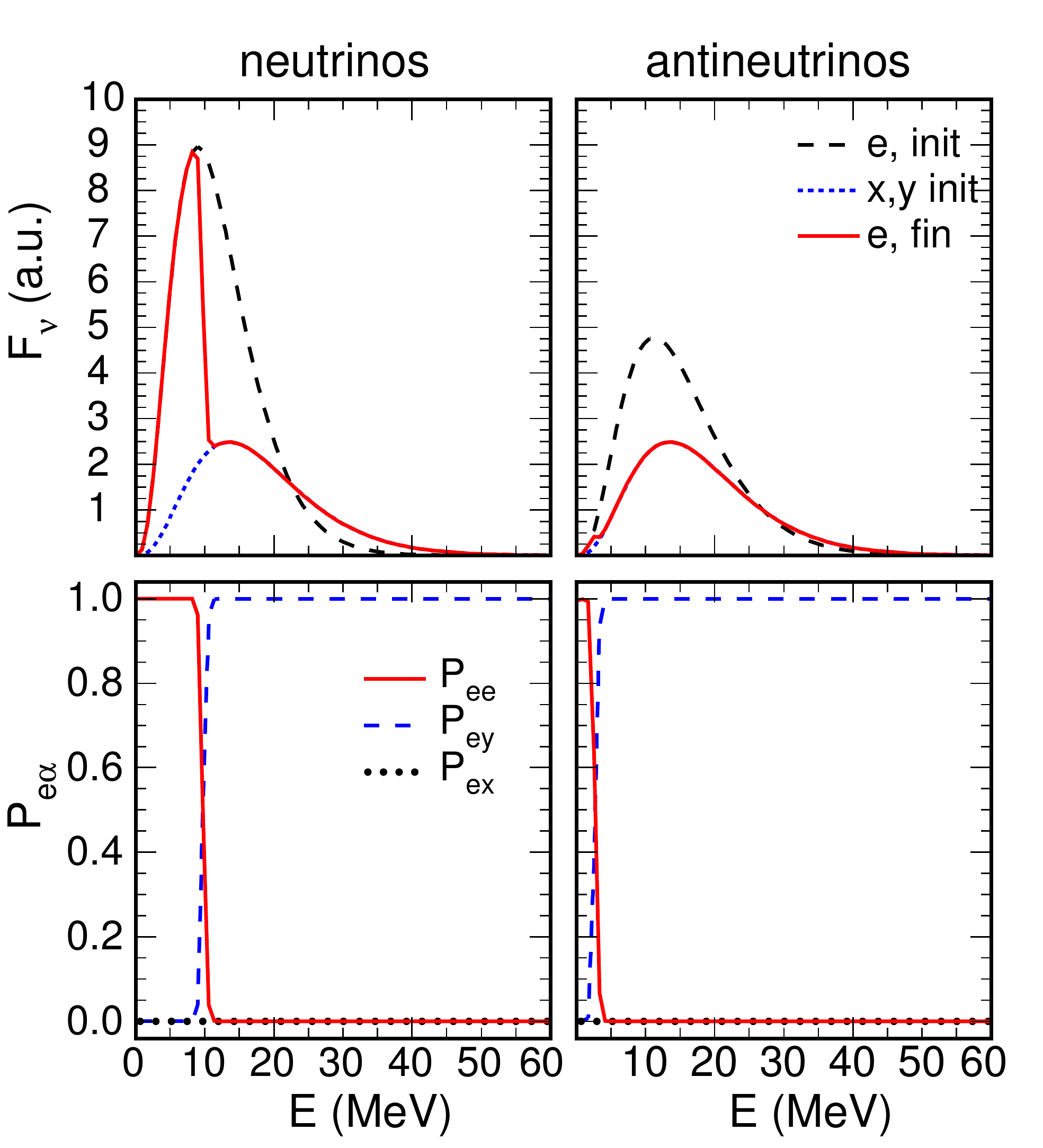}
\caption{Case with
$F^0_{\nu_e}: F^0_{\bar\nu_e}: F^0_{\nu_x}=2.40:1.60:1.0$. Three-flavor
 evolution in the single-angle case and in inverted mass hierarchy  for neutrinos (left panels) and antineutrinos (right panels). Upper panels: Initial energy spectra for $\nu_e$  (long-dashed curve) and $\nu_{x,y}$ (short-dashed curve) and for $\nu_e$ after collective oscillations (solid curve). Lower panels: Probabilities $P_{ee}$ (solid red curve), $P_{ey}$ (dashed blue curve), $P_{ex}$ (dotted black curve).
(Reprinted figure  from~\cite{Mirizzi:2010uz}; copyright (2011) by the American Physical Society.)
\label{accretsing}}
\vskip 1pt
\includegraphics[width=0.52\textwidth]{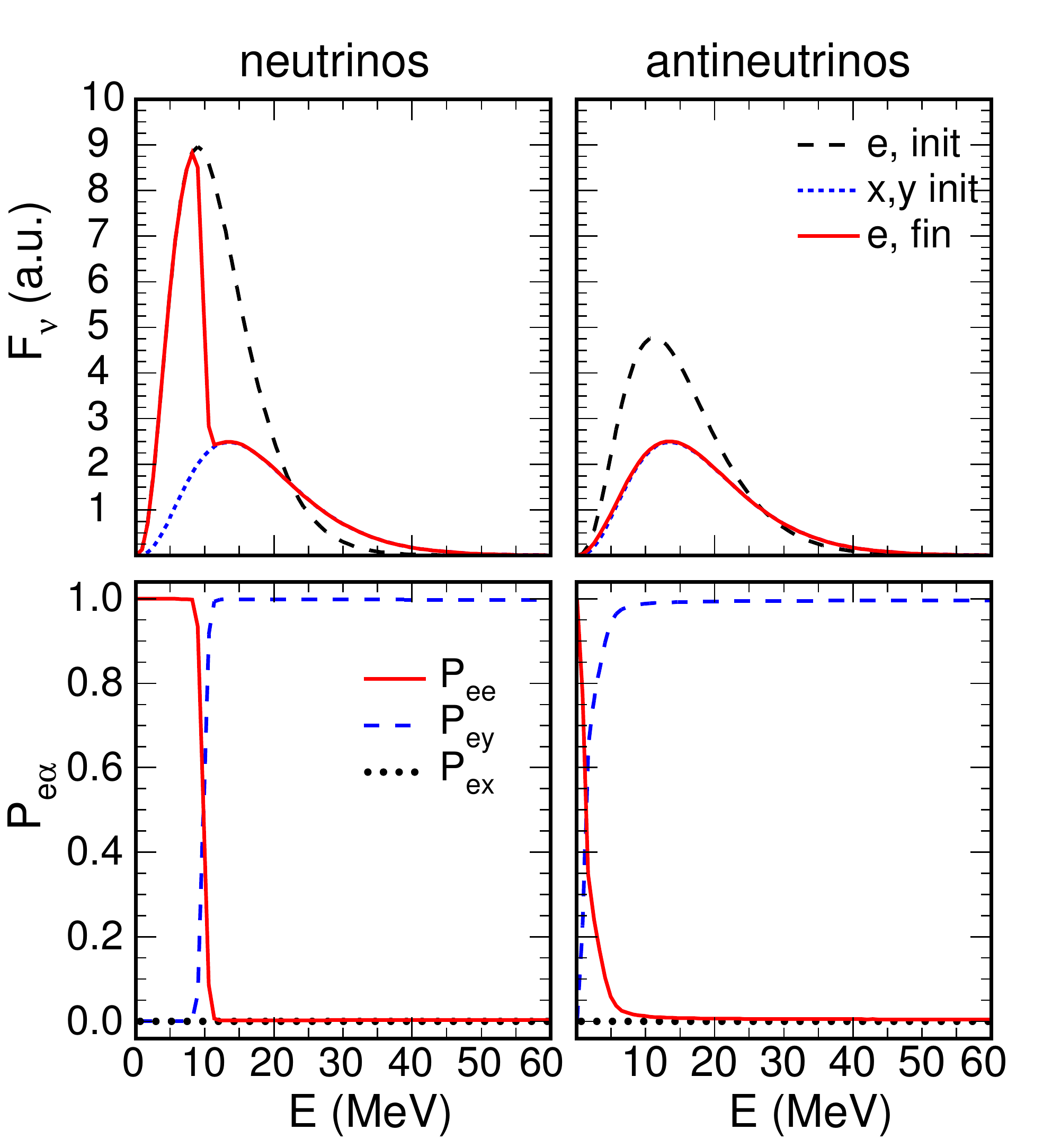}
\caption{The same as Fig.~\ref{accretsing} but in the multi-angle case. 
(Reprinted figure  from~\cite{Mirizzi:2010uz}; copyright (2011) by the American Physical Society.)
\label{accretmulti}}
\end{figure}

\item[-] $F^0_{\nu_e} \gg F^0_{\bar\nu_e} \gg F^0_{\nu_x}$. 
This  flux hierarchy is typically expected during the accretion phase (see Fig.~\ref{neutrinos-convection-s27}), where
the first part of the hierarchy is caused by the deleptonization of the collapsed core,
and the second is caused by the absence of charged-current interactions for neutrino species other than
$\nu_e$ and $\bar\nu_e$. 
Figures~\ref{accretsing} and \ref{accretmulti}, as examples of this configuration,   show the three-flavor multi-angle flavor evolution 
in IH for a flux ordering 
$F^0_{\nu_e}: F^0_{\bar\nu_e}: F^0_{\nu_x}=2.40:1.60:1.0$ (already used in Figs.~\ref{spectraaccr} 
and~\ref{splitaccr}).
In particular, Fig.~\ref{accretsing} refers to the single-angle evolution, while Fig.~\ref{accretmulti} is for the
multi-angle case.
The initial $\nu_e$  fluxes (dashed curves) and the final 
ones for $\nu_e$ and $\nu_{x,y}$ are  represented in the upper panels. Note that $\nu_{x,y}$ are linear combinations of the $\nu_{\mu,\tau}$
fluxes, defined as
$\nu_{x,y}
= \cos \theta_{23} \nu_{\mu} \mp \sin \theta_{23} \nu_{\tau}$~\cite{Di00}, with
$\theta_{23} \simeq \pi/4$ [see Eq.~(\ref{eq:parameters})].
 The conversion probabilities $P_{ee}, P_{ey}, P_{ex}$ are shown in the lower panels. 
As already discussed, one can see that the final $\bar\nu_e$ flux is almost completely \emph{swapped} with respect to the initial one,
while the final $\nu_e$ flux presents a peculiar spectral \emph{split} at $E_{\rm split} \simeq 10$ MeV, being swapped to
$\nu_y$ at higher energies. Flavor conversions occur in 
the $2\nu$ $(e-y)$ sub-system associated with $\Delta m^2$ and $\theta_{13}$ (as shown  in Fig.~\ref{splitaccr}). 
From the comparison with the multi-angle case, we see that these effects play a sub-leading role, being suppressed
by the large flavor hierarchy of the accretion phase~\cite{EstebanPretel:2007ec}. 
No self-induced flavor conversion occurs in NH, as expected. 
 
 The results presented in Figs.~\ref{accretsing} and \ref{accretmulti} have been obtained adopting an effective small mixing angle to simulate
 the matter effects; other multi-angle studies~\cite{Chakraborty:2011nf,Chakraborty:2011gd,Dasgupta:2011jf,Sarikas:2011am,Saviano:2012yh,Chakraborty:2014nma,%
Chakraborty:2014lsa},
 conducted under  a simplified setup, but including the matter background have shown as during the accretion phase the dense matter would dominate over the neutrino 
 density, generally suppressing the self-induced flavor conversions (see  Sec.~\ref{sec:polar}). 
 Therefore, it is not clear  wether this type of flavor evolution is realized in SNe, since probably the flux hierarchy is not large  during the cooling phase.

 \item[-] $F^0_{\nu_x} \gtrsim F^0_{\nu_e} \gtrsim  F^0_{\bar\nu_e}$.
 This spectral ordering with  a less pronounced flavor hierarchy among the different species is possible during the
cooling phase (see Fig.~\ref{neutrinos-convection-s27}).
For definitiveness we take  
$F^0_{\nu_e}: F^0_{\bar\nu_e}: F^0_{\nu_x}=0.85:0.75:1.0$.
Figures~\ref{coolsingl}--\ref{coolmulti} are in the same format adopted in Fig.~\ref{accretsing} and \ref{accretmulti}
 for the accretion phase. 
In particular, Fig.~\ref{coolsingl} refers to the single-angle case, while Fig.~\ref{coolmulti} is for the multi-angle evolution.
 Differently from the  accretion phase, \emph{multiple spectral splits} are present in both 
neutrino and antineutrino channels.
Three-flavor effects are observable in the single-angle scheme, while get suppressed in the multi-angle case.
   Moreover, the spectral swaps and splits are less pronounced,  due to 
some amount of multi-angle decoherence in the flavor conversions. In this regard,  complete decoherence could occur further 
 reducing the flavor asymmetry~\cite{Mirizzi:2010uz}.
Finally, we mention that for the flux ordering of the cooling phase  spectral splits and swaps would occur
also in NH.

\end{itemize}

In Table~I we summarize the results on the role of multi-angle effects, $3\nu$ effects  (associated with  $\delta m^2$) 
and spectral splits 
for different SN neutrino fluxes.
 Matter suppression effects (see Sec. \ref{sec:mattsupp}) should be mainly relevant during the accretion phase,  see two 
dedicated cases.

\begin{table}[!h]
\begin{center}
\caption{Summary of multi-angle effects, $3\nu$ effects and spectral splts for different SN neutrino fluxes,
assuming sub-leading matter effects. The cases ``accretion, $\lambda_r \ll \mu_r$'' (``accretion,  $\lambda_r \gg \mu_r$'') stand for 
absence (presence) of multi-angle matter suppression effects.}
\begin{tabular}{cccc}
\hline \hline
Initial spectral pattern & Multi-angle effects   & $\delta m^2$-effects & Spectral splits \vbox to15pt{}\\
\hline
$F^0_{\nu_e} \gg F^0_{\nu_x} \gg F^0_{\bar\nu_e}$ (neutronization) & no & no & no
\vbox to15pt{} \\
\hline
$F^0_{\nu_e} \gg F^0_{\bar\nu_e} \gg F^0_{\nu_x}$ (accretion, $\lambda_r \ll \mu_r$) & marginal & 
 absent & robust 
\vbox to15pt{} \\
\hline
$F^0_{\nu_e} \gg F^0_{\bar\nu_e} \gg F^0_{\nu_x}$ (accretion, $\lambda_r \gg \mu_r$) & relevant & 
 absent & no 
\vbox to15pt{} \\
\hline
$F^0_{\nu_x} \gtrsim F^0_{\nu_e} \gtrsim  F^0_{\bar\nu_e}$ (cooling) & relevant & 
 present/absent & smeared
\vbox to15pt{} \\
\hline
$F^0_{\nu_e} \simeq F^0_{\nu_x} \simeq F^0_{\bar\nu_e}$ (cooling) & strong & 
 present & washed-out 
\vbox to15pt{} \\
\hline
\end{tabular}
\label{tab:nuebar-effects}
\end{center}
\end{table}

\subsection{Multi-azimuthal-angle instability}
 \label{sec:MAA}
 
As discussed until now, numerical studies of self-induced flavor conversions have been performed within the bulb model.
However, it has been recently questioned if removing some of the symmetries assumed in  this model, this would trigger new instabilities
in the flavor evolution. 
A crucial  assumption of the bulb model is the spherical neutrino emission from the neutrinosphere, that  reduces to a cylindrical
symmetry along a given neutrino trajectory (see Fig.~\ref{nsphere}). 
Recently,  by means of a stability analysis of the linearized neutrino EoMs~\cite{Raffelt:2013rqa}, 
 it has been pointed out that removing the  assumption of axial symmetry in the $\nu$ propagation, a new multi-azimuthal-angle (MAA) instability
 could emerge in the flavor evolution of the dense SN neutrino  gas.
 The occurrence of this instability has been then clarified with simple toy 
models~\cite{Raffelt:2013isa,Duan:2013kba}. 
 The presence of MAA effects unavoidably implies the breaking of
 the spherical symmetry in the flavor evolution after the onset of the flavor conversions. 
 This would lead to a challenging multi-dimensional problem involving partial differential equations.
However, assuming that the variations of the global solution in the direction transverse to the radial one
 are small, one can still study the local solution  along a given line of sight, without 
worrying about its global behavior. 
This approach, even if  not completely self-consistent, allowed to obtain
the  first numerical solutions  of  the non-linear neutrino propagation equations in SNe, 
introducing
the azimuthal angle as angular variable in addition to the usual zenith angle in the multi-angle 
kernel [see Eq.~(\ref{eq:ham1})].
 
\begin{figure}[!t]
\begin{center}
 \includegraphics[angle=0,width=0.8\textwidth]{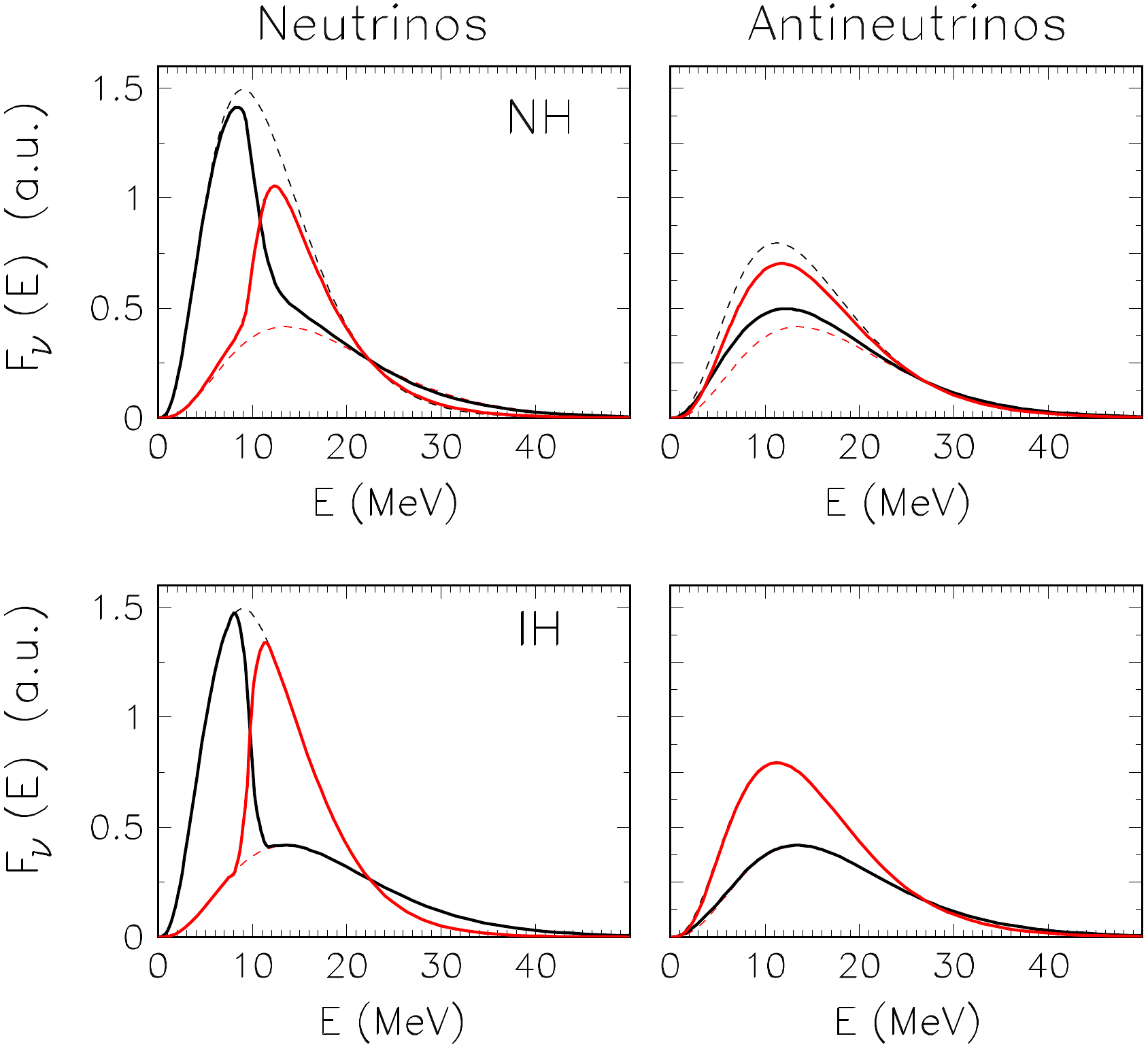} 
    \end{center}
\caption{Multi-azimuthal-angle flavor evolution for $\nu$'s
(left panel) and $\bar\nu$'s (right panel) in NH (upper panels) and IH (lower panels)
for fluxes with an initial ordering
$F^0_{\nu_e}:F^0_{\bar\nu_e}:F^0_{\nu_x}=2.40:1.60:1.0$.
 Energy spectra  for
$\nu_e$ (black dashed curves) and $\nu_x$ (light dashed curves) in absence of flavor oscillations
and  for $\nu_e$ (black continuous curves)
and $\nu_x$ (light continuous curves) after collective oscillations. 
(Reprinted figure  from~\cite{Mirizzi:2013wda}; copyright (2014) by the American Physical Society.)
} \label{accrMAA}
\end{figure}

 Adopting this generalized bulb model, it has been shown~\cite{Mirizzi:2013rla,Mirizzi:2013wda}  that the pattern of the spectral crossings (energies
where $F_{\nu_e} = F_{\nu_x}$, and  $F_{\bar\nu_e} = F_{\bar\nu_x}$) is crucial in determining the impact of MAA effects
on the flavor evolution. For neutrino spectra  with a strong excess of 
$\nu_e$ over $\bar\nu_e$, as expected during the accretion phase,  new flavor conversions  occur in NH.
This is represented in Fig.~\ref{accrMAA}, where we have considered initial fluxes as the ones of Fig.~\ref{splitaccr}.  In particular  the initial (dashed curves) and final fluxes (continuous curves) 
are shown for $e$ (black curves) and $x$ (light curves)  flavors, for neutrinos (left panels) and antineutrinos (right panels).
Upper panels show the NH case while lower panels refer to the IH one. One realizes that MAA effects produce 
 flavor conversions in NH, otherwise absent in the azimuthal symmetric model.
However, the dominance of matter terms during the accretion phase would typically suppress
also the MAA effects~\cite{Chakraborty:2014nma,Chakraborty:2014lsa}.
For spectra with a moderate flavor hierarchy, as the ones expected during the cooling phase,
  the growth of the MAA instability should be  inhibited. 
As a result, the oscillated spectra
are rather close to the ones seen in the azimuthal symmetric case; see  Fig.~\ref{coolmulti} for comparison (see also~\cite{Mirizzi:2013wda}).

\subsection{Spontanueous breaking of space-time symmetries}
 \label{sec:SSB}

The MAA  instability has shown as self-interacting neutrinos can 
\emph {spontaneously break}   the symmetries of the initial conditions.  
Such insight stimulated  further investigations  about the validity of the solution of the SN neutrino EoMs  worked out  within the
 bulb model (see Sec.~\ref{sec:eom}). 
 
 The  assumptions of azimuthal symmetry in neutrino propagation and quasi-stationary neutrino emission 
reduce the general partial
differential equations [see Eq.~(\ref{eq:eom})] into ordinary differential equations describing the stationary
 spatial evolution  of the dense neutrino gas along the radial direction.
 These
assumptions, although  adopted since the earliest papers on the subject,  are unjustified.
Indeed, it was tacitly assumed that  small deviations from them do not significantly perturb the flavor evolution.
However, it has been recently realized
 that instabilities may grow once initial symmetries   are relaxed, since self-interacting neutrinos can 
spontaneously break 
the
translation symmetries in time~\cite{Mangano:2014zda,Abbar:2015fwa,Dasgupta:2015iia} and space~\cite{Duan:2014gfa,Mirizzi:2015fva,Mirizzi:2015hwa,Duan:2015cqa,Chakraborty:2015tfa}.
This implies that the characterization of the self-induced effects obtained within the spherically symmetric bulb model
should be taken \emph{cum grano salis}.

 
 A self-consistent solution of the flavor evolution  would eventually  lead to solve the  complete space-time-dependent
problem described by the following partial differential equations~\cite{Sigl:1992fn}
\begin{equation}
\left(\frac{\partial}{\partial t} + \mathbf{v_p} \cdot \nabla_{\mathbf{x}}\right)  \varrho_{t,\mathbf{p, x}} = -i
 [\Omega_ {\mathbf{p,x}},  \varrho_{t,\mathbf{p,x}}] \,\ ,
\label{eq:eomcompl}
\end{equation} 
where $\Omega_ {\mathbf{p,x}}=\Omega^{\textrm{vac}}_ {\mathbf{p}}+\Omega^{\textrm{ref}}_ {\mathbf{p,x}}$, and an analogous equation exists for antineutrinos
[see Eq.~(\ref{eq:eom})]. This represents a formidable seven-dimensional problem. First attempts of solution have been recently
 presented in~\cite{Mangano:2014zda,Mirizzi:2015fva,Mirizzi:2015hwa} by Fourier transforming 
Eq.~(\ref{eq:eomcompl}).  For example, performing a spatial Fourier transform one finds 
\begin{equation}
\left(\frac{\partial}{\partial t} + \mathbf{v_p} \cdot {\bf k}\right)  \varrho_{t,\mathbf{p, k}} = -i
\int d^3 {\bf x} \,\ e^{-i {\bf k}\cdot{\bf x}} [\Omega_ {\mathbf{p,x}},  \varrho_{t,\mathbf{p,x}}] \,\ ,
\label{eq:eomfour}
\end{equation} 
which represents a tower of ordinary differential equations in $t$ for the different Fourier modes $ \varrho_{t,\mathbf{p, k}}$ with wavenumber ${\bf k}$,
which are coupled through the 
 interaction term.
This technique sheds the basis to study this challenging problem in the SN case. 

In this context, it has been recently pointed out    that even tiny space inhomogeneities could lead to new flavor  
instabilities~\cite{Duan:2014gfa,Mirizzi:2015fva}, developing at small scales, even at large neutrino densities, where oscillations are otherwise expected to be suppressed due to synchronization~\cite{Duan:2014gfa}.
 However, as shown in~\cite{Chakraborty:2015tfa},  large neutrino densities in a supernova are typically accompanied by a large matter density,
which produces the multi-angle matter effects (Sec.~\ref{sec:mattsupp}) that suppresses the low-radii small-scale instabilities.
Furthermore in~\cite{Dasgupta:2015iia}, it has been shown that  removing the assumption of the stationarity of the flavor evolution, these results can dramatically change. Indeed, the 
steady solution is not stable and self-interacting neutrinos can generate  a \emph{pulsating solution} with a frequency that 
   effectively compensates the phase dispersion associated with the large matter term, lifting the suppression of the space instabilities at small-scales. If these space-time instabilities develop fully and cascade, that paves the way for flavor conversions at large neutrino and matter densities. The flavor-content of SN neutrino fluxes will be strongly varying with time, perhaps leading to flavor-averaging and have profound consequences for supernova explosions and nucleosynthesis.

The above described very recent results add new challenges in the characterization of the self-induced SN neutrino flavor conversions. 
Remarkably,
the final goal would be a realistic treatment of  the self-induced flavor conversions within the SN environment, where large deviations from a spherical neutrino emission 
can be generated by  hydrodynamical instabilities such as SASI or LESA, especially during the accretion phase, as discussed in Sec.~\ref{sec:SNmodels}.

Another assumption that has recently  been questioned  is that neutrinos are free-streaming after the neutrinosphere~\cite{Cherry:2012zw}, as even a  fraction of neutrinos that occasionally scatter outside of the
 neutrinosphere should produce a small ``neutrino halo''~\cite{Cherry:2012zw,Cherry:2013mv}. On the basis of a stability analysis, it has been concluded 
  that the neutrino halo should not drastically affect flavor oscillations during the accretion phase~\cite{Sarikas:2012vb}; however a self-consistent numerical simulation is not available at the 
  moment.
The inclusion of the halo effect in the numerical simulations would change the nature of the 
flavor evolution,  turning it into a boundary value problem instead of a non-stationary initial value one. 
All these open issues  require further dedicated work to
fully clarify
the role of  flavor instabilities in the interacting
neutrino field.

\subsection{Mikheyev-Smirnov-Wolfenstein matter effect in the wake of the shock-wave}
 \label{sec:MSWshock}

Self-induced effects in SN neutrino oscillations, occurring in the deepest stellar regions
would eventually die out at $r\gtrsim {\mathcal O}(10^3)$~km. However, as neutrinos stream
through the outer layers of the stellar envelope,
they would feel ordinary matter effects. 
There is a wide literature on (SN) neutrino oscillations in matter to which we refer the interested
reader (see, e.g.,~\cite{Kuo:1989qe,Di00,Wolfenstein:1979ni,Reinartz:1983ba,Mikheev:1986if,Fogli:2001pm}).  
As the SN matter potential $\lambda_r =  \sqrt{2} G_F n_e(r)$ declines,
  neutrinos would eventually encounter  Mikheyev-Smirnov-Wolfenstein (MSW) \emph{resonances} when~\cite{Kuo:1989qe,Di00}
\begin{equation}
\lambda_r = \omega_{H,L} \,\ ,
\label{eq:resonance}
\end{equation}
corresponding to the atmospheric $\Delta m^2$ ($H$-resonance) and the solar $\delta m^2$ ($L$-resonance) mass-squared
differences, respectively.

\begin{figure}[t!]
\begin{center}
 \includegraphics[angle=0,width=0.95\textwidth]{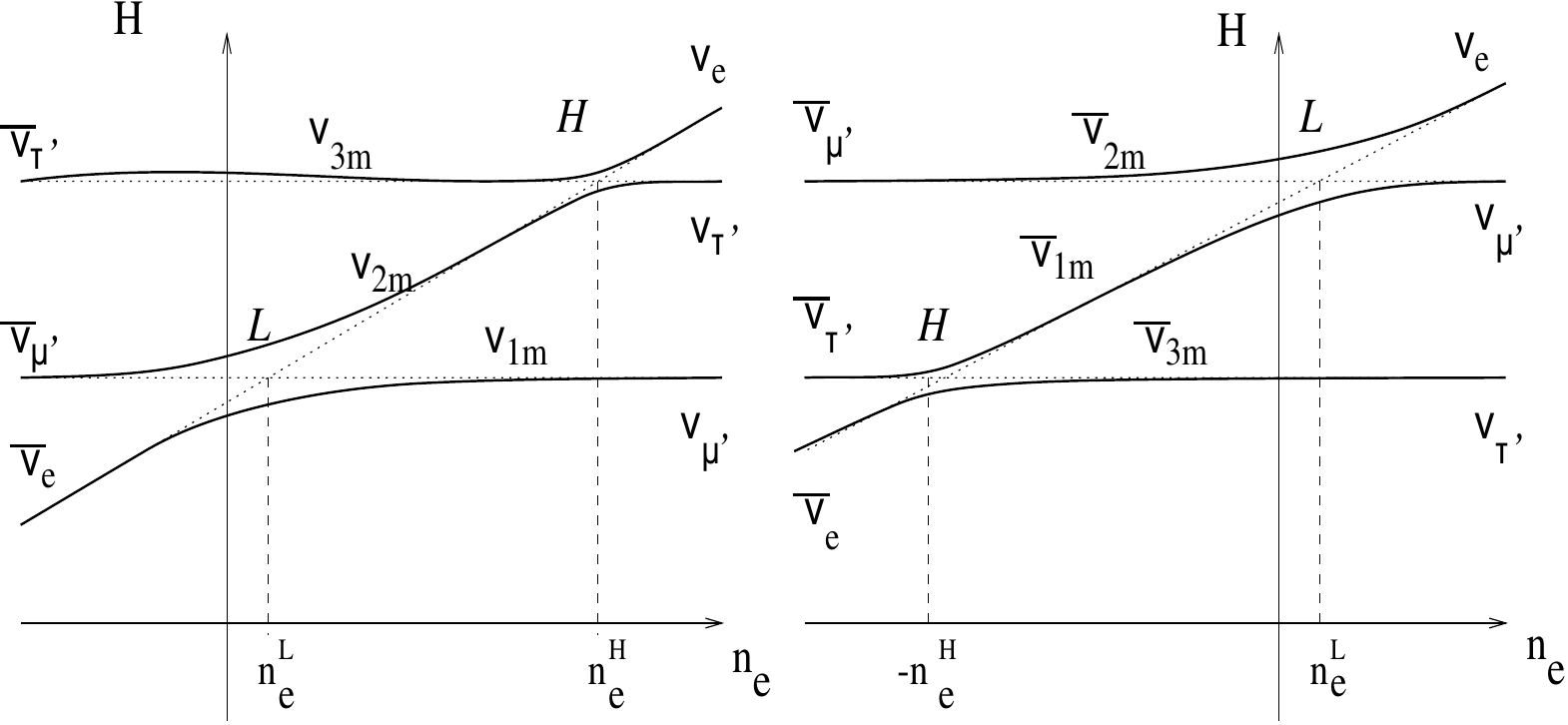} 
    \end{center}
\caption{Three-flavor level diagram for neutrino propagation eigenmodes,
relevant for neutrinos streaming from a SN core~\cite{Di00}
for normal hierarchy (left) and inverted hierarchy (right). (Figure taken from~\cite{Raffelt:2012kt} with permission.)
\label{level}}
\end{figure}

Typical regions where we expect $H$ and $L$ resonances have been shown in Fig.~\ref{densities}. We realize that
the two resonant regions are rather separated due to the hierarchy 
$\delta m^2 \ll \Delta m^2$, so that
one can typically factorize the dynamics studying the resonant effects in  $2\nu$ sub-sectors.  
 When one of the conditions in Eq.~(\ref{eq:resonance}) is fulfilled  a resonant amplification of the 
flavor conversions is expected. 
From the Hamiltonian (in absence of self-interactions), one can build the \emph{level crossing} diagrams for the two mass hierarchies
(see Fig.~\ref{level}) showing the 
neutrino propagation eigenstates from the regions at high-density to the vacuum, where
$\lambda_r=0$~\cite{Di00}.
These diagrams allow us to determine in which mass
eigenstate  will emerge a neutrino  produced in a given interaction eigenstate.
In the case of antineutrinos, the effective potential for $\bar\nu_e$ is
$\lambda_r = - \sqrt{2} G_F n_e(r)$. Antineutrinos can  be represented on the same level
crossing diagram, as neutrinos traveling through matter with ``effectively'' negative
$n_e(r)$. Given the resonance conditions in Eq.~(\ref{eq:resonance}), the sign of the 
matter potential $\lambda_r$ and of the mass-squared splittings, we realize that
the $H$-resonance can be satisfied by neutrinos in NH ($\lambda_r, \Delta m^2>0$)
or by antineutrinos in IH ($\lambda_r, \Delta m^2<0$). 
Therefore, in principle, the neutrino burst is sensitive to the neutrino mass hierarchy thanks to the matter effects, associated
with the $H$-resonance.
Conversely, the $L$-resonance
can be satisfied only by neutrinos in both the mass hierarchies (since $\delta m^2 >0$).

For typical SN  simulations, the matter density profile declines so slowly that the neutrino propagation is  \emph{adiabatic}, i.e.,
each mass eigenstate in Fig.~\ref{level} remains the same. However, this condition is violated 
at the (forward and reverse) shock-fronts where, due to the abrupt density change (see Fig.~\ref{densities}),  
strong non-adiabatic conversions  occur.
Neglecting self-induced flavor conversions, the neutrino flux arriving at Earth  can be expressed
in terms of energy-dependent $\nu_e$ survival probabilities $P_{ee}(E)$ at the shock-front~\cite{Di00}:
\begin{equation}
F_{\nu_e} =P_{ee}(E) F^0_{\nu_e}(E) + [1-P_{ee}(E)] F^0_{\nu_x}(E) \,\ .
\label{eq:fluxes}
\end{equation}
  An analogous expression exists for $\bar\nu_e$ with survival probabilities
  ${\bar P}_{ee}(E)$. In particular, considering for simplicity a static SN matter profile and  a complete adiabatic propagation we have
$(P_{ee},{\bar P}_{ee})= (0, \cos^2 \theta_{12})$ in normal mass hierarchy, and  
$(P_{ee},{\bar P}_{ee})= (\sin^2 \theta_{12}, 0)$ in inverted mass hierarchy~\cite{Di00}. 

Corrections to a pure adiabatic neutrino propagation are expressed in terms of  \emph{level crossing} probabilities
among the instantaneous eigenstates  in matter at the resonance point~\cite{Kuo:1989qe}. 
As shown in Fig.~\ref{densities}, the real SN density profile is non-monotonic and time-dependent, so that
multiple resonances would occur along it.
For $\theta_{13}$ as large as recently measured, the $H$-resonance is   adiabatic, 
except at the shock-fronts, where the crossing probability  $P_H = P_H(\Delta m^2, \theta_{13})$
between the  eigenmodes  $\nu_{3,m}$ and $\nu_{2,m}$ (see Fig.~\ref{level})
would be extremely non-adiabatic, giving $P_H \simeq 1$.
Then, as $\nu$'s propagate at larger radii, they would eventually    encounter the  $L$-resonance between
the  $\nu_{2,m}$ and $\nu_{1,m}$ states,
 associated with the  $(\delta m^2, \theta_{12})$ sub-sector. 
The $L$-resonance  intercepts the shock-front only at relatively late times
($t_{\rm pb}\gtrsim 10$~s) and it is never  strongly non-adiabatic.
Therefore,  sub-leading effects related 
a  level crossing probability
 $P_L \neq 0$ are typically neglected.

The survival probability $P_{ee}$ of SN neutrinos at Earth (neglecting Earth matter crossing) is related 
to the crossing probability $P_H$ at the shock-front~\cite{Fogli:2003dw}: 
\begin{equation}
\label{cases} P_{ee} \simeq  \left\{
\begin{array}{ll}
 \sin^2\theta_{12}\, P_H & (\nu,\;\mathrm{NH}), \\
 \cos^2\theta_{12}       & (\overline\nu,\;\mathrm{NH}), \\
 \sin^2\theta_{12}       & (\nu,\;\mathrm{IH}), \\
 \cos^2\theta_{12}\, P_H & (\overline\nu,\;\mathrm{IH}). \\
\end{array}\right.
\end{equation}
It is clear as  $P_H$ can modulate the (otherwise constant) survival probability  
of $\nu_e$ in NH and of $\bar\nu_e$ in IH, thus providing an important
handle to solve the current hierarchy ambiguity.

\begin{figure}[t!]
\begin{center}
 \includegraphics[angle=0,width=0.8\textwidth]{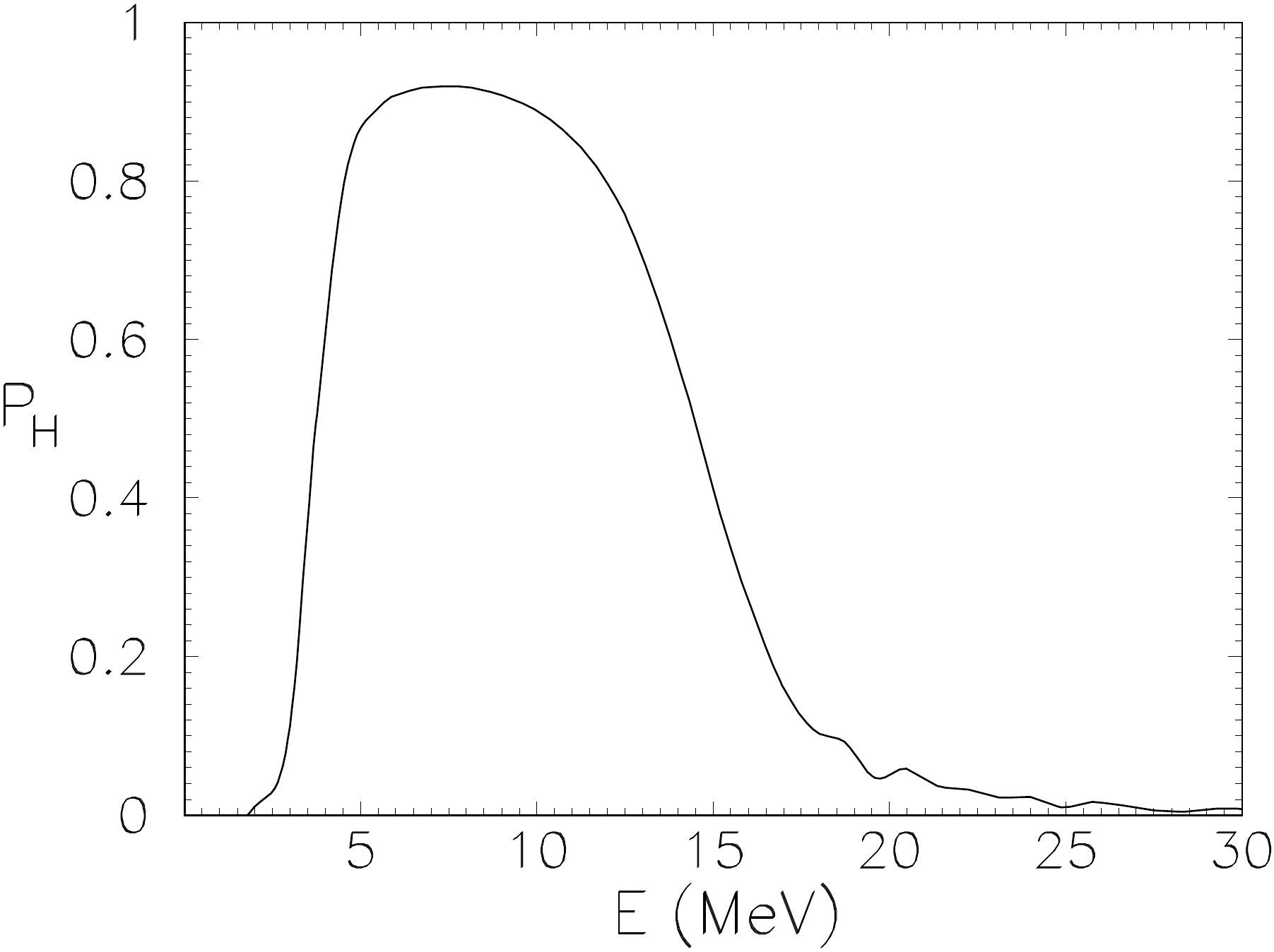} 
    \end{center}
\caption{Neutrino crossing probability
$P_H$ as function of the neutrino energy $E$
for the  SN matter potential of Fig.~\ref{densities} at post-bounce time $t_{\rm pb}=5$~s.
\label{PH}}
\end{figure}

In order to compute $P_H$, one has to  numerically integrate the  flavor evolution equations [Eq.~(\ref{eq:singleanglepolar})]
 in the basis of the instantaneous matter eigenstates. Analytical prescriptions  have been  presented in~\cite{Fogli:2003dw,Kneller:2005hf}.
In Fig.~\ref{PH}, we show the crossing probability
$P_H$ as a function of energy $E$
for the  SN matter density
profile of Fig.~\ref{densities} at post-bounce time $t_{\rm pb}=5$~s.  
We closely follow the approach of Ref.~\cite{Fogli:2003dw} to compute $P_H$ 
(the interested reader is referred to this reference for further details). 
The crossing probability has a typical top-hat structure, jumping from $P_H \sim 0$ (adiabatic regime)
to $P_H \sim 1$ (extremely non-adiabatic regime) when the resonance condition is satisfied across the forward shock-front. 
A more complicated pattern would emerge when resonances occur on both the forward and the reverse
 shock fronts~\cite{Fogli:2004ff,Tomas:2004gr}.
It is expected  that the transient violation of the adiabaticity condition, when neutrinos cross the shock-fronts, would 
 emerge as an observable
modulation of the neutrino signal. This signature could be particularly  useful to follow
in ``real-time'' the shock-wave propagation, as well as to probe the neutrino mass hierarchy.
This opportunity has been widely discussed in literature~\cite{Fogli:2004ff,Tomas:2004gr,Fogli:2003dw,Schi,%
Dasgupta:2005wn,Choubey:2006aq,Kneller:2007kg}.

A realistic characterization of matter effects, during neutrino propagation across the SN shock-wave, 
must also take into account stochastic density fluctuations, inhomogeneities
of various magnitudes as well as
correlation lengths in the ejecta layer in the wake of the shock front. 
These fluctuations are a result of hydrodynamic 
instabilities between the proto-neutron star and the SN shock 
during the very early stages of the SN explosion. They lead to 
large-scale explosion asymmetries and turbulence in a dense shell
of shock-accelerated ejecta, which subsequently also seed secondary 
instabilities in the outer shells of the exploding star (e.g., 
Refs.~\cite{Kifonidis:2003,Kifonidis:2006,Scheck:2006,Hammer:2010,Arcones:2011,Mueller:2012,Wongwathanarat:2013}).

Neutrino flavor conversions in a stochastic matter background have been subject of intense
investigations both in a general 
context~\cite{Schaefer:1987fr,Sawyer:1990tw,Loreti:1994ry,Nunokawa:1996qu,Balantekin:1996pp,Burgess:1996mz,torrente}
 and specifically in relation to SN neutrinos~\cite{Fogli:2006xy,Friedland:2006ta,Kneller:2010sc,Lund:2013uta,Loreti:1995ae,Choubey:2007ga,Benatti:2004hn,Kneller:2013ska}. 
It is expected that
stochastic matter fluctuations of sufficiently large amplitude may suppress flavor
conversions and lead to $P_H \simeq 1/2$ when the suppression is strong~\cite{Fogli:2006xy}. 
 Therefore, the spectral properties of the fluctuations
are very important for understanding the neutrino signal emerging from a core-collapse SN.
At the moment there is no unanimous consensus about the impact of matter fluctuations on the SN neutrino
flavor conversions. A recent study based on two-dimensional  SN simulations has shown 
a modest damping of the neutrino crossing probabilities~\cite{Borriello:2013tha}. However, further analysis would be
mandatory when high-resolution three-dimensional simulations will become available.

\subsection{Observable signatures of supernova neutrino flavor conversions}
 \label{sec:signatures}

After  discussing in detail how  self-induced and matter effects would process the 
SN neutrino fluxes, we  present some of the possible observable signatures
of flavor conversions in SNe and their sensitivity to the neutrino mass hierarchy. 
For definitiveness, we will refer to a Galactic SN at a distance $d=10$~kpc from the Earth.
We address the interested reader to Sec.~\ref{sec:detection} for a detailed discussion on SN neutrino detection
techniques.

\begin{figure}[t!]
\begin{center}
 \includegraphics[angle=0,width=0.8\textwidth]{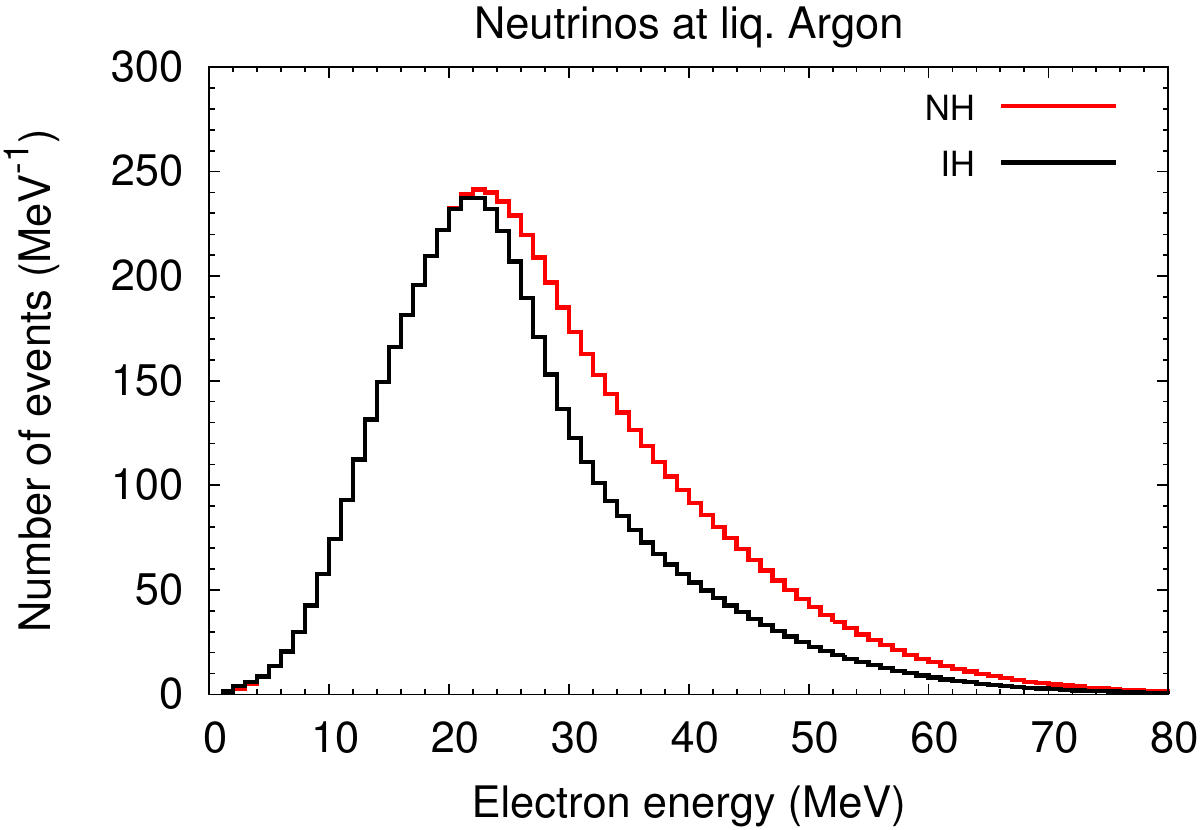} 
    \end{center}
\caption{Observable SN $\nu_e$ spectra in a 40-kton LAr TPC
for benchmark fluxes as in Fig.~\ref{coolsingl} in both the mass hierarchies.
(Figure adapted from~\cite{Choubey:2010up}; courtesy of B.~Dasgupta.)
\label{split_arg}}
\end{figure}

\begin{figure}[h!]
\begin{center}
 \includegraphics[angle=0,width=0.8\textwidth]{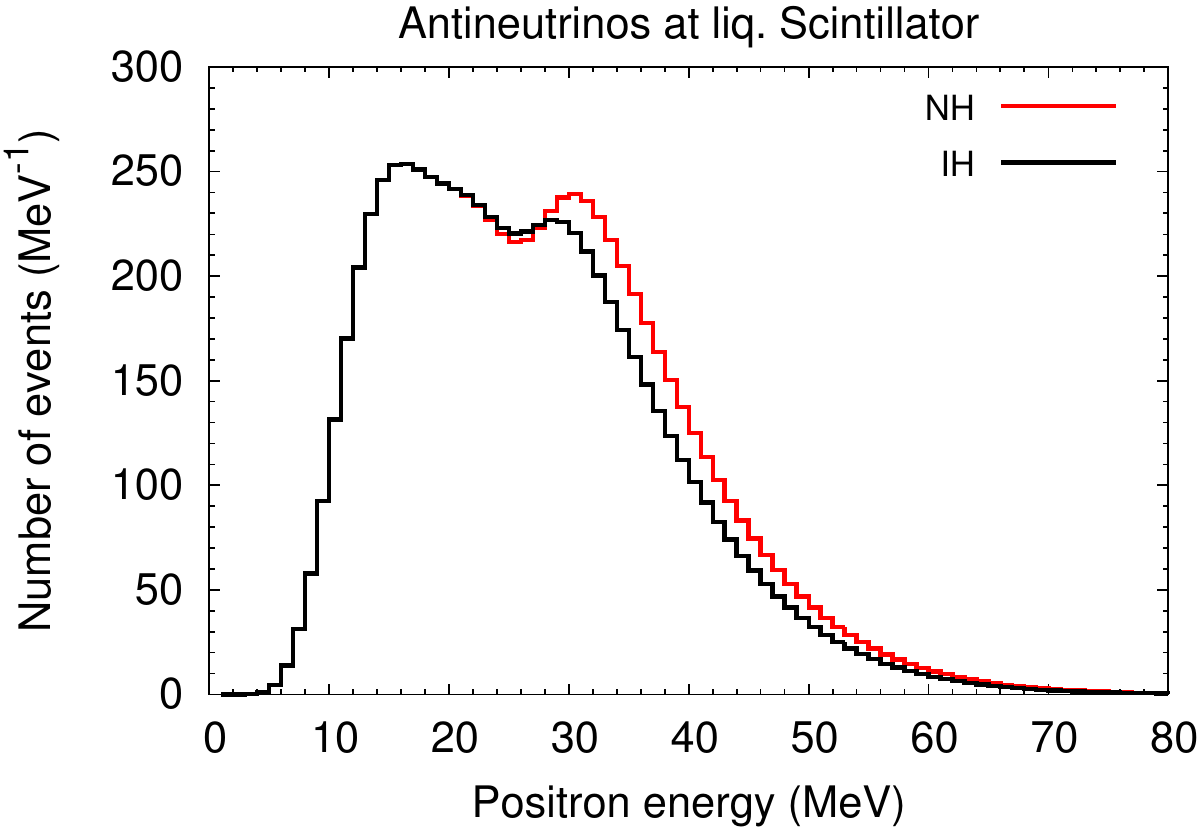} 
    \end{center}
\caption{Observable SN $\bar\nu_e$ spectra in a 20-kton liquid scintillator detector
for benchmark fluxes of Fig.~\ref{coolsingl} in both the mass hierarchies.
(Figure adapted from~\cite{Choubey:2010up}; courtesy of B.~Dasgupta.)
\label{split_scint}}
\end{figure}

\emph{(a) Self-induced spectral splits.}
As  discussed in this Section, in recent years the picture of SN neutrino oscillations, based
only on the MSW matter effects, has undergone  a change of paradigm
by the insight that the refractive effects of neutrinos on themselves are crucial.  Observationally, the most important consequence
of the self-induced flavor conversions 
is a swap of the $\nu_e$ and $\bar\nu_e$ spectrum with the non-electron species
$\nu_x$ and $\bar\nu_x$ in certain energy intervals, and the resultant spectral splits at
the edges of these swap intervals. 
Some of the
spectral splits could occur sufficiently close to the peak energies to produce significant distortions
in the  SN neutrino signal, observable in the large underground dectectors.
As an example, we show in Fig.~\ref{split_arg} the observable $\nu_e$ signal   
in  a 40-kton LAr TPC, while   Fig.~\ref{split_scint}
refers to the $\bar\nu_e$ signal in a 20-kton liquid scintillator detector. Both the mass hierarchy
cases are shown. The initial fluxes in these figures are the same as in Fig.~\ref{coolsingl} (during the cooling phase)
 where self-induced conversions are not matter-suppressed.
The self-induced flavor evolution has been characterized in a single-angle and three-flavors 
scenario (as in Fig.~\ref{coolsingl}).  
The MSW effect is calculated according to Eq.~(\ref{eq:fluxes})
and  (\ref{cases}), where it is assumed complete adiabatic propagation
($P_H=1$) neglecting possible shock-wave effects (see~\cite{Choubey:2010up} for details).

Concerning the electron spectrum produced by
$\nu_e$  in a LAr TPC (Fig.~\ref{split_arg}), the observable flux
is mostly due to the initial $F^0_{\nu_x}$ and then it does not present
any special spectral feature in both the mass hierarchies. 
Conversely, 
for the $\bar\nu_e$ signal  in a scintillator detector (Fig.~\ref{split_scint}), the observable positron spectrum
would be mostly due to $F^0_{\bar\nu_e}$ for $E<E_{\rm split}$ and  $F^0_{\nu_x}$ at higher
energies (see also Fig.~\ref{coolsingl}). 
This would produce a bimodal positron spectrum,
with two peaks produced  by the
two initial antineutrino distributions. The result is similar in both the mass hierarchies.
This example shows that spectral splits are potentially identifiable during the cooling phase, 
if  the average energies and
luminosities of non-electron fluxes are sufficiently large.

\emph{(b) Neutrinos from the SN neutronization burst.} 
The $\nu_e$ neutronization burst is a particularly interesting probe of flavor conversions, since it
can be considered almost as a ``standard candle'' being independent of the mass progenitor and nuclear EoS~\cite{Kachelriess:2004ds}.
As discussed in Sec.~\ref{sec:difffluxes}, self-induced effects are not operative on the SN $\nu_e$ neutronization  burst 
because of the large excess of $\nu_e$
due to the core deleptonization.
Therefore, the $\nu_e$ flux would be only affected by  MSW effects. At the very early post-bounce times relevant
for the prompt burst, MSW flavor conversions would occur along the  static progenitor matter density profile.
 We  expect  that the observable $\nu_e$ flux at Earth would be [Eq.~(\ref{eq:fluxes})]
\begin{eqnarray}  
 F_{\nu_e} &=& F^0_{\nu_x} \,\ \,\ \,\ \,\ \,\ \,\ \,\ \,\  \,\ \,\ \,\ \,\   \,\ \,\  \,\ \,\ \,\ \,\ \,\ \,\  \,\ \,\ \textrm{(NH)} \,\ , \\
 F_{\nu_e} &=&  \sin^2 \theta_{12} F^0_{\nu_e} +
\cos^2 \theta_{12} F^0_{\nu_x}  \,\ \,\ \,\ \,\ \textrm{(IH)} \,\ .
\end{eqnarray} 
 The observation of the neutronization
peak would indicate the inverted mass hierarchy, while we do not expect its detection in NH.

\begin{figure}[t!]
\begin{center}
 \includegraphics[angle=0,width=0.8\textwidth]{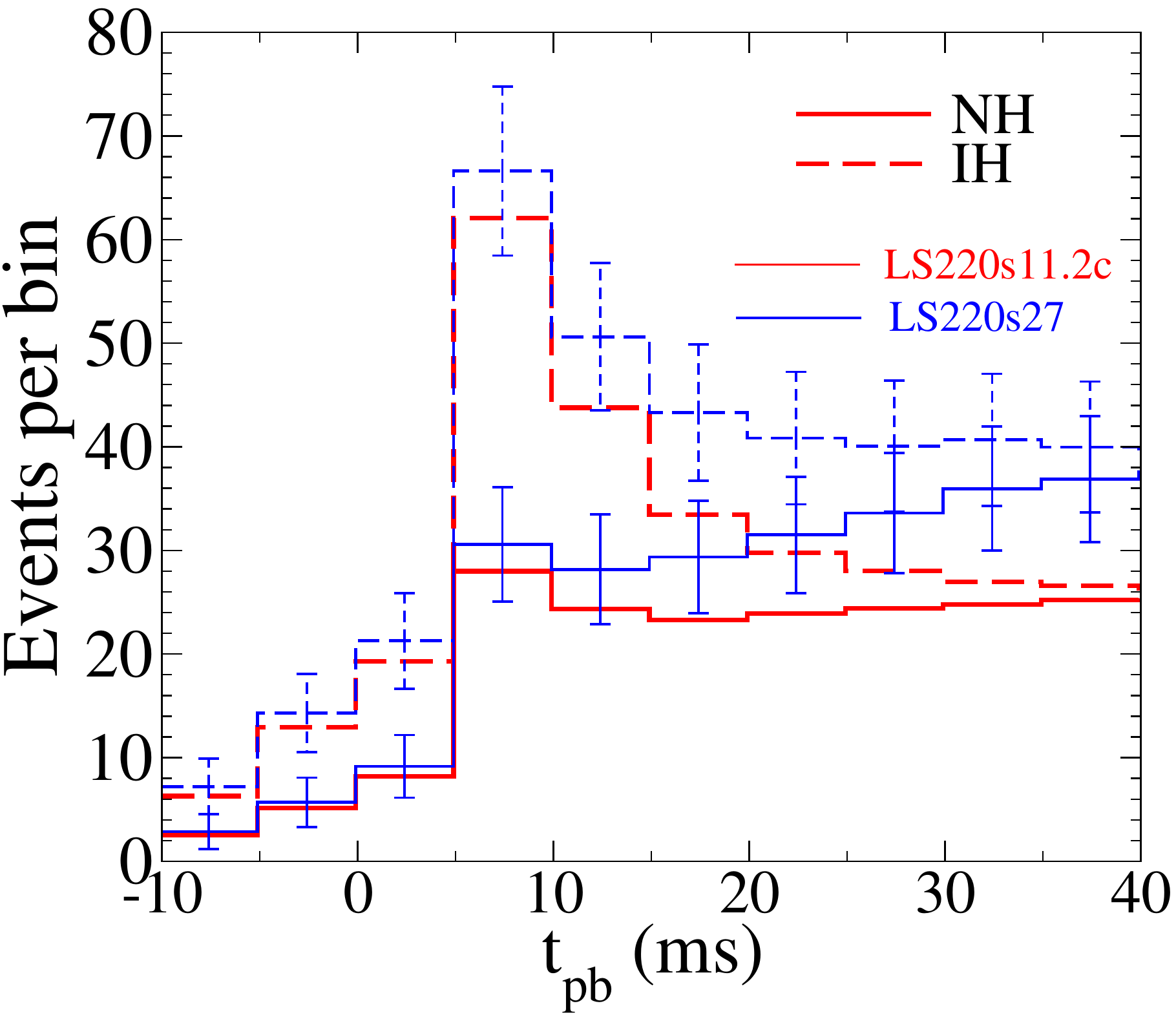} 
    \end{center}
 \caption{Neutronization events rate per time bin in a  560-kton WC detector
for 27~$M_{\odot}$ and 11~$M_{\odot}$ SN progenitors  (see Sec.~\ref{sec:protoneutronstars}) at $d=10$~kpc for both NH (continuous curve) and IH (dashed curve). 
\label{neutronmton}}
\end{figure}

\begin{figure}[h!]
\begin{center}
 \includegraphics[angle=0,width=0.8\textwidth]{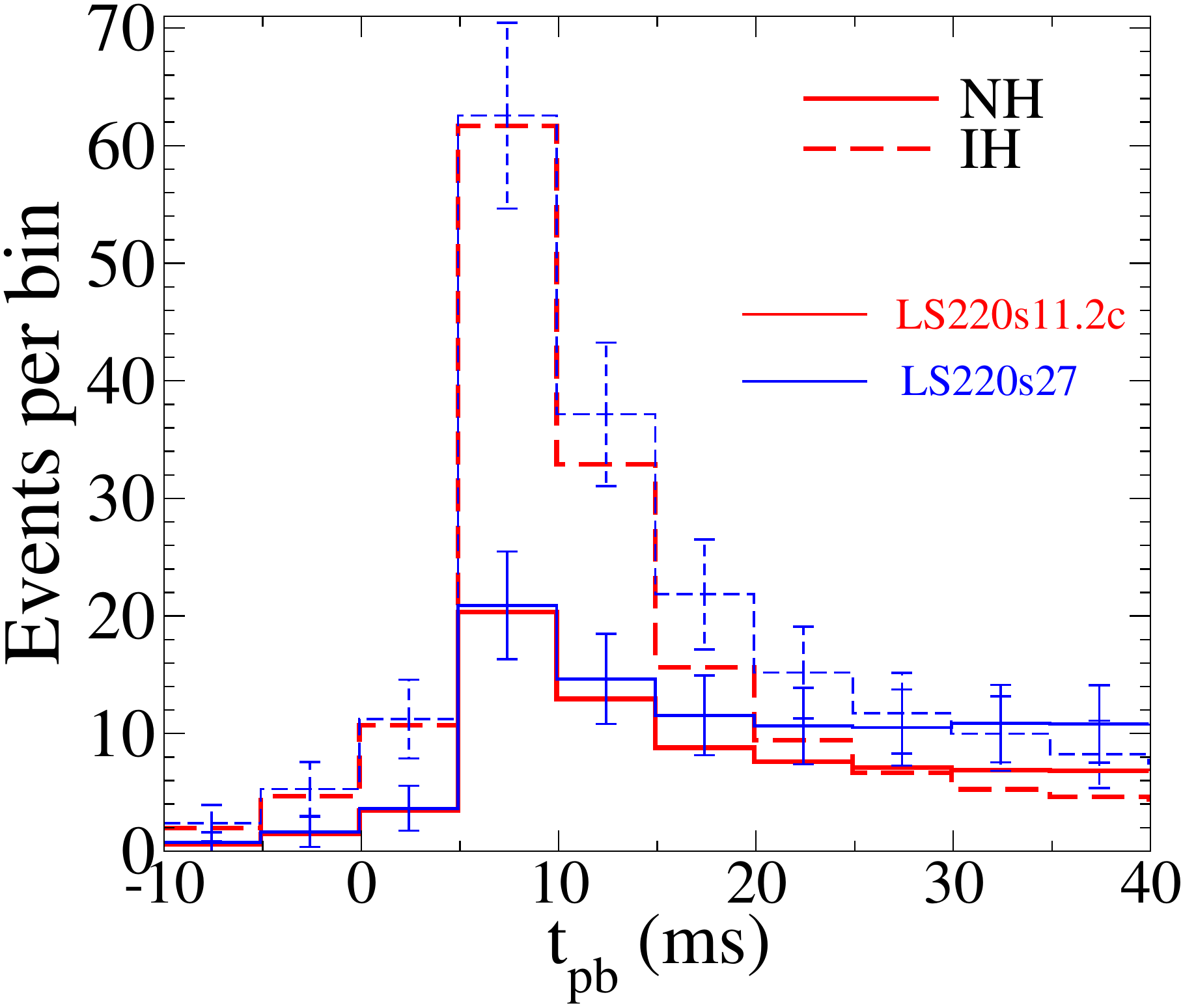} 
    \end{center}
 \caption{Neutronization events rate per time bin in a 40-kton LAr TPC
for 27~$M_{\odot}$ and 11~$M_{\odot}$ SN progenitors (see Sec.~\ref{sec:protoneutronstars})
 at $d=10$~kpc for both NH (continuous curve) and IH (dashed curve).
\label{neutronarg}}
\end{figure}

Concerning the possibility to detect the $\nu_e$ neutronization burst,
while Mton class WC detectors measure predominantly
the $\bar\nu_e$ flux using inverse beta decay,  they are also sensitive
to the subdominant $\nu_e$ channel via elastic scatterings on $e^-$ (see Sec.~\ref{sec:detection}). On the other hand,
a large LAr TPC will make the cleanest identification of the prompt $\nu_e$
burst through its unique feature of measuring $\nu_e$
charged-current interactions, enabling to
probe oscillation physics during the early stage of the SN
explosion. 
 We show the observable $\nu_e$ neutronization burst 
from  $11\ M_\odot$ and a $27\ M_\odot$ SN progenitors with LS EoS (see Sec.~\ref{sec:protoneutronstars})
in a 560-kton WC detector (Fig.~\ref{neutronmton}) and in a 40-kton  LAr TPC  
(Fig.~\ref{neutronarg});  the neutronization burst   peak is clearly visible in IH in both detectors. 
 Moreover, from these two figures one realizes that the variation in  the progenitor mass has a 
sub-leading impact in the features of the signal.

The detection of the SN neutronization burst has also been proposed to constrain
nonstandard scenarios, like Lorentz invariance violation~\cite{Chakraborty:2012gb}, 
neutrino decay~\cite{Ando:2004qe}, oscillations into light sterile neutrinos~\cite{Esmaili:2014gya},
neutrino-antineutrino oscillations mediated by a neutrino magnetic moment~\cite{Akhmedov:2003fu}. All these scenarios would lead to 
a suppression of the neutronization burst.

\begin{figure}[!t]  
\center
\includegraphics[angle=0,width=0.95\columnwidth]{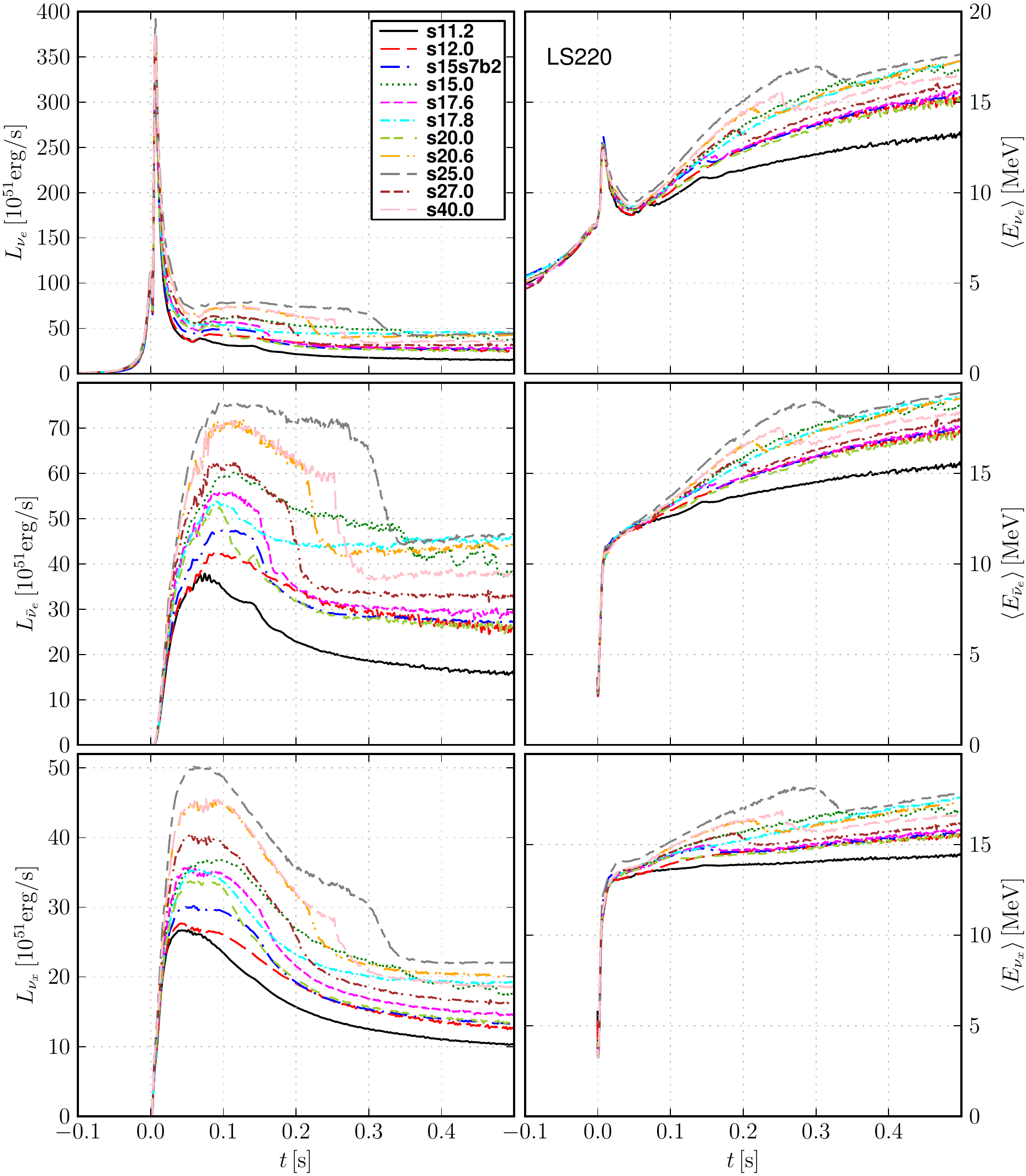} 
\caption{
Early postbounce evolution of luminosities (left panels), and mean energies (right panels)
for a set of eleven 1D simulations with progenitors of different masses  as obtained by the Garching group.
Quantities for  $\nu_e$,   $\overline\nu_e$, and  $\nu_x$   are shown in the top, middle and bottom panel, respectively. The vertical line
indicates the early timescale (100 ms). (Figure taken from~\cite{Janka:rise,Huedepohl:2013}.)
\label{rise}}  
\end{figure}  

\emph{(c) Rise time of the neutrino signal.} 
The rise time of a Galactic iron-core SN ${\overline\nu}_e$   light curve, observable
in large underground detectors,  can provide a diagnostic tool for the neutrino mass hierarchy.
Due to the  combination of  matter suppression of collective effects at early post-bounce times 
(see Sec.~\ref{sec:polar}) and the presence of the ordinary MSW effect in the outer layers of the SN, one expects that the observable $\bar\nu_e$ flux at Earth
would be given by
\begin{eqnarray}  
 F_{\bar\nu_e} &=& \cos^2 \theta_{12} F^0_{\bar\nu_e} + \sin^2 \theta_{12} F^0_{\bar\nu_x}   \,\   \,\ \,\  \,\ \,\ \,\ \,\ \,\ \,\  \,\ \,\ \textrm{(NH)} \,\ , \\
 F_{\bar\nu_e} &=&   F^0_{\bar\nu_x}  \,\ \,\ \,\ \,\ \,\ \,\ \,\ \,\ \,\ \,\ \,\ \,\ \,\ \,\ \,\ \,\ 
 \,\ \,\ \,\ \,\ \,\ \,\ \,\ \,\ \,\ \,\ \,\ \,\ \,\  \,\
\textrm{(IH)} \,\ .
\end{eqnarray} 
It is clear that the $F_{\bar\nu_e}$ 
flux at the Earth would basically reflect the original
$F^0_{\bar\nu_x}$ flux if IH occurs,
or closely match the $ F^0_{\bar\nu_e}$ flux in NH. 
As  from state-of-the-art  simulations~\cite{Janka:rise,Ott:2012jq},  temporal profiles of the original ${\overline\nu}_x$ and 
${\overline\nu}_e$ fluxes appear quite different in the accretion phase, and relatively model-independent. In particular,
  the ${\overline\nu}_x$ signal rises faster than the ${\overline\nu}_e$ one. Figure~\ref{rise}
shows the early post-bounce evolution of luminosities (left panels), and mean energies
(right panels) for a set  of eleven 1D simulations with progenitor of different masses  obtained
by the Garching group.
Due to this feature,  it would be possible to  distinguish the two mass hierarchies from the  rise time on ${\cal O}$(100) ms scale.

\begin{figure}[!t]  
\hspace{-5mm}
\includegraphics[angle=0,width=0.54\columnwidth]{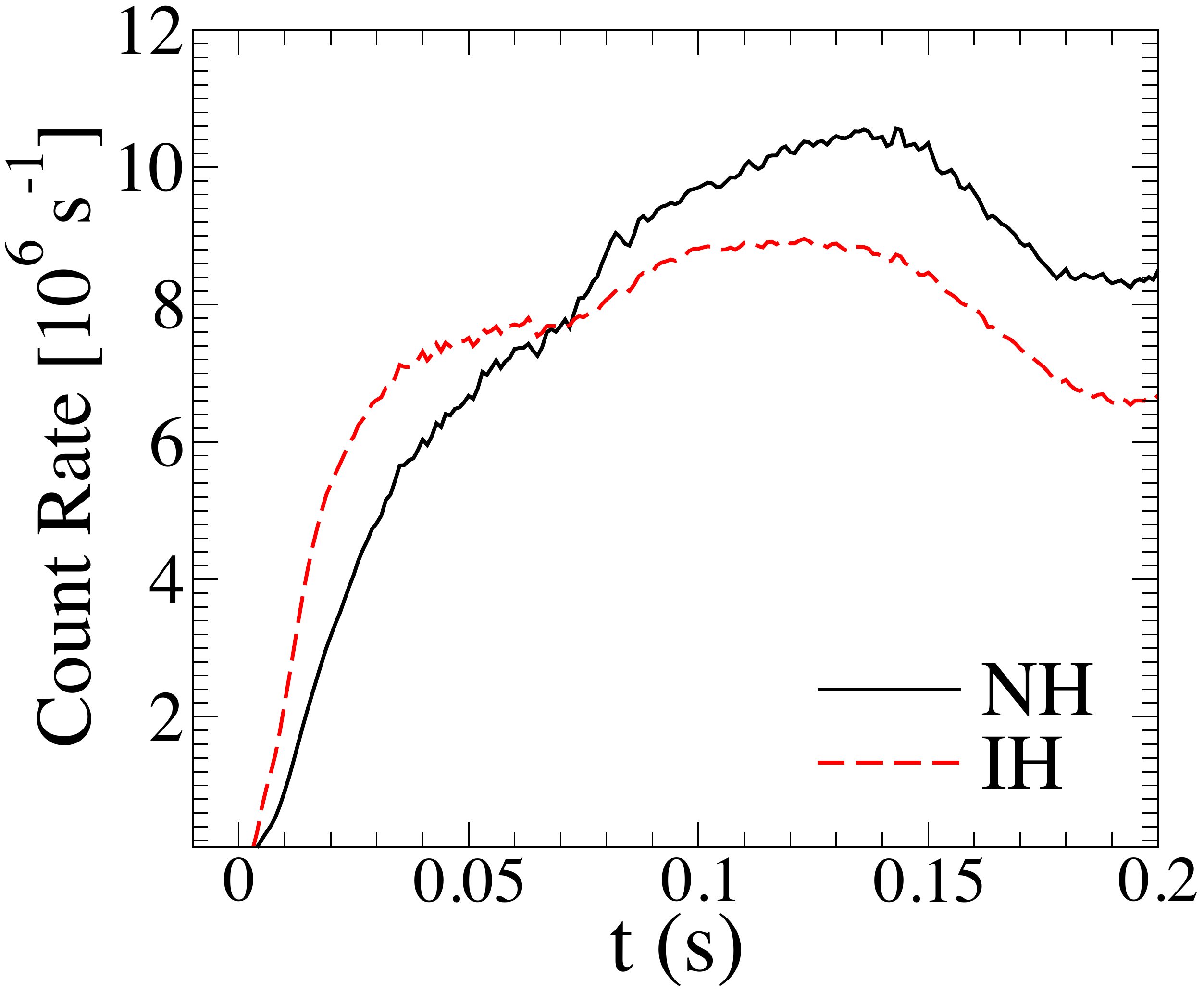} 
\includegraphics[angle=0,width=0.54\columnwidth]{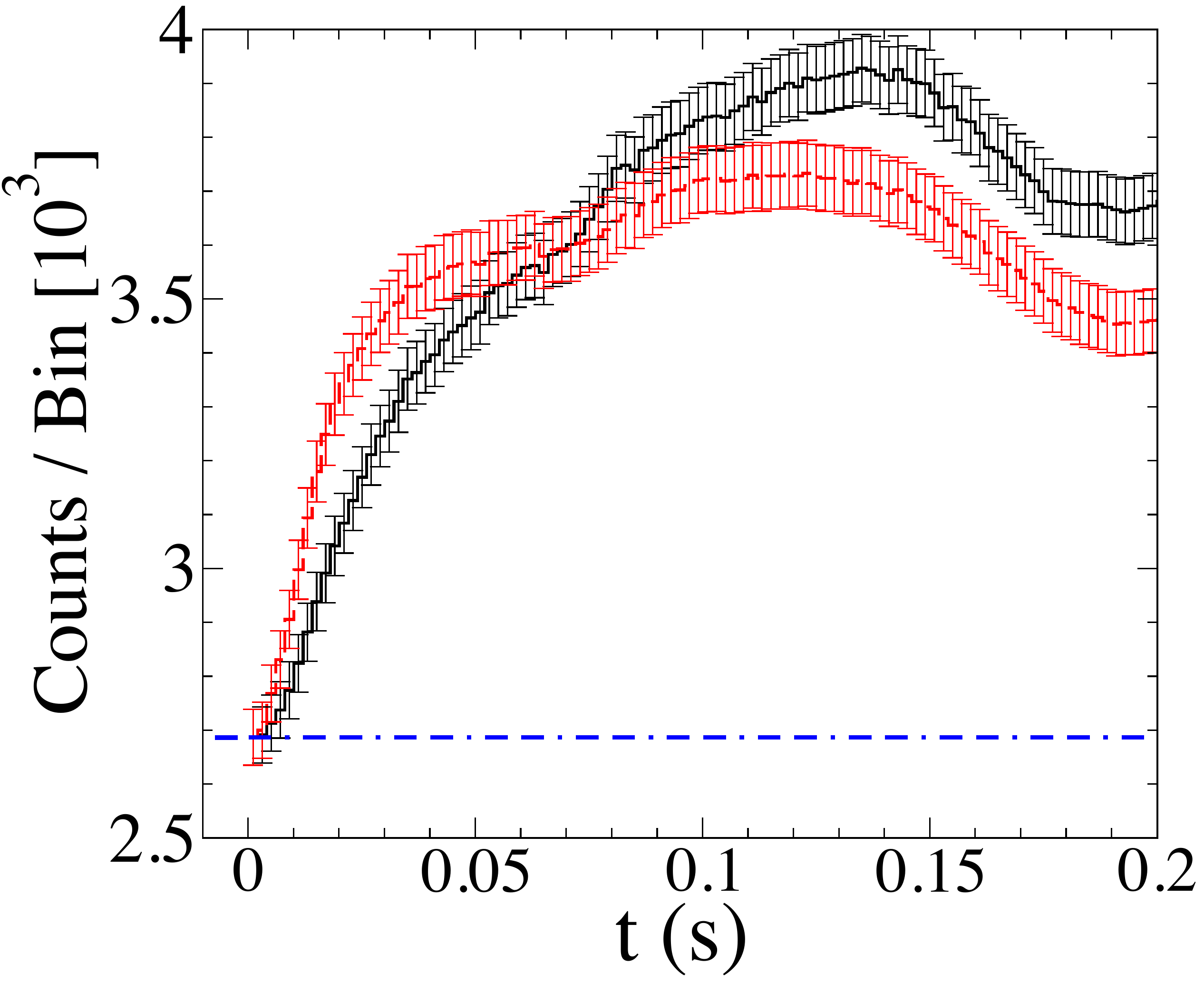} 
 \caption{Supernova signal in IceCube assuming a distance of 10 kpc,  
based on a simulation for a  15 $M_{\odot}$ SN progenitor from the Garching group.
In the right panel it is  illustrated the error size  using 2~ms bins with typical error estimates from the photomultiplier background noise. 
(Figure adapted from~\cite{Serpico:2011ir}.)
\label{rise_ice}}  
\end{figure}

Figure~\ref{rise_ice}  shows the expected {\it overall} signal rate $R(t)$ in the  IceCube detector for a Galactic SN 
(see Sec.~\ref{sec:detection} for details about the SN neutrino detection technique in IceCube).  We refer to the case of a 15 $M_{\odot}$ SN progenitor  also adopted for illustration in Fig.~\ref{rise}. The NH cases are shown with continuous curves, while the IH cases  with the dashed ones. The right panel overlies the error size  using 2~ms bins with typical error estimates from the photomultiplier background noise. 
The difference between the observed neutrino light curve in the NH and IH is evident. For the NH case, a relatively longer hump in the signal, associated with the accretion, is clearly visible. While in IH, the light curve has a  sudden rise. 

An important issue is the dependence of this signature is from theoretical uncertainties, most notably the progenitor
structure, EoS, and numerical schemes. The models tested in~\cite{Serpico:2011ir}  provide quite a satisfactory
test that the uncertainties associated with progenitor structure  do not spoil the viability of the method.
Concerning numerical schemes, the same signature has been found also within the  hydrodynamical simulations of the Basel/Darmstadt group~\cite{Fischer:2009af} 
and independently with the simulations of the Tapir group~\cite{Ott:2012jq}. 
 On the other
hand, a firm conclusion on the other uncertainties requires further studies. 
Therefore, given the potential importance  of the rise time signature in shedding light on the unknown neutrino mass
hierarchy,  it would be mandatory  to further explore the robustness of this feature with more  accurate simulations.

\emph{(d) Earth matter effect.}
It is widely known that  if  SN neutrinos should reach the
detector from  ``below''~\cite{Mirizzi:2006xx}, the Earth crossing would induce
an  energy-dependent modulation in the neutrino
survival probability~\cite{Di00,Lunardini:2001pb}. The appearance of the Earth effect depends on the 
neutrino fluxes and on the  mixing
scenario. The accretion phase  is particularly promising to detect Earth crossing signatures 
because the absolute 
SN $\nu$ flux
is large and the 
flavor-dependent 
flux differences are also large.
Due to the matter suppression  of the collective oscillations
during the accretion phase, SN neutrino initial fluxes will be processed by the
only MSW effect.
In this situation the oscillated SN $\nu$ fluxes at Earth (before Earth crossing) 
are given by Eq.~(\ref{eq:fluxes}) in terms of the 
energy-dependent $\nu_e$ survival probabilities $P_{ee}(E)$ (or ${\bar P}_{ee}(E)$ for $\bar\nu_e$).
The Earth effect can be taken into account by just mapping the non-vanishing 
$P_{ee} \to  1-P_{2e}$ and ${\bar P}_{ee} \to 1- {\bar P}_{2e}$, where $P_{2e}$ is the $\nu_2 \to \nu_e$
transition probability for neutrinos 
 propagating through the Earth, and analogously for ${\bar P}_{2e}$~\cite{Lunardini:2001pb}.
For large $\theta_{13}$,  the neutrino fluxes at Earth 
 for NH are~\cite{Di00}:
\begin{equation} 
F_{\bar\nu_e}^{\oplus} =(1 - \bar{P}_{2e}) F^0_{\bar\nu_e}+
\bar{P}_{2e} F^0_{\bar\nu_x} \,\,\,\,\,\,\,\,\,\,\,\, \textrm{and} \,\,\,\,\,\,\,\,\,\,\,\, F_{\nu_e}^{\oplus}=F^0_{\nu_x} \,\ ,
\label{eme-nh}  
\end{equation}  
while for IH 
\begin{equation} 
F_{\bar\nu_e}^{\oplus} =  F^0_{\bar\nu_x} \,\,\,\,\,\,\,\,\,\,\,\, \textrm{and} \,\,\,\,\,\,\,\,\,\,\,\,
F_{\nu_e}^{\oplus}  = (1-P_{2e}) F^0_{\nu_e}  +  P_{2e} F^0_{\nu_x} \,\ , 
\label{eme-ih}
\end{equation} 
where $F_{\nu}^{\oplus}$ indicates the neutrino fluxes after Earth crossing.

The analytical expressions for $P_{2e}$ and $\bar{P}_{2e}$ can be calculated
for the approximate two-density model of the Earth~\cite{Dighe:2003vm}.
When neutrinos traverse a distance $L$ through the mantle of the Earth,
these quantities  assume a very simple form~\cite{Di00,Lunardini:2001pb}
\begin{eqnarray}  
P_{2e} & = & \sin^2\theta_{12} + \sin2\theta^m_{12} \, \label{P2e}
 \sin(2\theta^m_{12}-2\theta_{12})  
\sin^2\left(  
\frac{\delta m^2 \sin2\theta_{12}}{4 E \,\sin2\theta^m_{12}}\,L  
\right)\,, 
\\
\bar{P}_{2e} & = & \sin^2\theta_{12} + \sin2\bar\theta^m_{12} \, \label{Pbar2e}  
 \sin(2\bar\theta^m_{12}-2\theta_{12})  
\sin^2\left(  
\frac{\delta m^2\,\sin2\theta_{12}}{4 E \,\sin2\bar\theta^m_{12}}\,L  
\right)\,,
\end{eqnarray}  
where $\theta^m_{12}$ and $\bar\theta^m_{12}$ are the effective
values of $\theta_{12}$ in the Earth matter for neutrinos and antineutrinos 
respectively~\cite{Fogli:2001pm}. 
The Earth crossing  induces a peculiar oscillatory signature in   
the neutrino energy spectrum. 
 From Eqs.~(\ref{eme-nh})--(\ref{eme-ih}),  it results that the Earth matter effect 
should be present for antineutrinos in NH and for neutrinos in 
IH, providing a potential tool to distinguish between these two cases. 

\begin{figure}[t!]
\begin{center}
 \includegraphics[angle=0,width=0.8\textwidth]{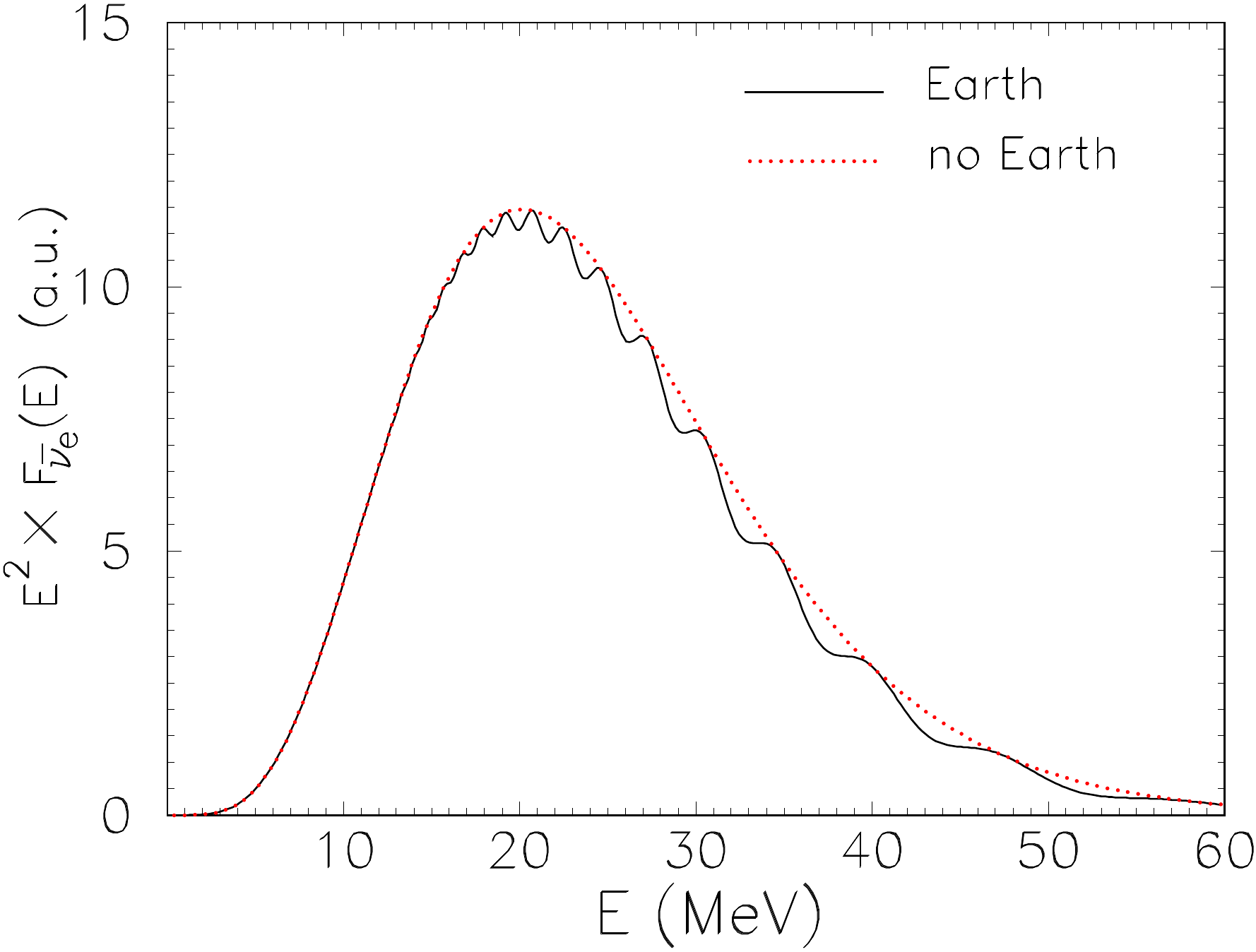} 
    \end{center}
\caption{Observable  signal $E^2 F_{\bar\nu_e}$ with
(continuous curve) and without (dotted curve) Earth crossing.
\label{earth}}
\end{figure}

Since the typical event rate associated with an inverse beta decay process  is 
$ \propto E^2 F_{\bar\nu_e}^{\oplus} (E)$---the detection cross section being $\sigma \propto E^2$---we show in Fig.~\ref{earth} the 
 $E^2 F_{\bar\nu_e}^{\oplus}$ as representative of the observable Earth-modulated signal,
 where we have taken as initial $\nu$ fluxes the ones in Fig.~\ref{coolsingl}.  
Earth matter effects could be measured in a single detector, if it has enough energy
resolution and statistics to track the wiggles in the observed energy spectrum, induced by the neutrino 
oscillations
in the Earth. In this context, a Fourier analysis of the SN neutrino signal  has been
proposed as a powerful tool to diagnose this modulation, identifying 
the peak associated with the Earth crossing in the power spectrum~\cite{Dighe:2003jg}.

The  observability of the Earth matter effect largely  depends on the neutrino  average energies  and on
 the flavor-dependent differences between
the primary spectra. In this regard, recent SN simulations indicate lower average energies
than previously expected~\cite{Janka:rise,Fischer:2009af} and a tendency towards the  equalization of the neutrino fluxes of different
flavors during the cooling phase (at  $t_{\rm pb} \gtrsim 1$~s). 
Motivated by these new inputs,  an updated study on the observability of this effect at large next-generation underground detectors (i.e., 0.4-Mton WC, 50-kton scintillation and 100-kton LAr detectors) was performed in Ref.~\cite{Borriello:2012zc}.
 It has been found that the detection of the SN neutrino Earth matter effect could be more challenging than expected from previous studies. Remarkably, it was argued that
 none of the proposed detectors shall be able to detect the Earth modulation for the neutrino signal of a typical Galactic SN at 10~kpc. This should be observable in a 100-kton LAr  detector for a SN at few kpc, while  all three detectors would clearly see the Earth signature for very close-by stars only ($d \sim 0.2$~kpc).

The Earth effect could also produce a modification in the  SN $\bar\nu_e$ light curve
measured by the IceCube neutrino telescope. Therefore, IceCube  could detect the Earth effect
by the relative  difference in the temporal signals   with a high-statistics Mton WC detector,
if only one of the two detectors is shadowed~\cite{Dighe:2003be}.
However, using the neutrino fluxes as from 
recent SN simulations also IceCube, used as co-detector together with a Mton-class WC detector,
seems  not to be able to detect any sizable variation in the SN neutrino event rate  
for any Galactic SN.

\emph{Synopsis of signatures of flavor conversions.}
To summarize, we present   a synopsis of the  discussed signatures of flavor conversions in SN neutrinos and 
their sensitivity to the neutrino mass hierarchy in Table~\ref{tab:signatures}. We focus on early time neutrino signal (neutronization
or accretion phase) where we assume that the self-induced effects are matter suppressed. Moreover, we take the MSW matter effects 
 along a progenitor static matter density profile.

\begin{table}[!h]
\begin{center}
\caption{Different detectable signatures of SN neutrino oscillations at early times and  sensitivity to the mass hierarchy.}
\begin{tabular}{cccccc}
\hline \hline
Mass Hierarchy & $P_{ee}$   & ${\bar P}_{ee}$ & $\nu_e$ burst & ${\bar\nu}_e$ rise time & Earth effects \vbox to12pt{}\\
\hline
NH & 0 & $\cos^2 \theta_{12}$ & absent & long & ${\bar \nu}_e$ 
\vbox to12pt{} \\
\hline
IH & $\sin^2 \theta_{12}$ & 0 & present & short & $\nu_e$
\vbox to12pt{} \\
\hline
\end{tabular}
\label{tab:signatures}
\end{center}
\end{table}

\subsection{Outlook}
\label{sec:conclu}

The dense SN core represents a unique laboratory  to probe neutrino flavor mixing
in high-density conditions. Indeed, within a radius of a few hundred kilometers,
the neutrino gas is so dense to become a ``background to itself.'' In these conditions,
we have shown how the neutrino flavor evolution equations become highly non-linear, leading to
surprising and counterintuitive  phenomena,
like neutrinos and antineutrinos of
different energies collectively oscillating to some other flavor. Another
surprising result is the partial or complete energy-dependent flavor
exchange, namely spectral swapping or spectral splits, as 
neutrinos travel out from a dense central region to the low neutrino
density of the outer layers.
During the past few years, our understanding of non-linear neutrino
oscillations has seen substantial progress.  Analytical and numerical works by several groups have shed light on 
the basic picture of non-linear neutrinos oscillations, although still many open issues remain to be clarified. 
In particular, it has been appreciated that  self-induced flavor conversions are related to 
instabilities in the flavor space.
While the self-induced flavor evolution  is a non-linear phenomenon, 
the onset of these conversions can be examined through a standard \emph{stability analysis} of the linearized
EoMs~\cite{Banerjee:2011fj} (see also~\cite{Sawyer:2008zs}).  Moreover, interesting conceptual studies  have been also carried on  to clarify  some of the aspects
related to the collective flavor dynamics in SNe~\cite{Raffelt:2010za,Raffelt:2011yb,Sarikas:2012ad,Pehlivan:2011hp,Hansen:2014paa}.

 The recent insight that interacting neutrino fields can \emph{spontaneously break}
symmetries inherent to the initial conditions in the flavor evolution together with hints of asymmetric neutrino emission
from the the most recent  SN simulations in 3D will require a critical re-investigation
of the previous results obtained within the bulb model. Seminal studies in this directions have just started.
Furthermore,  it has been speculated that in the transition region between the neutrino scattering-dominated regime (at high-densities) and the oscillation-dominated one (at lower densities)
spin-flavor oscillations could take place, allowing for  coherent transformations between neutrinos and antineutrinos~\cite{Vlasenko:2013fja,Cirigliano:2014aoa,Vlasenko:2014bva}. 
A possible role of  anomalous neutrino-antineutrino correlations in the development of the self-induced oscillations has also been discussed~\cite{Vaananen:2013qja,Serreau:2014cfa,Kartavtsev:2015eva}.
Remarkably, new physics scenarios could have a strong impact on the self-induced flavor conversions. 
In this regard,
 strong effects on collective oscillations have been  found  due to  spin-flip transitions~\cite{deGouvea:2012hg,deGouvea:2013zp},  triggered by a $\nu$ magnetic moment 100 times
larger than the Standard Model one, assuming typical magnetic fields in the SN envelope
($B\sim 10^{10}-10^{12}$~G). 
Moreover, the presence of non-standard flavor changing interactions among SN neutrinos and the background
fermions~\cite{EstebanPretel:2009is} or among neutrinos 
themselves~\cite{Blennow:2008er} would dramatically affect the self-induced flavor changes.
The putative existence of extra light sterile neutrino families is also expected to modify the active neutrino fluxes 
and the above described picture of neutrino self-interactions~\cite{Tamborra:2011is,Wu:2013gxa,Pllumbi:2014saa}.

An appropriate characterization  of these effects would motivate additional work
     to investigate their impact on the observable neutrino signal~\cite{Tamborra:2013,Wu:2014kaa,Gava:2009pj}, on the
			     r-process nucleosynthesis in SNe~\cite{Pllumbi:2014saa,Duan:2010af,Chakraborty:2009ej}, and on the
                 neutrino energy transfer to the stalled shock wave~\cite{Tamborra:2013,Tamborra:2014a,Tamborra:2014b}.

                 Matter effects on SN neutrino oscillations can imprint
peculiar signatures on the observable neutrino signal with a strong sensitivity to the mass hierarchy.
Neutrino flavor conversions on the wake of the shock-wave propagation
represent an intriguing possibility to follow the SN dynamics in real time through the neutrino signal.                  
 An accurate prediction of the observable oscillation signatures imprinted on the SN neutrino signal 
is extremely timely now, in relation to the intense
experimental activity for the realization of large-volume
detectors for low energy neutrino astronomy.

\newpage
\section{Diffuse Supernova Neutrino Background}
\label{sec:DSNB}
\noindent {\it Author: I.~Tamborra}
\\

While the occurrence rate of a Galactic SN is only $\le 3$ per century~\cite{vandenbergh:1991, Diehl:2006cf} and the probability of detecting neutrinos from the next Galactic explosion will be challenged by the location of the SN with respect to the detectors on Earth,  on average one SN is exploding   every second somewhere in the Universe. 
The Diffuse Supernova Neutrino Background (DSNB) is the total flux of neutrinos and antineutrinos with MeV energy emitted by  all SNe in 
our Universe. It is  expected to be isotropic and stationary and it will provide us with a unique portrait of the SN population.  Although not yet detected, the DSNB discovery prospects are excellent.  
The DSNB  detection will be  crucial to test our current understanding of the SN dynamics and SN redshift distribution. 
In this section, we report on the present status of the DSNB searches and its theoretical uncertainties. 

\subsection{Main ingredients and DSNB upper limits}
\label{sec:dsnbgeneral}

The DSNB flux  depends on the SN distribution according to their progenitor mass $M$ and redshift $z$ and on the neutrino energy spectra. For each neutrino flavor $\nu_\alpha$ ($\alpha = e, \bar{e}, \mu$ or~$\tau$) with observed energy $E$, the DSNB is defined as~\cite{Ando:2004hc}:
\begin{equation}
\Phi_{\nu_\alpha}(E)=\frac{c}{H_0} \int_{M_0}^{M_{\rm max} } dM \int_{0}^{z_{\rm max}}  dz\ \frac{\dot{\rho}_{SN}(z,M) F_{\nu_\alpha}(E (1+z),M)}{\sqrt{\Omega_M (1+z)^3+\Omega_\Lambda}}~,
\label{phiDSNB}
\end{equation}
where $c$ is the speed of light, $H_0 = 70\ \rm{km}~\rm{s}^{-1}~\rm{Mpc}^{-1}$ the Hubble constant,   $\Omega_M = 0.3$  and $\Omega_\Lambda = 0.7$ are  the matter and dark energy fractions of the cosmic energy density. $F_{\nu_\alpha}(E,M)$ is the oscillated $\nu_\alpha$ energy flux for a SN progenitor with mass $M$ introduced in Sec.~\ref{sec:polar}. The comoving SN rate (SNR) is labelled as $\dot{\rho}_{SN}$ and it is expected to be larger between $z=0$ and $z_{\rm max} \simeq 5$ (see Sec.~\ref{sec:SNR}).  The redshift correction of the energy in Eq.~(\ref{phiDSNB}) is due to the fact that each neutrino emitted from a SN at redshift $z$ will have energy $E^\prime = E (1+z)$ being $E$ the neutrino energy observed on Earth. As a consequence, neutrinos coming from high-$z$ are pushed towards lower energies; therefore the DSNB flux is dominated by  the $z\le 1$ contribution~\cite{Ando:2004hc}. 
 As it will be discussed in the next section, the SN mass distribution is such that the least massive stars (with $M \sim M_0 \simeq 8~ M_\odot$, being $M_0$ the minimum progenitor mass necessary to have a core-collapse SN)  give the larger contribution to the DSNB.  The DSNB is then weakly dependent  on $M_{\rm max} \simeq 125~ M_\odot$ (tentative upper limit for the occurrence of core-collapse SNe~\cite{Ando:2004hc}) and $z_{\rm max}$.

Since first studies on the DSNB and on its possible detection~\cite{Zeldovich:1965,Guseinov:1967,Ruderman:1965,Bisnovatyi:1982,Krauss:1984},  our knowledge  on the topic has increased considerably (see, e.g., Refs.\cite{Beacom:2010kk,Lunardini:2010ab} for dedicated review papers). 
From the experimental standpoint, a milestone was reached  when Super-Kamiokande placed an upper limit  on the $\bar{\nu}_e$ component of the flux~\cite{Malek:2002ns} ($\phi_{\bar{\nu}_e} \sim 0.1$--$1.~{\rm cm^{-2} s^{-1}}$ above 19.3 MeV threshold), excluding some theoretical models~\cite{Yuksel:2007mn}; while the upper limits on the other neutrino flavors above 19.3~MeV are less strong: $\phi_{\nu_e} < 73.30$--$154~{\rm cm^{-2} s^{-1}}$, $\phi_{\nu_\mu +\nu_\tau} < (1.$--$1.4) \times 10^3~{\rm cm^{-2} s^{-1}}$ and $\phi_{\bar{\nu}_\mu +\bar{\nu}_\tau} < (1.3$--$1.8) \times 10^3~{\rm cm^{-2} s^{-1}}$~\cite{Lunardini:2008xd}. The current most stringent upper limits have been placed  by the Super-Kamiokande experiment on the $\bar{\nu}_e$ component of the flux above 17.3 MeV threshold, $\phi_{\bar{\nu}_e} \le 2.8$--$3.0~{\rm cm^{-2} s^{-1}} $ at 90\% C.L.~\cite{Ikeda:2007sa,Bays:2011si,Zhang:2013tua}. More recently, Super-Kamiokande released a new upper limit by performing an analysis with a neutron-tagging technique:  $\Phi_{\bar{\nu}_e} < 5$--$30.~{\rm cm^{-2} s^{-1} MeV^{-1}}$ for neutrino energies between $13.3$~MeV and $17.3$~MeV~\cite{Zhang:2015zla}. As we will discuss later, the most promising energy window relevant for the DSNB detection against background is $11~{\rm MeV } \le E \le 40~{\rm MeV}$~\cite{Horiuchi:2008jz,Bays:2011si,Ando:2002ky,Ando:2002zj}.

\subsubsection{\emph{Cosmic supernova rate}}
\label{sec:SNR}
One of the key ingredients to compute the DSNB flux is the distribution of core-collapse SNe (i.e., Type II SNe and the subdominant Type Ib/c SNe) with the redshift $z$ and progenitor mass $M$. 
The SNR, $\dot{\rho}_{SN}(z, M)$, is proportional to the Star Formation Rate (SFR), $\dot{\rho}_{SF}(z)$, through the initial mass function (IMF), $\psi(M)$:
\begin{equation}
\dot{\rho}_{SN}(z,M) = \frac{\psi(M)}{\int_{0.5 M_\odot}^{M_{\rm max} }dM\ M \psi(M)}   \dot{\rho}_{SF}(z)\ ;
\label{snr}
\end{equation}
therefore normalization uncertainties on the IMF do not affect $\dot{\rho}_{SN}(z,M)$. 
 The distribution of stellar masses is assumed to follow a universal IMF, as the conventional
Salpeter scaling law:  $\psi(M) \propto M^{-2.35}$ for stellar masses $M$ between $0.1\ M_\odot$
and $100\ M_\odot$~\cite{Salpeter:1955it}, although even other  IMF scaling laws have been suggested, such as  an intermediate one ~\cite{Kroupa:2000iv} or a shallow one~\cite{Baldry:2003xi}.

\begin{figure}[t]
\centering
\includegraphics[width=0.6\textwidth]{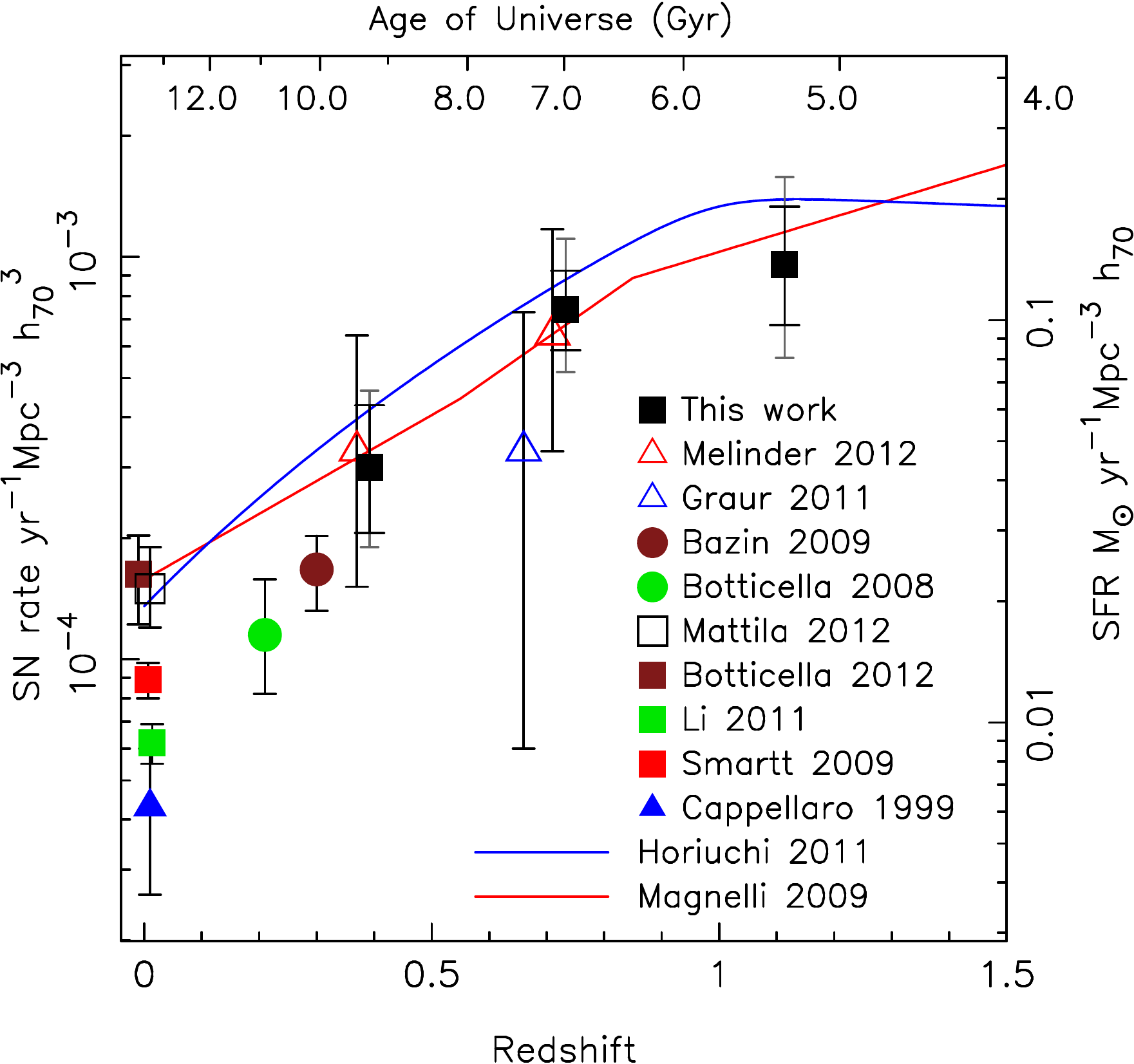}
 \caption{Cosmic supernova rate as a function of the redshift. The SFRs from \cite{Horiuchi:2011zz,Magnelli:2009}
 are shown as continue lines after extrapolating the SNR from the SFR as from Eq.~(\ref{snr}), assuming that progenitors with mass $8\ M_\odot \le M \le 50\ M_\odot$ produce optical SNe.  Measured SNRs from recent work are also shown.  (Reprinted  figure with permission from~\cite{Dahlen:2012cm}; copyright (2012) by the American Astronomical Society.)
}
 \label{plot:SNR}
\end{figure}

 Recent studies~\cite{Hopkins:2006bw,Yuksel:2008cu} suggest that the SFR can be fitted by a piecewise function of the redshift, for example~\cite{Hopkins:2006pr}: 
\begin{eqnarray}
\dot{\rho}_{SF}(z) \propto\Bigg\{ \begin{array}{lc}{(1+z)^{\delta}}\ \ \ \ \ \ \ z<1 \\
{(1+z)^{\alpha}}\  \ \ \ 1<z<4.5\ ,\\
{(1+z)^{\gamma}}\ \  4.5<z
\end{array} 
\label{snrparam}
\end{eqnarray}
with $\delta=3.28$, $\alpha=-0.26$, $\gamma=-7.8$, normalized such that $\int_{M_0}^{M_{\rm max} }dM\  \dot{\rho}_{SN}(0,M)  = 1.5 \times 10^{-4}~{\rm Mpc^{-3} yr^{-1}}$. 
Note that the DSNB dependence on $\dot{\rho}_{SF}(z > 1)$ is weak as the DSNB is mainly determined by the contribution coming from $z\le 1$.  The SNR defined as in Eq.~(\ref{snrparam}) is a growing function of the redshift implying that SNe were most frequent in the past (see also Fig.~\ref{plot:SNR}). 
Given the definition of the SNR (via the IMF), the DSNB  is dominated by the low-mass threshold $M_{0}$, although this quantity is difficult to predict accurately from theory since stellar properties change rapidly between $6$--$10\ M_\odot$. 
On the other hand, the upper limit mass $M_{\rm max}$ is less important because of the fast decline of the IMF with $M$.

In order to estimate the DSNB,  one could rely on direct measurements of the SNR, instead of using Eq.~(\ref{snr}).  However, while the shape of the SNR is basically well known,  
its measurements are governed  by normalization errors~\cite{Hopkins:2006bw,Horiuchi:2011zz}. Reference~\cite{Horiuchi:2011zz} pointed out that the normalization from direct SN observations was lower than that from SFR data by a factor $\sim 2$, this discrepancy is known as the ``supernova-rate problem.'' 
The reasons for such a mismatch may be manifold. It might be that a large number of SNe are actually dim  (either due to dust obscuration or being intrinsically weak), or that systematic changes are needed in our understanding of stars  and SN formation. 

Measurements of the SNR have greatly improved in the last few years~\cite{Li:2010kd,Botticella:2011nd,Mattila:2012zr,Dahlen:2012cm,Melinder:2012}. Figure~\ref{plot:SNR} shows the SNR from the most recent cosmic measurements in comparison with the one  predicted from the SFR assuming that all SNe with $M \ge 8\ M_\odot$ yield optical SNe~\cite{Hopkins:2006bw,Horiuchi:2011zz,Magnelli:2009}. It is clear as the SNR problem has been cured at low $z$ where we now have a better understanding of the dust extinction in the host galaxies and of the SN luminosity function, but the SNR is still affected by large  uncertainties at high $z$.  
It might also be that the fraction of ``invisible'' SNe (i.e., not optically visible SNe) increases with the redshift~\cite{Mattila:2012zr,Yuksel:2012zy}. 

The SNR estimation will hopefully  further improve within the next few years. In fact SNR measurements will come  from synoptic surveys which  will scan the full sky and probe the SN population with high sensitivity (see, e.g., discussion in Ref.~\cite{Lien:2010yb} and references therein). Such surveys are expected to pin down the error on the normalization of the SNR and, as a consequence, on the DSNB normalization. On the other hand, note that the forthcoming detection of the DSNB could provide independent constraints on the SFR~\cite{Ando:2004sb}.

\subsubsection{\emph{Time and progenitor dependence of the neutrino energy fluxes}}
\label{sec:timeSNprogenitor}
The dependence of the DSNB on neutrino oscillation physics and SN progenitors has been discussed in Ref.~\cite{Lunardini:2012ne}. 
As shown in Sec.~\ref{sec:SNmodels} and Fig.~\ref{rise},  the neutrino luminosities and spectra depend on the compactness of the progenitor star~\cite{O'Connor:2012am}. Moreover,  the luminosities of the different flavors are 
almost equal during the cooling phase, while $L_{\nu_e}, L_{\bar{\nu}_e} > L_{\nu_{\mu,\tau}}$ during the accretion phase. The mean energies are $\langle E_{\bar{\nu}_e,\nu_{\mu,\tau}}\rangle > \langle E_{\nu_e}\rangle$ during the cooling phase (see Fig.~\ref{neutrinos-convection-s27} and Sec.~\ref{sec:protoneutronstars}). 

Concerning the neutrino oscillation scenario (see Sec.~\ref{sec:oscillations}), while the MSW effect is well understood analytically and it occurs far away from the neutrinosphere,  neutrino self-interactions crucially depend on the initial hierarchy among neutrinos of different flavors, number of energy crossings and on the neutrino mass hierarchy. During the accretion phase ($t_{\rm pb} \le 1$~s),  complete or partial multi-angle matter suppression of neutrino self-interactions occurs because of the high matter potential~\cite{Chakraborty:2011nf,Chakraborty:2011gd,Saviano:2012yh,Sarikas:2011am}. While during the cooling phase, multiple spectral splits might appear according to the neutrino mass hierarchy and the number of crossings in the non-oscillated spectra. However, the shape of the spectral splits is smeared and their size reduced  due to the similarity of the non-oscillated flavor energy spectra and to $\nu$--$\nu$ multi-angle  effects, partial neutrino flavor conversions occur and the flavor transition become gentler as time increases~\cite{Lunardini:2012ne}.

As discussed in Sec.~\ref{sec:oscillations}, it is possible to factorize the effects of $\nu$--$\nu$ interactions and the MSW resonances for sake of simplicity,  as they occur in well separated spatial regions.  
Denoting with $F^{c}_{\nu_{\alpha}}$ the fluxes after the collective oscillations,  the unchanged $F_{\nu_\alpha}^0$ will undergo the traditional MSW conversions after $\nu$--$\nu$ interactions.
The fluxes ($F_{\nu_{e}}$ and $F_{\bar \nu_{e}}$) reaching the Earth after both the collective and MSW oscillations for NH and IH 
 are~\cite{Dasgupta:2007ws,Chakraborty:2008zp,Chakraboty:2010sz,Dighe:1999bi}:
\begin{eqnarray}
\label{fluxtable1}
F_{\nu_e}^{\rm NH}&=& \sin^2 \theta_{12} [1- P_{c}(F^{c}_{\nu_{e}},F^{c}_{\bar\nu_{e}},E)] (F^{0}_{\nu_e}-F^{0}_{\nu_y})  + F^{0}_{\nu_y}\ , \\ 
F_{\bar{\nu}_e}^{\rm NH}&=& \cos^2 \theta_{12} \bar P_{c}(F^{c}_{\nu_{e}},F^{c}_{\bar\nu_{e}},E) (F^{0}_{\bar{\nu}_e}-F^{0}_{\nu_y}) + F^{0}_{{\nu}_y}\ , \label{fluxtable2}\\ 
F_{\nu_e}^{\rm IH}&=& \sin^2 \theta_{12} P_{c}(F^{c}_{\nu_{e}},F^{c}_{\bar\nu_{e}},E) (F^{0}_{\nu_e}-F^{0}_{\nu_y}) + F^{0}_{{\nu}_y}\ , \\ \label{fluxtable3}
F_{\bar{\nu}_e}^{\rm IH} &=& \cos^2 \theta_{12}  [1-\bar P_{c}(F^{c}_{\nu_{e}},F^{c}_{\bar\nu_{e}},E)] (F^{0}_{\bar{\nu}_e}-F^{0}_{\nu_y}) + F^{0}_{\nu_y}\ . 
\label{fluxtable4}
\end{eqnarray}
In Eqs.~(\ref{fluxtable1})--(\ref{fluxtable4}) we have considered multi-angle 
$\nu$--$\nu$ interactions driven by $\Delta m^2$ and  $\theta_{13}$ between
$\nu_e$ and $\nu_y$, while the other flavor ($\nu_x$) does not evolve,   the effects on self-induced flavor conversions induced by 
the third flavor being a  negligible correction  for our purpose~\cite{Duan:2010bg,Friedland:2010sc,Dasgupta:2010ae,Dasgupta:2010cd}.

\subsection{Dependence of the DSNB from flavor oscillations and SN progenitor properties}

Adopting the SNR as from the SFR [Eq.~(\ref{snr})] and time-(and progenitor-) dependent oscillation physics described as 
in~Sec.\ref{sec:timeSNprogenitor}, the dependence of the DSNB on the neutrino oscillation physics and on the SN progenitors has been studied in Ref.~\cite{Lunardini:2012ne},  adopting  long-term SN simulations  for three progenitors with masses $M= 18,10.8,8.8\ M_\odot$~\cite{Fischer:2009af} (with Shen EoS~\cite{Shen:1998gq}). The features of the neutrino signal of these three progenitors have been extended to the whole estimated SN mass range   ($[M_0, M_{\rm max}] = [8\ M_\odot, 125\ M_\odot]$) and time-dependent oscillated neutrino spectra have been included in the DSNB computation implementing MSW and multi-angle $\nu$--$\nu$ interactions as a function of the post-bounce
\begin{table}[h!]
\center
\caption{Mean energies for the time-integrated neutrino flux $F_{\nu_\alpha}(E)$ in normal and inverted hierarchy. The listed cases are: (a) $F_{\nu_\alpha}(E)$ obtained considering $t_{\rm pb}=0.5$~s as representative post-bounce time and one  SN progenitor ($10.8\ M_\odot$) as representative of the whole stellar population, no oscillations; (b)  $F_{\nu_\alpha}(E)$ obtained including time-dependent fluxes for the $10.8\ M_\odot$ SN progenitor, no oscillations; (c) $F_{\nu_\alpha}(E)$ with time-dependent fluxes for the $10.8\ M_\odot$ SN progenitor, MSW effects included; (d) $F_{\nu_\alpha}(E)$ with time-dependent fluxes for the $10.8\ M_\odot$ progenitor, MSW and neutrino self-interactions included; (e) $F_{\nu_\alpha}(E)$ obtained including the whole stellar population, time-dependent fluxes, MSW and neutrino self-interactions. (Adapted from~\cite{Lunardini:2012ne})}
\begin{tabular}{llllll}
\hline
 &a & b & c & d & e \\
 \hline
$\langle E_{\nu_e} \rangle_{\rm NH}$~(MeV)  &$9.53$ & $8.96$ & $12.24$ & $11.90$& $11.96$\\
$\langle E_{\bar{\nu}_e} \rangle_{\rm NH}$~(MeV)  &$11.82$ & $10.84$ & $11.23$ & $11.50$& $11.62$\\
\hline
$\langle E_{\nu_e} \rangle_{\rm IH}$~(MeV)  &$9.53$ & $8.96$ & $10.83$ & $11.08$& $11.23$\\
$\langle E_{\bar{\nu}_e} \rangle_{\rm IH}$~(MeV)  &$11.82$ & $10.83$ & $12.24$ & $12.17$ &$12.18$\\
\hline
\end{tabular}
\label{tab:DSNB}
\end{table}
time.  A quantitative estimation of the dependence of the time-integrated $F_{\nu_\alpha}(E)$ on oscillation effects and on the stellar population  is provided in Table~\ref{tab:DSNB}~\cite{Lunardini:2012ne}. Notice that, as we will discuss later, the upper limit of the SN mass progenitor as well as the neutrino fluxes might change including failed SNe~\cite{Lunardini:2009ya}.

\begin{figure}[!h]
\centering
\includegraphics[width=0.95\textwidth]{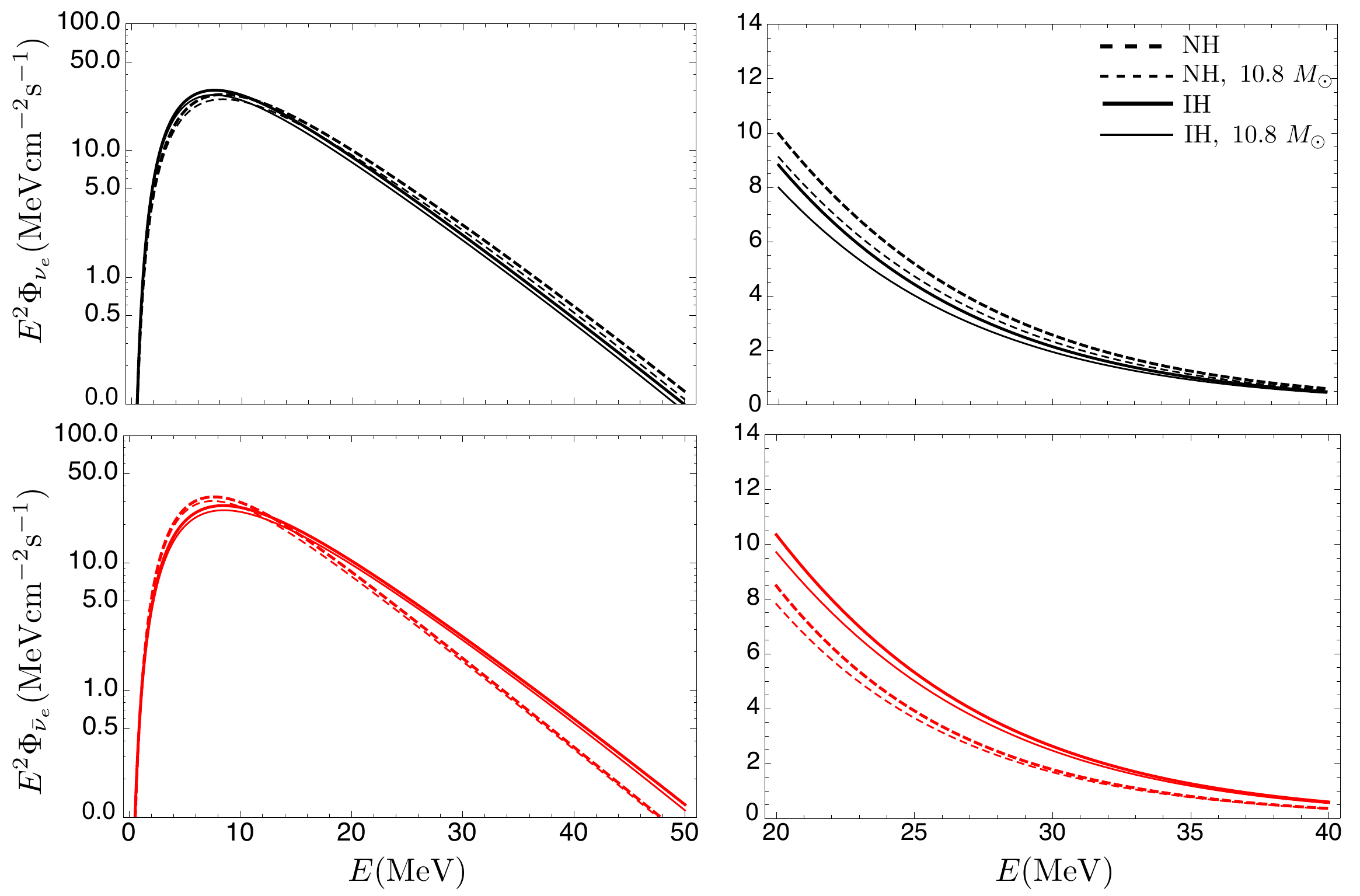}
 \caption{Left panels: DSNB as a function of the neutrino energy. Right panels: Zoom of the DSNB flux in the region of Super-Kamiokande detection ($E > 17.3$~MeV). The $\nu_e$ ($\bar{\nu}_e$) flux is plotted in black (red). The solid (dashed) line represents the DSNB in IH (NH). The thick lines represent the DSNB including the three SN progenitors and the thin line shows the DSNB obtained considering the $10.8\ M_\odot$ SN progenitor as representative of the whole stellar population. (Reprinted figure  from~\cite{Lunardini:2012ne};  copyright (2012) by the Institute Of Physics Publishing.)}
 \label{DSNB_plot}
\end{figure}
Figure~\ref{DSNB_plot} shows the DSNB, $E^2 \times \Phi_{\nu_\alpha}$ [with $\Phi_{\nu_\alpha}$ as in Eq.~(\ref{phiDSNB})], as a function of the energy for $\nu_e$ ($\bar{\nu}_e$) on the top (bottom).  In the plots on the right,  the region of interest for Super-Kamiokande detection ($E > 17.3$~MeV) is zoomed. A  variation of $10$--$20\%$ due to the neutrino mass hierarchy is visible at $E \simeq 20$~MeV.  
In order to investigate the DSNB dependence from the SN cooling and progenitor mass, a case using  $t_{\rm pb}=0.5$~s as representative of all post-bounce 
times and  the $10.8\ M_\odot$ SN progenitor as representative of the whole SN population is shown (thin lines) in comparison to the one computed 
including all SN mass models (thick lines). 

No signature due to $\nu$--$\nu$ interactions is visible in Fig.~\ref{DSNB_plot}, since neutrino self-interactions are suppressed during the accretion phase and they occur at different energies for each post-bounce time during the cooling~\cite{Lunardini:2012ne}. Note that Ref.~\cite{Fischer:2009af} does not include nucleon recoil effects  responsible for reducing the differences among the mean energies of different flavors during the cooling phase~\cite{Mueller:2012is,Mueller:2011aa,Huedepohl:2009wh}.  Therefore, the adopted SN inputs are useful to estimate the largest oscillation effects that could realistically affect the DSNB. Note that any flavor conversion occurring during the
 accretion phase will have a larger impact on the total DSNB, the luminosities during the accretion phase being larger than the cooling ones. While, during the cooling phase, the fluxes are  similar and any flavor conversion scarcely affect the DSNB.

Summarizing the impact of the different ingredients on the DSNB:  MSW effects are the largest source of variation of the DSNB with respect to the case without oscillations ($50$--$60 \%$),  the mass hierarchy induces variations of $\sim 20\%$; neutrino-neutrino interactions, time-dependence of the neutrino energy spectra over $\sim 10$~s and stellar population  are responsible for 
a smaller variations ($\sim 5$--$10 \%$)~\cite{Lunardini:2012ne}. 
Astrophysical uncertainties in the determination of the DSNB are instead larger: Besides the SN rate problem discussed above, the  SNR is affected by a normalization error of $\sim 25 \%$, although such error is expected to be dramatically reduced within the next decade. Another potentially small source of error on the DSNB is the  nuclear EoS~\cite{Mathews:2014qba}. Moreover, Refs.~\cite{Nakazato:2013maa,Nakazato:2015rya} pointed out as SN models with longer shock revival time produce a resultant higher DSNB event rate, since the larger the gravitational energy of the accreted matter, the larger is the resultant emitted neutrino energy. In principle, the shock revival time might be progenitor dependent and therefore introduce a non-negligible variation of the high-energy tail of the DSNB.

Although theoretical studies have pointed out that the DSNB is sensitive to the SN progenitor mass, 
EoS, variations of the shock revival time, neutrino mass hierarchy, and oscillation physics, we stress that the DSNB 
carries integrated population information. In view of our current knowledge of the flavor oscillation physics and stellar dynamics, it will 
therefore be extremely difficult to 
extract information on each of the mentioned aspects from a forthcoming DSNB detection. In this sense, the high statistics 
of neutrino events potentially provided by the next galactic explosion will be more useful to constrain the SN engine and to 
learn about its dynamics. As we will discuss in the next section, the DSNB might also be affected by the presence of invisible SNe 
(probably also responsible for the mismatch between the measured SNR and the one extracted from the SFR) and the DSNB detection
will be an extremely powerful  instrument to learn about the whole SN population.

\subsection{Contribution from invisible supernovae}
\label{sec:failed}

Besides stellar core-collapse events that trigger SN explosions, a significant fraction (some 10--30\%, possibly even more; e.g.,
~\cite{Horiuchi:2011zz,O'Connor:2010tk,Ertl:2015rga,Smartt:2008zd,Ugliano:2012kq,Horiuchi:2014ska,Kochanek:2013yca,Fryer:1999mi,Heger:2002by})
could lead to black holes (BH) ``directly,'' i.e., after a relatively short (fractions of a second to seconds) accretion 
phase of a transiently existing hot neutron star. Such ``failed SNe'' are expected to produce only faint electromagnetic emission 
by stripping their loosely bound hydrogen envelopes~\cite{Kochanek:2013yca,Lovegrove:2013ssa,Nadezhin:1980}.
The dim emission and red color make them hard to detect at extragalactic distances. Their neutrino emission, however, could account for an important contribution to the
DSNB. In Sec.~\ref{sec:blackholes} neutrino signals of such events were presented as predicted by recent hydrodynamic simulations.
The Super-Kamiokande Collaboration~\cite{Iida:2009qnz,Bays:2012wty} has provided  constraints on the neutrino flux from failed SNe, setting a limit about a factor
of four higher than predicted in Ref.~\cite{Lunardini:2009ya}. 

Black hole formation may also occur with considerably longer delay, if the  SN mechanism does not release enough energy to unbind the whole star.
In this case a fair fraction of the stellar matter can fall back to the neutron star (on time scales of minutes to hours) and can push the neutron
star beyond the mass limit for the black-hole formation. Such ``fallback SNe'' would radiate SN-like, though possibly sub-luminous, electromagnetic displays~\cite{Zhang:2007nw}.
Their neutrino signals, however, are initially indistinguishable from  those of  forming neutron stars in ordinary core-collapse SNe. Whether the potentially
massive fallback of matter to the neutron star produces observable neutrino emission is presently not clear because of the lack of detailed studies,
which are difficult due to the long time scales involved and the  likely importance of asymmetries and angular momentum (but see Ref.~\cite{Fraija:2015lya}). 
Since the rate of fallback SNe in the present-day Universe is predicted to be low~\cite{Ertl:2015rga},
we will not consider them in our discussion of the DSNB.

Taking into account the current uncertainties, we will focus on the contributions
from core-collapse and failed SNe to the DSNB in the following. We will see that the 
neutrino emission of failed SNe may indeed not be negligible but instead could 
make the more luminous component of the observable DSNB.

 Neglecting the progenitor mass dependence of the neutrino fluxes as well as of the SN rate, Eq.~(\ref{phiDSNB}) can be generalized as 
\begin{eqnarray}
\Phi_{\nu_\alpha}(E) &=&\Phi_{\nu_\alpha}^{\rm BH}(E) + \Phi_{\nu_\alpha}^{\rm NS}(E) = \\ \nonumber
&=& \frac{c}{H_0} \int_{0}^{z_{\rm max}}  dz\ \frac{\dot{\rho}_{\rm SN}(z) [f_{\rm BH} F_{\nu_\alpha}^{\rm BH}(E (1+z)) + 
f_{\rm NS} F_{\nu_\alpha}^{\rm NS}(E (1+z))]}{\sqrt{\Omega_M (1+z)^3+\Omega_\Lambda}}~,
\label{phifailed}
\end{eqnarray}
with $f_{\rm NS}$ ($f_{\rm BH}$) the fraction of core-collapse (failed) SN, such that $f_{\rm NS} + f_{\rm BH} = 1$, where we have subdivided the whole SN family in two major categories, neglecting the fallback SN class (that could easily fall in one or the other case according to the progenitor mass) and assumed that both these families follow the SFR redshift dependence.

The expected diffuse flux has been discussed in Ref.~\cite{Lunardini:2009ya}, assuming typical neutrino spectral parameters for the NS case as in Ref.~\cite{Keil:2002in} (i.e., $L_{\bar{\nu}_e} = L_{\bar{\nu}_x} = 5 \times 10^{52}$~erg/s, $\langle E_{\bar{\nu}_e} \rangle = 15$~MeV, $\langle E_{\bar{\nu}_x} \rangle = 18$~MeV, $\alpha_{\bar{\nu}_e} = 3.5$, $\alpha_{\bar{\nu}_e} = 2.5$) and modeling the BH case as in Ref.~\cite{Nakazato:2008vj} for a typical failed SN progenitor of $40\ M_\odot$ with stiffer Shen (S) EoS~\cite{Shen:1998gq} (i.e., $L_{\bar{\nu}_e} = 12.8 \times 10^{52}$~ergs, $L_{\bar{\nu}_x} = 4.9 \times 10^{52}$~erg/s, $\langle E_{\bar{\nu}_e} \rangle = 23.6$~MeV, $\langle E_{\bar{\nu}_x} \rangle = 24.1$~MeV) and softer Lattimer-Swesty (LS)  EoS~\cite{Lattimer:1991nc} (i.e., $L_{\bar{\nu}_e} = 4.5 \times 10^{52}$~ergs, $L_{\bar{\nu}_x} = 2.2 \times 10^{52}$~erg/s, $\langle E_{\bar{\nu}_e} \rangle = 20.4$~MeV, $\langle E_{\bar{\nu}_x} \rangle = 22.2$~MeV). 

\begin{figure}[t]
  \centering
 \includegraphics[width=0.49\textwidth]{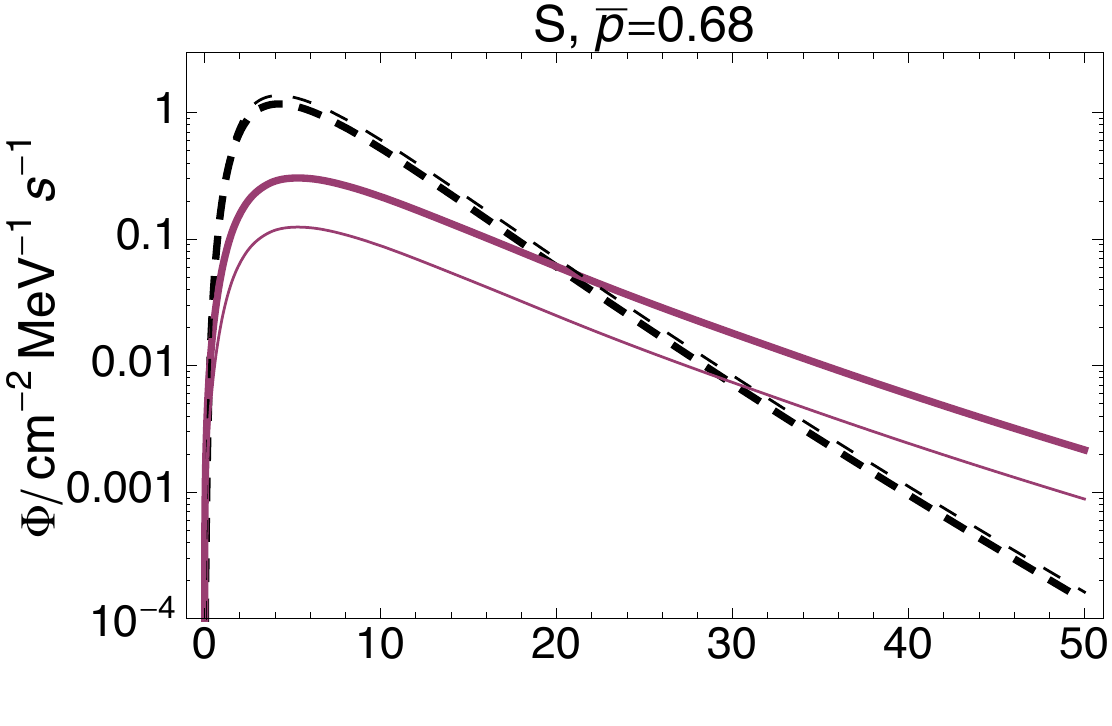}
 \includegraphics[width=0.49\textwidth]{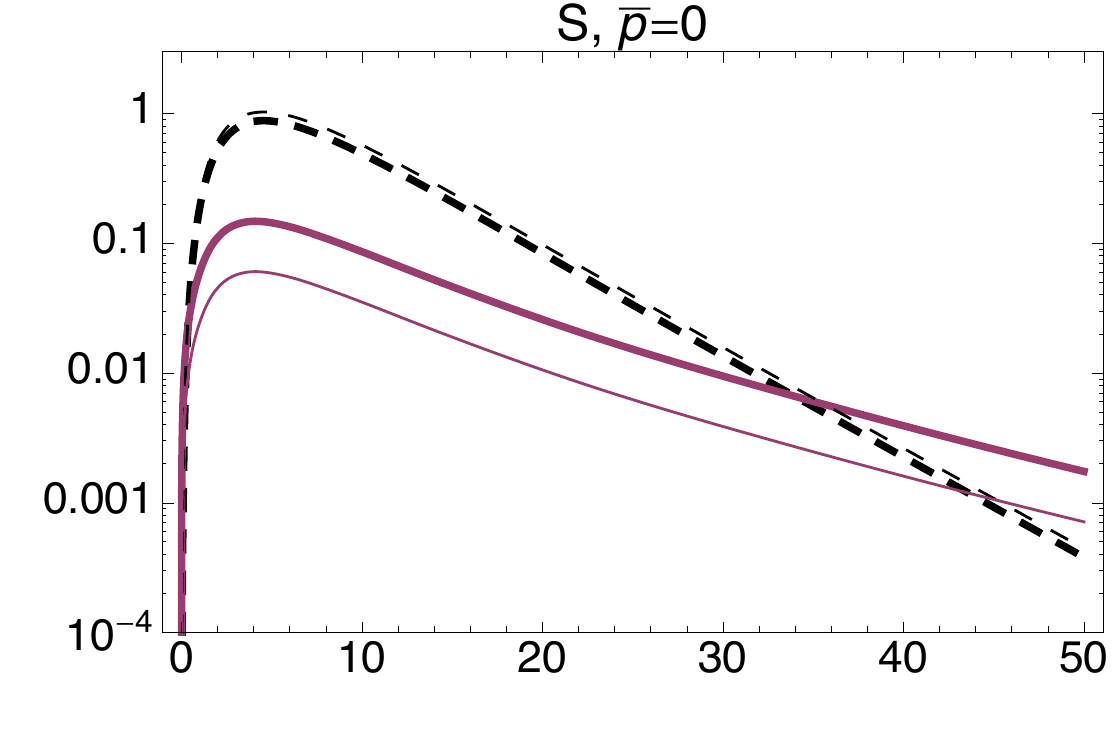}
 \includegraphics[width=0.49\textwidth]{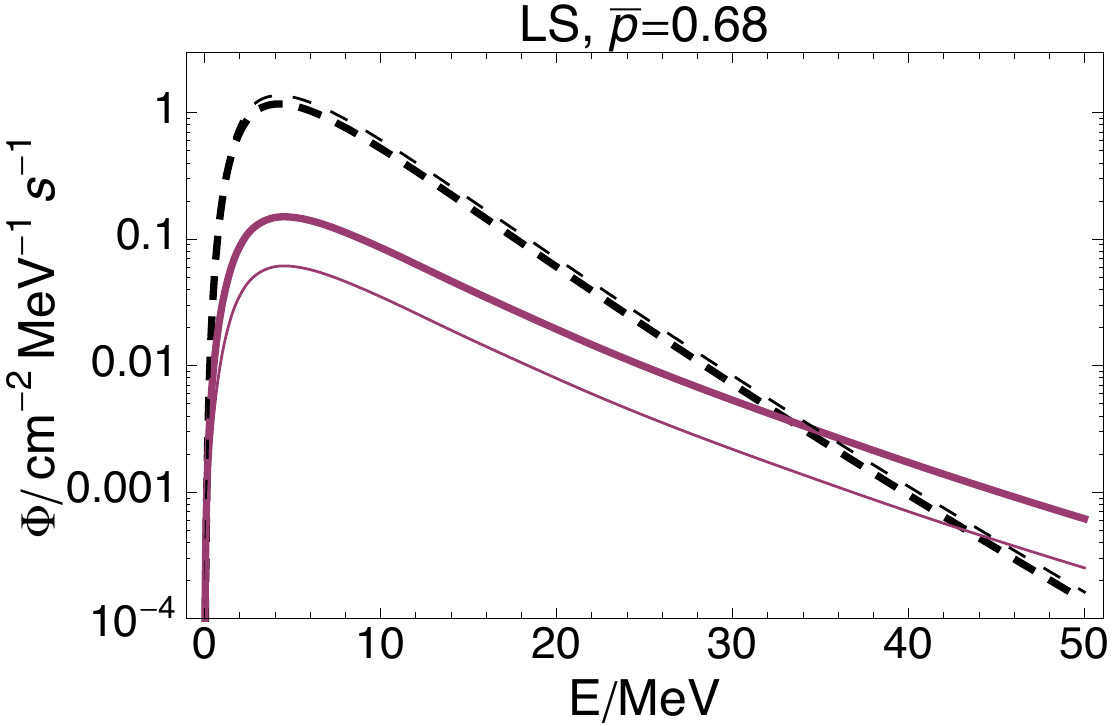}
 \includegraphics[width=0.49\textwidth]{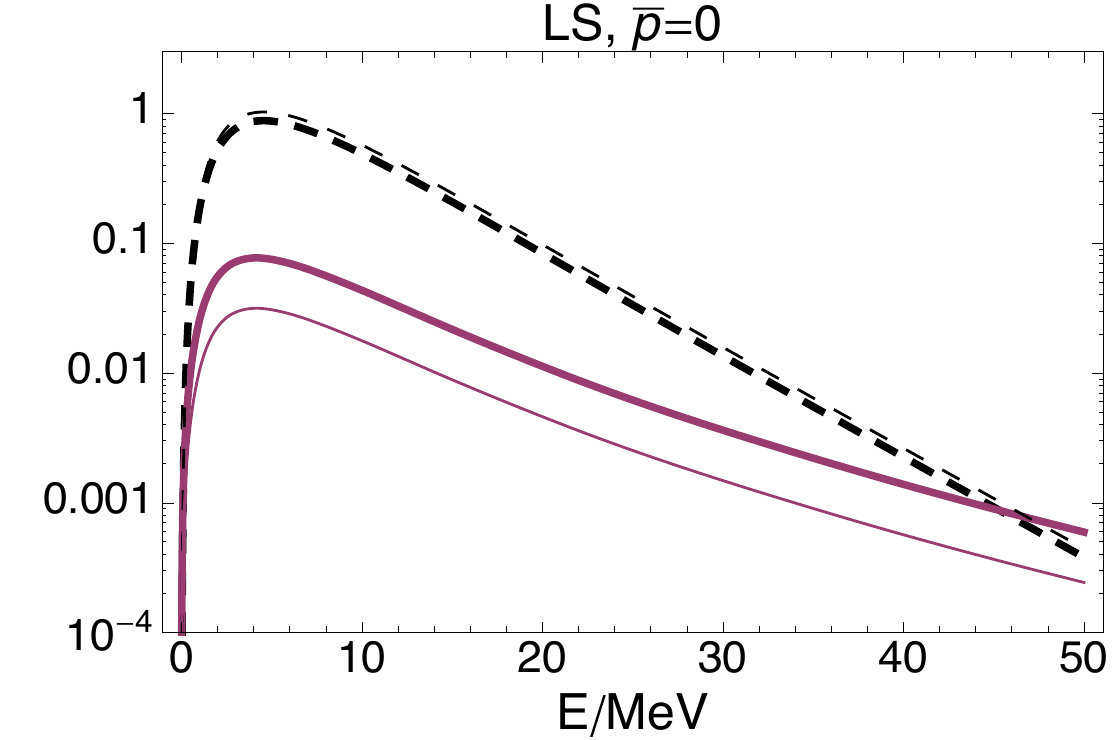}
   \caption{Diffuse flux of $\bar{\nu}_e$ from core-collapse SNe (failed SNe) as a function of the energy plotted with dashed (solid) curves for S and LS EoS and survival probability $\bar{P} = 0.68$ (including MSW effects, left panels) and $\bar{P}_{ee} = 0$ (no oscillations, right panels).  The fraction of failed SN is $f_{\rm BH} = 22\%$ (thick curves)  and $f_{\rm BH} = 9\%$  (thin curves). (Reprinted figure with permission from~\cite{Lunardini:2009ya}; copyright (2009) by the American Physical Society.)}
\label{failedspectra}
\end{figure}
The uncertainty of the fraction of failed SNe has been parametrized with  $f_{\rm NS} \in [0.78-0.91]$, corresponding to an upper limit for neutron-star-forming collapses varying in the interval $24$--$50\ M_\odot$.
Moreover, since the role of neutrino oscillations (especially of neutrino self-induced flavor conversions) has not been explored in this context, only MSW effects have been included:  $F_{\bar{\nu}_e}(E) = \bar{P}_{ee} F_{\bar{\nu_e}}^0(E) + (1 - \bar{P}_{ee}) F_{\bar{\nu_x}}^0(E)$, with $\bar{P}_{ee} = 0, 0.68$ (see Tab.~\ref{tab:signatures}, with $\bar{P}_{ee} = 0.68$ coming from the dependence of the survival probability from $\theta_{12}$). 
 The corresponding DSNB is shown in Fig.~\ref{failedspectra}.

Assuming flavor conversions play a non-negligible role (i.e., $\bar{P}_{ee} = 0.68$), the largest contribution from failed SNe comes from the stiffer EoS (S EoS case). In the latter case, the $\Phi_{\rm BH}$ enhances the DSNB  for $E \ge 20$~MeV, improving the DSNB detection chances (compare dashed with continue lines in Fig.~\ref{failedspectra})~\cite{Lunardini:2009ya}. 
More recently, similar conclusions have been reached by adopting the progenitor dependence as from  hydrodynamical simulations of a range of SN progenitors and  redshift dependence of the BH-formation rate  based on  the metallicity evolution of galaxies~\cite{Nakazato:2015rya}.

Although the largest contribution to the DSNB comes from $z \le 1$, 
the flux from failed SNe from higher redshifts might still be substantial~\cite{Keehn:2010pn,Mattila:2012zr,Yuksel:2012zy}. In this sense, the DSNB detection could directly allow to investigate the  black-hole forming collapse, constrain its energetics and cosmological rate  even beyond $z \simeq 1$.

If  the neutrino emission is larger for failed SNe than for ordinary SNe as expected, then
the signal increase in the detectors will be significantly larger in the high-energy tail of the DSNB spectrum~\cite{Horiuchi:2008jz,Lunardini:2009ya,Nakazato:2015rya,Sumiyoshi:2008zw,Sumiyoshi:2007pp,Nakazato:2008vj,Lunardini:2006pd}. This is promising in terms of neutrino detection chances and it would be a unique opportunity to study such SN progenitors, being otherwise optically invisible~\cite{Kochanek:2008mp,Gerke:2014ooa}.

As discussed in Ref.~\cite{Lien:2010yb}, we can already constrain the visible and invisible  SN rate at $z = 0$ on the basis of current data under the assumption that the shape of the SNR is known. It emerges
 that the allowed region for invisible SNe is indeed non-zero, but it cannot be arbitrarily large. Future observations will be extremely helpful on restricting the allowed region for visible SNe. 

\begin{figure}
  \centering
 \includegraphics[width=0.85\textwidth]{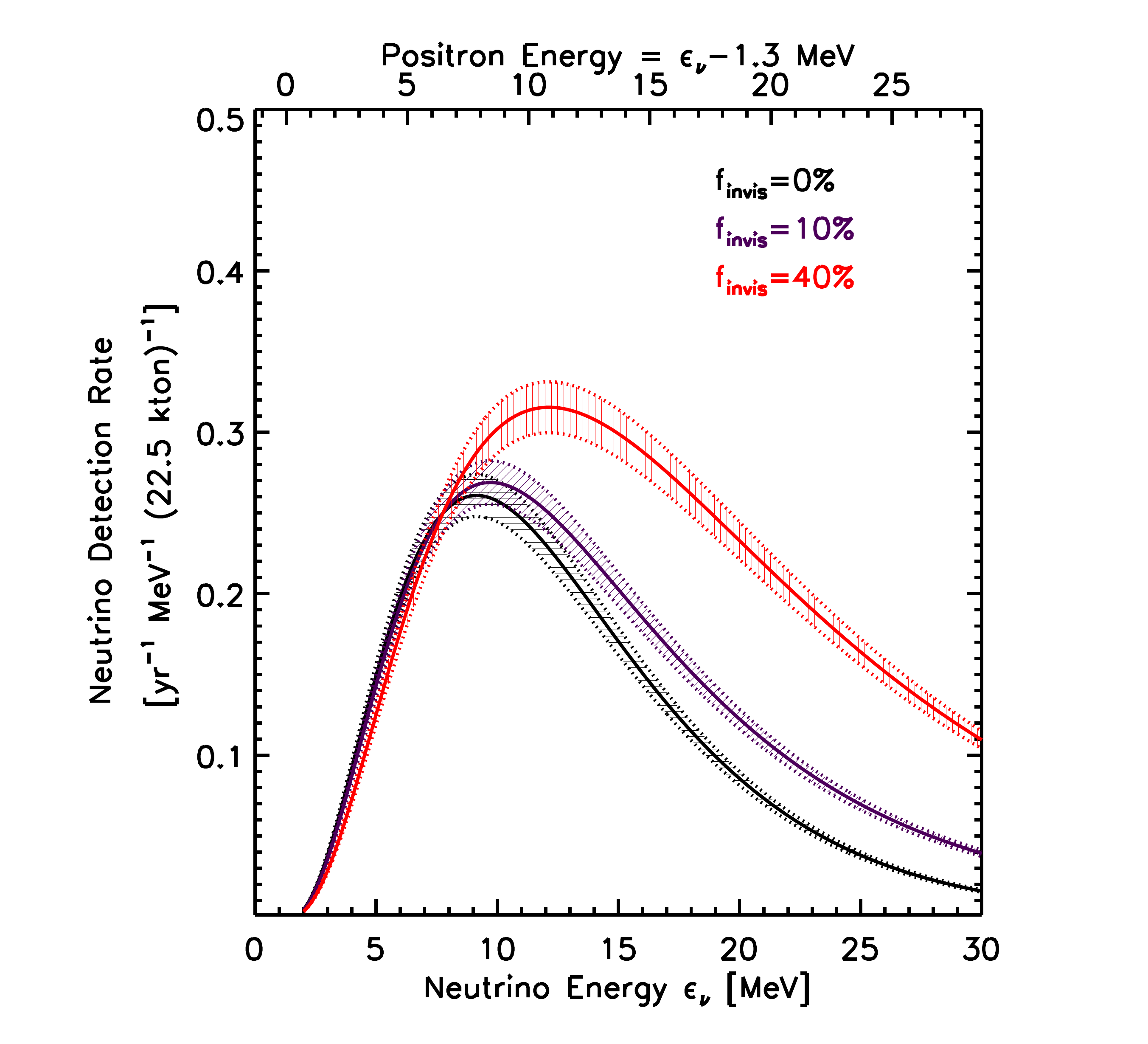}
   \caption{One-year neutrino detection rate in Super-Kamiokande as a function of the observed neutrino  energy for three different fractions of invisible SNe ($f_{\rm BH} =f_{\rm invis} = 0, 10, 40\%$). The band thickness of the curves is given by $5\%$ of the uncertainty on the SNR as expected from upcoming SN surveys. (Reprinted figure with permission from~\cite{Lien:2010yb};  copyright (2010) by the American Physical Society.)}
\label{failedpercentage}
\end{figure}

Assuming that in the next few years, upcoming SN surveys will be able to pin down to $5\%$ the uncertainty on the SNR, the variation of the one-year detection rate in Super-Kamiokande for different fractions of $f_{\rm BH}$ is discussed in Ref.~\cite{Lien:2010yb} by approximating  the neutrino energy spectrum with a Fermi-Dirac distribution with zero chemical potential and total energy emitted for each flavor  to $L_{\nu_{\alpha}} = 0.5 \times 10^{53}$~erg, $\langle E_{\alpha} \rangle = 12.6$~MeV for the failed SNe, while   $L_{\nu_{\alpha}} = 0.9 \times 10^{53}$~erg, $\langle E_{\nu_{\alpha}} \rangle = 23.6$~MeV  for core-collapse SNe.
Figure~\ref{failedpercentage} shows the expected one-year detection rate in Super-Kamiokande for $f_{\rm BH} = f_{\rm invis} = 0, 10, 40\%$. It is evident as, increasing the fraction of invisible SNe, a higher detection rate is expected especially in the high energy tail of the energy spectrum. Note as, any indirect experimental constraint of this kind on the the NS/BH formation ratio will be also useful to constrain the still disputed mechanism(s) by which stars explode as SNe.

We  assumed that ``failed'' SNe are a fraction of the total SNR through this Section. However, a first attempt to model the rate of failed  SNe as a function of the redshift has been done in Ref.~\cite{Yuksel:2012zy}, based on gamma-ray burst observations which could originate from core collapses yielding  rapidly rotating BHs. Reference~\cite{Yuksel:2012zy} seems indeed to find a much higher fraction of failed SNe at $z \ge 1$ than locally. See also Refs.~\cite{Nakazato:2015rya,Ertl:2015rga} for dedicated discussions.

\subsection{Detection perspectives}
\begin{figure}[t]
  \centering
 \includegraphics[width=0.85\textwidth]{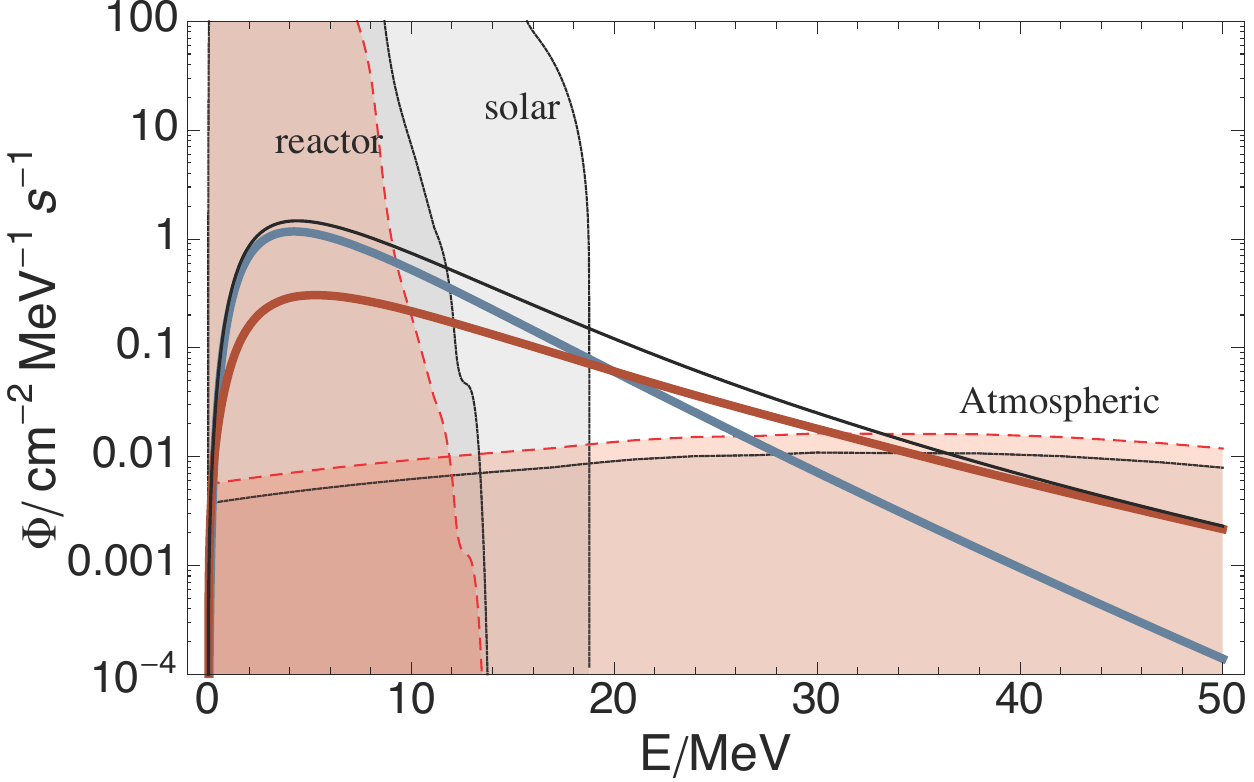}
   \caption{Expected diffuse flux of $\bar{\nu}_e$ from core-collapse SNe in  blue and failed SNe (S EoS, $\bar{P}_{ee} = 0.68$ and $f_{\rm NS} = 0.78$) in  brown. Their sum is in  black (see Sec.~\ref{sec:failed}). The atmospheric
   and reactor fluxes are shown for the Kamioka (solid, gray) and Homestake (dashed, red) sites. Since the $\nu_e$ and $\bar{\nu}_e$ atmospheric fluxes 
   are very similar, only one is plotted.
   (Reprinted figure with permission from~\cite{Keehn:2010pn}; copyright (2012) by the American Physical Society.)}
\label{spectradiff}
\end{figure}

The DSNB detection is one of the main goals of neutrino astrophysics. However, the DSNB detection is strongly affected by backgrounds, which reduce the detectable energy window  to $11~{\rm MeV } \le E \le 40~{\rm MeV}$~\cite{Horiuchi:2008jz,Bays:2011si}.
Figure~\ref{spectradiff} shows the expected DSNB, in the presence of standard core-collapse (blue line) and failed (brown line) SNe and their sum (black line), as discussed in Sec.~\ref{sec:failed}. The expected  atmospheric, reactor and solar neutrino backgrounds are also plotted.

Water Cherenkov detectors, such as Super-Kamiokande, being mostly sensitive to $\bar{\nu}_e$, have  atmospheric 
and reactor neutrinos as major backgrounds. The atmospheric neutrinos fall within the same DSNB energy range and have an isotropic distribution in space, 
therefore it is very difficult to discriminate the DSNB signal from the background.  However, the DSNB is larger than the atmospheric background around  $30$--$40$ MeV (see Fig.~\ref{spectradiff}), thus
restricting the experimental detection window to this range~\cite{Fogli:2004ff}.  Reactor neutrino events could    instead be distinguishable 
from their direction~\cite{Hochmuth:2005nh}, but no dedicated study concerning the DSNB exists yet. Liquid argon detectors
are mostly sensitive to $\nu_e$, and therefore solar and atmospheric neutrino 
fluxes are the main backgrounds~\cite{Cocco:2004ac}. 
Solar neutrinos prevent studying the DSNB below $18$~MeV for all the above-mentioned detectors (see Fig.~\ref{spectradiff}); however they can be subtracted very effectively since their direction and flux can be well determined\footnote{Note that solar neutrinos
might constitute a non-minor impediment to the DSNB detection in LAr detectors possibly due to uncertainties in the cross section and poor pointing.}.
Clearly the detection-energy windows would be larger for an enhanced DSNB (i.e., including the contribution from failed SNe), and therefore a larger $f_{\rm BH}$ would be advantageous for signal to background discrimination.  

An enhanced signal discrimination over the background for  the DSNB detection in Super-Kamiokande  will be reached  by dissolving gadolinium 
 in water~\cite{Beacom:2003nk} (see Sec.~\ref{sec:detection}); this option is currently being tested through the 
Evaluating Gadolinium's Action on Detector Systems
 (EGADS) facility~\cite{VaginsHAnSE}  and the Super-Kamiokande Collaboration has just approved the future enrichment of Super-Kamiokande with
 Gd, the so-called SuperK-Gd project~\cite{SuperK-Gd}. 
This would result in a reduction of the background  by a factor of $\sim 5$ for invisible muons and by at least an order of magnitude for spallation~\cite{Beacom:2003nk,Fogli:2004ff,Li:2014sea,Li:2015kpa}.
The planned liquid scintillator detector, JUNO, should also be able to separate DSNB neutrino events from the background very efficiently, offering a great opportunity to detect the DSNB~\cite{Li:2014qca,An:2015jdp,Mollenberg:2014pwa}.

\begin{figure}
\includegraphics[width=0.9\textwidth]{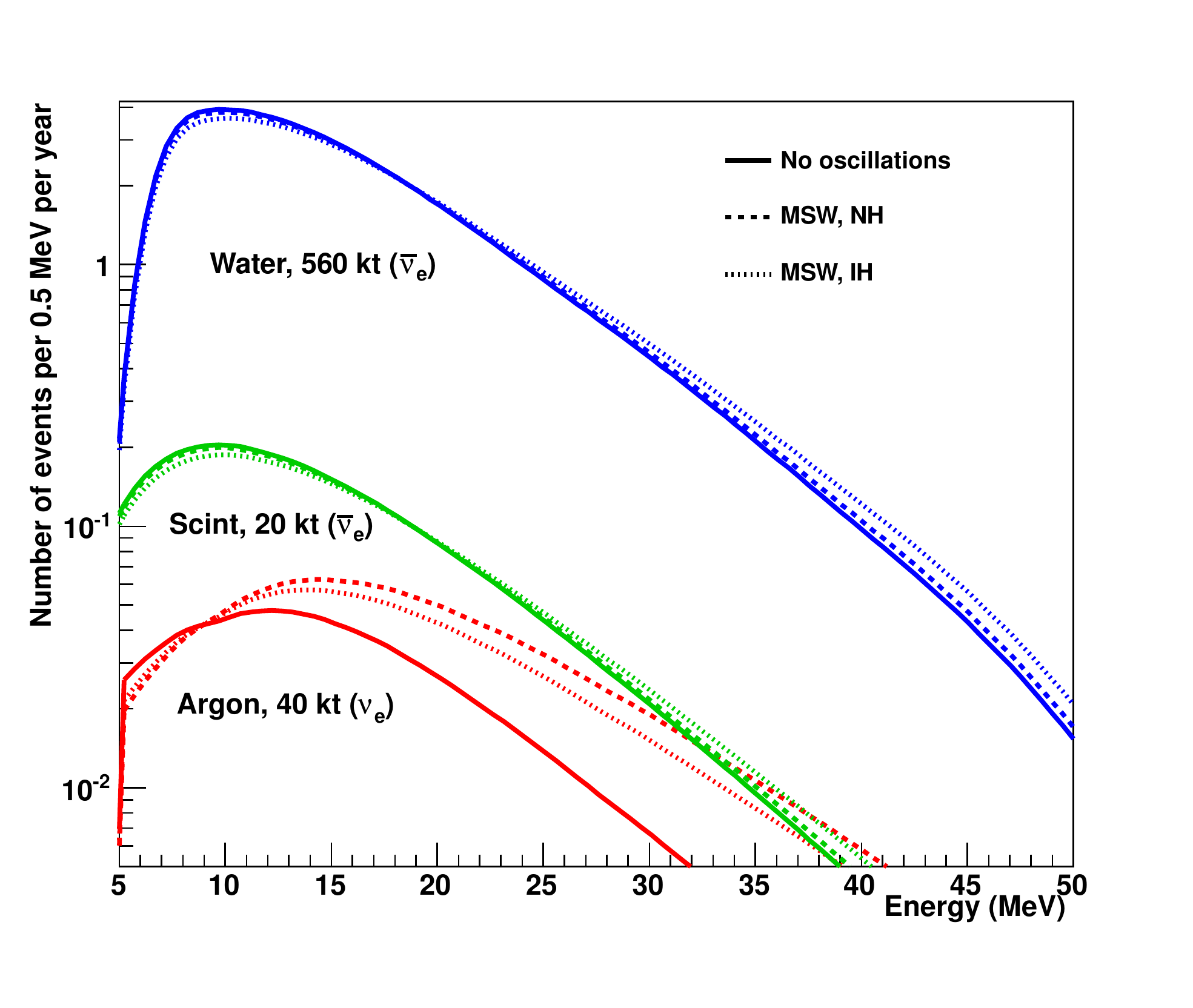}     
\centering
\caption{Smeared event rates as a function of detected energy based on an optimistic estimation of the DSNB obtained assuming a $27\ M_\odot$ SN progenitor 
(see Sec.~\ref{sec:protoneutronstars}) as representative of the SN population and MSW effects only. The rates are shown for a 560-kton WC detector (top, blue curves), a 20-kton liquid scintillator (center, green curves) and a 40-kton LAr detector (bottom, red curves). MSW oscillations under NH and IH assumptions are included.}\label{fig:dsnbrate}
\end{figure}
In order to give an idea of the expected number of events for the detectors described above (see also Sec.~\ref{sec:detection}), 
Fig.~\ref{fig:dsnbrate} shows the smeared event rates as a function of the detected energy for a 560-kton WC detector (blue curve), a 20-kton liquid scintillator (green curve) and a 40-kton LAr   detector (red curve). The rates have been computed including the dominant  reaction channels (i.e., IBD in water and scintillator detectors, and $\nu_e$ CC interactions on $^{40}$Ar for the LAr  detector) and assuming the $27\ M_\odot$ SN progenitor introduced in  Sec.~\ref{sec:protoneutronstars} as representative of the whole SN population; for simplicity, $\nu$--$\nu$ interactions have been neglected. Note as the  presented rates may be optimistic because of the high NS mass of the adopted  $27\ M_\odot$ SN progenitor.

\subsection{Outlook}
The study of the DSNB is still affected by several theoretical uncertainties, preventing us from a precise forecasting of the expected signal. Assuming that synoptic surveys will be able to pin down the uncertainty on the SNR up to $5\%$ and that the neutrino mass hierarchy will be known within the next decade, what could we learn from the DSNB detection?
\begin{itemize}
\item Constraints on the stellar population. The DSNB receives contributions from all families of SN progenitors and it will be an independent test of the global SN rate.
 \item Constraints on the fraction of core-collapse and failed SNe. The DSNB will be a precious instrument to test our current knowledge of the core-collapse physics, especially for the failed SN class only detectable through its neutrinos. Constraints on the NS/BH formation ratio may also indirectly help to decipher the still disputed mechanism(s) by which stars explode as supernovae.
 \item Constraints on the neutrino emission properties. As discussed in Ref.~\cite{Yuksel:2005ae}, with the existing current upper limits on the ${\bar{\nu}_e}$ diffuse background from Super-Kamiokande and the few events from the SN 1987A, it is possible to exclude some 
 regions of the $L_{\bar{\nu}_e}$--$\langle E_{\bar{\nu}_e}\rangle$ parameter space. Such bounds will surely be  improved by real data.
\item A precise estimation of the DSNB is also relevant for the background modeling in direct dark matter searches~\cite{Billard:2013qya,Grothaus:2014hja}. 
\item The DSNB detection could be useful to constrain neutrino decay models~\cite{Ando:2003ie,Fogli:2004gy}, neutrino electric or magnetic transition moments~\cite{Raffelt:2009mm} and the existence of light scalar bosons~\cite{Goldberg:2005yw}. 
\end{itemize}

\clearpage

\section{Conclusions and Perspectives}\label{sec:conclusions}

\begin{figure}[htpb]
\centering
\includegraphics[width=1.0\textwidth]{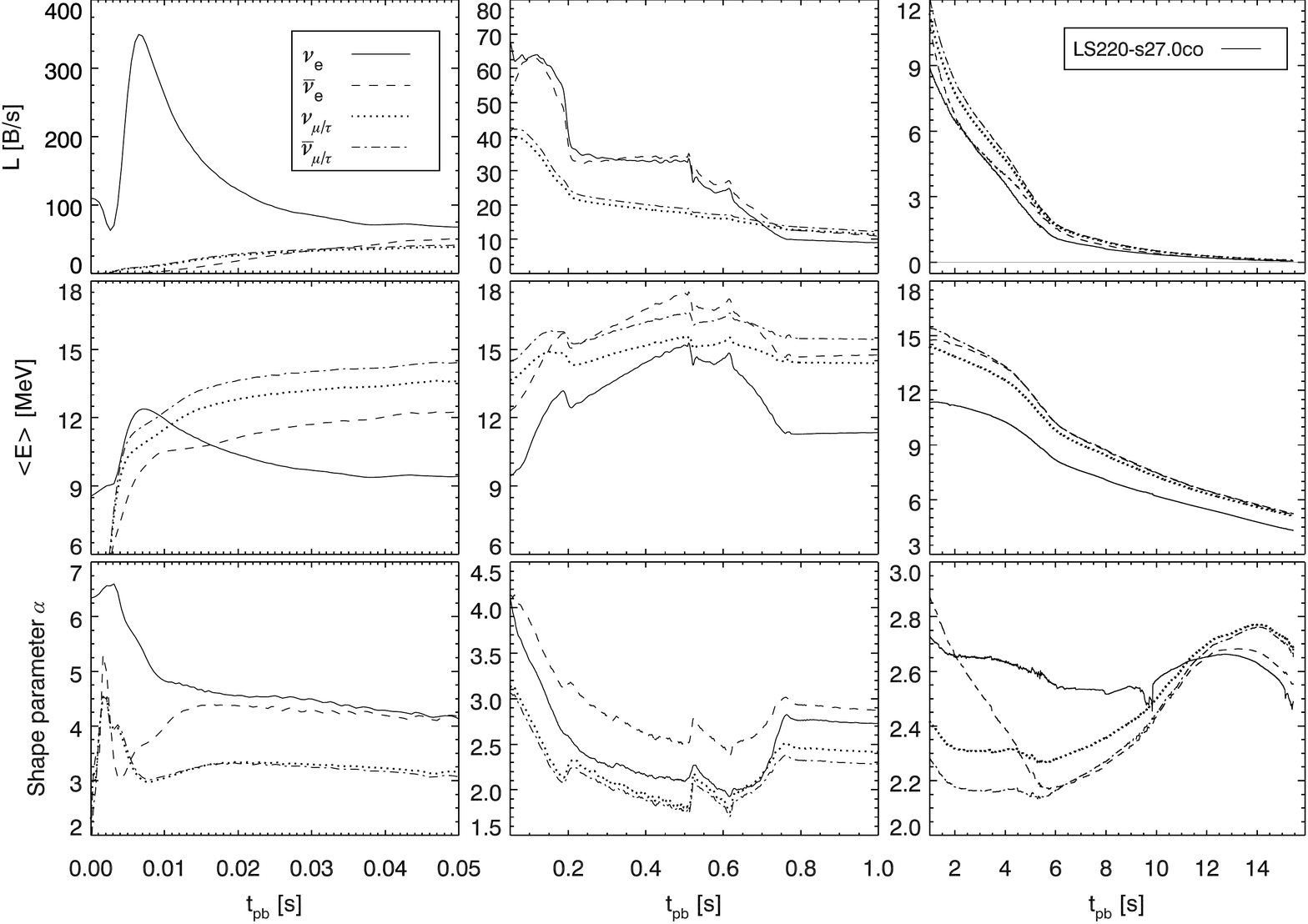}
\caption{Post-bounce evolution of the neutrino signal from a
neutron star with baryonic (final gravitational) mass of $\sim$1.776
($\sim$1.592)\,$M_\odot$ formed in the explosion of a 27\,$M_\odot$
progenitor. The panels show the luminosities ({\em top}, in units
of B\,s$^{-1} = 10^{51}$\,erg\,s$^{-1}$), mean energies (ratios
of energy fluxes to number fluxes; {\em middle}),
and spectral shape parameters ({\em bottom})
for $\nu_e$ (solid lines), $\bar\nu_e$ (dashed), $\nu_{\mu,\tau}$ 
(dotted) and $\bar\nu_{\mu,\tau}$ (dash-dotted) during the 
shock-breakout burst of $\nu_e$ ({\em left column}), accretion phase
({\em middle column}), and proto-neutron star cooling phase
({\em right column}). The 1D simulation was performed with the 
nuclear EoS of Lattimer \& Swesty~\cite{Lattimer:1991nc} 
(using the incompressibility
modulus of $K = 220$\,MeV) and took into account self-energy shifts in 
the $\beta$-interactions of free nucleons and a mixing-length treatment
of proto-neutron star convection.
The steep decline of the $\nu_e$ and $\bar\nu_e$
luminosities at $\sim$0.2\,s is associated with the decrease of the
mass-accretion rate when the edge of the stellar silicon layer falls
through the stalled shock. The explosion was artificially
initiated at 0.5\,s and accretion finally ends at $\sim$0.8\,s
after bounce.
\label{neutrinos-s27}}
\end{figure}

Neutrinos play a crucial role during all stages of stellar collapse and explosion. 
The emission of electron neutrinos produced by electron captures accelerates the 
implosion of the unstable, degenerate stellar core and mediates the neutronization 
of its matter. The absorption of electron neutrinos and antineutrinos above the 
neutrinospheric layers initiates and powers the blast wave of the explosion and
determines the trans-iron nucleosynthesis by setting the neutron-to-proton ratio 
in the innermost, neutrino-heated ejecta. The emission of neutrinos and 
antineutrinos of all flavors drives the cooling evolution of the newly formed 
neutron star that is left behind at the center of the supernova explosion.
In order to summarize the expected neutrino emission characteristics, 
we show in Fig.~\ref{neutrinos-s27} the source luminosities $L_{\nu}$ (upper panels), 
average energies $\langle E_\nu \rangle$ (middle panels), and spectral shape 
parameters $\alpha$ (lower panels) of $\nu_e$, $\bar\nu_e$ and $\nu_x$ for 
the shock-breakout, accretion and cooling phases of a 27 $M_{\odot}$ supernova
progenitor simulated by the Garching group.

The detection of two dozen electron antineutrinos from SN 1987A in three 
underground experiments splendidly confirmed our basic theoretical picture of 
stellar collapse and the birth of neutron stars. However, the signal statistics 
were too poor to yield detailed information of the explosion mechanism, which is 
still one of the most nagging problems in stellar astrophysics. The solution of 
this riddle is of fundamental importance for a better understanding of the origin 
of neutron stars and black holes and for the definition of the role of supernovae 
in the cosmic cycle of element formation. In this context only neutrinos (and gravitational waves) 
can serve as direct probes of the physics taking place at the center of the 
explosion. A well resolved time and energy dependent neutrino signal from a 
future Galactic event will therefore provide a benchmark of unprecedented 
value for supernova physics.

Hydrodynamical supernova simulations have now reached the multi-dimensional
front including a full and sophisticated treatment of the neutrino reactions  
that play a role in the stellar core. The first three-dimensional supernova models
of that sort have shown that neutrinos carry imprints of hydrodynamic processes
occurring around the newly formed neutron star during the accretion phase.
In particular, quasi-periodic modulations of the neutrino emission reflect
dynamical mass motions that precede and enable the onset of the explosion.
Rapid declines of the neutrino luminosity and non-monotonicities in the 
evolution of the spectral parameters may reveal the
composition-shell structure of the progenitor and can indicate the end of the 
accretion phase (see Fig.~\ref{neutrinos-s27}).
The spectral evolution of the neutrino signal is determined 
by the neutrinospheric conditions and thus carries radius and mass information 
of the contracting proto-neutron star.   The late-time evolution of the 
neutrinos radiated during the cooling phase has a great potential to yield 
important insights into the still incompletely understood properties of matter 
at supranuclear densities in the interior of the nascent neutron star.

While multi-dimensional supernova modeling seems to be on a good track to confirm 
the viability of the neutrino-driven explosion mechanism within the present 
uncertainties of input physics and initial conditions, an empirical validation 
of the theoretical concept is still missing. A Super-Kamiokande-size detector 
for electron neutrinos like DUNE, in combination with existing (Super-Kamiokande, 
IceCube, etc.) and proposed (Hyper-Kamiokande, JUNO) water and liquid 
scintillator detectors for electron antineutrinos, would 
provide excellent time and/or spectral resolution for a Galactic supernova. Therefore, they
would offer a fundamentally new quality for supernova research through neutrino 
measurements. Resolving the prompt shock-breakout burst of electron neutrinos 
would allow to determine the distance to the supernova even without 
electromagnetic detection and would define a precise reference point for the 
instant of core bounce. The simultaneous information in both electron neutrino 
and antineutrino sectors could set constraints on the equation of state (e.g.,
the baryonic symmetry energy) at neutrinospheric conditions and, in particular, 
it would help to answer the long-standing question whether supernovae are 
sources of r-process elements or of certain proton-rich isotopes produced 
by the neutrino-proton process. The direct comparison of electron neutrino 
and antineutrino measurements could also serve to clarify the existence of 
a dipolar lepton-emission asymmetry (LESA) that is a recent, unexpected 
(though yet unconfirmed) discovery by the first three-dimensional supernova 
models with important implications for supernova nucleosynthesis and pulsar 
kicks. Moreover, a multitude of complex and not yet fully understood  flavor  transformation effects can modify the neutrino 
signal on the way from the source at the supernova center to the detectors 
on Earth. Supernova neutrino data would 
provide extremely valuable information to decipher the evolution of the 
neutrino flavor field, which is indispensable for disentangling the 
source properties from effects imprinted on the signal by flavor  
transformations.

The role of astrophysical messengers
played by supernova neutrinos  is largely associated
with the signatures imprinted on the observable neutrino burst by the supernova
core dynamics and by the flavor conversions occurring deep inside the star. 
Within a radius of a few hundred kilometers
from the neutrinosphere,
the neutrino field is so dense to become a ``background to itself,''
making the neutrino flavor evolution highly non-linear and leading
to surprising and counterintuitive collective phenomena, when the
entire neutrino system oscillates coherently as a single collective mode. 
The rich phenomenology associated with these non-linear flavor dynamics is still
in its infancy and many unexpected results have been found in the past years.
It is therefore mandatory to further investigate the role of flavor instabilities
in the interacting neutrino field. Directions of further  
studies include: The characterization of the self-induced oscillations in 
models with reduced symmetries
compared to what is usually assumed in the neutrino light-bulb model;
investigations of the role of flavor conversions in the light of multi-dimensional
hydrodynamic supernova simulations and their impact on the explosion dynamics 
and nucleosynthesis; or the role of residual scatterings and the possibility 
of self-induced spin-flavor transitions.

Matter effects on supernova neutrino oscillations could imprint
peculiar signatures on the observable neutrino signal with a strong 
sensitivity to the still-unknown neutrino mass hierarchy.
Neutrino flavor conversions in the wake of the shock-wave propagation 
could also represent an intriguing possibility to follow the supernova
dynamics in real time through measurements of the neutrino signal.
The impact of the matter turbulence on the
shock-wave signature would require 
further investigations. Studies in these directions will be   mandatory once future,
high-resolution supernova simulations will become available.

Current and planned large underground neutrino detectors offer 
unprecedented opportunities to study supernova and neutrino properties
through high-statistics signals from different interaction channels.
However, the exciting possibilities associated with the next Galactic explosion
are in tension with the low Galactic supernova rate. The supernova rate in our
Galaxy and within the Magellanic Cloud is estimated to be between one and 
several per century, but
even with a high assumed rate we might have to wait a long time for the
next Galactic supernova. In contrast, the DSNB, i.e., the flux of neutrinos 
and antineutrinos coming from all past core-collapse supernovae in our 
Universe, is a guaranteed signal to study supernovae by their neutrinos
in the forseeable future. The perspectives for a detection of the DSNB
within the next decade are promising, especially with  the just approved SuperK-Gd project (i.e., gadolinium-enhanced 
Super-Kamiokande detector) and with JUNO as a planned liquid 
scintillator experiment. Such a measurement would push the frontier 
of neutrino astronomy literally towards the edges of our Universe.

In conclusion,
the detection of supernova neutrinos represents the next frontier of 
low-energy neutrino astrophysics. Supernovae are celestial laboratories
where neutrinos, after escaping from the highly opaque core, play a crucial 
role in the mechanism of the stellar explosion and where they might allow
us to gain new insights into fundamental neutrino properties.
In spite of remarkable progress, a satisfactory understanding of the 
formation of the neutrino emission as well as  of the flavor   conversions 
in this exotic environment  is still lacking. 
Further theoretical and experimental work is therefore needed in order to
get prepared for exploiting the wealth of information that the next
Galactic supernova explosion is going to provide.

\acknowledgments
A.M.~thanks A.~Bettini and G.L.~Fogli for proposing this review project
during the ISAPP School 2013 in Carfranc (Spain).
We thank Basudeb Dasgupta and Antonio Marrone for providing adapted versions
of some of the plots shown in this review.
The supernova modeling at Garching acknowledges
input by Ale\-xander Lohs, Andreas Marek, and Bernhard M\"uller. It 
received financial support by the Deutsche Forschungsgemeinschaft through
the Cluster of Excellence EXC~153 ``Origin and Structure of the Universe''
(http://www.universe-cluster.de) and by the European Research Council 
through ERC-AdG No.\ 341157-COCO2CASA. The numerical simulations became
possible by high performance computing resources
(Tier-0) provided by PRACE on SuperMUC (GCS@LRZ, Germany), CURIE~TN
(GENCI@CEA, France), and MareNostrum (BSC, Spain), and by
the Gauss Centre for Supercomputing on SuperMUC (LRZ, Germany).
We also thank the Max Planck Computing and Data Facility (MPCDF) for computing
time on the IBM iDataPlex system \emph{hydra}.
The work of A.M.~is supported by the Italian Ministero dell'Istruzione,
Universit\`a e Ricerca (MIUR) and Istituto Nazionale
di Fisica Nucleare (INFN) through the ``Theoretical Astroparticle Physics'' projects.
I.T.~acknowledges support from
the Netherlands Organization for Scientific Research (NWO).
N.S.~thanks
the European Union FP7 ITN INVISIBLES (Marie Curie
Actions, PITN- GA-2011- 289442).
  The research of
K.S.~has been supported by the US Department of Energy and the National Science Foundation.
S.C.~acknowledges the European Union through a Marie Curie Fellowship, Grant No.
PIIF-GA-2011-299861 and through the ITN Invisibles, Grant No. PITN-GA-2011-289442.

\newpage 

\section*{Acronyms}

\begin{table}[h]
  \caption{List of the most used acronyms.}
  \label{tab:acron}
  \begin{tabular}{llc}
    \hline
    {\bf 1D} & One-dimensional & \\
   {\bf 2D} & Two-dimensional & \\
   {\bf 3D} & Three-dimensional & \\
  {\bf BH} & Black Hole & \\
    {\bf CC} & Charged Current & \\
    {\bf CE$\nu$NS} & Coherent Elastic Neutrino-Nucleus Scattering & \\
    {\bf DSNB} & Diffuse Supernova Neutrino Background  & \\
    {\bf EoM} & Equations of Motion & \\
    {\bf EoS} & Equation of State & \\
       {\bf ES} & Elastic scattering & \\
	{\bf IBD} & Inverse Beta Decay & \\
	{\bf IMB} & Irvine-Michigan-Brookhaven & \\
	{\bf  IMF} & Initial Mass Function & \\
    {\bf IH} & Inverted Hierarchy & \\
    {\bf LAr TPC} & Liquid Argon Time Projection Chamber & \\
     {\bf LESA} & Lepton Emission Self-sustained Asymmetry & \\
    {\bf l.h.s.} & left-hand-side & \\
    {\bf LSD} & Large Scintillator Detector & \\
    {\bf LVD} & Large Volume Detector & \\
     {\bf LS EoS} & Lattimer-Swesty Equation of State &  \\
{\bf MAA}  & Multi-azimuthal-angle &  \\
      {\bf MSW}    & Mikheyev-Smirnov-Wolfenstein &       \\
		{\bf NC} & Neutral Current & \\
		{\bf NH} & Normal Hierarchy & \\
		{\bf NS} & Neutron Star & \\
		{\bf r.h.s.} & right-hand-side & \\
		{\bf S EoS} & Shen et al.~Equation of State &  \\
		{\bf SASI} & Standing Accretion Shock Instability & \\
	{\bf SFR} & Star Formation Rate & \\
	{\bf SN} & Supernova &  \\  
	{\bf SNe} & Supernovae &  \\  
	{\bf SNEWS} & SuperNova Early Warning System & \\
	 {\bf SNR} & Supernova Rate & \\
	{\bf WC} & Water Cherenkov & \\
	\hline
  \end{tabular}
\end{table}

\end{document}